\documentclass[a4paper, 11pt, twoside]{book}
\usepackage[DIV=16,BCOR=1.8cm,headinclude=true]{typearea}

\usepackage{layout}

\usepackage[titletoc]{appendix}

\usepackage[utf8]{inputenc}

\usepackage[T1]{fontenc}

\usepackage{CJKutf8}
\usepackage[encapsulated]{CJK}

\usepackage{prettyref}
\usepackage[backend=bibtex,style=phys, biblabel=brackets]{biblatex}
\usepackage{amsmath}
\usepackage{amssymb}
\usepackage{amsthm}

\usepackage[hidelinks,colorlinks,allcolors=black]{hyperref} 
\usepackage[noabbrev]{cleveref}

\usepackage{bbm}

\usepackage{color}
\usepackage{float}
\usepackage{graphicx}
\usepackage{mathtools}
\usepackage[dvipsnames]{xcolor}
\usepackage[font=small,labelfont=bf,labelsep=space]{caption}
\usepackage{wrapfig}
\usepackage{subcaption}
\usepackage{array}
\usepackage{dsfont}
\usepackage{makecell}
\usepackage[babel]{csquotes}
\usepackage{tikz-cd}

\usepackage{braket}

\usepackage{enumitem}

\usepackage{fancyhdr}

\usepackage{etoolbox}

\usepackage[export]{adjustbox}

\AtBeginEnvironment{proof}{\parskip=0pt}

\renewcommand{\chaptermark}[1]%
         {\markboth{\thechapter.\ #1}{}}
\renewcommand{\sectionmark}[1]%
         {\markright{\thesection\ #1}}
\usepackage[symbols,nogroupskip,sort=none]{glossaries-extra}

\usepackage{accents}

\usetikzlibrary{3d} 
\usetikzlibrary{calc}

\newenvironment{Abstract}[1]
    {\begin{center}\Large \bfseries{#1} \end{center} 
      \begin{changemargin}{.7cm}{.7cm}
      \begingroup\normalsize}
    {\endgroup \end{changemargin}}

\newtheoremstyle{plainfont}  
  {8pt}                      
  {8pt}                      
  {\normalfont}              
  {}                         
  {\bfseries}                
  {.}                        
  { }                        
  {}                         

\newtheorem{thm}{Theorem}[chapter]
\newtheorem{lem}[thm]{Lemma}
\newtheorem{cor}[thm]{Corollary}

\newtheorem{propo}[thm]{Proposition}

\newtheorem{conj}{Conjecture}[chapter]

\newtheorem{defn}{Definition}[chapter]

\newtheorem{problem}{Problem}[chapter]

\theoremstyle{plainfont}
\newtheorem{ex}{Example}[chapter]

\newtheorem{rem}{Remark}[chapter]

\let\oldlem\lem
\renewcommand{\lem}{%
  \crefalias{thm}{lem}
  \oldlem}

\let\oldpropo\propo
\renewcommand{\propo}{%
  \crefalias{thm}{propo}
  \oldpropo}

\let\oldcor\cor
\renewcommand{\cor}{%
  \crefalias{thm}{cor}
  \oldcor}

\crefname{thm}{Theorem}{Theorems}
\crefname{lem}{Lemma}{Lemmas}
\crefname{propo}{Proposition}{Propositions}
\crefname{cor}{Corollary}{Corollaries}

\crefname{defn}{Definition}{Definitions}

\crefname{ex}{Example}{Examples}
\crefname{rem}{Remark}{Remarks}

\crefname{figure}{Fig.}{Figs.}

\crefname{table}{Table}{Tables}

\crefname{chapter}{Chapter}{Chapters}
\crefname{section}{Section}{Sections}

\crefname{conj}{Conjecture}{Conjecture}

\DeclareMathOperator{\Tr}{Tr}

\newcommand{\conv}{\mathrm{conv}}
\newcommand{\aff}{\mathrm{aff}}
\newcommand{\FHK}{\mathcal{F}_{HK}}

\newcommand{\HN}{\mathcal{H}_{N}}
\newcommand{\Ho}{\mathcal{H}_{1}}
\newcommand{\VFactor}[1]{L_{#1}}

\newcommand{\ketbrap}[1]{\ket{#1}\!\bra{#1}}
\newcommand{\ketbra}[2]{\ket{#1}\!\bra{#2}}

\newcommand{\doverd}[1]{\frac{\mathrm{d}}{\mathrm{d}#1}}

\newcommand{\dstraight}{\mathrm{d}}
\newcommand{\Dstraight}{\mathrm{D}}
\newcommand{\HdR}{H_{\mathrm{dR}}}

\newcommand{\proj}[1]{\mathbf{p}_{#1}}

\DeclareMathOperator{\ad}{ad}

\DeclareMathOperator{\Ad}{Ad}

\DeclareMathOperator{\sgrad}{sgrad}

\DeclareMathOperator{\dom}{dom}
\DeclareMathOperator{\im}{im}

\DeclareMathOperator{\relint}{relint}

\DeclareMathOperator*{\argmin}{arg\,min}

\let\Re\relax
\DeclareMathOperator{\Re}{Re}
\let\Im\relax
\DeclareMathOperator{\Im}{Im}

\newcommand*\xbar[1]{\hbox{\vbox{
       \hrule height 0.6pt 
       \kern0.3ex
       \hbox{%
         \kern-0.2em
         \ensuremath{#1}%
         \kern 0.0em
         }}}}

\newcommand*\xxbar[1]{\hbox{\vbox{
       \hrule height 0.6pt 
       \kern0.3ex
       \hbox{%
         \kern-0.0em
         \ensuremath{#1}%
         \kern 0.0em
         }}}}

\newcommand{\Fp}{\mathcal{F}_{\!p}}
\newcommand{\Fe}{\mathcal{F}_{\!e}}

\newcommand{\nalpha}{m^{(\alpha)}}
\newcommand{\nbeta}{m^{(\beta)}}
\newcommand{\HNP}{\mathcal{H}_N^{(P)}}

\newcommand{\FpP}{\mathcal{F}_p^{(P)}}

\newcommand{\cldmap}{\mathcal{A}}

\newcommand{\exterior}{\mathop{\scalebox{0.95}{$\bigwedge$}}\nolimits}

\begin{document}

\pagenumbering{roman}
\setcounter{page}{1}

\begin{titlepage}
\begin{center}
\begin{LARGE}
\textsc{Ludwig-Maximilians-Universit{\"a}t M{\"u}nchen}\\
\end{LARGE}
\begin{Large}
\textsc{Faculty of Physics}\\
\end{Large}
\vspace{1cm}
\begin{LARGE}
\textsc{Master's thesis}\\
\end{LARGE}
\vspace{1cm}
    {\parindent0cm
    \rule{\linewidth}{.5ex}}  
\begin{flushright}
    \centering
\begin{Huge}
\textbf{Geometry of Generalized Density Functional Theories}\\
\end{Huge}
\end{flushright}
\rule{\linewidth}{.5ex}
\vspace{0cm}
\end{center}
\begin{center}
\vspace{0cm}
\begin{LARGE}
Chih-Chun Wang\\
\end{LARGE}
\vspace{1cm}
\includegraphics[width=2in]{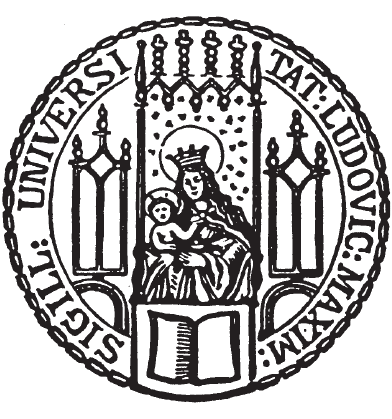}\\
\vspace{.5cm}
\begin{Large}
Supervised by \\
Dr. Christian Schilling\\
\end{Large}
\vspace{.5cm}
\begin{Large}
Munich, October 5, 2025
\end{Large}
\end{center}
\end{titlepage}

\clearpage{\pagestyle{empty}\cleardoublepage}
\cleardoublepage

\begin{titlepage}
\setcounter{page}{3}
\begin{center}
\begin{LARGE}
\textsc{Ludwig-Maximilians-Universit{\"a}t M{\"u}nchen}\\
\end{LARGE}
\begin{Large}
\textsc{Fakult{\"a}t f{\"u}r Physik}\\
\end{Large}
\vspace{1cm}
\begin{LARGE}
\textsc{Masterarbeit}\\
\end{LARGE}
\vspace{1cm}
    {\parindent0cm
    \rule{\linewidth}{.5ex}}  
\begin{flushright}
    \centering
\begin{Huge}
\textbf{Geometrie der verallgemeinerten Dichtefunktionaltheorien}\\
\end{Huge}
\end{flushright}
\rule{\linewidth}{.5ex}
\vspace{0cm}
\end{center}
\begin{center}
\vspace{0cm}
\begin{LARGE}
Chih-Chun Wang\\
\end{LARGE}
\vspace{1cm}
\includegraphics[width=2in]{graphics/lmu_siegel.pdf}\\
\vspace{.5cm}
\begin{Large}
Betreut von \\
Dr. Christian Schilling\\
\end{Large}
\vspace{.5cm}
\begin{Large}
M{\"u}nchen, den 5.~Oktober 2025
\end{Large}
\end{center}
\end{titlepage}

\clearpage{\pagestyle{empty}\cleardoublepage}
\cleardoublepage

%
%
%

\thispagestyle{empty}

\begin{Abstract}{Abstract}
 Density functional theory (DFT) is an indispensable \textit{ab initio} method
in both quantum chemistry and condensed matter physics. Based on
recent advancements in reduced density matrix functional theory
(RDMFT), a variant of DFT that is believed to be better suited for
strongly correlated systems, we construct a mathematical framework
generalizing all ground state functional theories, which in particular
applies to fermionic, bosonic, and spin systems. We show that within
the special class of such functional theories where the space of
external potentials forms the Lie algebra of a compact Lie group,
the $N$-representability problem is readily solved by applying techniques from
the study of momentum maps in symplectic geometry, an approach complementary
to Klyachko's famous solution to the quantum marginal problem. The ``boundary
force'', a diverging repulsive force from the boundary of the functional's
domain observed in previous works but only qualitatively understood in
isolated systems, is studied quantitatively and extensively in our work.
Specifically, we present a formula capturing the exact behavior of the
functional close to the boundary. In the case where the Lie algebra is
abelian, a completely rigorous proof of the boundary force formula based on
Levy-Lieb constrained search is given. Our formula is a first step towards
developing more accurate functional approximations, with the potential of
improving current RDMFT approximate functionals such as the Piris natural
orbital functionals (NOFs). All key concepts and ideas of our work are
demonstrated in translation invariant bosonic lattice systems.
\end{Abstract}

\clearpage

\thispagestyle{empty}

\begin{Abstract}{Zusammenfassung}

 Die Dichtefunktionaltheorie (DFT) ist eine unverzichtbare \textit{ab initio}-Methode
 sowohl in der Quantenchemie als auch in der Festkörperphysik. 
 Basierend auf
 jüngsten Fortschritten in der Funktionaltheorie der reduzierten Dichtematrix
 (RDMFT), einer Variante der DFT, die als besser geeignet für stark korrelierte
 Systeme gilt, entwickeln wir einen mathematischen Rahmen, der alle
 Grundzustands-Funktionaltheorien verallgemeinert und insbesondere auf
 fermionische, bosonische und Spin-Systeme anwendbar ist.
 Wir zeigen, dass innerhalb der speziellen Klasse solcher Funktionaltheorien,
 in denen der Raum der externen Potentiale die Lie-Algebra einer kompakten
 Lie-Gruppe bildet, das $N$-Repräsentierbarkeitsproblem sich leicht durch die
 Anwendung von Methoden aus der Untersuchung der Impulsabbildungen in der
 symplektischen Geometrie lösen lässt – ein Ansatz, der komplementär zu
 Klyachkos berühmter Lösung des Quantenmarginalproblems ist.
 Die ``Randkraft'', eine divergierende abstoßende Kraft von dem Rand des
 Funktionaldefinitionsbereichs, die in früheren Arbeiten beobachtet, aber in
 isolierten Systemen nur qualitativ verstanden wurde, wird in unserer Arbeit
 quantitativ und umfassend untersucht.
 Genauer gesagt stellen wir eine Formel vor, die das exakte Verhalten des
 Funktionals nahe des Rands erfasst. 
 Im Fall, dass die Lie-Algebra abelsch
 ist, wird ein vollständig strenger Beweis der Randkraftformel auf Basis der
 Levy-Lieb'schen eingeschränkten Suche präsentiert.
 Unsere Formel stellt einen ersten Schritt zur Entwicklung genauerer
 Funktionalapproximationen dar, mit dem Potenzial, aktuelle
 RDMFT-Funktionalapproximationen wie 
 Piris' Funktionale der natürlichen Orbitale (NOFs)
 zu
 verbessern. 
 Alle zentralen Konzepte und Ideen unserer Arbeit werden an
 translationsinvarianten bosonischen Gittersystemen veranschaulicht.

\end{Abstract}

\cleardoublepage

\tableofcontents

\glsxtrnewsymbol[description={Hohenberg-Kohn functional}]{FHK}{\ensuremath{\FHK}}
\glsxtrnewsymbol[description={pure functional}]{Fp}{\ensuremath{\Fp}}
\glsxtrnewsymbol[description={ensemble functional}]{Fe}{\ensuremath{\Fe}}

\glsxtrnewsymbol[description={set of pure states}]{pureset}{\ensuremath{\mathcal{P}}}
\glsxtrnewsymbol[description={set of ensemble states}]{ensset}{\ensuremath{\mathcal{E}}}
\glsxtrnewsymbol[description={projective Hilbert space}]{proj_hilb_space}{\ensuremath{\mathbb{P}(\mathcal{H})}}

\glsxtrnewsymbol[description={annihilator of a vector or a subset $w$}]{annihil}{\ensuremath{w^\circ}}
\glsxtrnewsymbol[description={dual map of a linear map $f$}]{dual_map}{\ensuremath{f^*}}

\glsxtrnewsymbol[description={inner product or dual pairing}]{bracket_inner_product}{\ensuremath{\braket{\cdot,\cdot}}}

\glsxtrnewsymbol[description={convex hull}]{conv}{\ensuremath{\conv}}
\glsxtrnewsymbol[description={affine hull}]{aff}{\ensuremath{\aff}}
\glsxtrnewsymbol[description={relative interior}]{relint}{\ensuremath{\relint}}
\glsxtrnewsymbol[description={translation space of an affine space $A$}]{transl_aff}{\ensuremath{\overrightarrow{A}}}
\glsxtrnewsymbol[description={standard $n$-simplex}]{nsimplex}{\ensuremath{\Delta_n}}
\glsxtrnewsymbol[description={standard simplex over a finite set $X$, i.e., $\{t\in \mathbb{R}_{\ge 0}^X \mid \sum_{i\in X}t_i = 1\}$}]{Ssimplex}{\ensuremath{\Delta_X}}

\glsxtrnewsymbol[description={smooth functions on a smooth manifold $M$}]{smoothfunctions}{\ensuremath{C^\infty(M)}}
\glsxtrnewsymbol[description={vector fields on $M$}]{vectfields}{\ensuremath{\mathfrak{X}(M)}}
\glsxtrnewsymbol[description={tangent space of $M$ at a point $x$}]{tangent_space}{\ensuremath{T_xM}}
\glsxtrnewsymbol[description={de Rham cohomology of $M$}]{derham}{\ensuremath{\HdR^\bullet(M)}}

\glsxtrnewsymbol[description={derivative of a smooth map $f$}]{derivative}{\ensuremath{Df}}
\glsxtrnewsymbol[description={derivative of $f$ at a point $x$}]{derivative_at_point}{\ensuremath{D_xf}}

\glsxtrnewsymbol[description={exterior derivative of a differential form $\alpha$}]{exterior_derivative}{\ensuremath{\dstraight\alpha}}
\glsxtrnewsymbol[description={contraction of a vector field $X$ with $\alpha$}]{contraction}{\ensuremath{\iota_X\alpha}}
\glsxtrnewsymbol[description={Lie derivative of $\alpha$ along $X$}]{Lie_derivative}{\ensuremath{L_X\alpha}}


\setlength{\glsdescwidth}{.75\textwidth}

\printunsrtglossary[type=symbols,style=long,title={Notation and Symbols}]

\clearpage{\pagestyle{empty}\cleardoublepage}
\cleardoublepage


\pagenumbering{arabic}

\setcounter{page}{1}



\chapter{Introduction}
\label{chap:intro}

One fundamental problem in the physical sciences is that of understanding
systems consisting of many constituents. In quantum chemistry, one is
interested in the behavior of the electrons in an atom or molecule. A large
part of solid-state physics is dedicated to the study of how electrons
propagate and scatter in a lattice. In the realm of ultracold atoms, one
studies systems of interacting bosonic or fermionic atoms trapped in optical
lattices. If there is no time dependence, the properties of all of these
systems, in particular those of the ground state, are described by the
Schr\"odinger equation
\begin{equation}
\label{eq:schrodinger}
H\ket{\Psi} = E\ket{\Psi}.
\end{equation}
In principle, then, we know everything about the system once we know the
Hamiltonian $H$, provided that we have the computational power to solve the
eigenvalue equation \eqref{eq:schrodinger}. The problem is, of course, that for
the vast majority of interesting systems, and by ``interesting'' we mean, among
other things, that the number of constituents is large, directly solving
Eq.~\eqref{eq:schrodinger} is practically impossible even numerically.

Various methods have been devised to overcome this problem by employing
sophisticated and physically motivated approximations, each with its own
merits. Density matrix renormalization group (DMRG)
\cite{whiteDensityMatrixFormulation1992,hallbergNewTrendsDensity2006,schollwockDensitymatrixRenormalizationGroup2011,ciracMatrixProductStates2021}
is particularly well-suited for one-dimensional systems.
Dynamical mean-field theory (DMFT) 
is especially effective with strongly correlated 
electronic systems
\cite{georgesDynamicalMeanfieldTheory1996,zgidDynamicalMeanfieldTheory2011,vollhardtDynamicalMeanfieldTheory2012}.
In density functional theory (DFT)
\cite{hohenbergInhomogeneousElectronGas1964,kohnSelfConsistentEquationsIncluding1965},
the role of the electronic wave function as the basic quantity is replaced by
the electron density $\rho(\mathbf{r})$, and ground state properties are
encoded in the density functional $\mathcal{F}(\rho)$. The functional is
\textit{universal} in the sense that it depends only on the kinetic energy of
the electrons and the Coulomb interaction, but not on the applied external
potential $v(\mathbf{r})$, which, for example, could be the electron-nucleus
interaction. It has been an indispensable \textit{ab-initio} numerical method
in both solid-state physics and modern chemistry
\cite{burkePerspectiveDensityFunctional2012,beckePerspectiveFiftyYears2014,hasnipDensityFunctionalTheory2014}.
Practical implementations of DFT are typically based on the Kohn-Sham scheme
\cite{kohnSelfConsistentEquationsIncluding1965} together with suitable
approximations of the so-called exchange-correlation functional
\cite{kohnSelfConsistentEquationsIncluding1965,
leeDevelopmentColleSalvettiCorrelationenergy1988,
beckeDensityfunctionalThermochemistryIII1993,
perdewGeneralizedGradientApproximation1996,toulouseReviewApproximationsExchangeCorrelation2023}.

Reduced density matrix functional theory (RDMFT)
\cite{gilbertHohenbergKohnTheoremNonlocal1975,donnellyElementaryPropertiesEnergy1978,pirisNaturalOrbitalFunctional2007,sharmaReducedDensityMatrix2008,pernalReducedDensityMatrix2015}
has established itself as an alternative to DFT. 
With the one-particle reduced density matrix (1RDM) treated as the basic
variable, the ground state problem is similarly reduced to that of determining
the density \textit{matrix} functional $\mathcal{F}(\gamma)$. 
Since the kinetic energy is a simple function of the 1RDM, the density matrix
functional depends solely on the particle-particle interaction, which suggests
that a fictitious noninteracting Kohn-Sham system does not need to be
introduced. Moreover, it has been shown
\cite{sharmaReducedDensityMatrix2008,lathiotakisDiscontinuitiesChemicalPotential2010}
that RDMFT performs better for transition metal oxides as compared to DFT with
the local density approximation, which tends to predict metallic ground states
as opposed to Mott insulators, indicating that RDMFT might be better suited for
strongly correlated systems.

Nevertheless, as with DFT, the exact density matrix functional is usually
unknown apart from few small systems
\cite{lopez-sandovalDensitymatrixFunctionalTheory2000,cohenLandscapeExactEnergy2016,benavides-riverosReducedDensityMatrix2020}.
Numerous proposed approximate functionals exist in the literature
\cite{goedeckerNaturalOrbitalFunctional1998,buijseApproximateExchangecorrelationHole2002,gritsenkoImprovedDensityMatrix2005,
marquesEmpiricalFunctionalsReduceddensitymatrixfunctional2008,
sharmaReducedDensityMatrix2008,
leivaAssessmentNewApproach2005,pirisNewApproachTwoelectron2006a,pirisNaturalOrbitalFunctional2011,pirisGlobalMethodElectron2017,
pirisGlobalNaturalOrbital2021,pirisAdvancesApproximateNatural2023}, most of
which are based on modifying M\"uller's reconstruction of the two-particle
reduced density matrix (2RDM) in terms of the 1RDM
\cite{mullerExplicitApproximateRelation1984}. Since RDMFT is formally exact (as
is DFT), it is important to take into account known exact properties of the
functional when proposing approximations. One such possibility is the behavior
of the functional near the boundary of its domain. Recently, it has been
observed that the functional exhibits a diverging repulsive force from the
boundary in various fermionic and bosonic systems
\cite{schillingDivergingExchangeForce2019,benavides-riverosReducedDensityMatrix2020,liebertFunctionalTheoryBoseEinstein2021,maciazekRepulsivelyDivergingGradient2021},
effectively preventing the 1RDM from reaching the boundary. In fermionic
systems, this has been termed the \textit{exchange force}, whereas in bosonic
systems the same phenomenon has been referred to as the \textit{BEC force},
since previous work on bosonic systems mostly focused on investigating the
functional near the Bose-Einstein condensate (BEC) vertex.
To be more precise, the observed force is always repulsive, and takes the form
\begin{equation}
  \doverd{\epsilon}\mathcal{F}(\rho_* + \epsilon \eta)
   \approx  -\mathcal{G}\frac{1}{\sqrt{\epsilon}}
\end{equation}
for some prefactor $\mathcal{G}$, where $\rho_*$ is a density on the boundary
of $\dom\mathcal{F}$, $\eta$ is an inward-pointing vector, and $\epsilon$ is small.
Not only is such a force intimately related to the presence or absence of
\textit{(quasi-) pinning}
\cite{cioslowskiUnoccupiedNaturalOrbitals2006,benavides-riverosQuasipinningEntanglementLithium2013,
schillingQuasipinningItsRelevance2015,tenniePinningFermionicOccupation2016,tennieInfluenceFermionicExchange2017,
schillingImplicationsPinnedOccupation2020,maciazekImplicationsPinnedOccupation2020},
understanding quantitatively why and how the derivative of the functional
diverges as the density approaches the boundary of the domain could potentially
lead to more accurate approximate functionals. However, these questions have
not been addressed in the literature in a systematic way; the only rigorous
quantitative demonstration which is not restricted to particular small systems
is Ref.~\cite{maciazekImplicationsPinnedOccupation2020}, where a bound of the
functional $\mathcal{F}$ near the boundary has been found, which in turn implies that the
derivative must diverge. Yet, there is no known formula for computing the
prefactor $\mathcal{G}$.

In this thesis, we show that this diverging behavior of the derivative of the
functional is entirely general and does not depend on the specific details of
each system. In light of this, we often refer to the phenomenon as the
\textit{boundary force} to emphasize its geometric origin. More importantly, as
will become apparent in the course of this thesis, the fact that such
divergence exists does not even depend on the ``type'' of the functional theory
itself, so boundary forces should also exist in ordinary DFT. To make these
claims more precise, then, we shall set up a framework of functional theories
which generalizes both DFT and RDMFT, which we do in
\cref{chap:generalized_ft}. It turns out, however, that generality comes at the
cost of tractability: in order to obtain useful statements, it seems necessary
to restrict oneself to subclasses of functional theories satisfying certain
``niceness conditions''. We will explore two such classes: the \textit{abelian}
functional theories and \textit{momentum map} functional theories, for each of
which we derive a concrete formula for the boundary force prefactor
$\mathcal{G}$.

The rest of this thesis is organized as follows. After a brief review of DFT
and RDMFT in \cref{sec:dft_intro}, we present a general framework of
functional theories in \cref{chap:generalized_ft}, defining in the abstract
setting concepts and objects such as density maps, representability, and
universal functionals, which will be of central importance throughout the
thesis. \cref{chap:generalized_ft} is also where basic results
like the weak Hohenberg-Kohn theorem and the relations between the universal
functionals are established. 
In \cref{chap:momentum_rdmft}, the abstract framework is instantiated by a
functional theory describing translation invariant bosonic lattice systems. We
show that the translation symmetry can be exploited to simplify RDFMT, with
momentum space occupation numbers playing the role of density in the new
functional theory. Having worked out concrete examples in
\cref{chap:momentum_rdmft}, we move again to the abstract realm in
\cref{chap:abelian_functional_theory} to develop the theory of what we call
abelian functional theories, which include both (finite-dimensional) DFT and
the momentum space functional theory in \cref{chap:momentum_rdmft} as special
cases and constitute arguably the simplest class of generalized functional
theories. The central result in this chapter is
\cref{thm:boundary_force_formula}, which establishes the existence of a
diverging boundary force and provides a formula to compute the prefactor
$\mathcal{G}$. Roughly two thirds of \cref{chap:abelian_functional_theory} is devoted to
proving \cref{thm:boundary_force_formula}.

The last two chapters are dedicated to the study of a class of functional
theories built upon representations of compact Lie groups, to which
(finite-dimensional) RDMFT, DFT, and the functional theory of
\cref{chap:momentum_rdmft} all belong.
We will refer to these as momentum map functional theories.
The rigid structure allows us to translate the representability
problem to that of characterizing images of \textit{momentum maps} from symplectic
geometry,  an approach heavily explored and developed in quantum information
theory, in particular in the quantum marginal problem
(note: the word ``momentum'' in ``momentum map'' has nothing to do, for our
purpose, with the familiar physical concept, although the two are etymologically related). 
In
\cref{chap:interlude}, we present a short introduction to the theory of
momentum maps. The most relevant results from this chapter are Kirwan's theorem
(\cref{thm:kirwan}), which is a statement about certain convexity properties of the image
of a momentum map, and \cref{thm:characterization_of_kirwan_polytope}, which
gives a way to characterize the image. Almost all content of
\cref{chap:interlude}, including these two theorems, is drawn from existing
work in the literature, though largely unfamiliar to the functional theory
community. In \cref{chap:momentum_map_functional_theories}, we apply the
machinery set up in the previous chapter to discuss momentum map functional
theories themselves, with primary focus on the boundary behavior of the
functional. The main result is \cref{conj:boundary_force_formula_nonabelian},
which is the counterpart of \cref{thm:boundary_force_formula}. We put forward
an argument for the conjecture based on perturbative calculations, although we
do not have a rigorous proof at the moment.

\section{Background: DFT and RDMFT}
\label{sec:dft_intro}

We quickly review the fundamentals of DFT and RDMFT, starting with the former.
For DFT, nice introductions can be found in
Refs.~\cite{vanleeuwenDensityFunctionalApproach2003,lewinUniversalFunctionalsDensity2023}.
Although there are extensions which accommodate finite temperatures and time
dependence
\cite{merminThermalPropertiesInhomogeneous1965,rungeDensityFunctionalTheoryTimeDependent1984},
we will only be concerned with the ground state theory here and throughout the
thesis. Subtleties of dealing with infinite-dimensional Hilbert spaces are
swept aside, since we will only consider finite-dimensional ones from
\cref{chap:generalized_ft} onwards.

Consider a system of $N$ electrons interacting via a pair interaction $U$ (the
exact form of $U$ will not matter, but the most common choice is the Coulomb
interaction), under the influence of an external potential $v(\mathbf{r})$. 
The Hamiltonian is
\begin{equation}
\label{eq:quantum_chemistry_hamiltonian}
H(v) = \sum_{i=1}^N \frac{p_i^2}{2m} + \sum_{1\le i<j\le N} U(\mathbf{r}_i-
\mathbf{r}_j) + \sum_{i=1}^Nv(\mathbf{r}_i).
\end{equation}
For example, the external potential $v(\mathbf{r})$ could come from a
collection of nuclei around the electrons, in which case the Hamiltonian
\eqref{eq:quantum_chemistry_hamiltonian} would be that of an atom or molecule
after the Born-Oppenheimer approximation. Moving the nuclei then corresponds to
changing the external potential $v(\mathbf{r})$, and for each $v(\mathbf{r})$
there corresponds a ground state $\ket{\Psi(v)}$ (assuming no degeneracy for
simplicity), from which we can compute the electron density $\rho(\mathbf{r})$
by
$$
\rho(\mathbf{r}) = N\sum_{\sigma_1, \dotsb, \sigma_N\in \{\uparrow,\downarrow\}}
\int \braket{\mathbf{r}, \sigma_1, \mathbf{r}_2, \sigma_2, \dotsb, \mathbf{r}_N, \sigma_N|\Psi}
\braket{\Psi|\mathbf{r}, \sigma_1, \mathbf{r}_2, \sigma_2, \dotsb \mathbf{r}_N, \sigma_N} \dstraight \mathbf{r}_2 \dotsb \dstraight \mathbf{r}_N.
$$
We then have a map
$$
v \mapsto \ket{\Psi(v)} \mapsto \rho.
$$
Hohenberg and Kohn \cite{hohenbergInhomogeneousElectronGas1964} showed that
this map is in fact invertible, in the sense that given the ground state
electron density $\rho(\mathbf{r})$, the external potential $v(\mathbf{r})$ can
be recovered up to a constant. Consequently, the ground state $\ket{\Psi}$ is a
functional of $\rho$, and so is every property of the ground state. The
\textit{Hohenberg-Kohn functional} is defined as
\begin{equation}
  \FHK (\rho) := \braket{\Psi(\rho)|T+U|\Psi(\rho)},
\end{equation}
where $T$ is the kinetic energy. One basic relation is
\begin{equation}
\label{eq:intro_dft_legendre}
E(v) = \min_\rho \left[\int v(\mathbf{r})\rho(\mathbf{r})\dstraight\mathbf{r}+ \FHK(\rho)\right],
\end{equation}
where $E(v)$ is the ground state energy corresponding to the external potential $v$.
Indeed, if $\rho$ is the ground state density for $v$, then $E(v) = \int
v(\mathbf{r})\rho(\mathbf{r})\dstraight \mathbf{r} + \FHK(\rho)$. For any other density $\rho'$, we have
$$
\int v(\mathbf{r}) \rho(\mathbf{r}) \dstraight \mathbf{r}  + \FHK(\rho)
\le \braket{\Psi(\rho')|H(v)|\Psi(\rho')}
= \int v(\mathbf{r})\rho'(\mathbf{r})\dstraight \mathbf{r}
+ \FHK(\rho').
$$

Instead of $\FHK$, one can define the \textit{constrained search functionals}
\cite{levyUniversalVariationalFunctionals1979,liebDensityFunctionalsCoulomb1983}:
\begin{equation}
\begin{aligned}
&\mathcal{F}_p(\rho) := \min_{\ket{\Psi}\mapsto \rho}\braket{\Psi|T+U|\Psi}\\
&\mathcal{F}_e(\rho) := \min_{\Gamma\mapsto \rho}\Tr(\Gamma(T+U)),
\end{aligned}
\end{equation}
where the minimizations are over $N$-particle pure states $\ket{\Psi}$ and ensemble states
$\Gamma$ which yield $\rho$ as the density respectively.
Eq.~\eqref{eq:intro_dft_legendre} still holds if we replace $\FHK$ with $\Fp$
or $\Fe$.

The domain of $\FHK$ is the set of densities $\rho$ that are (pure) ground
state electron densities for some external potential $v$, a condition on $\rho$
known as $v$\textit{-representability}. Similarly, the domain of $\Fp$ ($\Fe$),
are the sets of densities $\rho$ such that there exists a pure (ensemble)
state which yields $\rho$ as the density, a condition called pure (ensemble) state  
$N$-representability.

For RDMFT, one considers a family of Hamiltonians of the form
$$
H(h) = h + U,
$$
where $h$ is a single-particle operator, and $U$ is any interaction. For atoms
or molecules, $h$ consists of both the kinetic energy term and the
electron-nucleus interaction, and $U$ is the electron-electron repulsion.
However, we imagine that we are free to choose any single-particle operator
$h$, so in particular we may alter the kinetic term. This is physically
relevant, for example, in ultracold atom experiments, where the hopping rates
of atoms in an optical lattice can be tuned. 

For each $h$, there corresponds a ground state $\ket{\Psi}$ and a 1RDM $\gamma
= N\mathrm{Tr}_{N-1}(\ketbrap{\Psi})$. A generalization of the Hohenberg-Kohn theorem
\cite{gilbertHohenbergKohnTheoremNonlocal1975} implies that the map
$$
    h \mapsto \ket{\Psi} \mapsto \gamma
$$
is invertible (up to shifting $h$ by constants). We can then similarly define the  
Hohenberg-Kohn functional  (also called the Gilbert functional)
$$
\FHK(\gamma) := \braket{\Psi(\gamma)|U|\Psi(\gamma)},
$$
which satisfies
$$
E(h) = \min_{\gamma} \left[\Tr(h \gamma) + \FHK(\gamma)\right],
$$
where $E(h)$ is the ground state energy of $H(h)$. The constrained search
functionals can also be defined \cite{valoneConsequencesExtending1matrix1980}:
$$
\begin{aligned}
&\mathcal{F}_p(\gamma) := \min_{\ket{\Psi}\mapsto \gamma}\braket{\Psi|U|\Psi}\\
&\mathcal{F}_e(\gamma) := \min_{\Gamma\mapsto \gamma}\Tr(\Gamma U)
\end{aligned}
$$
$v$-representability, pure state $N$-representability, and ensemble state
$N$-representability are defined analogously.

\chapter{Generalized Functional Theory}
\label{chap:generalized_ft}


\section{Motivation}

In 
DFT, the electron density $\rho(\mathbf{r})$ is the
variable of the universal functional, and the external potential $v(\mathbf{r})$
can be seen as a parameter of the Hamiltonian. In RDMFT, the 1RDM $\gamma$
plays the role of $\rho(\mathbf{r})$, and the single-particle operator
$h$ plays the role of $v(\mathbf{r})$. There is,
in fact, nothing special about the electron density or the 1RDM. The underlying
mathematical structure of both functional theories depends solely on the fact
that there is a pair of conjugate variables, in the sense of Legendre-Fenchel
transformations, in each case: $(\rho, v)$ in DFT and $(\gamma, h)$ in RDMFT. 
In both functional theories, we aim to obtain the ground state energies of a
family of Hamiltonians, each consisting of a fixed part, which we shall also
call the ``interaction'' in the abstract language that follows, and a variable
part. For example, in DFT the Hamiltonian has the form $H(v) = \sum_i
v(\mathbf{r}_i) + W$, where $W = T+U$, and in RDMFT one has $H(h) = \sum_ih_i +
U$.

Each state $\Gamma$ on the Hilbert space, pure or mixed, is assigned a
``reduced quantity'', which is $\rho$ or $\gamma$ respectively in DFT and
RDMFT, with which the expectation value of the variable part of the Hamiltonian
can be computed. For example, $\Tr[\left(\sum_iv(\mathbf{r}_i)\right) \Gamma] =
\int  v(\mathbf{r}) \rho(\mathbf{r})\dstraight\mathbf{r}$ in DFT and
$\Tr[(\sum_ih_i)\Gamma] = \Tr(h\gamma)$ in RDMFT. 
The reduced quantity, $\rho$ or $\gamma$, is then the simplest object obtained from the
full description $\Gamma$ of the quantum mechanical system with which it is still possible to
recover the expectation value of the variable part of the Hamiltonian. Indeed,
the point of view that the full electronic wave function $\Psi(\mathbf{r}_1,
\dotsb, \mathbf{r}_N)$ might contain too many complicated and unnecessary
details for determining the physical properties of the ground state was a
significant driving force of the development of both DFT and RDMFT
\cite{lowdinQuantumTheoryManyParticle1955,hohenbergInhomogeneousElectronGas1964,liebDensityFunctionalsCoulomb1983},
going back to Thomas and Fermi
\cite{thomasCalculationAtomicFields1927,fermiMetodoStatisticoDeterminazione1927}.

Mathematically, then, we see that there is a dual relation, in the
algebraic sense, between the reduced quantity and the variable part of the
Hamiltonian. 
By considering different scopes within which the latter is allowed to vary, we
obtain distinct functional theories. This idea has been explored and exploited
extensively in the literature, expanding the decades-old framework of Hohenberg
and Kohn to an ever more diverse range of systems, sometimes even finding
applications and connections to other branches of quantum many-body physics.
Here, we give a list of examples of such developments, apologizing in advance
for not being exhaustive due to the myriad of possible modifications and
extensions of DFT that have been proposed in the literature.

\begin{enumerate}[label={(\arabic*)},leftmargin=*]

  \item Already in 1972, von Barth and Hedin devised a spin-polarized
    variant of DFT \cite{barthLocalExchangecorrelationPotential1972}, called spin
    density functional theory (SDFT), taking $\rho_{\sigma\sigma'}(\mathbf{r}):=
    \left\langle\psi_\sigma^\dagger(\mathbf{r})\psi_{\sigma'}(\mathbf{r})\right\rangle$ as the
    reduced quantity, coupled to an external potential $v_{\sigma\sigma'}(\mathbf{r})$
    now sensitive to the spin degree of freedom. Equivalently, the electron density
    $\rho(\mathbf{r})$ is augmented with the magnetization, with the
    external potential being of magnetic nature. 
    The Hohenberg-Kohn theorem
    famously fails in the sense that a one-to-one correspondence between
    $\rho_{\sigma\sigma'}(\mathbf{r})$ and $v_{\sigma\sigma'}(\mathbf{r})$ cannot be established
    \cite{barthLocalExchangecorrelationPotential1972,ullrichNonuniquenessSpindensityfunctionalTheory2005}.

\item In 1988, Oliveira, Gross, and Kohn
proposed a functional theory for superconductors
\cite{oliveiraDensityFunctionalTheorySuperconductors1988}, where the reduced
    variable is the pair $(\rho, \Delta)$, where $\rho(\mathbf{r}):=
    \left\langle\sum_\sigma\psi_\sigma^\dagger(\mathbf{r}) \psi_\sigma(
    \mathbf{r})\right\rangle$ is the usual electron density and $\Delta(\mathbf{r}, \mathbf{r}'):=
    \left\langle\psi_{\uparrow}(\mathbf{r})\psi_{\downarrow}(\mathbf{r}')\right\rangle$ is
    the nonlocal gap function. The resulting functional then depends on both $\rho$ and
    $\Delta$, and a procedure analogous to the Kohn-Sham scheme can be carried out
    to compute the ground state properties. Subsequent works led by L\"uders,
    Marques, and Lathiotakis
    \cite{ludersInitioTheorySuperconductivity2005,marquesInitioTheorySuperconductivity2005}
    incorporated the full many-body nuclear density as an extra density component.

\item Higuchi and Higuchi \cite{higuchiArbitraryChoiceBasic2004} considered a
modification of the constrained search, introducing an arbitrary local reduced quantity
    $X(\mathbf{r})$ associated to each quantum state, thus yielding a constrained
    search functional $\mathcal{F}(\rho, X):= \min_{\Psi\mapsto (\rho, X)}\braket{\Psi|T +
      U|\Psi}$. However, in this formulation, $X(\mathbf{r})$ is treated as a purely
    auxiliary quantity and does not appear in the Hamiltonian. That is, the
    family of Hamiltonians of interest is still $H(v) = T + U + \sum_{i}v(\mathbf{r}_i)$.

\item L\'ibero and Capelle
  \cite{liberoSpindistributionFunctionalsCorrelation2003} proposed a functional
    theory for spin lattice systems, in particular for the inhomogeneous Heisenberg
    model, with Hamiltonian $H = J\sum_{\braket{i,j}}\mathbf{S}^{(i)}\cdot \mathbf{S}^{(j)}+
    \sum_i \mathbf{B}^{(i)}\cdot \mathbf{S}^{(i)}$. Here, the reduced quantity is the list of
    all single-site spin expectation values $\braket{\mathbf{S}^{(i)}} =
    (\braket{S^{(i)}_x}, \braket{S^{(i)}_y}, \braket{S^{(i)}_z})$, which are coupled to a
    site-dependent magnetic field $\{\mathbf{B}^{(i)}\}$. 
    Later, in a similar vein,
    Alcaraz and Capelle \cite{alcarazDensityFunctionalFormulations2007}
    described a functional theory for quantum chains with Hamiltonian of the
    form $H = H_0 + \sum_i v_i N_i$, where each $N_i$ is the ``spin'' operator
    associated to site $i$ and takes discrete values. In both works, a variant
    of the local density approximation is employed to estimate the correlation
    functional, which is by definition the difference between the exact
    functional and the one obtained from mean-field theory.
\item 
    Quantum phase transitions (QPTs) refer to nonanalyticity of properties of
    a system at zero temperature caused by tuning control parameters of the Hamiltonian
    \cite{sachdevQuantumPhaseTransitions2011,vojtaQuantumPhaseTransitions2003}.
    Wu et al. \cite{wuLinkingEntanglementQuantum2006} established a
    general functional-theoretic framework to investigate QPTs with Hamiltonian
    $H=H_0 + \sum_l \lambda_l A_l$, where $H_0$ and the $A_l$ are arbitrary
    operators and the reduced quantity is the collection of expectation values
    $\{\braket{A_l}\}$. The authors were then able to apply results from DFT,
    such as the HK theorem, to examine the dependence of an entanglement
    measure on the control parameters $\lambda_l$. See also
    Refs.~\cite{nagyDensityFunctionalTheory2013,nagyQuantumPhaseTransitions2013,guDensityFunctionalFidelityApproach2009}
    for later refinements.

\item More recently, Penz and van Leeuwen
  \cite{penzDensityfunctionalTheoryGraphs2021} studied a system consisting of
  spinless fermions placed on a finite graph. There, the Hamiltonian is $H = h
  + U$, where $U$ is a fixed operator and $h$ is a single-particle operator
  containing a vertex-dependent external potential together with hopping terms
  specified by the edges of the graph. The reduced quantity is simply the
  discrete analog of the electron density, i.e., the electron number on each
  vertex, and the external potential is allowed to vary. The pure state
  $N$-representability problem is solved without complication, and the
  mechanism behind potential violation of the HK theorem in the same sense as
  in (1) is discussed in detail. 
  Note that an intermediate case between this setting and ordinary DFT, where
  electrons move in a continuum, is discussed by Chayes et al.
  \cite{chayesDensityFunctionalApproach1985}, in which infinite but discrete
  lattice systems are considered.

\item Within RDMFT, Liebert et al.
  \cite{liebertRefiningRelatingFundamentals2023} demonstrated the
  construction of new functional theories by further reducing the 1RDM $\gamma$
  in case the single-particle Hamiltonian $h$ is allowed to vary within a
  restricted subspace. In this context, the reduced quantity corresponds to the
  orthogonal projection of $\gamma$ onto said subspace.
\end{enumerate}

In each variant or modification of the original DFT listed above, often
nontrivial portions of the work are dedicated to presenting an adapted proof of
(some version of) the HK theorem. It is almost a tradition at this point in the
field that whenever a novel variant of functional theory is proposed, the main
mathematical results are to be re-proved so as to reassure the reader that
replacing the electron density $\rho(\mathbf{r})$ by some other reduced quantity
does not break the mathematical foundations that underlie a functional theory.

The goal of this chapter is thus to extract the abstract structure constitutive
to (1)--(7), 
thereby establishing a very general mathematical framework in which
notions like ``universal functionals'', ``$v$-representability'', and
``$N$-representability'' can be defined. 
The advantage of doing so is that each
time we prove a mathematical statement, 
it will be simultaneously true for all functional theories as long as a certain
set of minimal assumptions are satisfied.

Before going into formal mathematics, three remarks are in order: 
(a) No attempt
will be made to broaden the mathematical theory to cover excited states or
finite temperatures; we focus only on functional theories for the ground state
problem. This is not because generalizations are impossible, but because we
will anyway be concerned with only the ground state and its properties in the
subsequent chapters. 
(b) For technical reasons, we always work with
finite-dimensional vector spaces. This means that terms like ``Hilbert space''
or ``vector space'' without modifiers are understood to be finite-dimensional.
The generalization to infinite dimensions should be at least intuitively clear. 
(c) As already hinted at in Examples (1)--(7) above, the HK theorem in its strongest
form, shown in the original paper
\cite{hohenbergInhomogeneousElectronGas1964}, does not hold in general. This is
because the proof relies on the hypothesis that the ground state wave function
does not vanish on a set of positive measure (for a mathematically rigorous
discussion, see Ref.~\cite{liebDensityFunctionalsCoulomb1983}), a condition
that is prohibitively strong when the Hilbert space is finite-dimensional.
However, as will become clear in \cref{sec:hk_theorem}, there is a weaker
version of the HK theorem that is good enough for defining a universal
functional. The failure of the strong form of the HK theorem is intimately tied
to the existence of ground state degeneracies, and a satisfactory investigation
of this phenomenon can be found in
Refs.~\cite{penzStructureDensityPotentialMapping2023,penzGeometryDegeneracyPotential2023,penzGeometricalPerspectiveSpin2024}.
In the following, we will not have much to say about this topic.

\section{Definitions}

Recall that if $\mathcal{H}$ is a Hilbert space, the Lie algebra
$\mathfrak{u}(\mathcal{H})$ of the unitary group consists of all skew-Hermitian
endomorphisms (operators). Hence, we will write $i\mathfrak{u}(\mathcal{H})$
for the real vector space of Hermitian operators.

\begin{defn}
  \label{defn:generalized_ft}
  A \textbf{generalized functional theory} is a tuple $(V, \mathcal{H},
  \iota, W)$, where $V$ is a real vector space, $\mathcal{H}$ a Hilbert space,
  $\iota: V\rightarrow i\mathfrak{u}(\mathcal{H})$ a linear map and $W\in
  i\mathfrak{u}(\mathcal{H})$ is any Hermitian operator. 
  The map $\iota$ is called the \textbf{potential map}, and $W$ is called the
  \textbf{interaction} or the \textbf{fixed part} of the generalized functional
  theory. An element of $V$ is called an \textbf{external potential}.
\end{defn}

We are interested in determining the ground state energy of Hamiltonians of the
form $H(v) = \iota(v) + W$, where $v\in V$ is any external potential. Thus, we
define the \textit{(ground state) energy function} to be
\begin{equation}
  E(v) := \min_{\substack{\ket{\Psi}\in \mathcal{H}: \\ \lVert \Psi\rVert=1}} \braket{\Psi|\iota(v) +W|\Psi}.
\end{equation}
For any $v\in V$, let $\mathbf{G}_p(v)$ denote the set of all pure ground
states $\ketbrap{\Psi}$ of $\iota(v) + W$ and let $\mathbf{G}_e(v)$ denote the set of all
ensemble ground states. 
Clearly, (1) $\mathbf{G}_e(v)$ is the convex hull of $\mathbf{G}_p(v)$,
(2) $\mathbf{G}_p(v) = \mathbf{G}_e(v) \cap \mathcal{P}$, and (3) $E(v) =
\Tr(\Gamma H(v)) = \Tr(\Gamma(\iota(v)+W))$ for any $\Gamma \in
\mathbf{G}_e(v)$, in particular for any $\Gamma \in \mathbf{G}_p(v)$.

\begin{propo}
  \label{thm:energy_is_concave}
  The energy function $E: V\rightarrow\mathbb{R}$ is concave.
  \begin{proof}
    Take $v_1, v_2\in V$ and $t\in [0,1]$. Then 
    $$
    \begin{aligned}
      &E(tv_1 + (1-t)v_2) 
    = \min_{\Gamma\in \mathcal{P}} 
      \Tr[\Gamma (t\iota(v_1) + (1-t)\iota(v_2) + W)]\\
      &= \min_{\Gamma\in \mathcal{P}} \Big[
        t\Tr[\Gamma(\iota(v_1)+ W)]
        +(1-t)\Tr[\Gamma(\iota(v_2)+ W)]
        \Big]\\
      & \ge t\min_{\Gamma\in \mathcal{P}}\Big[
        \Tr[\Gamma(\iota(v_1)+W)]
        \Big]
      + (1-t)\min_{\Gamma\in\mathcal{P}}\Big[
        \Tr[\Gamma(\iota(v_2)+W)]\Big]\\
      &= tE(v_1) + (1-t)E(v_2).
    \end{aligned}
    $$
  \end{proof}
\end{propo}

\begin{ex}
  \label{ex:discrete_DFT}
  Consider an $N$-electron system on discrete lattice sites in any spatial
  dimension labeled by $i=1,2,\dotsb, d$. The
  single-particle Hilbert space is 
  $$
    \mathcal{H}_1 = \mathbb{C}^d =
    \mathrm{span}_{\mathbb{C}}\{\ket{1}, \ket{2}, \dotsb ,\ket{d}\},
  $$
  where $\ket{i}$ is the single-particle state where an electron is located at
  site $i$. The $N$-particle Hilbert space is $\mathcal{H} =
  \exterior^N\mathcal{H}_1$. Take $V=\mathbb{R}^d$ and $\iota: V\rightarrow
  i\mathfrak{u}(\mathcal{H}), v \mapsto \sum_{i=1}^dv_ia_i^\dagger
  {a_i^{\phantom{\dagger}}}$. Let $T$ be a suitable discretized kinetic energy
  operator, and $U$ the discretized Coulomb interaction, and take $W = T+U$.
  Then the generalized functional theory defined by $(V, \mathcal{H}, \iota,
  W)$ is the ordinary DFT (on a finite lattice).
\end{ex}

\begin{ex}
  \label{ex:rdmft}
  Consider a system of $N$ indistinguishable particles, either bosons or fermions. The
  $N$-particle Hilbert space is $\mathcal{H} = \exterior^N\mathcal{H}_1$ for
  fermions and $\mathcal{H}_N = \mathrm{Sym}^N\mathcal{H}_1$ for bosons. Take
  $V = i\mathfrak{u}(\mathcal{H}_1)$ and $\iota: V\rightarrow i\mathfrak{u}(\mathcal{H})$
  to be the map that sends a single-particle operator to the corresponding 
  operator on $\mathcal{H}$, i.e., 
  $$
    \iota\left(u\ketbra{i}{j} + \bar u \ketbra{j}{i}\right) = ua_i^\dagger a_j^{\phantom{\dagger}} 
    + \bar u a_j^\dagger a_i^{\phantom{\dagger}},
  $$
  where $(\ket{i})_{i=1}^{\dim\mathcal{H}_1}$ is any orthonormal basis for $\mathcal{H}_1$
  and $a_i^\dagger, a_i^{\phantom{\dagger}}$ are the respective creation and annihilation operators.
  Take $W$ to be an
  arbitrary interaction, e.g. $W = \sum_{ijkl}w_{ijkl}a^\dagger_i a^\dagger_j
  a_k^{\phantom\dagger} a_l^{\phantom{\dagger}}$ for a two-particle interaction. The generalized functional theory
  defined by $(V, \mathcal{H}, \iota, W)$ is RDMFT.
\end{ex}

\begin{ex}
  \label{ex:nonabelian_spin_chain}
  Consider a chain of $N$ two-level spins, whose Hilbert space is $\mathcal{H} =
  (\mathbb{C}^2)^{\otimes N}$. 
  Take $V = (\mathbb{R}^3)^N$ and 
  $$
  \begin{aligned}
    &\iota: (\mathbb{R}^3)^N\rightarrow i\mathfrak{u}(\mathcal{H})\\
    &(v^1,\dotsb, v^N) \mapsto 
    \sum_{i=1}^N (v^i_x X_i + v^i_y Y_i + v^i_z Z_i),
  \end{aligned}
  $$
  where $X_i, Y_i, Z_i$ are the Pauli operators of the $i$-th spin.
  We can take the interaction to be, for example
  $$
      W = -\sum_{i}^N(X_i X_{i+1} + Y_i Y_{i+1} + Z_i Z_{i+1}).
  $$
  Then $(V,  \mathcal{H},\iota, W)$ recovers the
  spin-$\frac{1}{2}$ case of the spin functional theory discussed by L\'ibero and
  Capelle \cite{liberoSpindistributionFunctionalsCorrelation2003}.
  Generalization to arbitrary spin is straightforward.
\end{ex}

\begin{rem}
  A note on terminology:
  calling $W$ the ``interaction'' comes from RDMFT, where $W$ contains 
  only the particle-particle interaction term and \textit{not} the kinetic term.
  This piece of terminology, however, need not make any physical sense in other
  systems. For example, in DFT, we want $W$ to be the sum of both
  the electron-electron interaction \textit{and} the kinetic energy. Hence, the word
  ``interaction'' should be regarded merely as a reminder of how RDMFT is a
  special case of \cref{defn:generalized_ft}. Whenever there is an
  opportunity for confusion, we shall call $W$ the ``fixed part'' of the
  functional theory.
  Similarly, the term ``external potential'', which now refers to any $v\in V$ in
  our abstract setting, comes from the common nomenclature used in DFT. The
  corresponding term in RDMFT would be ``single-particle operator'', which has a very
  specific definition, and consequently is not chosen for the general
  terminology. Sometimes, by abuse of language, we call $\iota(v)$, instead of
  $v$, an external potential.
\end{rem}

\begin{rem}
  In many cases $V$ will be a subspace of $i\mathfrak{u}(\mathcal{H})$ and
  $\iota: V\hookrightarrow i\mathfrak{u}(\mathcal{H})$ is the inclusion. However,
  we allow the possibility that $\iota$ is not injective.
\end{rem}

In the following, we fix a generalized functional theory $(V, \mathcal{H},
\iota, W)$. Recall that an \textit{ensemble state} (or \textit{density operator}) on
$\mathcal{H}$ is a positive semidefinite operator $\Gamma\in
i\mathfrak{u}(\mathcal{H})$ such that $\Tr(\Gamma) = 1$. The set of all
ensemble states forms a convex compact subset of $i\mathfrak{u}(\mathcal{H})$,
whose extreme points are exactly the pure states, i.e., rank-one projectors. We
will use $\mathcal{E}$ and $\mathcal{P}$ to denote the sets of all ensemble
states and pure states on $\mathcal{H}$ respectively. 

It turns out to be more convenient to think of ensemble states not as operators
on $\mathcal{H}$, but rather as linear functionals on
$i\mathfrak{u}(\mathcal{H})$. This means that we will think of an ensemble state
$\Gamma$ as belonging to $(i\mathfrak{u}(\mathcal{H}))^*$ via the canonical 
isomorphism $i\mathfrak{u}(\mathcal{H})\rightarrow
(i\mathfrak{u}(\mathcal{H}))^*, \Gamma \mapsto \mathrm{Tr}(\Gamma(\cdot))$.

Take any ensemble state $\Gamma\in (i\mathfrak{u}(\mathcal{H}))^*$ and any
external potential $v\in V$. The expectation value of the operator $\iota(v)\in
i\mathfrak{u}(\mathcal{H})$ with respect to $\Gamma$ is $\braket{\Gamma,
\iota(v)}$, where $\braket{\cdot,\cdot}: (i\mathfrak{u}(\mathcal{H}))^* \times
i\mathfrak{u}(\mathcal{H})\rightarrow \mathbb{R}$ is the natural pairing. Then,
we have by duality 
\begin{equation}
  \label{eq:state_potential_duality}
  \braket{\Gamma, \iota(v)} = \braket{\iota^*(\Gamma), v}.
\end{equation}
In this equation, the brackets $\braket{\cdot,\cdot}$ on the
right hand side refer to the pairing 
$V^*\times V\rightarrow \mathbb{R}$.
For a fixed ensemble state $\Gamma$, Eq.~\eqref{eq:state_potential_duality}
holds for any external potential $v\in V$. In other words, $\iota^*(\Gamma)\in V^*$ contains
all the information needed to compute the expectation value of any $v\in V$.
Hence, the dual vector $\iota^*(\Gamma)\in V^*$ 
corresponds to exactly the electron density $\rho(\mathbf{r})$ in DFT, which
can be used to compute the expectation value of any external potential
$v(\mathbf{r})$, or the 1RDM $\gamma$ in RDMFT, which is sufficient to compute
the expectation value of any single-particle operator $h$. This motivates the
following:
\begin{defn}
  The dual map $\iota^*: (i\mathfrak{u}(\mathcal{H}))^* \rightarrow V^*$ is the
  \textbf{density map}. A \textbf{density} is a dual vector $\rho \in V^*$. 
\end{defn}

Thus, for example, $\iota^*(\mathbf{G}_p(v))$ is the set of all densities in
$V^*$ that come from a pure ground state of $\iota(v) + W$ for each external
potential $v\in V$.

We discuss now the notion of \textit{representability} of densities.
Roughly speaking, this has to do with whether a density $\rho\in V^*$ comes
from a quantum state $\Gamma$. In RDMFT, for example, one speaks of the
\textit{$N$-representability problem}, which asks whether a given
single-particle density operator $\gamma$ is the 1RDM of some pure or ensemble
$N$-particle state. Depending on whether we allow $\Gamma$ to be an ensemble
state, and whether $\Gamma$ has to be a ground state of $\iota(v) + W$ for some
external potential $v\in V$, there are in total four types of representability
that can be defined. The importance of this notion will become clear when we
define the various universal functionals in \cref{sec:universal_functionals}: we will see that the domain of each
of the universal functionals is the set of those densities $\rho\in V^*$ that are
representable in one of the following four senses.

\begin{defn}
  \label{def:representability}
  A density $\rho\in V^*$ is: 
  \vspace{-1em}
  \begin{itemize}
    \item \textbf{pure (resp. ensemble) state representable} if there exists a
      pure (resp. ensemble) state $\Gamma \in (i\mathfrak{u}(\mathcal{H}))^*$
      such that $\iota^*(\Gamma) = \rho$. 
    \item \textbf{pure (resp. ensemble) state $v$-representable} if there exists
      an external potential $v\in V$ and a pure (resp. ensemble) state
      $\Gamma \in (i\mathfrak{u}(\mathcal{H}))^*$ such that $\Gamma$ is a ground state of $\iota(v)+W$ and
      $\iota^*(\Gamma) = \rho$. 
  \end{itemize}
\end{defn}

That is, the set of pure (resp. ensemble) state representable densities is
$\iota^*(\mathcal{P})$ (resp. $\iota^*(\mathcal{E})$), and the set of pure
(resp. ensemble) state $v$-representable densities is $\bigcup_{v\in
V}\iota^*(\mathbf{G}_p(v))$ (resp. $\bigcup_{v\in V}\iota^*(\mathbf{G}_e(v))$).
We will refer to the problem of determining and characterizing the sets
$\iota^*(\mathcal{P})$ and $\iota^*(\mathcal{E})$ as the
\textit{representability problem}.

Applied to DFT or RDMFT, the notion of pure/ensemble state representability
coincides exactly with that of pure/ensemble state $N$-representability. Since
the present framework (Definition~\ref{defn:generalized_ft}) generalizes beyond
$N$-particle systems, we drop the prefix ``$N$-'' in the abstract terminology.

\begin{ex}
  \label{ex:2bodyqma}
  Consider a system of $N$ electrons. Let $V$ consist of all two-particle
  operators, i.e., those of the form $a_i^\dagger a_j^\dagger a_k^{\phantom{\dagger}} a_l^{\phantom{\dagger}}$, 
  and let $\iota: V\hookrightarrow i\mathfrak{u}(\mathcal{H}_N)$ be the
  inclusion. The density map $\iota^*$ is, up to normalization, the partial
  trace $\Tr_{N-2}$ over all but two particles, and $\iota^*(\Gamma)$
  is proportional to the 2RDM. The problem of determining whether a
  two-particle operator $\rho$ is in the set $\iota^*(\mathcal{E})$ is QMA
  complete \cite{liuQuantumComputationalComplexity2007}.
\end{ex}

The preceding example shows that determining the sets $\iota^*(\mathcal{E})$
and $\iota^*(\mathcal{P})$ can be extremely difficult in general. Indeed, if
the Hamiltonian $H$ of a system contains at most two-particle interactions,
then the energy expectation value $\braket{\Psi|H|\Psi}$ depends linearly on
the 2RDM. Consequently, an efficient
characterization of the set of (ensemble state) representable 2RDMs would lead
to an efficient algorithm, which works \textit{independently of the
Hamiltonian}, for obtaining the exact ground state energy and ground state 2RDM
using standard convex optimization techniques. This serves to warn us against
being overly ambitious about the representability problem, and motivates us to
single out particular classes of generalized functional theories for which the
representability problem is comparatively more tractable. 

With the notion of representability defined, we can now address a technical
subtlety that we left out in \cref{defn:generalized_ft}: when
defining the potential map $\iota: V\rightarrow i\mathfrak{u}(\mathcal{H})$, we
could have chosen to exclude the possibility that a nonzero element $v\in V$ is
sent to a multiple of $\mathbbm{1} \in i\mathfrak{u}(\mathcal{H})$.
Evidently, this does not affect the physics, since adding a multiple of the
identity to the Hamiltonian merely shifts the energy levels (in particular the
ground state energy $E(v)$) by a constant. 
However, due to our definition of densities as elements of $V^*$, the existence
of $v\neq 0$ such that $\iota(v)\propto \mathbbm{1}$ will reduce the dimension
of the set of representable densities in $V^*$. In RDMFT, for example, this
reduction of dimension corresponds to the normalization of the trace of the
1RDM: $\Tr(\gamma)=N$.

Recall that an \textit{affine combination} of finitely many vectors is a linear
combination with coefficients summing to one. The \textit{affine hull} of a
subset $X$ of a vector space, denoted as $\aff(X)$, is the set of all affine
combinations of vectors in $X$. If $A$ is an affine set, we write $\vec
A$ to denote the vector space of translations, i.e., $\vec A:= \{a-b \mid a,b\in
A\}$. Also recall that the \textit{annihilator} $v^\circ$ of a vector $v$
is the vector space of all dual vectors that vanish on $v$.

\begin{lem}
  \label{lem:canonical_dual_space}
  $\overrightarrow{\aff(\iota^*(\mathcal{P}))}= \overrightarrow{\aff(\iota^*(\mathcal{E}))}\cong
  \left(\frac{V}{\iota^{-1}(\mathrm{span}\{\mathbbm{1}\})}\right)^*$ canonically.
  \begin{proof}
    The first equality is obvious.

    $\overrightarrow{\aff(\iota^*(\mathcal{E}))} =
    \overrightarrow{\iota^*(\aff(\mathcal{E}))}=
    \iota^*\left(\overrightarrow{\aff(\mathcal{E})}\right)$ by linearity.
    But $\overrightarrow{\aff(\mathcal{E})}= \mathbbm{1}^\circ\subset (i\mathfrak{u})^*$.
    Indeed, ``$\subset$'' is obvious due to normalization of density operators, and
    for each traceless Hermitian operator $B$, we see from the spectral decomposition
    $B = \sum_{n} \lambda_n\ket{n}\!\bra{n}$, $\sum_n \lambda_n=0$, that $B\in
    \overrightarrow{\aff(\mathcal{E})}$ (here we identify
    $(i\mathfrak{u}(\mathcal{H}))^*\cong i\mathfrak{u}(\mathcal{H})$ by the
    Hilbert-Schmidt inner product), showing ``$\supset$''. Thus, proving the
    lemma is equivalent to showing 
    $$
      \iota^*(\mathbbm{1}^\circ) {\cong} \left(\frac{V}{\iota^{-1}(\mathrm{span}\{\mathbbm{1}\})}\right)^*,
    $$
    which is true because
    $$
    \left(\frac{V}{\iota^{-1}(\mathrm{span}\{\mathbbm{1}\})}\right)^*
    \cong (\iota^{-1}(\mathrm{span}\{\mathbbm{1}\}))^\circ 
    = \iota^*(\mathbbm{1}^\circ)
    $$
    canonically.
  \end{proof}
\end{lem}

\begin{propo}
  \label{thm:dimension_of_ens_rep}
  The dimension of the set $\iota^*(\mathcal{E})$ of ensemble state representable
  densities is equal to $\dim V - \dim
  \iota^{-1}(\mathrm{span}\{\mathbbm{1}\})$. In particular,
  $\iota^*(\mathcal{E})$ has full dimension if and only if the potential map $\iota$ is injective 
  and $\mathbbm{1}\in i\mathfrak{u}(\mathcal{H})$ is not contained in the image of $\iota$.
  To be more precise,
  $$
    \aff(\iota^*(\mathcal{E})) = V^* \Leftrightarrow \iota^{-1}(\mathrm{span}\{\mathbbm{1}\}) = 0.
  $$
  \begin{proof}
    This is a direct consequence of Lemma~\ref{lem:canonical_dual_space}.
  \end{proof}
\end{propo}

\begin{ex}
  Consider the functional theory defined by $(i\mathfrak{u}(2), \mathbb{C}^2,
  \mathrm{Id}_{i\mathfrak{u}(2)}, W)$, where the interaction $W$ is
  irrelevant. We have $\dim \mathrm{Id}^{-1}(\mathrm{span}\{\mathbbm{1}\}) =
  \dim \mathrm{span}\{\mathbbm{1}\} =1$, so Proposition~\ref{thm:dimension_of_ens_rep}
  states that $\mathrm{Id}^*(\mathcal{E})$ has dimension $\dim(i\mathfrak{u}(2))-1=3$, 
  which is expected since $\mathcal{E}$ is just the Bloch ball.
\end{ex}

It will be convenient to have a formula for the derivative of $\iota^*|_{\mathcal{P}}$. 
For any pure state $\ketbrap{\Psi}\in \mathcal{P}$, we may identify the
tangent space $T_{\ketbrap{\Psi}}\mathcal{P}$ with $\ket{\Psi}^\perp =
\{\ket{\phi}\in \mathcal{H} \mid \braket{\phi|\Psi}=0\}$ (to get the
actual identification, note that the derivative of the map
$\mathcal{H}\rightarrow \mathcal{P}, \ket{\Psi'}\mapsto \ketbrap{\Psi'}$
at $\ket{\Psi}$ has kernel $\mathbb{C}\ket{\Psi}$, thus
$T_{\ketbrap{\Psi}}\mathcal{P}\cong \mathcal{H} / \mathbb{C}\ket{\Psi}\cong
\ket{\Psi}^\perp$ canonically).

\begin{lem}
  \label{lem:derivative_of_density_map}
  The derivative of the density map $\iota^*$ along the set of pure states
  $\mathcal{P}$ is characterized by
  \begin{equation}
    \left\langle
        D_{\ketbrap{\Psi}}\iota^*|_{\mathcal{P}}(\ket{\phi}), v 
        \right\rangle
     = 2\mathrm{Re}\braket{\phi|\iota(v)|\Psi}
  \end{equation}
  for any $\ket{\Psi}\in \mathcal{H}$, $\lVert \Psi\rVert^2=1$,
  $\ket{\phi}\in T_{\ket{\Psi}\!\bra{\Psi}}\mathcal{P}$, and $v\in V$. 
  \begin{proof}
    $$
    \begin{aligned}
      &\left\langle D_{\ketbrap{\Psi}}\iota^*|_{\mathcal{P}}(\ket{\phi}), v\right\rangle\\
      &=\doverd{t}\Big|_{t=0} \braket{\iota^*(\bra{\Psi}+t\bra{\phi})(\ket{\Psi}+t\ket{\phi}), v}\\
      &=\doverd{t}\Big|_{t=0} (\bra{\Psi}+t\bra{\phi})\iota(v)(\ket{\Psi}+t\ket{\phi})\\
      &= \braket{\phi|\iota(v)|\Psi} + \braket{\Psi|\iota(v)|\phi}.
    \end{aligned}
    $$
  \end{proof}
\end{lem}

\section{The Hohenberg-Kohn Theorem}
\label{sec:hk_theorem}
The conceptual basis of DFT is the so-called
Hohenberg-Kohn theorem \cite{hohenbergInhomogeneousElectronGas1964}, which
establishes a one-to-one correspondence between external potentials and
densities in the case of the inhomogeneous electron gas (that is, a system of
$N$ electrons moving in $\mathbb{R}^3$ interacting with each other via the
Coulomb interaction and subject to an external potential $v(\mathbf{r})$), 
which in turn allows us to define the \textit{Hohenberg-Kohn functional}.
Actually proving this statement in full mathematical rigor is very subtle, and
it does not apply to finite-dimensional systems (see discussions below).
Nevertheless, for the definition of the functional, it turns out to be
sufficient to prove a weaker version of the Hohenberg-Kohn theorem. 

\begin{thm}[weak Hohenberg-Kohn; see, for example, Theorem~7 of Ref.~\cite{penzDensityfunctionalTheoryGraphs2021}]
  \label{thm:weak_HK}
  For any $v, v'\in V$, if $\rho \in \iota^*(\mathbf{G}_p(v))\cap
  \iota^*(\mathbf{G}_p(v'))$, then $(\iota^*)^{-1}(\rho)\cap \mathbf{G}_p(v) =
  (\iota^*)^{-1}(\rho)\cap \mathbf{G}_p(v')$. In other words, if $\iota(v)+W$
  and $\iota(v')+W$ both share a ground state density $\rho$, then any pure
  ground state $\Gamma$ $(=\ketbrap{\Psi})$ of $\iota(v)+W$ whose
  density is $\rho$ is also a ground state of $\iota(v')+W$ and vice versa.
  The same statement holds if we replace ``$p$'' with ``$e$'' (and ``pure''
  with ``ensemble'').
  \begin{proof}
    Assume $\rho \in \iota^*(\mathbf{G}_p(v))\cap \iota^*(\mathbf{G}_p(v'))$.
    Take $\Gamma\in (\iota^*)^{-1}(\rho)\cap \mathbf{G}_p(v)$ and $\Gamma' \in
    (\iota^*)^{-1}(\rho)\cap \mathbf{G}_p(v')$. In other words, we take a pure
    ground state $\Gamma$ of $\iota(v)+W$ and a pure ground state $\Gamma'$ of $\iota(v')+W$ 
    such that $\iota^*(\Gamma) = \iota^*(\Gamma') = \rho$. Then
    \begin{equation}
      \begin{aligned}
        &\braket{\Gamma,\iota(v)+W} \le \braket{\Gamma', \iota(v)+W}\\
        &= \braket{\Gamma', \iota(v)} + \braket{\Gamma', W}
        =\braket{\iota^*(\Gamma'), v} + \braket{\Gamma', W}\\
        &= \braket{\rho, v} + \braket{\Gamma', W}\\
        &=\braket{\rho, v-v'} + \braket{\rho, v'} + \braket{\Gamma', W}\\
        &= \braket{\rho, v-v'} + \braket{\iota^*(\Gamma'), v'} + \braket{\Gamma', W}\\ 
        &= \braket{\rho, v-v'} + \braket{\Gamma', \iota(v') + W},
      \end{aligned}
    \end{equation}
    implying $\braket{\Gamma, \iota(v) + W} + \braket{\rho, v'-v}\le \braket{\Gamma', \iota(v')+W}$. 
    Therefore,
    \begin{equation}
      \braket{\Gamma', \iota(v')+W} \le 
      \braket{\Gamma, \iota(v')+W}
      = \braket{\Gamma, \iota(v)+W} + \braket{\rho, v'-v}
      \le \braket{\Gamma', \iota(v')+W}. 
    \end{equation}
    Both ends of this chain of inequalities agree, so the inequalities must
    have been equalities. In particular, $\braket{\Gamma', \iota(v')+W} =
    \braket{\Gamma, \iota(v')+W}$, showing that $\Gamma$ is also a ground state
    of $\iota(v')+W$. Similarly, $\Gamma'$ is also a ground state of $\iota(v)
    + W$. 

    It is easy to check that everything carries through if we replace ``$p$'' with``$e$'' and ``pure''
    with ``ensemble''.
  \end{proof}
\end{thm}

In contrast to this weakened version, the original Hohenberg-Kohn theorem
claims that from the ground state density $\rho$, one can already recover the
external potential $v$ up to a constant. More precisely:

\begin{thm}[strong Hohenberg-Kohn, \textbf{not true in general!}]
  \label{thm:strong_HK}
  If two external potentials $v, v'\in V$ are such that $\iota(v) + W$ and $\iota(v') + W$
  have respective pure ground states $\Gamma, \Gamma'$ that share the same density, i.e.,
  $\iota^*(\Gamma) = \iota^*(\Gamma')$, then $\iota(v)-\iota(v')$ is
  proportional to the identity.
\end{thm}

At this point, it is convenient to introduce the notion of \textit{unique $v$-representability}:
\begin{defn}
    \label{def:unique_vrep}
    A density $\rho\in V^*$ is \textbf{uniquely (pure state) $v$-representable} if there
    exists an external potential $v\in V$ such that $\rho \in
    \iota^*(\mathbf{G}_p(v))$, and any other external potential $v'\in V$ for
    which $\rho \in \iota^*(\mathbf{G}_p(v'))$ satisfies $\iota(v') - \iota(v) \propto \mathbbm{1}$.
\end{defn}

The strong Hohenberg-Kohn theorem then claims that all pure state $v$
representable densities are uniquely $v$-representable.

This stronger version holds, with some mathematical subtleties that were
settled rather recently
\cite{liebDensityFunctionalsCoulomb1983,pinoRestatementHohenbergKohn2007,garrigueUniqueContinuationManybody2018},
for $N$-electron systems in $\mathbb{R}^3$, which was the context in Hohenberg
and Kohn's seminal paper. 
In finite systems, counterexamples are found and discussed in
Refs.~\cite{penzDensityfunctionalTheoryGraphs2021,penzGeometryDegeneracyPotential2023,penzGeometricalPerspectiveSpin2024},
where a functional theory for fermions on finite graphs are studied. Here, we
will construct an extremely simple example that violates
Theorem~\ref{thm:strong_HK}.

\begin{ex}[counterexample to the strong Hohenberg-Kohn theorem]
  \label{ex:HK_counterexample}
  Take $V= \mathbb{R}$, $\mathcal{H} = \mathbb{C}^2 =
  \mathrm{span}\{\ket{\uparrow}, \ket{\downarrow}\}$, 
  $\iota(v) = v Z$,
  where $Z=\mathrm{diag}(1, -1)$ is the Pauli Z operator, and $W=0$.
  Physically, we might want to think of a spin-$\frac{1}{2}$ particle fixed at
  the origin and subject to an external tunable magnetic field in the $z$
  direction. 
  We claim that in the generalized functional theory given by $(V,
  \mathcal{H}, \iota, W)$, the density $1\in V^* = \mathbb{R}$ violates the
  strong HK theorem.
  Take distinct external potentials $v,v'<0$. Then both $\iota(v)+W=vZ$
  and $\iota(v')+W=v'Z$ share the same ground state $\ket{\uparrow}$,
  hence the same density, which can be readily computed:
  $$
  \begin{aligned}
    \rho = \iota^*(\ketbrap{\uparrow})
     = (\braket{\uparrow|\cdot |\uparrow}) \circ (v \mapsto v Z)
    = (v \mapsto v\braket{\uparrow|Z|\uparrow}) = (v \mapsto v)
    =  1.
  \end{aligned}
  $$
\end{ex}

Note that this is the minimal example where the strong Hohenberg-Kohn theorem
can be violated: if $\mathcal{H}$ is one-dimensional, then $\iota(V)\subset
\mathbbm{1}\cdot \mathbb{R}$, so the conclusion of the theorem trivially holds.

Even though the strong Hohenberg-Kohn theorem does not hold in general, something useful
can still be said about unique $v$-representability. Recall that if $f: M\rightarrow
N$ is a smooth map of smooth manifolds, a point $q\in N$ is called a \textit{regular
value} if the derivative $D_pf: T_pM\rightarrow T_qN$ is surjective for all
$p\in f^{-1}(q)$. Otherwise, $q\in N$ is said to be a \textit{critical value}
(if $q$ lies outside of the image of $f$, it is a regular value vacuously).

\begin{propo}
  \label{thm:unique_vrep_regular_value}
  Let $\rho\in V^*$ be a pure state $v$-representable density. If $\rho$ is a
  regular value of the map $\iota^*|_{\mathcal{P}}: \mathcal{P}\rightarrow
  \aff(\iota^*(\mathcal{P}))$, then $\rho$ is uniquely $v$-representable.
  \begin{proof}
    Suppose $v,v'\in V$ are such that $\rho$ is a (pure) ground state density for both
    $\iota(v)+W$ and $\iota(v')+W$.
    Then by the weak Hohenberg-Kohn theorem (Theorem~\ref{thm:weak_HK}), there
    is a state $\ket{\Psi}$ such that $\iota^*(\ketbrap{\Psi}) = \rho$ and
    $$
    \begin{aligned}
      &(\iota(v)+W)\ket{\Psi}= E\ket{\Psi}\\
      &(\iota(v')+W)\ket{\Psi} = E'\ket{\Psi}
    \end{aligned}
    $$
    for some $E,E'$. Taking the
    difference, we get $\iota(v-v')\ket{\Psi} = (E-E')\ket{\Psi}$.

    Suppose $\iota(v - v') \not\propto\mathbbm{1}$, then there is some
    $\gamma\in \mathbbm{1}^\circ$ such that $\braket{\iota^*(\gamma), v-v'} =
    \braket{\gamma, \iota(v-v')}\neq 0$. But $\iota^*(\mathbbm{1}^\circ) =
    \overrightarrow{\aff(\iota^*(\mathcal{P}))}$ (see the proof of
    Lemma~\ref{lem:canonical_dual_space}), so $\alpha:=\iota^*(\gamma)$
    is a vector in $\overrightarrow{\aff(\iota^*(\mathcal{P}))}$ such that
    $\braket{\alpha, v-v'}\neq 0$.

    Since the map
    $D_{\ket{\Psi}\!\bra{\Psi}}\iota^*|_{\mathcal{P}}: \ket{\Psi}^\perp
    \rightarrow \overrightarrow{\aff(\iota^*(\mathcal{P}))}$,
    $\ket{\phi}\mapsto (v\mapsto 2\mathrm{Re}\braket{\phi|\iota(v)|\Psi})$ is
    surjective (where we used Lemma~\ref{lem:derivative_of_density_map}), we
    can find a $\ket{\phi}\in \ket{\Psi}^\perp$ such that $(v\mapsto
    2\mathrm{Re}(\braket{\phi|\iota(v)|\Psi}) = \alpha$. But then
    $$
      2\mathrm{Re}\braket{\phi|\iota(v-v')|\Psi} = \braket{\alpha, v-v'}.
    $$
    The right hand side is nonzero, while the left hand side vanishes since
    $\braket{\phi|\iota(v-v')|\Psi}=(E-E')\braket{\phi|\Psi} = 0$. Hence, we
    have a contradiction and it must have been that $\iota(v-v')\propto \mathbbm{1}$.
  \end{proof}
\end{propo}

\begin{rem}
  In the statement of Proposition~\ref{thm:unique_vrep_regular_value}, we have
  to restrict the codomain of $\iota^*|_{\mathcal{P}}$ to
  $\aff(\iota^*(\mathcal{P}))$, which is fine because we know the image of
  $\iota^*|_{\mathcal{P}}$ is contained in the latter set anyway. We do this to
  prevent the condition that a density be a regular value from being too restrictive: indeed, if
  $\aff(\iota^*(\mathcal{P}))$ has positive codimension in $V^*$, then every
  density will be a critical value of the map $\iota^*|_{\mathcal{P}}:
  \mathcal{P}\rightarrow V^*$! Of course, in this case, the conclusion of
  Proposition~\ref{thm:unique_vrep_regular_value} would be still be true, albeit
  vacuously, but then we would obtain a much weaker statement.
  In light of Proposition~\ref{thm:dimension_of_ens_rep}, we see that this sort
  of inconvenience is the price we pay for allowing the potential map $\iota$
  to map nonzero elements $v\in V$ to multiples of $\mathbbm{1}$.
\end{rem}

Applying Sard's theorem, which states that the set of critical values has
measure zero for any smooth map, we get:
\begin{cor}
  In any generalized functional theory, the set of pure state $v$-representable
  densities which are not uniquely $v$-representable, i.e., densities which
  violate the strong Hohenberg-Kohn theorem, has measure zero in
  $\aff(\iota^*(\mathcal{P}))$.
\end{cor}

Thus, we conclude that the strong HK theorem is not violated \textit{too drastically}.

The study of the properties of set of (uniquely) pure state $v$-representable
densities has a history in the literature. For more detail, see 
Refs.~\cite{katrielMappingLocalPotentials1981,kohn$v$RepresentabilityDensityFunctional1983,ullrichDegeneracyDensityFunctional2002,ullrichNonuniquenessSpindensityfunctionalTheory2005,penzDensityfunctionalTheoryGraphs2021,penzGeometryDegeneracyPotential2023,penzStructureDensityPotentialMapping2023}.

%
%

\section{Universal Functionals}
\label{sec:universal_functionals}
In this section, we will define a total of three functionals within a
generalized density functional theory $(V,\mathcal{H}, \iota, W)$. In practice,
that is, in physically interesting and large systems, none of the functionals
can be exactly obtained due to either the non-constructive nature of the
definition (for the Hohenberg-Kohn functional) or the sheer complexity of the
optimization problems involved (for the constrained search functionals).
However, as we will see, each of these three functionals plays a significant
theoretical role in understanding functional theories in general, and they are
related in precise ways discussed in the second half of this section.

\begin{defn}
  The \textbf{Hohenberg-Kohn functional} $\FHK$ is the real-valued
  function defined on the set of pure state $v$-representable densities in the following
  way: given a $v$-representable density $\rho\in V^*$, take an external
  potential $v\in V$ so that $\rho \in \iota^*(\mathbf{G}_p(v))$. Then define
  $\FHK(\rho)=\braket{\Gamma, W}$ for any $\Gamma\in \mathbf{G}_p(v)\cap
  (\iota^*)^{-1}(\rho)$.
\end{defn}

The definition of the value of the Hohenberg-Kohn functional at $\rho$ depends
on two choices: first, one has to choose an external potential $v\in V$ such
that $\rho$ is a ground state density of $\iota(v)+W$, then one specific pure
ground state $\Gamma\in \mathbf{G}_p(v)$ such that $\iota^*(\Gamma) = \rho$ is
chosen. It is precisely the weak Hohenberg-Kohn theorem
(\cref{thm:weak_HK}) that guarantees that $\FHK(\rho)$ does
not depend on these two choices.

\begin{propo}
  The Hohenberg-Kohn functional is well-defined.
  \begin{proof}
    We need to show that $\FHK(\rho)$ does not depend on the choice of
    $v$ and $\Gamma$. 
    Suppose $v'\in V$ also satisfies $\rho\in\iota^*(\mathbf{G}_p(v'))$.
    Take any $\Gamma'\in \mathbf{G}_p(v')\cap (\iota^*)^{-1}(\rho)$. By
    \cref{thm:weak_HK}, we have $\Gamma'\in \mathbf{G}_p(v)\cap (\iota^*)^{-1}(\rho)$.
    Therefore
    \begin{equation}
      \braket{\Gamma', \iota(v)+W} = \braket{\Gamma, \iota(v)+W}
      \Rightarrow \braket{\Gamma',W} = \braket{\Gamma,W}.
    \end{equation}
  \end{proof}
\end{propo}


\begin{defn}
  \label{def:pure_ensemble_functional}
  The \textbf{pure functional} $\Fp$ and the \textbf{ensemble
  functional} $\Fe$ are defined as
  \begin{equation}
    \label{eq:def_pure_functional}
    \begin{aligned}
      \Fp(\rho)&: \iota^*(\mathcal{P})\rightarrow \mathbb{R}\\
      &\rho \mapsto \min_{\Gamma \in (\iota^*)^{-1}(\rho)\cap\mathcal{P}}
    \braket{\Gamma, W}
    \end{aligned}
  \end{equation}
  \begin{equation}
    \label{eq:def_ensemble_functional}
    \begin{aligned}
      \Fe(\rho)&: \iota^*(\mathcal{E})\rightarrow \mathbb{R}\\
      &\rho \mapsto \min_{\Gamma \in (\iota^*)^{-1}(\rho)\cap\mathcal{E}}
    \braket{\Gamma, W}.
    \end{aligned}
  \end{equation}
  That is, $\Fp(\rho)$ (resp. $\Fe(\rho)$) is defined to be
  the minimum of the expectation value $\braket{\Gamma, W}$ over all pure (resp.
  ensemble) states $\Gamma$ such that $\iota^*(\Gamma)=\rho$.
\end{defn}

\begin{rem}
  The density map $\iota^*$ is linear, in particular continuous, so both
  $(\iota^*)^{-1}(\rho)\cap \mathcal{P}$ and $(\iota^*)^{-1}(\rho)\cap
  \mathcal{E}$ are compact. Since the function $\Gamma \mapsto \braket{\Gamma,
  W}$ is continuous, the infima in Eqs.~\eqref{eq:def_pure_functional} and
  \eqref{eq:def_ensemble_functional} are attained. This justifies writing
  $\min$ instead of $\inf$.
\end{rem}

\begin{rem}
  In the literature, these are also often called \textit{constrained search}
  functionals because of the constrained minimizations appearing in their
  definitions. In the DFT literature, the pure functional is sometimes also
  called the \textit{Levy-Lieb functional}, and the ensemble functional is
  sometimes called the \textit{Lieb functional}. In the RDMFT literature, the
  Hohenberg-Kohn functional is called the \textit{Gilbert
  functional}, due to Gilbert \cite{gilbertHohenbergKohnTheoremNonlocal1975}, 
  and the ensemble functional is often called the \textit{Valone
  functional}, due to Valone \cite{valoneConsequencesExtending1matrix1980}.
  Throughout this thesis, we will stick to the more descriptive names
  given in Definition~\ref{def:pure_ensemble_functional}.
  We will also refer to both $\Fp$ and $\Fe$ as 
  the constrained search functionals, and to the minimizations
  \eqref{eq:def_pure_functional} and \eqref{eq:def_ensemble_functional}
  as the \textit{constrained search}.
\end{rem}

\begin{rem}
  A brief note on the word ``functional'': in DFT, a suitable function space
  which is an infinite-dimensional Banach space (see, for example, Section 2 of
  Ref.~\cite{vanleeuwenDensityFunctionalApproach2003}) plays the role of $V$,
  so $\FHK, \Fp, \Fe$ are functionals in the usual sense. In our setting,
  $V$ is finite-dimensional, so it may be more natural to call them
  ``functions''. However, we choose to adopt the traditional terminology.
\end{rem}

\begin{ex}
  \label{ex:qubit_functional_theory}
  We construct a simple functional theory and compute its functionals. Let
  $V=\mathbb{R}, \mathcal{H}=\mathbb{C}^2=\mathrm{span}\{\ket{\uparrow},
  \ket{\downarrow}\}, \iota(v) = vZ,$ and $W = \lambda X$, where
  $\lambda$ is a parameter. Note that the functional theory defined by
  $(V,\mathcal{H},\iota,W)$ is identical to the one in
  Example~\ref{ex:HK_counterexample} when $\lambda=0$. It is easy to verify that 
  $\iota^*(\mathcal{E}) = \iota^*(\mathcal{P}) = [-1, 1]\subset V^* = \mathbb{R}$.

  We consider first the pure functional $\Fp(\rho):=
  \min_{\ket{\Psi}:\braket{\Psi|Z|\Psi} = \rho}\braket{\Psi|\lambda X|\Psi}$,
  where $\rho \in [-1,1]$ is some fixed density. Every state $\ket{\Psi}$ which
  satisfies $\braket{\Psi|Z|\Psi} = \rho$ is of the form 
  $$
    \ket{\Psi} = \frac{1}{\sqrt{2}}\Big(\sqrt{1+\rho}\ket{\uparrow} +
    e^{i\phi}\sqrt{1-\rho}\ket{\downarrow}\Big),
  $$
  where $e^{i\phi}$ is any phase. Hence,
  $$
  \Fp(\rho) = \min_{\phi\in[0,2\pi]}\lambda\sqrt{1+\rho}\sqrt{1-\rho}\cos\phi
   = -|\lambda|\sqrt{1+\rho} \sqrt{1-\rho}.
  $$
  Next, we determine the ensemble functional $\Fe(\rho):=\min_{\Gamma:
  \braket{\Gamma,Z}=\rho}\lambda\Tr(\Gamma X)$. Recall that any ensemble state $\Gamma$ on
  $\mathbb{C}^2$ can be written as $\Gamma =
  \frac{1}{2}(\mathbbm{1}+a_xX + a_yY + a_zZ)$, where $a = (a_x,a_y,a_z)\in \mathbb{R}^3$
  satisfies $\lVert a\rVert \le 1$. Clearly, the condition
  $\braket{\Gamma, Z}=\rho$ is the same as $a_z=\rho$, so 
  $$
  \Gamma = \frac{1}{2}(\mathbbm{1} + \rho Z+ a_x X+ a_y Y),
  $$
  where $a_x^2 + a_y^2 \le 1-\rho^2$. Since $\braket{\Gamma, X} = a_x$, we have 
  $$
  \Fe(\rho) = \min_{(a_x,a_y): a_x^2+a_y^2 \le 1-\rho^2} \lambda a_x
   = -|\lambda|\sqrt{1-\rho^2}.
  $$
  So we see that $\Fe(\rho) = \Fp(\rho)$. We can visualize the constrained
  search for both the pure functional and the ensemble functional as shown in
  Fig.~\ref{fig:blochball_constrained_search}. Since the Hilbert space
  $\mathcal{H}$ is two-dimensional, the set of all pure states $\mathcal{P}$ is
  the Bloch sphere, and the set of ensemble states $\mathcal{E}$ is the convex
  hull, i.e., the ball. For any ensemble (in particular pure) state $\Gamma$,
  the density map $\iota^*$ sends $\Gamma$ to its $z$-coordinate on the Bloch
  ball. The preimage $(\iota^*)^{-1}(\rho)$ then intersects $\mathcal{P}$ and
  $\mathcal{E}$ in a circle and a disk respectively. 

  \begin{figure}
    \centering
    \includegraphics[width=.64\textwidth]{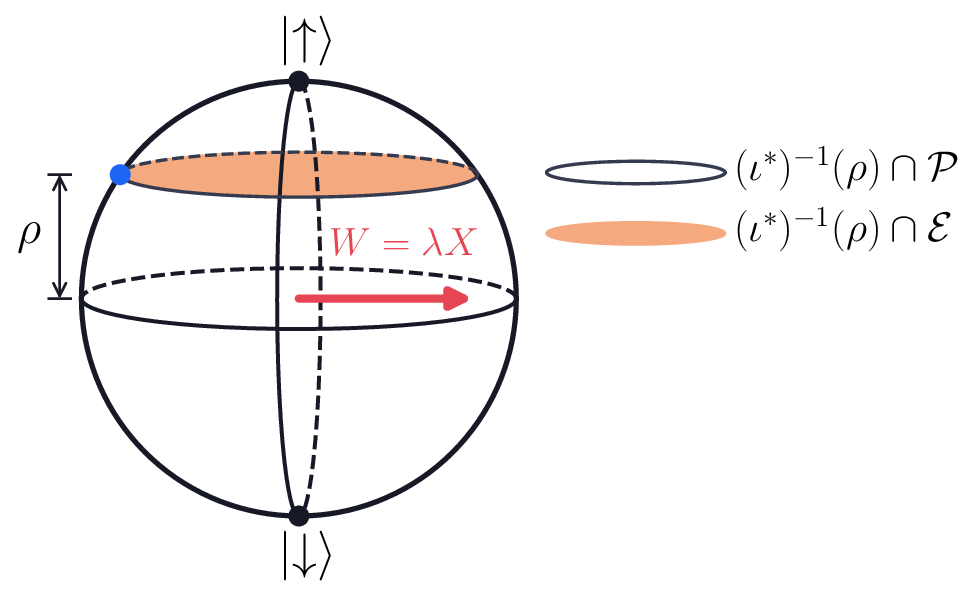}
    \caption{Illustration of the geometry of computing the pure and ensemble
    functionals. The sphere represents the set $\mathcal{P}$ of all pure states, and the ball
    is the set $\mathcal{E}$ of ensemble states. 
    The value of the pure functional $\Fp$ at a
    density $\rho\in[-1,1]$ is defined to be the minimum of $\braket{\Gamma, W}$,
    where $\Gamma$ is a pure state mapping to $\rho$. The set of all admissible
    pure states $\Gamma$ is $(\iota^*)^{-1}(\rho)\cap \mathcal{P}$, which is
    nothing but the horizontal circle at height $\rho$ in the figure.
    Similarly, for the ensemble functional, the set of admissible ensemble
    states is the orange disk. We may think of the fixed interaction $W=\lambda
    X$ as a vector in $\mathbb{R}^3$, and in this case it points towards the
    positive $x$ axis if $\lambda > 0$ (red arrow). The minimizer of
    $\braket{\Gamma, W}$ is then the blue dot on the left for both pure and ensemble
    states, which explains why $\Fp$ and $\Fe$ agree.
    }

    \label{fig:blochball_constrained_search}
  \end{figure}

  Of course, since $\mathcal{H} = \mathbb{C}^2$, we can also directly
  calculate the ground state energy $E(v)$ and the ground state $\ket{\Psi(v)}$
  for each external potential $v\in\mathbb{R}$. 
  The Hamiltonian is $H(v) = vZ+ \lambda X$,
  so $E(v) = -\sqrt{\lambda^2 + v^2}$ and, for $\lambda\neq 0$,
  $$
    \ket{\Psi(v)}\!\bra{\Psi(v)} =
    \frac{1}{2}\left(\mathbbm{1} - \frac{v}{\sqrt{\lambda^2+v^2}}Z -
    \frac{\lambda}{\sqrt{\lambda^2+v^2}} X\right).
  $$
  It follows that the Hohenberg-Kohn functional satisfies 
  $$
  \begin{aligned}
    &\FHK(\braket{\Psi(v)|Z|\Psi(v)}) = \lambda\braket{\Psi(v)|X|\Psi(v)}\\
    \Rightarrow &
    \FHK\left(-\frac{v}{\sqrt{\lambda^2 + v^2}}\right)
    = -\frac{\lambda^2}{\sqrt{\lambda^2 + v^2}}.
  \end{aligned}
  $$
  Inverting $\rho = -v/\sqrt{\lambda^2 + v^2}$ then gives
  $$
  \FHK(\rho) = -\frac{\lambda^2}{\sqrt{\lambda^2+\frac{\lambda^2\rho^2}{1-\rho^2}}}
  = - |\lambda| \sqrt{1-\rho^2}\hspace{2em} \rho\in (-1,1).
  $$
  Note that $1$ and $-1$ are not pure state $v$-representable (unless
  $\lambda=0$), so $\FHK$ is not defined there.

  If $\lambda=0$, then $\FHK(\rho) =0$ by definition, so the formula above
  still holds. We conclude that the three functionals in the functional theory
  $(V,\mathcal{H},\iota,W)$ are identical on $(-1,1)$ for all $\lambda$.
\end{ex}

\section{Properties of the Functionals}
So far, we have defined three functionals. These are the Hohenberg-Kohn
functional $\mathcal{F}_{HK}$ and the pure and ensemble functionals
$\Fp,\Fe$. The practical usefulness of these functionals
lies in the fact that they all determine the ground state energy $E(v)$ for
\textit{all} external potentials $v\in V$.

\begin{thm}
  \label{thm:energy_functional_legendre_transform}
  For any external potential $v\in V$, we have
  \begin{equation}
    \begin{aligned}
      E(v) &= 
    \min_{\rho\in \bigcup_{v\in V}\iota^*(\mathbf{G}_p(v))}[\braket{\rho, v} + \mathcal{F}_{HK}(\rho)] 
      = \min_{\rho\in \iota^*(\mathcal{P})}[\braket{\rho, v}+ \mathcal{F}_p(\rho)]
      = \min_{\rho\in \iota^*(\mathcal{E})}[\braket{\rho, v} + \mathcal{F}_e(\rho)].
    \end{aligned}
  \end{equation}
\begin{rem}
  The sets involved in the three minimizations are exactly the sets of pure
  state $v$-representable, pure state representable and ensemble state
  representable densities respectively (see comments after
  Definition~\ref{def:representability}). 
\end{rem}
  \begin{proof}
    Fix an arbitrary external potential $v\in V$.

    For any pure state $v$-representable density $\rho'$, let $\Gamma'$
    be a pure ground state of $\iota(v')+ W$ for some $v'\in V$ such that
    $\iota^*(\Gamma')=\rho'$.
    Then
    $$
      \braket{\rho', v} +\mathcal{F}_{HK}(\rho')
      = \braket{\rho', v} +\braket{\Gamma', W} 
      = \braket{\Gamma', \iota(v) + W}
      \ge E(v).
    $$
    On the other hand, let $\Gamma$ be a pure ground state of $\iota(v)+W$.
    Then $E(v) = \braket{\iota^*(\Gamma), v} + \braket{\Gamma, W}$. This shows
    the first equality.

    The second equality follows from splitting the minimization in the
    Rayleigh-Ritz variational principle into two steps:
    $$
    \begin{aligned}
      &E(v) = \min_{\Gamma \in \mathcal{P}} \braket{\Gamma, \iota(v) + W}
      = \min_{\rho\in \iota^*(\mathcal{P})} \min_{\Gamma\in (\iota^*)^{-1}(\rho)\cap \mathcal{P}}\braket{\Gamma, \iota(v)+W}\\
      &=\min_{\rho\in \iota^*(\mathcal{P})}\Big[\braket{\rho, v}+
      \min_{\Gamma\in (\iota^*)^{-1}(\rho)\cap \mathcal{E}}\braket{\Gamma,W}
      \Big].
    \end{aligned}
    $$
    The third equality can be shown in exactly the same way.
  \end{proof}
\end{thm}

Since the three functionals yield the same ground state energy in the
precise sense of Theorem~\ref{thm:energy_functional_legendre_transform}, one
might hope to establish simple relations among them, such as one being an extension of another. 
As a first step towards understanding what potentially could be said, note that
the domains of these functionals sit inside each other:
$$
  \dom\FHK \subset \dom\Fp \subset\dom \Fe,
$$
which are the sets of pure state $v$-representable, pure state representable
and ensemble state representable densities respectively. Hence, we may ask
whether $\Fp$ and $\Fe$ are extensions of $\FHK$, the answer to which is
positive:

\begin{propo}
  \label{thm:functionals_agree_on_v_rep}
  The functionals $\mathcal{F}_{HK}, \mathcal{F}_p, \mathcal{F}_e$ agree on
  their common domain, which is the set of pure state $v$-representable
  densities $\bigcup_{v\in V}\iota^*(\mathbf{G}_p(v))$.
  \begin{proof}
    It is clear that $\bigcup_{v\in V}\iota^*(\mathbf{G}_p(v))\subset
    \iota^*(\mathcal{P})\subset \iota^*(\mathcal{E})$: the first inclusion
    follows from $\mathbf{G}_p(v)\subset \mathcal{P}$, and the second from
    $\mathcal{P}\subset \mathcal{E}$. Therefore, the common domain is the set
    of pure state $v$-representable densities.

    Take a pure state $v$-representable density $\rho\in V^*$ and let $v\in V$
    be an external potential with pure ground state $\Gamma$ such that $\rho =
    \iota^*(\Gamma)$. Then $\mathcal{F}_{HK}(\rho) = \braket{\Gamma,W}$ by
    definition. We will show that $\mathcal{F}_p(\rho) = \braket{\Gamma, W}$.

    Take any pure state $\Gamma'$ such that $\iota^*(\Gamma')=\rho$, then
    $$
    \begin{aligned}
      &\braket{\Gamma', \iota(v) + W} \ge \braket{\Gamma, \iota(v)+W}\\
      &\Rightarrow \braket{\rho, v} + \braket{\Gamma', W} \ge \braket{\rho, v} + \braket{\Gamma, W}\\
      &\Rightarrow \braket{\Gamma', W} \ge \braket{\Gamma, W},
    \end{aligned}
    $$
    implying $\Fp(\rho)\ge \braket{\Gamma ,W}$. But clearly
    $\Fp(\rho)\le \braket{\Gamma, W}$, so we conclude
    $\Fp(\rho) = \braket{\Gamma, W}$.

    It is straightforward to show $\mathcal{F}_e(\rho)=\braket{\Gamma,W}$ by
    repeating the same argument.  
  \end{proof}
\end{propo}

The preceding proposition does not say anything about the values of the
functionals outside the set of pure state $v$-representable densities. In
particular, it does \textit{not} claim (nor does it rule out) that the pure and
ensemble functionals agree on their common domain, which is
$\iota^*(\mathcal{P})$. In fact, there are cases where the two do not agree
\cite{schillingCommunicationRelatingPure2018}.

Another important relation among $\FHK, \Fp$, and $\Fe$ can be established
using some machinery from convex analysis. Recall that if $g: \mathcal{V}\rightarrow
\mathbb{R}\cup\{\pm\infty\}$ is any function on a vector space $\mathcal{V}$, the
\textit{convex conjugate} or \textit{Legendre-Fenchel transform} of $g$ is the
function $g^*: \mathcal{V}^*\rightarrow \mathbb{R}\cup\{\pm\infty\}$ defined by $g^*(y) =
\sup_{x\in U}(\braket{x,y} - g(x))$. If a function $g$ is defined on a proper
subset of $\mathcal{V}$, we extend it by setting $g(x)=+\infty$ for $x$ outside the
domain and define its convex conjugate as that of the extension.

A basic result is that $g^*$ is convex for any $g$, and $g^{**}$ is the 
lower convex envelope of $g$. Moreover, $g^{**} = g$ if and only if $g$ is
itself convex, proper and lower semicontinuous (or $g$ is identically $+\infty$
or $-\infty$) by the Moreau-Fenchel theorem. We now show that the ensemble
functional satisfies this condition:

\begin{propo}
  \label{thm:ensemble_functional_is_convex}
  The ensemble functional $\Fe: \iota^*(\mathcal{E})\rightarrow \mathbb{R}$ is
  convex and lower semicontinuous. 
  \begin{proof}
    Take ensemble state representable densities $\rho_1,
    \rho_2\in \iota^*(\mathcal{E})$, and $t\in [0,1]$. Then the density
    $\rho:= t\rho_1 + (1-t)\rho_2$ is also ensemble state representable (because
    $\mathcal{E}$ is convex and $\iota^*$ is linear), hence
    $\mathcal{F}_e(\rho)$ is defined. We have
    $$
    \begin{aligned}
      &\Fe(\rho)
     = \min_{
       \Gamma \in (\iota^*)^{-1}(\rho)\cap \mathcal{E}
     }
       \braket{\Gamma, W}.\\
      \end{aligned}
    $$
    Since the set $\mathcal{E}$ of all ensemble states on $\mathcal{H}$ is
    convex, the set $\{\Gamma \in \mathcal{E}\mid \iota^*(\Gamma)=\rho\}$
    contains $\{t\Gamma_1 + (1-t)\Gamma_2 \mid \Gamma_i\in \mathcal{E}, \iota^*(\Gamma_1)+\rho_1,
    \iota^*(\Gamma_2)=\rho_2\}$ as a subset. It follows that we can bound the minimum from
    above by restricting to the latter set, obtaining
    $$
    \begin{aligned}
    \Fe(\rho) \le 
      &\min_{\Gamma_1\in (\iota^*)^{-1}(\rho_1)\cap\mathcal{E}}
    \min_{\Gamma_2\in (\iota^*)^{-1}(\rho_2)\cap\mathcal{E}}
      \braket{t\Gamma_1 + (1-t)\Gamma_2, W}\\
      &= t\min_{\Gamma_1\in (\iota^*)^{-1}(\rho_1)\cap \mathcal{E}}
      \braket{\Gamma_1, W}
      + (1-t)\min_{\Gamma_2\in (\iota^*)^{-1}(\rho_2)\cap \mathcal{E}}
      \braket{\Gamma_2, W}\\
      &\le t\Fe(\rho_1) + (1-t)\Fe(\rho_2).
    \end{aligned}
    $$
    So $\Fe$, and therefore also $\mathcal{\tilde F}_e$, is convex.

    Let $\rho \in \iota^*(\mathcal{E})$ be any ensemble representable density.
    Suppose $(\rho_i)_i$ is a sequence in $\iota^*(\mathcal{E})$ converging to
    $\rho$. Let $\Gamma \in (\iota^*)^{-1}(\rho)\cap \mathcal{E}$ be a
    minimizer of the constrained search, so that $\Fe(\rho) = \braket{\Gamma, W}$.
    For each $i$, let $\Gamma_i'\in (\iota^*)^{-1}(\rho_i)\cap \mathcal{E}$ be
    a minimizer of the constrained search at $\rho_i$. We can extract a
    subsequence $(i_j)_j$ such that $\Fe(\rho_{i_j})\xrightarrow{j\rightarrow\infty}
    \liminf_{i\rightarrow\infty}\Fe(\rho_{i})$ and
    $\Gamma_{i_j}'\xrightarrow{j\rightarrow\infty} \Gamma'$ for some
    $\Gamma'\in \mathcal{E}$ by the compactness of $\mathcal{E}$. 
    Then
    $$
    \iota^*(\Gamma') = \lim_{j\rightarrow\infty}\iota^*(\Gamma_{i_j}')
    = \lim_{j\rightarrow\infty} \rho_{i_j} = \rho,
    $$
    from which it follows that
    $$
    \liminf_{i\rightarrow\infty} \Fe(\rho_i)
    = \lim_{j\rightarrow\infty} \Fe(\rho_{i_j})
    = \lim_{j\rightarrow\infty} \braket{\Gamma_{i_j}', W}
    = \braket{\Gamma', W} \ge \Fe(\rho).
    $$
    This shows that the ensemble functional $\Fe$ is lower semicontinuous.
  \end{proof}
\end{propo}

Next, note that
Theorem~\ref{thm:energy_functional_legendre_transform} is the same as saying
that the ground state energy function $E$ is, up to some minus signs, the
convex conjugate of the three functionals. More precisely, one has
\begin{equation}
  \label{eq:energy_functional_legendre_transform}
  E(v) = -\mathcal{F}_{HK}^*(-v) = -\mathcal{F}_p^*(-v) = -\mathcal{F}_e^*(-v).
\end{equation}
This implies immediately that the ground state energy function $E$ is concave,
giving an alternative proof of Proposition~\ref{thm:energy_is_concave}.

Multiplying Eq.~\eqref{eq:energy_functional_legendre_transform} by $-1$ and taking the convex conjugate gives
\begin{equation}
  (-E)^*(-\rho) = \mathcal{F}_{HK}^{**}(\rho)
   = \mathcal{F}_p^{**}(\rho) = \mathcal{F}_e^{**}(\rho) = \mathcal{F}_e(\rho),
\end{equation}
where the last equality follows from the fact that the ensemble functional
$\mathcal{F}_e$ is convex (more precisely, that its extension is convex, proper,
and lower semicontinuous) due to
Proposition~\ref{thm:ensemble_functional_is_convex}. This proves the
following:

\begin{propo}
  \label{thm:ensemble_functional_is_lower_convex_envelope}
  The ensemble functional $\mathcal{F}_e$ is the lower convex envelope of both
  the Hohenberg-Kohn functional $\mathcal{F}_{HK}$ and the pure functional $\mathcal{F}_{p}$.
\end{propo}

%

\section{The Purification Trick}
\label{sec:purification_trick}
  Recall that any ensemble state $\Gamma$ on $\mathcal{H}$ is the marginal of
  some pure state on $\mathcal{H}\otimes \mathcal{H}$. In other words, there
  always exists a pure state $\bar\Gamma$ on $\mathcal{H}\otimes \mathcal{H}$,
  called a \textit{purification} of $\Gamma$, such that $\Gamma =
  \Tr_2(\bar\Gamma)$. More precisely, if $\Gamma=\sum_i t_i \ketbrap{\Psi_i}$
  is any ensemble state, we may take $\ket{\bar \Psi} = \sum_i
  \sqrt{t_i}\ket{\Psi_i}\otimes \ket{i}$, where $(\ket{i})_{i=1}^{\dim
  \mathcal{H}}$ is any orthonormal basis, and $\bar\Gamma :=
  \ketbrap{\bar\Psi}$, for which it holds that $\Gamma = \Tr_2(\bar \Gamma)$.
  Based on this idea, we now show that the ensemble functional of any
  generalized functional theory is, in fact, a pure functional of a suitably
  constructed functional theory. 

  This has both conceptual and practical appeal. Conceptually, this means that
  Definition~\ref{defn:generalized_ft} gives rise to a class of functional
  theories large enough such that the notion of ``ensemble functional'' is a
  redundant or automatic one as soon as that of ``pure functional'' is defined.
  Practically, this result is extremely useful for proving properties of the
  ensemble functional: any general statement which is true about the pure
  functional also applies to the ensemble functional, precisely because the
  latter is a special case of the former. As an example, when we derive a
  formula for the boundary force via the pure state constrained search
  (Eq.~\eqref{eq:def_pure_functional}) in
  Chapter~\ref{chap:abelian_functional_theory}, the same formula will apply to
  the ensemble functional after replacing the Hilbert space with a suitably
  larger one.
 
  Even though we will primarily be interested in the ensemble functional, which
  is the lower convex envelope of the pure functional
  (Proposition~\ref{thm:ensemble_functional_is_lower_convex_envelope}), we
  present here a slightly more general construction, which will result in a
  family of not necessarily convex functionals, with the pure and ensemble
  functionals being two extreme cases.

\begin{defn}
  The \textbf{$k$-convex hull} of a subset $X$ of a vector space is
  $\mathrm{conv}_k(X) := \{\sum_{i=1}^k t_i s_i\mid s_i\in X, \sum_{i=1}^kt_i=1, t_i\ge 0\}$. 
  The \textbf{$k$-convexification} of a function $f: X\rightarrow \mathbb{R}$ is the function
  $\mathrm{conv}_k(f): \mathrm{conv}_k(X)\rightarrow \mathbb{R},$ 
  $s \mapsto \inf_{\{t_i\}, \{s_i\}: \sum_i t_is_i = s} \sum_it_if(s_i)$
  (where the conditions $\sum_i t_i =1$ and $t_i\ge 0$ are left implicit).
\end{defn}

\begin{rem}
$\conv_k(\cdot)$ is not idempotent in general, in contrast to
$\conv(\cdot)$. However, if $\conv(X)$ is at most
$(k-1)$-dimensional, then $\conv(X) = \conv_k(X)$ by Carath\'eodory's theorem.
\end{rem}

Let $(V, \mathcal{H}, \iota, W)$ be a generalized functional theory. 
\begin{defn}
  \label{defn:k_convex_functional_theory}
  For a
  positive integer $k$, the associated \textbf{$k$-convexification} or
  \textbf{$k$-convexified functional theory} is the generalized functional
  theory constructed from the data $(V, \mathcal{H}\otimes \mathbb{C}^k,
  \bar\iota, W\otimes \mathbbm{1})$, where $\bar\iota(v) := \iota(v)\otimes
  \mathbbm{1}$.
\end{defn}

The intuition behind Definition~\ref{defn:k_convex_functional_theory} is that
we want to represent ensemble states of rank at most $k$ on the ``primal''
Hilbert space $\mathcal{H}$ as pure states on the ``primal + ancillas'' Hilbert
space $\mathcal{H}\otimes \mathbb{C}^k$. If $k=\dim\mathcal{H}$, we will simply
refer to the functional theory constructed this way as the \textit{convexified
functional theory}.

Objects and notions like the pure functional, pure state representability, etc.
in this newly constructed functional theory are then automatically defined. We
will typically use a bar (\;$\bar{}$\;) to indicate that an object belongs to
the $k$-convexified functional theory, so, for example,
$\mathcal{\bar P}$ refers to the set of pure states on
$\mathcal{H}\otimes \mathbb{C}^k$, and $\mathcal{\bar F}_p$ denotes the pure
functional of the $k$-convexification.

\begin{lem}
  \label{lem:k_convex_hull_pure_repr}
  The $k$-convex hull of $\iota^*(\mathcal{P})$ is equal to the set of pure
  state representable densities in the $k$-convexification. That is,
  \begin{equation}
    \mathrm{conv}_k(\iota^*(\mathcal{P})) = {\bar\iota}^*(\mathcal{\bar P}).
  \end{equation}
  \begin{proof}
    Take a density $\rho \in V^*$ in the $k$-convex hull of
    $\iota^*(\mathcal{P})$. Then $\rho = \sum_{i=1}^k t_i\iota^*(\Gamma_i)$,
    where $\sum_it_i =1$, $t_i\ge 0$, and the $\Gamma_i = \ketbrap{\Psi_i}$ are
    pure states on $\mathcal{H}$.  Let $(\ket{i})_{i=1}^k$ be the standard
    orthonormal basis for $\mathbb{C}^k$. Define $\ket{\bar \Psi}:= \sum_k
    \sqrt{t_i}\ket{\Psi_i}\otimes\ket{i}$ and $\bar \Gamma :=
    \ket{\bar\Psi}\!\bra{\bar\Psi}$. Then for all $v\in V$
    $$
    \begin{aligned}
      &\left\langle \bar\iota^*(\bar\Gamma), v\right\rangle
    = \braket{\bar\Psi|\iota(v)\otimes \mathbbm{1}|\bar\Psi}
      =\sum_{i,j=1}^k
      \sqrt{t_it_j}
      (\bra{\Psi_i}\otimes \bra{i})(\iota(v)\otimes \mathbbm{1})
      (\ket{\Psi_j}\otimes \ket{j})\\
      &=\sum_{i,j=1}^k\sqrt{t_it_j}\braket{\Psi_i|\iota(v)|\Psi_j}\delta_{ij}
      =\sum_{i=1}^k t_i\braket{\Psi_i|\iota(v)|\Psi_i}
      = \sum_{i=1}^k t_i\left\langle \iota^*(\Gamma_i), v\right\rangle\\
      &= \left\langle\sum_{i=1}^k t_i\iota^*(\Gamma_i), v\right\rangle
      =\braket{\rho, v}
    \Rightarrow  \bar\iota^*(\bar\Gamma) = \rho.
    \end{aligned}
    $$
    This shows ``$\subset$''. 

    To show inclusion in the other direction, take any $\ket{\bar\Psi}\in \mathcal{H}\otimes
    \mathbb{C}^k$. Consider the Schmidt decomposition $\ket{\bar\Psi} =
    \sum_{i=1}^k\lambda_i \ket{\Psi_i}\otimes \ket{\tilde i}$, where
    $(\ket{\tilde i})_{i=1}^k$ is an orthonormal basis for $\mathbb{C}^k$. By
    the same calculation, we see that
    ${\bar\iota}^*(\ketbrap{\bar\Psi})=
    \sum_{i=1}^k|\lambda_i|^2\iota^*(\ketbrap{\Psi_i}$. This shows ``$\supset$''.
  \end{proof}
\end{lem}

\begin{cor}
  \label{cor:ensemble_image_purification}
  The set of ensemble state representable densities is equal to the set of pure
  state representable densities of the convexified functional theory.
\end{cor}

\cref{lem:k_convex_hull_pure_repr} is equivalent to saying that the
domains of the $k$-convexified functional and the pure functional in the
$k$-convexified functional theory coincide. We now show that the two
functionals themselves are indeed the same. In loose terms, ``$k$-convexifying commutes
with constructing the pure functional''.

\begin{propo}
  For all $k$, the $k$-convexification of the pure functional of a functional
  theory is the pure functional of the $k$-convexification of the functional
  theory. That is,   
  \begin{equation}
    \conv_k(\Fp) = \mathcal{\bar F}_p.
  \end{equation}
  \begin{proof}
    Fix a density $\rho\in
    \conv_k(\iota^*(\mathcal{P}))={\bar\iota}^*(\bar{\mathcal{P}})$.
    Unpacking the definitions, we get
    $$
    \begin{aligned}
      &\conv_k(\Fp)(\rho) 
     = \inf\Big\{
       \sum_{i=1}^k t_i
     \min \{\braket{\Gamma, W}| \iota^*\Gamma = \rho_i, \Gamma\in \mathcal{P}\}
      \mid \sum_i t_i\rho_i =\rho, \rho_i\in \iota^*(\mathcal{P}), t\in \Delta_{k-1}
     \Big\}\\
   &=\inf\Big\{
      \min \big\{\sum_{i=1}^k t_i\braket{\Gamma_i, W}| \iota^*\Gamma_i = \rho_i, \Gamma_i\in \mathcal{P}\big\}
      \mid \sum_i t_i\rho_i =\rho, \rho_i\in \iota^*(\mathcal{P}), t\in \Delta_{k-1}
     \Big\}\\
   &= \inf\Big\{
     \sum_{i=1}^k
      t_i\braket{\Psi_i|W|\Psi_i}\mid 
      \iota^*(\ketbrap{\Psi_i}) = \rho_i,
      \sum_i t_i\rho_i = \rho, \rho_i\in \iota^*(\mathcal{P}), \braket{\Psi_i|\Psi_i}=1,
      t\in \Delta_{k-1}
     \Big\}\\
   &=\inf \Big\{
     \sum_{i=1}^kt_i \braket{\Psi_i|W|\Psi_i}\mid
      \sum_i t_i \iota^*(\ketbrap{\Psi_i}) = \rho, 
      \braket{\Psi_i|\Psi_i}=1, t\in \Delta_{k-1}
   \Big\},
    \end{aligned}
    $$
    where we have introduced the standard $(k-1)$-simplex $\Delta_k:= \{t\in
    \mathbb{R}^k \mid \sum_{i=1}^k t_i=1, t_i\ge 0\}$ to make the constraints look simpler.
    Also note that the set of admissible $((\ket{\Psi_i})_{i=1}^k, t)$ in the
    last line is compact, so we may replace $\inf$ with $\min$.

    We turn our attention to the pure functional of the $k$-convexified
    functional theory. It is defined (see
    Definition~\ref{def:pure_ensemble_functional}) as a constrained minimization on
    $\mathcal{H}\otimes \mathbb{C}^k$:
    $$
    \begin{aligned}
      & \mathcal{\bar F}_p (\rho)
     = \min\Big\{
       \braket{\bar\Gamma, W\otimes \mathbbm{1}} \mid {\bar\iota}^*(\bar\Gamma) = \rho, \bar\Gamma\in \mathcal{\bar P}
       \Big\}\\
     &=
      \min\Big\{
       \braket{\bar\Psi| W\otimes\mathbbm{1}|\bar \Psi} \mid 
       {\bar\iota}^*(\ketbrap{\bar\Psi}) = \rho, \braket{\bar\Psi|\bar\Psi}=1
       \Big\}.
    \end{aligned}
    $$
    Each pure state $\ket{\bar\Psi}\in \mathcal{H}\otimes \mathbb{C}^k$ can be
    written as $\sum_i \sqrt{t_i}\ket{\Psi_i}\otimes \ket{i}$, where
    $\braket{\Psi_i|\Psi_i}=1$ and $\sum_i t_i =1$ (note that the
    $\ket{\Psi_i}$ are not required to be orthogonal). It follows that
    $$
    \mathcal{\bar F}_p(\rho)
      = \min\Big\{
        \sum_i t_i\braket{\Psi_i|W|\Psi_i} 
        \mid \sum_i t_i \iota^*(\ket{\Psi_i}\!\bra{\Psi_i}) = \rho,
        \sum_i t_i = 1, t_i\ge0, \braket{\Psi_i|\Psi_i} = 1
       \Big\}.
    $$
    Therefore $\mathcal{\bar F}_p(\rho) = \conv_k(\Fp)$.
  \end{proof}
\end{propo}

\begin{cor}
  The ensemble functional of a functional theory is the pure functional of the
  convexification of the functional theory.
\end{cor}

\chapter{Translation Invariant RDMFT for Bosons}
\label{chap:momentum_rdmft}

In Chapter~\ref{chap:generalized_ft}, we presented an abstract framework of
functional theories. Apart from the motivating examples and a few rather
artificial and simple constructions such as those presented in
Examples~\ref{ex:HK_counterexample} and \ref{ex:qubit_functional_theory}, the
focus was on extracting the essence shared by all functional theories (for
finite-dimensional systems at zero temperature) and exploring the consequences
of our abstract definitions. In particular, we did not address in depth how
functional theories can arise ``in nature''.

In this chapter, we discuss specific instances of functional theories that
occur naturally when one considers a system of interacting bosons with
translation invariance. Recall that in RDMFT, one is concerned with a family of
Hamiltonians $H(h) = h + W$, where $h$ is a single-particle operator and $W$ is
a fixed interaction. The assumption that the system respects translational
symmetry then has two implications: (1) The fact that $H(h)$ commutes with the
group of translations means that each Hamiltonian is block-diagonal
with respect to the orthogonal decomposition of the Hilbert space into
subspaces labeled by irreducible representations of the group of translations. 
(2) Since not all
single-particle operators are translation invariant, the space of allowed $h$
is significantly reduced. In the language of Chapter~\ref{chap:generalized_ft},
the space of external potentials is replaced by a subspace of much lower
dimension. Consequently, the dimension of the density space, which is the dual,
is also reduced.

Compared to \cref{chap:generalized_ft}, the present one will have
more emphasis on physical ideas and less on mathematical rigor. In
Chapter~\ref{chap:abelian_functional_theory}, we will define and discuss a
class of functional theories known as \textit{abelian functional theories}, of
which the present bosonic momentum space functional theory is a special case.
Hence, this chapter serves also as a precursor to the next one, providing
concrete examples to guide our intuition when developing the abstract theory.

\textit{The contents of this chapter are based on a forthcoming paper in
collaboration with Christian Schilling.}

\section{Motivation}
Following the experimental realization of Bose-Einstein condensation (BEC)
\cite{andersonObservationBoseEinsteinCondensation1995a,
bradleyEvidenceBoseEinsteinCondensation1995,
davisBoseEinsteinCondensationGas1995}, the field of ultracold atoms has
undergone rapid advancements
\cite{blochUltracoldQuantumGases2005,blochManybodyPhysicsUltracold2008},
providing a versatile platform for quantum simulations
\cite{blochQuantumSimulationsUltracold2012,grossQuantumSimulationsUltracold2017,schaferToolsQuantumSimulation2020a},
among other applications. 
Atoms loaded into an optical lattice generated by laser beams can simulate
many-body systems with accurately controllable parameters
\cite{jakschColdBosonicAtoms1998,jakschColdAtomHubbard2005,takahashiQuantumSimulationQuantum2022},
facilitating the study of the bosonic superfluid-Mott insulator transition
\cite{greinerQuantumPhaseTransition2002,blochSuperfluidtoMottInsulatorTransition2022}.
Therefore, it is of interest to develop a functional theoretic description of bosonic
lattice systems, which was proposed in
Ref.~\cite{benavides-riverosReducedDensityMatrix2020}. Since the particle
density alone contains insufficient information for the characterization of BEC
in terms of Penrose and Onsager's off-diagonal long-range order criterion
\cite{penroseBoseEinsteinCondensationLiquid1956}, RDMFT is naturally
advantageous over (the bosonic variant of) DFT. The theory put forward in
Ref.~\cite{benavides-riverosReducedDensityMatrix2020}, however, neither assumes
nor takes advantage of symmetries.

In the following, a functional theory (strictly speaking, a family of
functional theories) for bosonic lattice systems with translation invariance
is constructed by carrying out a two-step reduction of RDMFT based on symmetry, resulting in
a simple characterization of the domains of the pure functional and
the ensemble functional. This then allows us to derive an exact functional
form, which is in turn intimately tied to the domain's geometry. 
Unlike previous works, we establish an exact formula for the 
boundary force, a diverging repulsive force from the boundary of the
functional domain, a phenomenon previously observed in a broad range of systems
\cite{benavides-riverosReducedDensityMatrix2020,liebertFunctionalTheoryBoseEinstein2021,
schillingDivergingExchangeForce2019,maciazekRepulsivelyDivergingGradient2021},
generalizing in particular the notion of the BEC force
\cite{benavides-riverosReducedDensityMatrix2020,liebertFunctionalTheoryBoseEinstein2021}.

%

\section{Incorporating Translational Symmetry in RDMFT}
\label{sec:bosonic_momentum_symmetry}

Consider a system of $N$ spinless bosons on a one-dimensional periodic lattice
with $d$ sites. The group of translations is then the cyclic group
$\mathbb{Z}/d\mathbb{Z}= \{e, g, g^2, \dotsb, g^{d-1}\}$, where the generator $g$ acts on
$\mathcal{H}_1$ by shifting site $i$ to site $i+1$. In momentum space, we have
$g\ket{k} = e^{\frac{2\pi i}{d}\cdot k}\ket{k}$, where $\ket{0}, \dotsb,
\ket{d-1}$ denote the momentum space basis for $\Ho$. Let $\ket{m_0, \dotsb,
m_{d-1}}\propto \prod_k(b_k^\dagger)^{m_k}\ket{}$ denote the normalized
$N$-particle state with $m_k$ bosons having momentum $k$ for each $k$, where
the $b_k^\dagger$ and $b_k^{\phantom{\dagger}}$ are the creation and annihilation operators in
momentum space respectively. Such a state will be referred to as a
\textit{permanent} with \textit{occupation number vector} $m = (m_0, \dotsb,
m_{d-1})\in \mathbb{R}^d$.

In terms of the permanents, the action of the translation group on $\HN$ is
given by
\begin{equation}
  \label{eq:translation_g_eigenvalue}
  g\ket{m_0, \dotsb, m_{d-1}} = e^{\frac{2\pi i}{d} \sum_{k=0}^{d-1} k
  m_k}\ket{m_0, \dotsb, m_{d-1}}.
\end{equation}
The Hilbert space $\HN$ decomposes into the direct
sum of isotypic components 
with respect to the $\mathbb{Z}/d\mathbb{Z}$-representation:
$$
\HN = \bigoplus_{P=0}^{d-1} \HN^{(P)},
$$
where the total momentum $P$ labels the isomorphism classes of the irreducible
representations of $\mathbb{Z}/d\mathbb{Z}$. For each $P$, the subspace $\HNP$
is spanned by all permanents with total momentum $P$:
\begin{equation}
  \HN^{(P)} = \mathrm{span}_{\mathbb{C}}
  \left\{\ket{m_0, \dotsb, m_{d-1}} \;\Big\vert \;\sum_{k=0}^{d-1}km_k \equiv P \pmod d\right\}.
\end{equation}

As an example, suppose there are $N=3$ bosons on $d=3$ lattice sites. The
$N$-particle Hilbert space decomposes into $d=3$ subspaces, labeled by the total
momentum $P=0,1,2$. For $P=0$, we have $\mathcal{H}_3^{(0)} =
\mathrm{span}_{\mathbb{C}}\{\ket{3,0,0} ,\ket{0,3,0}, \ket{0,0,3}, \ket{1,1,1}\}$.

Note that the condition $\sum_k km_k\equiv P\pmod d$ is equivalent to $e^{\frac{2\pi i}{d}
\sum_kkm_k} = e^{2\pi i P /d}$, so we are simply decomposing $\HN$ into
eigenspaces of the generator $g$ of the translation group
$\mathbb{Z}/d\mathbb{Z}$. Since the Hamiltonian is block-diagonal with respect
to this decomposition, the ground state problem can be considered separately in
each symmetry subspace. Hence, we fix a specific total momentum $P\in\{0,\dotsb,
d-1\}$ and restrict our attention to the subspace $\HNP$ in the following.

The second part of the construction of the symmetry-adapted functional theory
starts with the fact that any one-particle operator $h$ which commutes with
$\mathbb{Z}/d\mathbb{Z}$ must be of the form $h = \sum_{k=0}^{d-1} v_k n_k$,
where $n_k := b_k^\dagger b_k^{\phantom{\dagger}}$ is the number operator in
momentum space. Hence, the space of allowed $h$ is parametrized by $v = (v_0,
\dotsb, v_{d-1})\in \mathbb{R}^d$. To be precise, we have a map
$$
\begin{aligned}
  \iota: \;& \mathbb{R}^d \rightarrow i\mathfrak{u}(\mathcal{H}_N^{(P)})\\
  & v\mapsto \sum_{k=0}^{d-1} v_k n_k.
\end{aligned}
$$
Note that in the RDMFT literature $v$ is usually used to denote the actual
external potential, which is diagonal in position space, and does not include
the kinetic operator. Here, however, we choose to be as consistent as possible
with the notation used in \cref{chap:generalized_ft}.

\begin{defn}
  \label{defn:bosonic_momentum_space_functional_theory}
  For particle number $N$, number of lattice sites $d$, total momentum
  $P\in\{0, \dotsb, d-1\}$, and interaction $W$, the \textbf{bosonic momentum
  space functional theory at $(d, N, P)$} is the generalized functional theory
  defined by the data $(\mathbb{R}^d, \mathcal{H}_N^{(P)}, \iota,
  \tilde W)$, where $\tilde W$ is the restriction
  of $W$ to $\HNP$.
\end{defn}

In this chapter, we will often simply write $W$ instead of $\tilde W$ for the
restriction. Also note that although the potential map $\iota$ depends on $P$, this
dependence is suppressed.

One can check by direct calculation that the density map is given by
(recall that $\braket{\Gamma, n_i} = \Tr(\Gamma n_i)$)
$$
\begin{aligned}
  &\iota^*: \mathcal{E}^{(P)}\rightarrow \mathbb{R}^d \\
  & \Gamma \mapsto \left(\braket{\Gamma, n_0}, \dotsb, \braket{\Gamma, n_{d-1}}\right),
\end{aligned}
$$
where $\mathcal{E}^{(P)}$ denotes the set of ensemble states on $\HNP$.
Similarly, $\mathcal{P}^{(P)}$ will denote the set of pure states. Let
$\FHK^{(P)}, \Fp^{(P)},$ and $\Fe^{(P)}$ be the Hohenberg-Kohn functional, pure
functional, and ensemble functional of the bosonic momentum space functional
theory respectively. Unpacking the definitions, we have
\begin{equation}
  \label{eq:symmetry_sector_functional_def}
  \FpP(m)  =
      \min_{
        \substack{
          \ket{\Psi}\in \HNP:\\
          \braket{\Psi|n_k|\Psi} = m_k \forall k
        }
      }
  \braket{\Psi|W|\Psi},
\end{equation}
and similarly for $\mathcal{F}_e^{(P)}$. The domain $\dom \FpP$, resp. $\dom
\mathcal{F}_e^{(P)}$, is then the set of all momentum occupation number vectors
$m\in \mathbb{R}^d$ for which there exists a pure, resp. ensemble, $N$-particle
state $\Gamma$ on $\HNP$ such that $\braket{\Gamma, n_k} = m_k$ for all $k$ (c.f.
\cref{def:pure_ensemble_functional}). This representability problem will be
addressed in \cref{sec:bosonic_momentum_functional_domain}.

Starting from Eq.~\eqref{eq:symmetry_sector_functional_def}, it is a
straightforward exercise to verify that the ground state energy of
$\sum_{k=0}^{d-1} v_k n_k +W$ on all of $\mathcal{H}$ is given by $\min_P
E^{(P)}(v)$, where $E^{(P)}$ is the ground state energy function of the
momentum space functional theory at $(d,N,P)$. Although all results in this
chapter are general, we will often choose a specific interaction $W$ for
definiteness. A convenient and also physically relevant choice is the
two-particle interaction of the Bose-Hubbard model, which is given by
\begin{equation}
  \label{eq:hubbard_W}
  W = \frac{1}{d}\sum_{k_1,k_2,k_3,k_4=0}^{d-1}
  \delta_{k_1+k_2, k_3+k_4} b_{k_1}^\dagger b_{k_2}^\dagger b_{k_3}^{\phantom{\dagger}} b_{k_4}^{\phantom{\dagger}},
\end{equation}
where the Kronecker delta is evaluated modulo $d$.

\section{Functional Domains}
\label{sec:bosonic_momentum_functional_domain}
The goal of this section is to determine the domains of the functionals $\FpP,
\mathcal{F}_e^{(P)}$ in each subspace $\HNP$, which amounts to characterizing
the sets $\iota^*(\mathcal{P}^{(P)})$ and $\iota^*(\mathcal{E}^{(P)})$.

Obvious necessary representability conditions on $m\in\mathbb{R}^d$ are
$\sum_k m_k = N$ and $m_k \ge 0$ for all $k\in \{0, \dotsb, d-1\}$, 
which cut out a $(d-1)$-simplex
in $\mathbb{R}^d$. This is expected: the identity $\mathbbm{1} =
N^{-1}\sum_{k=0}^{d-1} n_k$ is in the image of the potential map $\iota$, so
the codimension of $\iota^*(\mathcal{E}^{(P)})$ must be at least one by
\cref{thm:dimension_of_ens_rep}. 
These constraints are, however, not sufficient in general. As we will show in a
moment, for most choices of $(d, N, P)$, the domains $\dom \Fp^{(P)},
\dom\Fe^{(P)}$ are simplices with certain ``forbidden regions'' removed around
the vertices, characterized by an additional set of inequality constraints.
These forbidden regions strongly constrains the kinematics of the system, not
only preventing the quantum state (strictly speaking, its momentum occupation
number vector $m$) from exiting the domain, but actually exerting a
\textit{repulsive force} that drives the state away from the boundary. This
latter claim, which is at this point not at all obvious, will be made precise
and shown in \cref{sec:generalized_bec_force}, and provides a solid motivation
for investigating the geometric structure of $\dom\Fp^{(P)}$ and
$\dom\Fe^{(P)}$.

\begin{thm}
  \label{thm:domain_of_F}
  For each value of the total momentum $P\in \{0, \dotsb, d-1\}$, the domains of
  $\Fp^{(P)}$ and $\Fe^{(P)}$ coincide, and are both given by
  \begin{equation}
    \label{eq:boson_domain_convex_hull}
      \mathrm{conv}\underbrace{\left\{
      m \in \mathbb{N}_0^d \;\Big\vert\; \sum_{k=0}^{d-1} m_k = N,\; \sum_{k=0}^{d-1}km_k \equiv P \pmod d
      \right\}
        }_{=:\Omega^{(P)}},
  \end{equation}
  the convex hull of the occupation number vectors of all permanents with total
  momentum $P$.
  \begin{proof}
      For any $N$-particle state $\ket{\Psi}$ with momentum $P$, i.e.,
      $\ket{\Psi}\in \HNP$, we have $\ket{\Psi} = \sum_{\alpha=1}^{\dim \HNP}
      c_\alpha \ket{m^{(\alpha)}}$, where each $m^{(\alpha)}$ satisfies
      $\sum_{k=0}^{d-1}km^{(\alpha)}_k \equiv P \pmod d$. Hence, the momentum
      occupation number vector $m$ of $\ket{\Psi}$ is given by
      \begin{equation}
        \label{eq:n_vector_from_state_coefficients}
        \begin{aligned}
          m_k = \braket{\Psi|n_k|\Psi}
          = \sum_{\alpha,\beta=1}^{\dim\HNP} \bar c_\alpha c_\beta \braket{\nalpha|n_k|\nbeta}
          = \sum_\alpha |c_\alpha|^2 m^{(\alpha)}_k,
        \end{aligned}
      \end{equation}
      where we have used $\braket{\nalpha|n_k|\nbeta}
      =m^{(\alpha)}_k\delta_{\alpha\beta}$. Looking at
      Eq.~\eqref{eq:n_vector_from_state_coefficients}, we immediately see that
      the occupation number vector $m$ sweeps through the entirety of the
      convex hull of the $\nalpha$ as the wave function coefficients $c_\alpha$
      are varied. This shows $\iota^*(\mathcal{P}^{(P)}) = \conv(\Omega^{(P)})$.

      $\iota^*(\mathcal{E}^{(P)})$ coincides with $\iota^*(\mathcal{P}^{(P)})$
      because the former is the convex hull of the latter, which is already
      convex.
  \end{proof}
\end{thm}

\begin{figure}
  \centering
  \subfloat[$d=2,N=4,P=0$\label{fig:d2N4P0}]{
    \includegraphics[width=.31\textwidth]{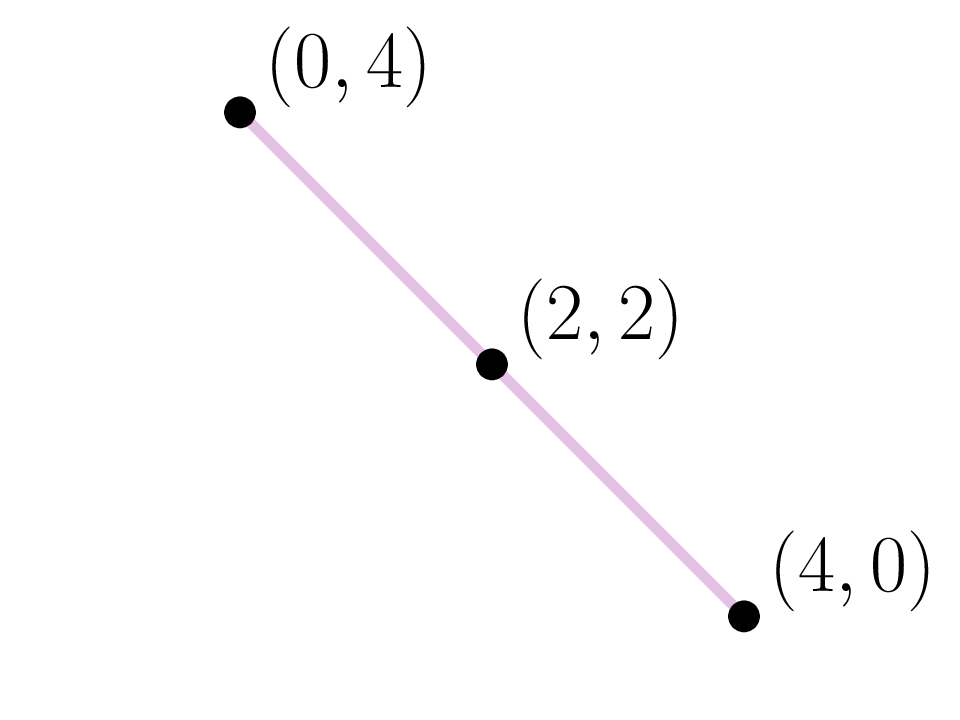}
  }
  \subfloat[$d=2,N=4,P=1$\label{fig:d2N4P1}]{
    \includegraphics[width=.31\textwidth]{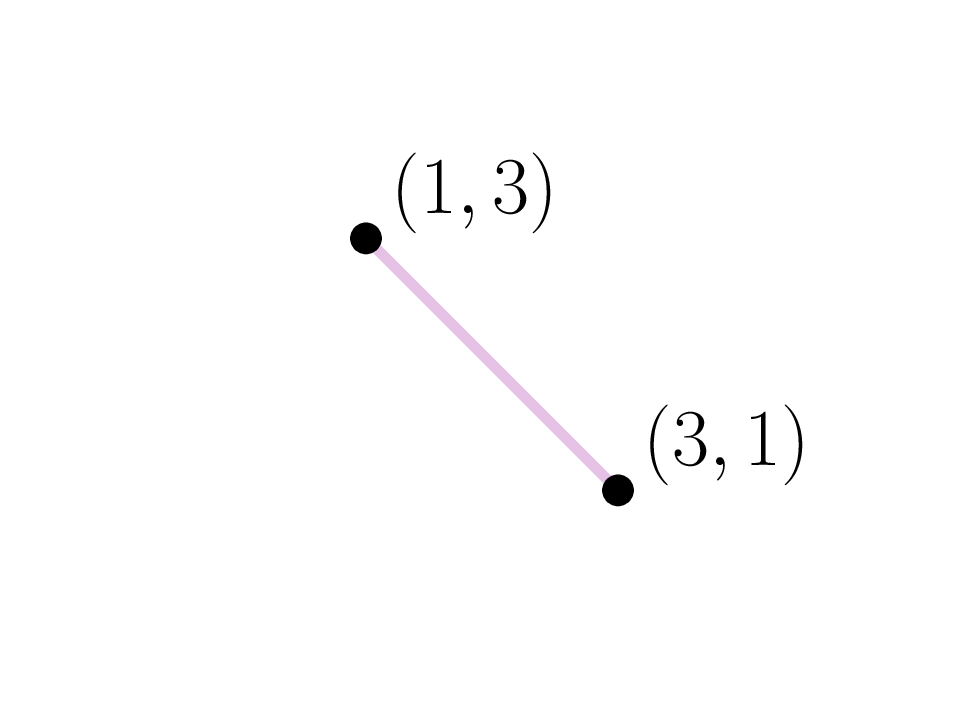}
  }

  \subfloat[$d=3,N=3,P=0$\label{fig:d3N3P0}]{
    \includegraphics[width=.31\textwidth]{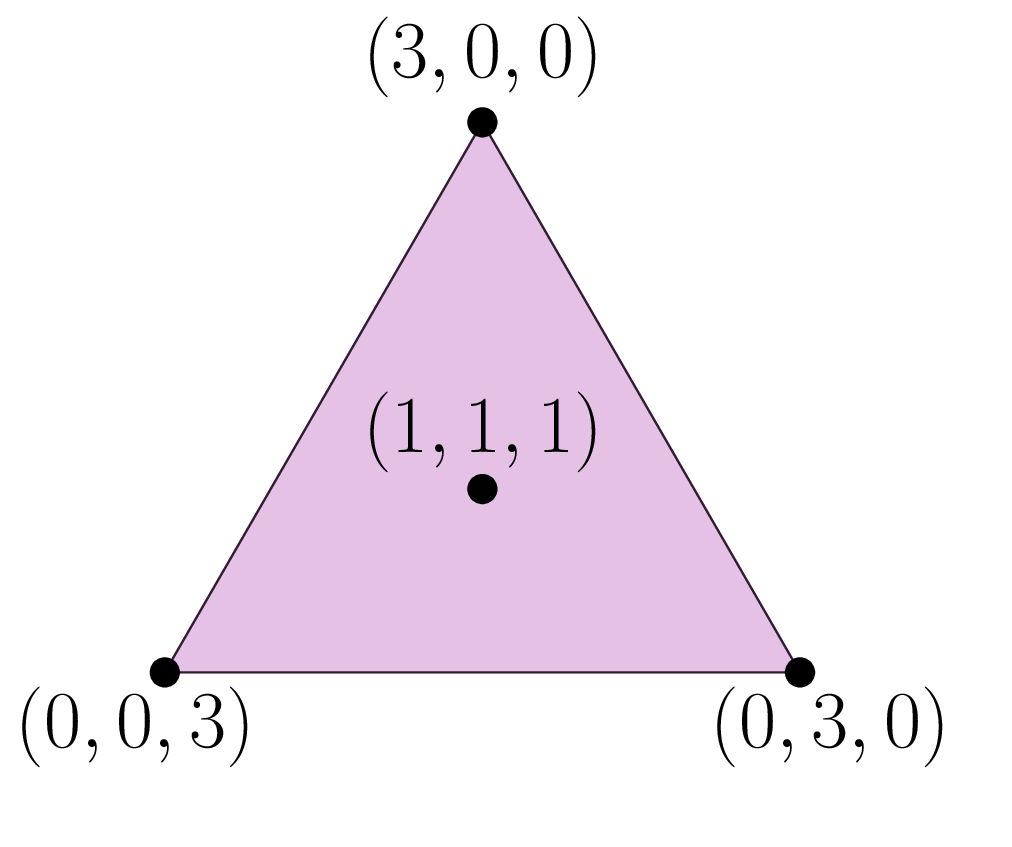}
  }
  \subfloat[$d=3,N=3,P=1$\label{fig:d3N3P1}]{
    \includegraphics[width=.31\textwidth]{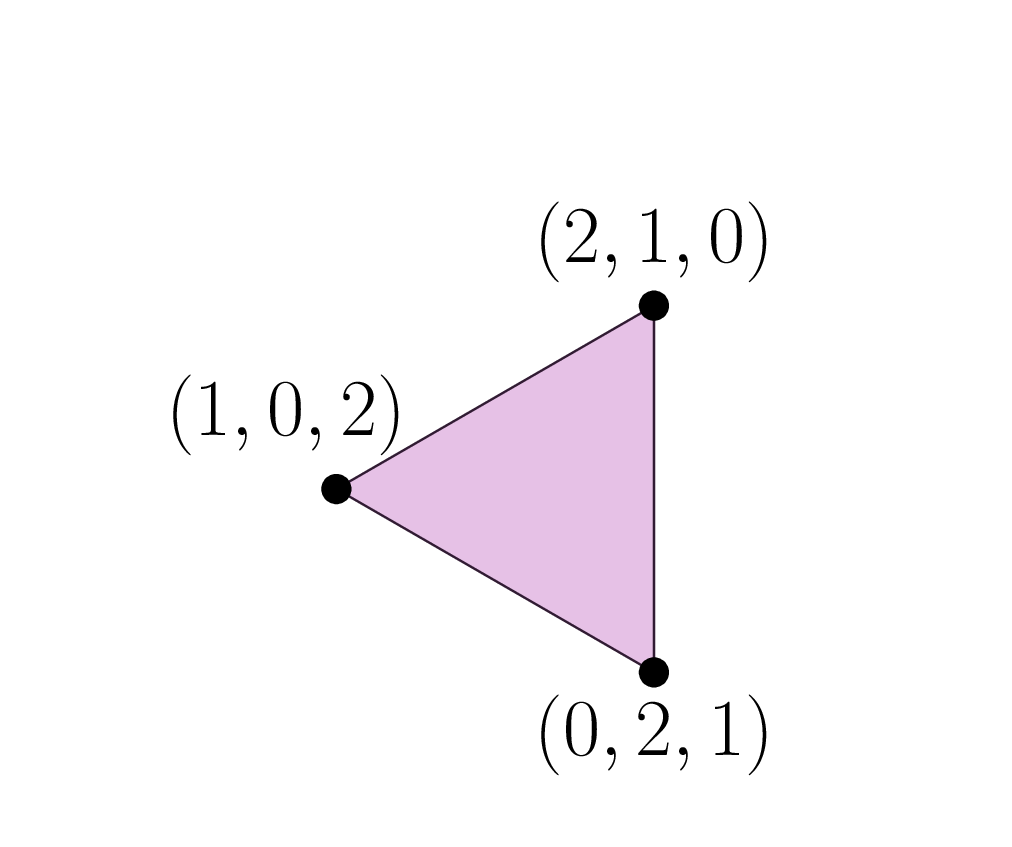}
  }

  \subfloat[$d=3,N=12,P=0$\label{fig:d3N12P0}]{
    \includegraphics[width=.36\textwidth]{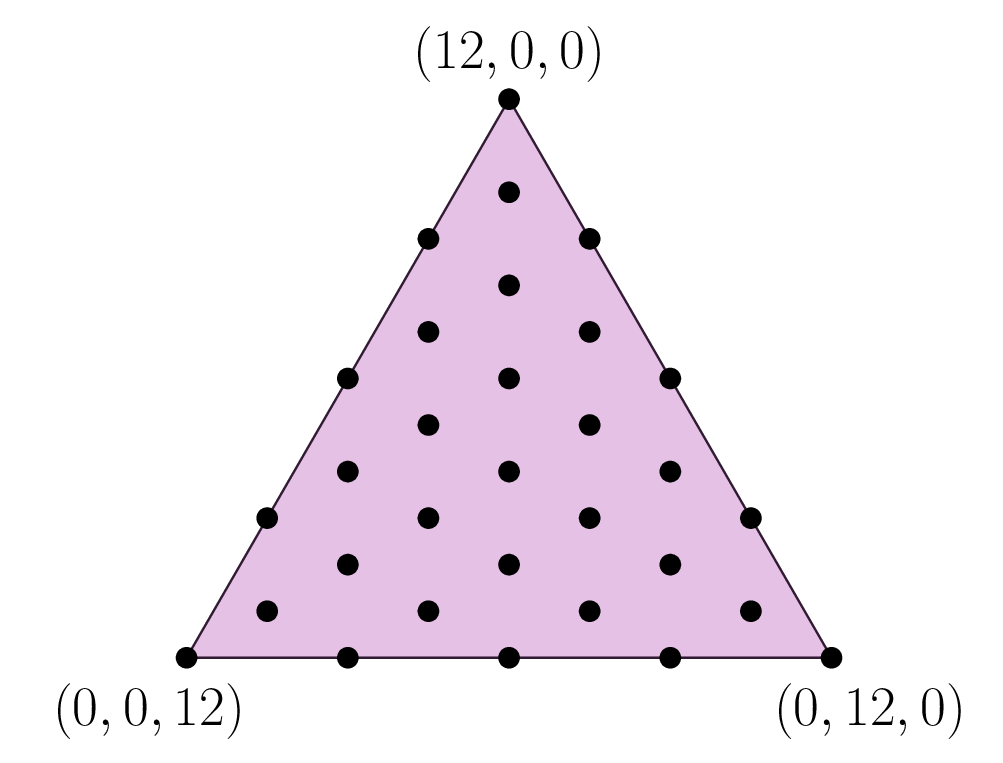}
  }
  \subfloat[$d=3,N=12,P=1$\label{fig:d3N12P1}]{
    \includegraphics[width=.36\textwidth]{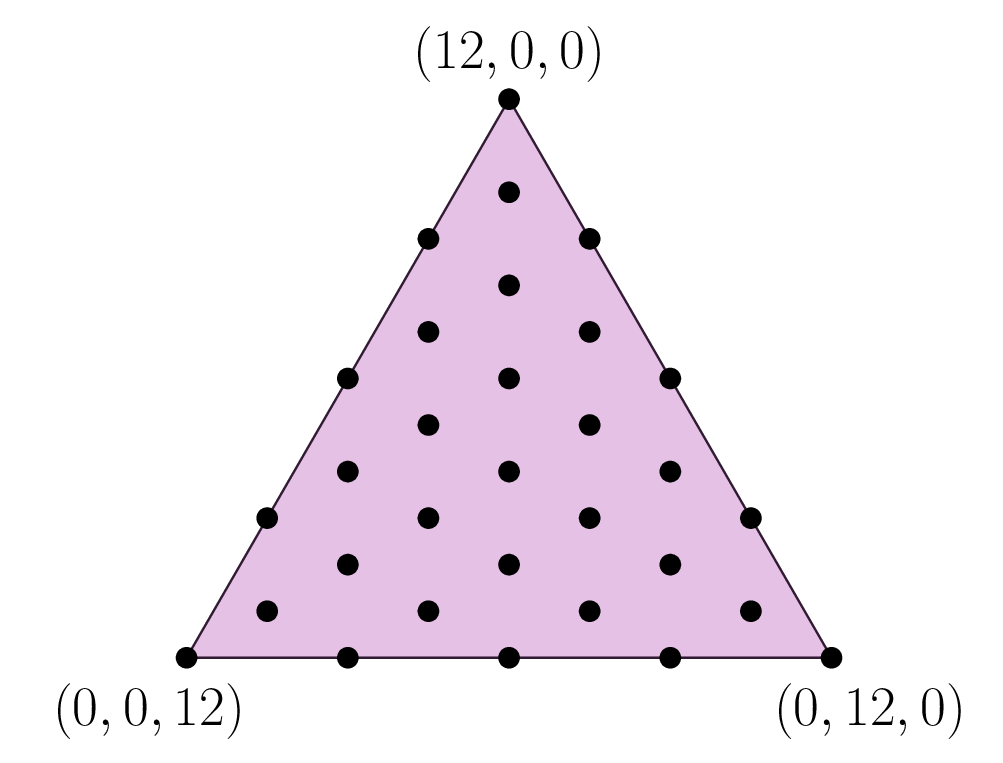}
  }

  \subfloat[$d=4,N=4,P=0$\label{fig:d4N4P0}]{
    \includegraphics[width=.36\textwidth]{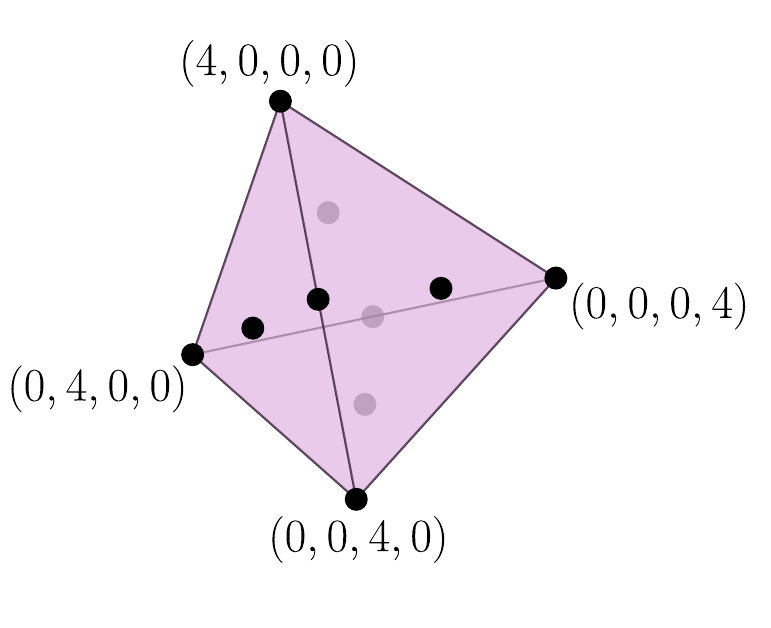}
  }
  \subfloat[$d=4,N=4,P=1$\label{fig:d4N4P1}]{
    \includegraphics[width=.36\textwidth]{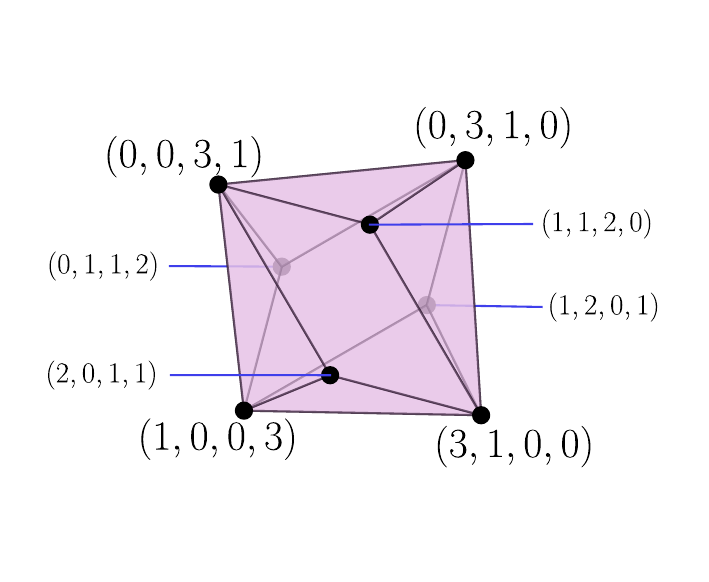}
  }

  \caption{Structure of $N$-particle momentum space permanents and domains
  of the functional.}
  \label{fig:domains_dNP}
\end{figure}

\begin{ex}
For $d=2$ lattice sites, the $N$-particle Hilbert space is spanned by the
permanents 
$$
    \ket{N,0}, \ket{N-1,1}, \dotsb, \ket{0, N}.
$$
There are
two possible values of the total momentum $P$, namely $0$ and $1$. Assuming $N$
is even, then the $P=0$ subspace is spanned by $\ket{N,0}, \ket{N-2, 2},
\dotsb, \ket{0, N}$ and the $P=1$ subspace by $\ket{N-1, 1}, \ket{N-3, 3},
\dotsb, \ket{1, N-1}$. According to Theorem~\ref{thm:domain_of_F}, we have
\begin{equation}
  \begin{aligned}
    &\mathrm{dom}(\Fe^{(0)}) = \mathrm{dom}(\Fp^{(0)}) = \{(n_0, n_1)\in \mathbb{R}^2\mid n_0+n_1= N, 0 \le n_1\le N\}\\
    &\mathrm{dom}(\Fe^{(1)}) = \mathrm{dom}(\Fp^{(1)}) = \{(n_0, n_1)\in \mathbb{R}^2\mid n_0+n_1= N, 1 \le n_1\le N-1\}.
  \end{aligned}
\end{equation}
  The domains for $N=4$ are illustrated in \cref{fig:d2N4P0} and \cref{fig:d2N4P1}.
\end{ex}

For general $(d, N, P)$, the domain will be more complicated. However, we can
make several observations:
\begin{enumerate}
  \item Whenever $d$ divides $N$, the states $\ket{N,0,\dotsb, 0},
    \ket{0,N,0,\dotsb, 0}, \dotsb$ all belong to the zero-momentum subspace. Therefore,
    $\dom\Fe^{(0)} = \dom \Fp^{(0)}$ is a simplex 
    (see \cref{fig:d2N4P0,fig:d3N3P0,fig:d3N12P0,fig:d4N4P0}).

  \item In the limit $N\rightarrow \infty$ at fixed $d$ and $P$, each vertex
    of the simplex $\{m\in \mathbb{R}^d \mid \sum_k m_k = N, m_k \ge 0\}$
    becomes closer (relative to $N$) to the nearest occupation number vector
    $m^{(\alpha)}$ with total momentum $P$, so the domain approaches a simplex 
    (see \cref{fig:d3N12P1}).
\end{enumerate}

\cref{fig:d2N4P0,fig:d3N3P0,fig:d3N12P0,fig:d3N12P1,fig:d4N4P0}
also show a distinctive feature of the bosonic setting in contrast to fermions
(see Ref.~\cite{schillingDivergingExchangeForce2019}): in general, there are
occupation number vectors that are not extreme points. As we will see in
Section~\ref{sec:generalized_bec_force}, both the shape of the domain and the
distribution of the occupation number vectors of the permanents therein will
influence the behavior of the function near the boundary.

\section{Form of the Universal Functional}
\label{sec:bosonic_momentum_functional_form}
In usual RDMFT without incorporating symmetries, as outlined
in \cref{chap:intro}, various approximate forms of the
universal functional have been proposed in the literature, mostly for
electronic systems
\cite{mullerExplicitApproximateRelation1984,marquesEmpiricalFunctionalsReduceddensitymatrixfunctional2008,pirisNaturalOrbitalFunctional2011,pirisGlobalNaturalOrbital2021,schmidtMachineLearningUniversal2021,sharmaReducedDensityMatrix2008}.
In isolated cases, exact analytical functionals of particular model systems
have been derived and studied
\cite{cohenLandscapeExactEnergy2016,benavides-riverosReducedDensityMatrix2020}.
However, the presence of symmetries is rarely exploited in the way it is in
\cref{sec:bosonic_momentum_symmetry,sec:bosonic_momentum_functional_domain}.
These simplifications due to translational symmetry were explored in
Ref.~\cite{schillingDivergingExchangeForce2019}, where the pure functional
$\Fp$ was found to take a form that heavily depends on the geometry of the
domain. The goal of this section is to adapt the results of
Ref.~\cite{schillingDivergingExchangeForce2019} to bosons. 
In light of the purification trick (see \cref{sec:purification_trick}), we will
focus on the pure functional from now on, and the superscript on $\Fp^{(P)}$
will be dropped, with the understanding that $\Fp$ always refers to the pure
state functional of the bosonic momentum space functional theory at some fixed $(d, N,
P)$.


As is apparent in \cref{sec:bosonic_momentum_functional_domain}, the domain of
$\Fp$ is a subset of the hyperplane $\{m \in \mathbb{R}^n \mid \sum_k m_k
= N\}$ bounded by a list of $J = J(d, N, P)$ affine constraints 
\begin{equation}
	\label{eq:Dj_constraints}
	D^{(j)}(m) = \braket{m, S^{(j)}} - \nu^{(j)} \ge 0,
\end{equation}
where $S^{(j)} \in \mathbb{R}^d$, $\nu^{(j)}\in \mathbb{R}$, and  $j=1, \dotsb,
J$ labels the facets. Note that there is some ambiguity in the choice of
$S^{(j)}$ and $\nu^{(j)}$. For instance, scaling them or replacing them with 
$$
\begin{aligned}
	&S^{(j)}\rightarrow  S^{(j)} + (\mu,\dotsb, \mu)\\
	&\nu^{(j)}\rightarrow \nu^{(j)} + \mu N
\end{aligned}
$$ 
for any $\mu$ results in an equivalent constraint. This ambiguity, however,
will not affect any physical result in this section.

\begin{ex}
	Take $(d,N,P) = (3,3,0)$ as an example. As one can see in \cref{fig:d3N3P0},
	the domain is $\{m \in \mathbb{R}^3 \mid m_k\ge0,
m_1+m_2+m_3= 3\}$, whence 
\begin{equation}
	\label{eq:d3N3P0_constraints}
	\begin{aligned}
		&D^{(1)}(m) = m_1\\
		&D^{(2)}(m) = m_2\\
		&D^{(3)}(m) = m_3.
	\end{aligned}
\end{equation}
	That is, we can take $S^{(j)}$ to be the $j$-th standard basis vector $e_j$
	and $\nu^{(j)}=0$.
\end{ex}

The idea is to express $\Fp(m)$ in terms of $D^{(j)}(m), j=1, \dotsb, J$, with
$D^{(j)}(m)$ having the geometric interpretation of being the distance from $m$
to the $j$-th facet (up to a multiplicative factor). 
We proceed in two steps:
first, we treat a simplified scenario, the so-called ``simplex setting'' (see
below for precise definition). Although a general combination of $(d, N, P)$
will not satisfy the simplex assumption, this idealized setting will allow us
to derive the functional in a straightforward way, illustrating the essential
features without encountering technical complications. Next, in
\cref{sec:F_beyond_simplex}, armed with the intuition obtained in the simplex
setting, we will drop this assumption and derive a functional form in full
generality.

\subsection{The Simplex}
\label{sec:F_simplex}
In this section, we will assume that the
domain is a simplex with exactly $\dim \HN^{(P)}$ vertices, meaning in
particular that the momentum occupation number vector $\nalpha$ of each
permanent $\ket{\nalpha}$ has to be a vertex. This is an extremely
strong condition to impose on $(d, N, P)$, and only specific combinations of
low dimension and low particle number will result in such a scenario. For example, 
\cref{fig:d2N4P1,fig:d3N3P1} satisfy this assumption, while
the remaining do not. Note that although the functional domains in 
\cref{fig:d2N4P0,fig:d3N3P0,fig:d3N12P0,fig:d4N4P0}
are simplices geometrically, they are excluded from the present section because
of the extra $\nalpha$ vectors which are not extreme points. 
Hereafter, \textit{the simplex assumption} or \textit{simplex
setting} will always refer to this stronger condition than merely that the
domain be a simplex.

Let $(\nalpha)_{\alpha=1}^{\dim\HNP}$ be the list of vertices. The Hilbert
space $\HNP$ is then spanned by the permanents $\ket{m^{(1)}}, \ket{m^{(2)}},
\dotsb$. Since there is a one-to-one correspondence between vertices and
facets, the number $J$ of facets is equal to $\dim\HNP$, 
and we will use Greek
letters $\alpha,\beta$ to label the constraints instead of $j$, as we did in
Eq.~\eqref{eq:Dj_constraints}. To be more precise, we label the constraints so
that $D^{(\alpha)}=0$ corresponds to the facet opposite the vertex $\nalpha$.

Suppose we are interested in the value
of $\Fp$ at some $m\in \dom \Fp$. 
If $\ket{\Psi}=\sum_\alpha c_\alpha\ket{\nalpha}$ 
is a state such that $\iota^*(\ketbrap{\Psi}) = m$, 
i.e., $\braket{\Psi|n_k|\Psi} = m_k$ for all $k$, 
then 
$$
	\sum_\alpha |c_\alpha|^2 \nalpha = m 
$$
(see Eq.~\eqref{eq:n_vector_from_state_coefficients}). Applying $D^{(\beta)}$
to both sides, we get $\sum_{\alpha} |c_\alpha|^2
D^{(\beta)}(\nalpha) = D^{(\beta)}(m)$. But $D^{(\beta)}(\nalpha) = \VFactor{\alpha} \delta_{\alpha\beta}$, 
where $\VFactor{\alpha} := D^{(\alpha)} (\nalpha)$, 
because every vertex except $\nbeta$ lies on the hyperplane $D^{(\beta)} = 0$. 
It follows that
\begin{equation}
	\label{eq:simplex_coeff_D_relation}
	|c_\beta|^2 = \frac{D^{(\beta)}(m)}{\VFactor{\alpha}},
\end{equation}
which has a geometric interpretation as a distance ratio. Therefore
\begin{equation}
	\label{eq:F_simplex}
	\Fp(m) = \sum_{\alpha,\beta=1}^{\dim \HN^{(P)}} 
	W_{\alpha\beta}\bar\xi_\alpha \xi_\beta
	\sqrt{\frac{D^{(\alpha)}(m)}{\VFactor{\alpha}}}\sqrt{\frac{D^{(\beta)}(m)}{\VFactor{\beta}}},
\end{equation}
where $W_{\alpha\beta}:=\braket{\nalpha|W|\nbeta}$ and the $\xi_\alpha$
are phases chosen so that Eq.~\eqref{eq:F_simplex} is minimized.

\begin{ex}
	Consider $N=3$ bosons on $d=3$ lattice sites with total momentum
	$P=1$ (\cref{fig:d3N3P1}). 
	The spanning permanents are 
	$$
		\ket{m^{(1)}} = \ket{2,1,0}, \ket{m^{(2)}}=\ket{0,2,1}, \ket{m^{(3)}} = \ket{1,0,2}.
	$$ 
	The domain of the functional is given by the three inequality constraints
	\begin{equation}
  	\begin{aligned}
  	&D^{(1)}(m) = m_0 - m_2 + 1 \ge 0 \\ 
  	&D^{(2)}(m) = m_1 - m_0 + 1 \ge 0 \\ 
  	&D^{(3)}(m) = m_2 - m_1 + 1 \ge 0,
  	\end{aligned}
	\end{equation}
	together with the equality constraint $m_0+m_1+m_2=3$.
	We have $L_1=L_2=L_3 = 3$, so Eq.~\eqref{eq:F_simplex} becomes
\begin{equation}
  \begin{aligned}
    &\Fp(m) = \\
    &U_0+\frac{2}{3}U_1\Re\left[
    	 \bar\xi_1\xi_2 \sqrt{D^{(1)}(m) D^{(2)}(m)}
     + \bar\xi_2\xi_3 \sqrt{D^{(2)}(m) D^{(3)}(m)}
     + \bar\xi_3\xi_1 \sqrt{D^{(3)}(m) D^{(1)}(m)} 
     \right],
  \end{aligned}
\end{equation}
where we have assumed $W_{11}=W_{22}=W_{33}=U_0$ and $W_{\alpha\neq \beta} =
U_1$, which is the case for the Hubbard model. If $U_1$ is positive,
minimization over the phases $\xi_\alpha$ is not trivial but still
relatively simple. A straightforward calculation shows that $\Fp(m)= U_0 -
U_1$ if 
$\sqrt{D^{(1)}(m)}, \sqrt{D^{(2)}(m)}, \sqrt{D^{(3)}(m)}$
satisfy the triangle inequality (that is, $\sqrt{D^{(1)}(m)}\le
\sqrt{D^{(2)}(m)} + \sqrt{D^{(3)}(m)}$ plus cyclic permutations),
otherwise 
$$
	\Fp(m) = U_0 + \frac{2}{3}U_1\left(\sqrt{D^{(1)}(m)D^{(2)}(m)}-\sqrt{D^{(2)}(m)D^{(3)}(m)}
	- \sqrt{D^{(3)}(m)D^{(1)}(m)}\right) 
$$
	if $D^{(3)}(m)\ge D^{(1)}(m), D^{(2)}(m)$ (and similarly for the other two possibilities).
\end{ex}

\subsection{Beyond Simplex}
\label{sec:F_beyond_simplex}
Outside of the simplex setting, that is, whenever the domain is not a simplex
with $\dim \HNP$ vertices, it is no longer possible to recover
$(|c_\alpha|^2)_{\alpha=1}^{\dim\HNP}$ uniquely from $(D^{(j)}(m))_{j=1}^J$ as
we did in Eq.~\eqref{eq:simplex_coeff_D_relation}.
Instead, the collection of numbers
$D^{(j)}(m)$ depend linearly on the squared moduli of the wave function
coefficients $|c_\alpha|^2$, with the associated linear map having a nontrivial
kernel. Hence, the procedure to derive an equation analogous to
Eq.~\eqref{eq:F_simplex} involves partially inverting said linear map, a method
elaborated in the following.

Fix some $m\in \dom \Fp$, and suppose $\ket{\Psi}=\sum_{\alpha}
c_\alpha\ket{\nalpha}$ is a state in $\HNP$ with momentum occupation number
vector $m$. Then, by Eq.~\eqref{eq:n_vector_from_state_coefficients},
\begin{equation}
  \label{eq:D_from_c_temp}
  \begin{aligned}
    &D^{(j)}(m) := 
    \braket{m, S^{(j)}}- \nu^{(j)}
    =\braket{\Psi|\sum_{k=0}^{d-1} S^{(j)}_k n_k - \nu^{(j)}\mathbbm{1}|\Psi}\\
    &=\sum_{\alpha} |c_\alpha|^2 \left(\braket{S^{(j)}, \nalpha}- \nu^{(j)}\right)\\
    &=\sum_{\alpha} |c_\alpha|^2 D^{(j)} (\nalpha).
  \end{aligned}
\end{equation}
Eq.~\eqref{eq:D_from_c_temp} can be written more compactly as 
\begin{equation}
  \label{eq:D_from_y_temp}
			D(m) = Ty,
\end{equation}
where $D(m) := (D^{(1)}(m), \dotsb, D^{(J)}(m))$,
$T$ is the $J\times \dim \HNP$ matrix defined by $T_{j\alpha}:= 
D^{(j)}(\nalpha)$, and $y\in \mathbb{R}^{\dim\HNP}$ is the vector with components $y_\alpha
:=|c_\alpha|^2$. The situation is as follows: we know $m$, and hence
$D(m)$ (remember the components of $D(m)$ are, up to scaling, the
distances from $m$ to the $J$ facets), and would like to find all $y$ that
solve Eq.~\eqref{eq:D_from_y_temp} (and have nonnegative entries).

Recall that if $A=U\Sigma V^\top$ is the singular value decomposition of a real
matrix $A$, the \textit{Moore-Penrose inverse} of $A$ is defined by $A^+ = V
\Sigma^+ U^\top$, where $\Sigma^+$ is obtained from taking the transpose of
$\Sigma$ and inverting its nonzero entries. The Moore-Penrose inverse satisfies
$AA^+ A = A$ and $A^+AA^+ = A^+$.

Let $T^+$ denote the Moore-Penrose inverse of $T$. Then
Eq.~\eqref{eq:D_from_y_temp} is solved by 
\begin{equation}
  \label{eq:y_from_D_by_pseudoinverse}
		y = T^+ D(m) + x,
\end{equation}
 where $x\in\mathbb{R}^{\dim\HNP}$ is any vector such that 
 (1) $Tx=0$ and (2) the resulting $y$ has nonnegative components, with condition (2) depending on
	$m$. 
	To see that Eq.~\eqref{eq:y_from_D_by_pseudoinverse} really
	solves Eq.~\eqref{eq:D_from_y_temp}, let $T$ act on both sides of
	Eq.~\eqref{eq:y_from_D_by_pseudoinverse} to get $Ty = TT^+ D(m)$. 
	By pure state representability of $m$, there exists a state $\ket{\Psi'} =
	\sum_\alpha c_\alpha'\ket{\nalpha}$ such that $\iota^*(\ketbrap{\Psi'}) = m$,
	implying $D(m) = Ty'$, where $y'_\alpha := |c_\alpha'|^2$. So we have
	$Ty = TT^+ T y' = Ty' = D(m)$.

Therefore
\begin{equation}
  \ket{\Psi} = \sum_{\alpha=1}^{\dim \HNP} \xi_\alpha \sqrt{
  	(T^+D(m))_\alpha + x_\alpha}
  \ket{\nalpha},
\end{equation}
where the $\xi_\alpha$ are phases, which at this point are arbitrary. As a
result,
	\begin{equation}
		\label{eq:general_F_form}
    \Fp(m) = \sum_{\alpha,\beta=1}^{\dim\HNP} W_{\alpha\beta}\bar\xi_\alpha \xi_\beta
      \sqrt{(T^+D(m))_\alpha + x_\alpha \vphantom{(T^+D(m))_\beta + x_\beta}}
      \sqrt{(T^+D(m))_\beta + x_\beta}.
	\end{equation}
For each $m\in \dom\Fp$, the phases $\xi_\alpha$ and the vector $x$ are to be
chosen so that Eq.~\eqref{eq:general_F_form} is minimized. If, in addition to
translation invariance, we require that the total Hamiltonian $H(v)$ also be
invariant under combined action of lattice inversion and time-reversal, then we
may take the phases $\xi_\alpha$ to be real-valued, which would lead to a
different but equivalent functional $\mathcal{\tilde F}_p$ (in the sense of yielding the same ground
state energy, see
Ref.~\cite{liebertRefiningRelatingFundamentals2023} for more detail). One
approximation strategy for $\mathcal{\tilde F}_p$ would then be to divide the
domain into disjoint regions, and fixing a sign pattern
$(\xi_\alpha)_\alpha$ for all $m$ in each region. We do not pursue this further
here.

Eq.~\eqref{eq:general_F_form} reveals a remarkable relation between the
functional $\Fp$ and its domain. We see that the ``weight'' of each
permanent $\ket{\nalpha}$ is controlled by $D(m)$, 
which quantifies the distances from $m$ to each facet. Moreover, all quantities
in the two square roots in Eq.~\eqref{eq:general_F_form} are purely geometric:
they have nothing to do with the interaction $W$ (of course, the minimization over
$x$ will, in general, depend on $W$).
To see how Eq.~\eqref{eq:general_F_form} reduces to Eq.~\eqref{eq:F_simplex} in
the simplex setting (remember that by ``simplex setting'' we really mean
\textit{two} conditions: that the domain is a simplex and that $\dim \HNP$ is
equal to the number of vertices), observe that $T_{\beta\alpha} =
L_\alpha\delta_{\beta\alpha}$ by the definition of $L_\alpha$, and that the
kernel of $T$ is trivial. Therefore, the extra degrees of freedom parametrized by $x$
disappear from Eq.~\eqref{eq:general_F_form}, and $T^+_{\beta\alpha}$ is simply
$L_\beta^{-1}\delta_{\beta\alpha}$. Consequently, we recover Eq.~\eqref{eq:F_simplex}.

\begin{ex}
Take $d=2, N=4, P=0$. The domain of $\Fp$ is given by
	the convex hull of $m^{(1)} = (4,0)$, $m^{(2)} = (2,2)$, $m^{(3)} = (0,4)$ (see Fig.~\ref{fig:d2N4P0}). 
	There are two facets, which
	in this case are simply the vertices $m^{(1)}$ and $m^{(3)}$ since
	the domain is one-dimensional, leading to two respective constraints
	$D^{(1)}(m) = m_0$ and $D^{(2)}(m) = m_1$. Thus 
	\begin{equation}
		T = (D^{(j)}(\nalpha))_{j\alpha} = 
		\begin{pmatrix}
			4 & 2 & 0\\
			0 & 2 & 4
		\end{pmatrix},
	\end{equation}
	the Moore-Penrose inverse of which is
	\begin{equation}
			T^+ = \frac{1}{24}\begin{pmatrix}
		5 & -1\\
		2 & 2\\
			-1 & 5
		\end{pmatrix}.
	\end{equation}
The kernel of $T$ is spanned by the vector $(-1,2,-1)$, so
	$x = q(-1,2,-1)$ with $q \in \mathbb{R}$. It follows that the radicands in
Eq.~\eqref{eq:general_F_form} are 
\begin{equation}
	\begin{aligned}
		&T^+D(m) + x = \frac{1}{24}
		\begin{pmatrix}
		5 & -1\\
		2 & 2\\
			-1 & 5
		\end{pmatrix}
		\begin{pmatrix}
			m_0 \\ m_1
		\end{pmatrix} + q\begin{pmatrix}-1\\2\\-1\end{pmatrix}\\
		 	&= \frac{1}{24}\begin{pmatrix}
					5m_0 - m_1\\ 2m_0 + 2m_1 \\ -m_0 + 5m_1
		 \end{pmatrix}
			+ q \begin{pmatrix}-1\\2\\-1\end{pmatrix}
			=
\begin{pmatrix}
	\frac{1}{3} + \frac{1}{8}(m_0-m_1) - q\\ \frac{1}{3}+2 q\\\frac{1}{3} - \frac{1}{8}(m_0-m_1) - q
		 \end{pmatrix},
		\end{aligned}
\end{equation}
	where we have used $m_0+m_1= 4$.
	If we assume $W_{13}=0$, $W_{11}=W_{33}$, and $W_{12}=W_{23} =: w$, which
	is true for the Hubbard model (Eq.~\eqref{eq:hubbard_W}), and set
	$W_{11}=W_{33}=0$ without loss of generality (we can always do this by
	shifting the energy by a constant), then a choice of minimizing phases is
	$(q_1, q_2, q_3) = (1,-1,1)$ and Eq.~\eqref{eq:general_F_form} gives
	\begin{equation}
		\begin{aligned}
			\Fp&(m) =  W_{22}\left(\frac{1}{3}+2q\right) -2|w| \sqrt{\frac{1}{3}+2q}
				\left[\sqrt{\frac{1}{3} + \frac{1}{8}(m_0-m_1) - q} +
		\sqrt{\frac{1}{3} - \frac{1}{8}(m_0-m_1) - q}
		\right],
		\end{aligned}
	\end{equation}
	minimized over $q$.
	\end{ex}

	We would like to point out that Eq.~\eqref{eq:general_F_form} is an
	\textit{exact} expression of the pure functional $\Fp$.
	Of course, expressing $\Fp$ in 
	this form
does not trivialize in any way the task of computing $\Fp$: for
a general $W$, there is no simple way of determining $x$ and the phases
$\xi_\alpha$ (which will depend on $m$), and for interesting systems
where $\dim \HNP \gg d$, the nullity of the matrix $T$, which is 
the dimension of the space of candidates for $x$, will be extremely large.
Nonetheless, Eq.~\eqref{eq:general_F_form} establishes an important link
between $\Fp$ and the geometry of its domain, and hints
at a potential approximation scheme for the functional through a suitable estimate of $x(m)$. 

\section{Generalized BEC Force}
\label{sec:generalized_bec_force}
In this section, we carry out a quantitative investigation of the behavior of
the functional close to the boundary of its domain.

\begin{figure}
  \centering
  \includegraphics[width=.95\textwidth]{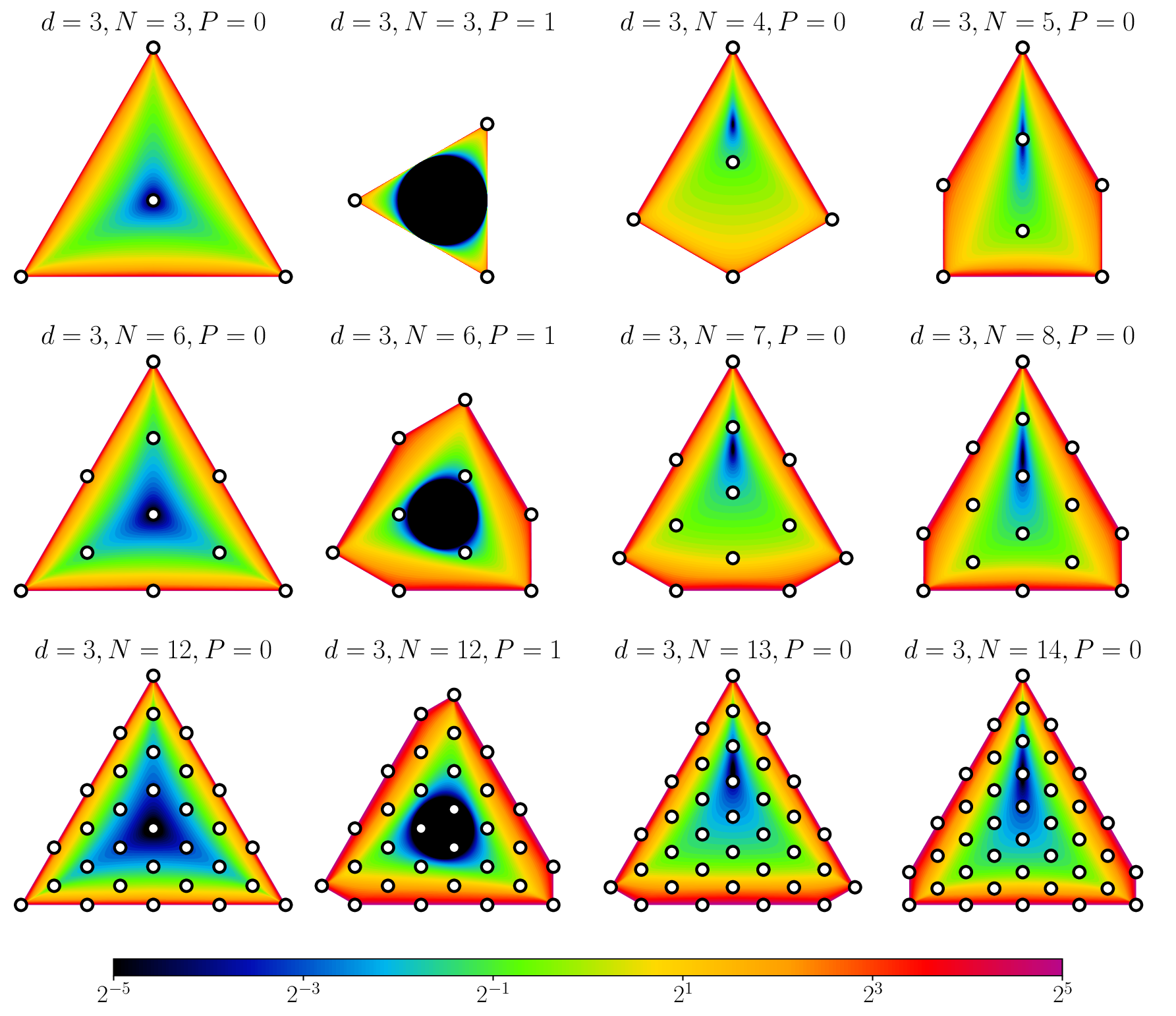}
  \caption{Plots of the magnitude of the derivative of the functional for various
  combinations of $(N, P)$ with $d=3$. Whenever $N$ is not a multiple of $3$,
  all momenta are equivalent, so only the case $P=0$ is presented. If $N$ is a
  multiple of $3$, only $P=1$ and $P=2$ are equivalent, thus we present plots
  for $P=0$ and $P=1$.
  For $N=3,6,12$, the functional is affine on the black disc in the center of
  its domain, which is a degeneracy region (see Refs.~\cite{penzGeometryDegeneracyPotential2023,penzGeometricalPerspectiveSpin2024}
  for more detailed discussions).
  }
  \label{fig:gradient_F_plots}
\end{figure}

Observe that in Eq.~\eqref{eq:F_simplex} the functional takes the approximate
form $\Fp(m)\sim F_0 - \mathcal{G} \sqrt{D^{(\omega)}(m)}$ whenever $D^{(\omega)}(m)$
approaches zero, where $F_0$ and $\mathcal{G}$ are constants.
Consequently, the derivative $\dstraight_m\Fp$ diverges as 
$1/\sqrt{D^{(\omega)}(m)}$ as $D^{(\omega)}(m)\rightarrow 0$. The same phenomenon is demonstrated in
Fig.~\ref{fig:gradient_F_plots}, where we plot the quantity 
$\lVert\dstraight_m\Fp\rVert$ for the Hubbard model with $d=3$ (so that the domain is
$2$-dimensional) and several combinations of $(N,P)$. 
%
One may infer by inspecting the plots in Fig.~\ref{fig:gradient_F_plots} that
$\dstraight\Fp$ always diverges near the boundary in all cases. 


This turns out to be a commonly observed feature in RDMFT: whenever the
variable playing the role of a ``density'' approaches the boundary of the
domain of the functional, the derivative diverges as one over the square root
of the distance to the boundary \cite{schillingDivergingExchangeForce2019,
benavides-riverosReducedDensityMatrix2020,
liebertFunctionalTheoryBoseEinstein2021,maciazekRepulsivelyDivergingGradient2021}.
In fermionic systems \cite{schillingDivergingExchangeForce2019}, this has been
dubbed the ``exchange force'', while for bosons analogous behavior of the
functional near the BEC vertex $(N,0,\dotsb)$ is called the ``BEC force''
\cite{liebertFunctionalTheoryBoseEinstein2021,maciazekRepulsivelyDivergingGradient2021}.

As we will demonstrate in this section, this force exists not only in the
vicinity of the BEC vertex, but more generally near any facet of $\dom(\Fp)$,
which is why we will refer to it as the \textit{generalized BEC force}, or
simply \textit{boundary force} when we move to the more general setting of
abelian functional theories in \cref{chap:abelian_functional_theory}.
It is
exactly the presence of such a force that prohibits the so-called
\textit{pinning}, which refers to situations where the 1RDM (or other
generalized notions of density) of the ground state lies precisely on the
boundary of the set of pure $N$-representable 1RDMs
\cite{maciazekImplicationsPinnedOccupation2020,schillingImplicationsPinnedOccupation2020,schillingHubbardModelPinning2015},
i.e., the domain of $\Fp$, and is related to the question of whether natural
occupation numbers of the ground state can vanish
\cite{cioslowskiNaturalOrbitalsTheir2024}.

The existence of this repulsive force in various fermionic and bosonic systems
has been quantitatively studied in
Ref.~\cite{maciazekRepulsivelyDivergingGradient2021} by constructing a family of
suitably approximated wave functions, giving an upper bound of the pure
functional $\Fp$, which then implies a divergence of the derivative near the
boundary. The constructed wave functions also serve as a means to obtain a
functional approximation. Here, in contrast, we will provide an \textit{exact}
formula for the ``repulsion strength'', a quantity which will be defined below, of the
diverging generalized BEC force. Such a formula will not only provide a
criterion for the presence or absence of pinning, but will also contribute an
exact relation the functional of any translation invariant bosonic system has
to satisfy, independently of the interaction $W$. Moreover, it will be evident
from our derivation that repulsive forces from the boundary are a common
feature in general functional theories, and we will keep coming back to this phenomenon
throughout this thesis as we explore functional theory classes of different
levels of generality.

Before proceeding, a brief note on convention: recall that in determining the
facet-defining inequalities, there is an ambiguity in $S^{(j)}\in \mathbb{R}^d$
and $\nu^{(j)}\in\mathbb{R}$, as we remarked shortly after
Eq.~\eqref{eq:Dj_constraints} in
Section~\ref{sec:bosonic_momentum_functional_form}. Throughout the rest of this
section, we will fix the normalization of the constraint coefficients $S^{(j)},
\nu^{(j)}$ by requiring that $S^{(j)}$ be tangent to $\dom\Fp$ and
that $\lVert S^{(j)}\rVert$ be unity. 
If $\iota$ is injective, which is true for most $(d,N,P)$, then $\dom\Fp$ has
codimension one in $\mathbb{R}^d$, so the former condition is equivalent to
imposing $\sum_\alpha S^{(j)}_\alpha = 0$. 
For example, the first constraint in Eq.~\eqref{eq:d3N3P0_constraints} for
$(d,N,P)=(3,3,0)$ would become $D^{(1)}(m) = (2m_1 - m_2 - m_3 +
3)/\sqrt{6}$, for which $S^{(1)} = \frac{1}{\sqrt{6}}(2,-1,-1)$ and
$\nu^{(1)} = -\frac{3}{\sqrt{6}}$. This is merely a convenient choice that will
simplify the formulas that follow.

\subsection{The Simplex}
\label{sec:BECforce_simlex}

As in Section~\ref{sec:F_simplex}, a clearer understanding can
be gained by first analyzing the problem under the assumption that $\dom(\Fp)$
is a simplex with $\dim \HNP$ vertices. Although a heuristic argument for a
diverging repulsive force in the simplex setting is given at the beginning of
this section, here we will derive a formula for the repulsion strength
(defined in Eq.~\eqref{eq:repulsion_strength_def}).
Recall (see Section~\ref{sec:F_simplex})
that we have occupation number vectors $\nalpha$, $\alpha=1,\dotsb,
\dim\HNP$ corresponding to the permanents, 
with the respective constraint associated to the facet opposite
$\nalpha$ denoted as $D^{(\alpha)}(m)$. 
Pick any $\omega \in \{1,\dotsb, \dim\HNP\}$, and let us focus on the
values of the functional $\Fp$ in a small region near a point $m_*$ on the facet labeled by $\omega$,
i.e., the one opposite the vertex $m^{(\omega)}$.  
That is to say, $D^{(\omega)}(m_*) = 0$ by assumption and we shall
investigate $\Fp(m_* + \epsilon \eta)$ for some inward-pointing vector $\eta$. We
will also assume that $m_*$ does not simultaneously lie on another facet. In
other words, $D^{(\alpha)}(m_*) > 0$ whenever $\alpha \neq \omega$.

In principle, we could carry out the calculations for an arbitrary direction
$\eta$ so long as it is inward-pointing. However, the outcome should not
differ much: if the diverging boundary force exists, which we are yet to
establish, then, intuitively, moving in any direction other than the one
perpendicular to the facet should result in only minute changes in the value of
$\Fp$. For this reason, we will take here $\eta = S^{(\omega)}$,
where $S^{(\omega)}$ consists of the coefficients of the linear part of the
constraint $D^{(\omega)}$ as defined in Eq.~\eqref{eq:Dj_constraints}.
Then, almost by definition, the vector $S^{(\omega)}$ is the inward
unit vector perpendicular to the facet $\omega$.

We will make use of Eq.~\eqref{eq:F_simplex} to examine $\Fp(m_* +\epsilon
S^{(\omega)})$. More specifically, the derivative 
\begin{equation}
  \label{eq:repulsion_strength_def}
  \doverd{\sqrt{\epsilon}}
  \Big\vert_{\epsilon=0}\Fp(m_* + \epsilon S^{(\omega)})
\end{equation}
will be computed, and this quantity will be referred to as the
\textit{repulsion strength} of the boundary force at $m_*$.
If one computed $\dstraight\Fp / \dstraight\epsilon$ near but not at $m_*$ instead, 
one would get 
$$
\frac{1}{2\sqrt{\epsilon}}\left[
  \doverd{\sqrt{\epsilon}}
  \Big\vert_{\epsilon=0}\Fp(m_* + \epsilon S^{(\omega)})
\right]
$$ 
(plus other terms that are insignificant in the limit $\epsilon\rightarrow 0$).
Therefore, the repulsion strength is, up to a factor of $2$, the prefactor in
front of $1/\sqrt{\epsilon}$ of the diverging derivative.

It is straightforward to check
$$
  \label{eq:D_from_epsilon}
D^{(\alpha)}(m_* + \epsilon  S^{(\omega)}) = 
  D^{(\alpha)}(m_*) + 
  \epsilon \braket{S^{(\alpha)}, S^{(\omega)}},
$$
which, for $\alpha = \omega$, yields 
\begin{equation}
  \label{eq:D_omega_is_epsilon}
  D^{(\omega)}(m_* + \epsilon S^{(\omega)}) = \epsilon.
\end{equation}
Of course, Eq.~\eqref{eq:D_omega_is_epsilon} is unsurprising:
$\epsilon$ parameterizes the distance to $m_*$, whereas $D^{(\omega)}$ is
the distance to the facet.
Nonetheless, Eq.~\eqref{eq:D_omega_is_epsilon} confirms that $
S^{(\omega)}$ is indeed inward-pointing as we had required. 
Now, if we
plug Eq.~\eqref{eq:D_omega_is_epsilon} into Eq.~\eqref{eq:F_simplex}, separate
the terms involving $\omega$, and perform the minimization over the phase
$\xi_\omega$, we get
\begin{equation}
  \label{eq:complicated_mess}
  \begin{aligned}
    \Fp(m_* + \epsilon S^{(\omega)})
    &= \min_{(\xi_{\alpha})_{\alpha\neq \omega}}
    \Bigg[\sum_{\alpha\neq \omega, \beta\neq\omega}
        \bar \xi_\alpha \xi_\beta W_{\alpha\beta}
    \sqrt{
      \frac{D^{(\alpha)}(m_* + \epsilon S^{(\omega)})}{L_\alpha}
    }
    \sqrt{
      \frac{D^{(\beta)}(m_* + \epsilon S^{(\omega)})}{L_\beta}
    }\\
    &\hspace{4em}
    -2\sqrt{\frac{\epsilon}{L_\omega}}\left|\sum_{\alpha\neq \omega}\xi_\alpha W_{\omega\alpha}
    \sqrt{\frac{D^{(\alpha)}(m_* + \epsilon S^{(\omega)})}
    {L_\alpha}}\right|  + \frac{\epsilon}{L_\omega} W_{\omega\omega}
    \Bigg].
  \end{aligned}
\end{equation}
The goal is to take the derivative with respect to $\sqrt{\epsilon}$ at
$\epsilon=0$ (Eq.~\eqref{eq:repulsion_strength_def}). This can be accomplished
via Danskin's theorem \cite{danskinTheoryMaxMinApplications1966}, which asserts that it is
valid to interchange the order of minimization and taking directional
derivatives (which is what we need here; in particular, it does not imply that
the pure functional $\Fp$ is differentiable),
while restricting the range of the new minimization to the set of minimizers of
the original one. That is, if we define
$$
X_0 := \argmin\limits_{(\xi_{\alpha})_{\alpha\neq\omega}} 
\sum_{\alpha\neq\omega, \beta\neq \omega} \bar\xi_\alpha \xi_\beta
W_{\alpha\beta} \sqrt{\frac{D^{(\alpha)}(m_*)}{L_\alpha}} \sqrt{\frac{D^{(\beta)}(m_*)}{L_\beta}},
$$
then
\begin{equation}
  \label{eq:BECforce_simplex}
  \doverd{\sqrt{\epsilon}}\Big|_{\epsilon=0}\Fp(m_* + \epsilon S^{(\omega)})
  =-\frac{2}{\sqrt{L_\omega}}\max_{(\xi_\alpha)_{\alpha\neq \omega} \in X_0} \left|\sum_{\alpha\neq \omega}\xi_\alpha W_{\omega\alpha}
    \sqrt{\frac{D^{(\alpha)}(m_*)}{L_\alpha}}\right|.
\end{equation}
Let $\ket{\Phi}$ denote the state $\sum_{\alpha\neq
\omega}\xi_\alpha\sqrt{D^{(\alpha)}(m_*)/L_\alpha}\ket{\nalpha}$ 
with the phases $\xi_\alpha$ chosen so that the maximum in
Eq.~\eqref{eq:BECforce_simplex} is achieved.
Then we can write Eq.~\eqref{eq:BECforce_simplex} more compactly as
\begin{equation}
  \label{eq:BECforce_simplex_2}
  \doverd{\sqrt{\epsilon}}\Big|_{\epsilon=0}\mathcal{F}(m_* + \epsilon S^{(\omega)})
  = -2\frac{\left|\braket{m^{(\omega)}|W|\Phi}\right|}{\sqrt{L_\omega}}.
\end{equation}
First, note that the minus sign indicates repulsion from the facet.
The geometric factor $\sqrt{L_\omega}$ in the denominator implies that the
repulsion strength is weaker as the distance from the vertex to the facet
increases, which gives us the following physical interpretation: the farther
away the vertex $m^{(\omega)}$ is from the facet $\braket{m, S^{(\omega)}} -
\nu^{(\omega)}=0$, the less relevant the state $\ket{m^{(\omega)}}$ is for
describing the physics for densities near the facet, since having a nonzero
overlap with $\ket{m^{(\omega)}}$ comes at the cost of deviating from the facet
at a rate proportional to the distance $L_\omega = D^{(\omega)}(m^{(\omega)})$.
(This is also why we imposed $\left\Vert S^{(\omega)}\right\Vert=1$: so that
$D^{(\omega)}(m)$ is not only proportional to the distance from $m$ to the
facet, but actually \textit{is} the distance.)



\subsection{Beyond Simplex}
\begin{figure}
  \centering
  \includegraphics[width=.84\columnwidth]{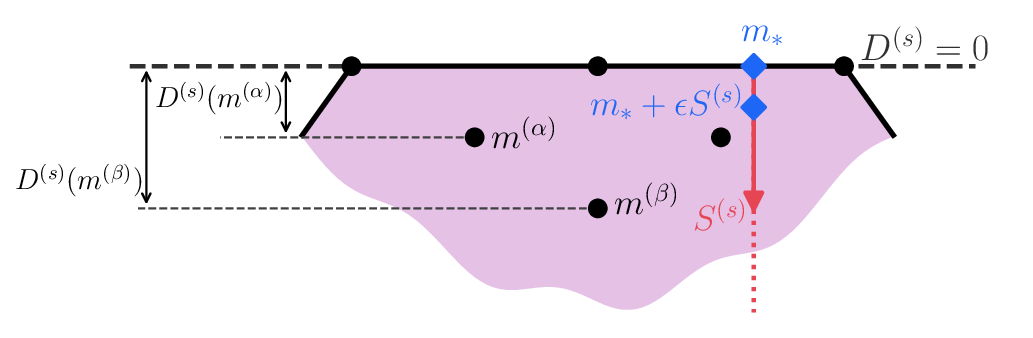}
  \caption{Illustration of the geometric objects involved in the analysis of the
  generalized BEC force. The upper facet is the facet labeled by $s$, i.e., the
  points on it satisfy $D^{(s)} = 0$. In the diagram, $\nalpha$ and $\nbeta$
  are the occupation number vectors of two permanents (arbitrarily
  chosen for the purpose of illustration) not lying on the facet. Their
  distances $D^{(s)}(\nalpha)$ and $D^{(s)}(\nbeta)$ are indicated on the left.
  The repulsion strength is
  defined in terms of the values of the functional $\Fp$ along the straight
  path starting at a point $m_*$ on the facet (blue diamond marker on top)
  and extending inwards perpendicularly to the facet (red dotted line),
  i.e., in the direction of the normal vector $S^{(s)}$ (red arrow).
  }
  \label{fig:facet_diagram}
\end{figure}
We now drop the simplex assumption and derive a formula for the repulsion
strength for any $(d, N, P)$. 
As is often the case, abandoning the simplex assumption generates substantial
complications in establishing parallel results. 
First and foremost, 
there is no longer a one-to-one correspondence between the facets and the
occupation number vectors $\{\nalpha\}$ of the permanents, and instead we have $J$
constraints $D^{(j)}, j=1,\dotsb, J$ and $\dim \HNP$ occupation number vectors
$\nalpha$, $\alpha=1,\dotsb, \dim\HNP$. 
For the notation, this means that we
will use Latin letters $j,s, \dotsb$ to label facets, rather than Greek letters
$\alpha,\omega, \dotsb$, which we reserve for labeling permanents.
We proceed as before, choosing a specific $s\in \{1, \dotsb, J\}$ and fixing a
density $m_*$ on the facet labeled by $s$, so that $D^{(s)}(m_*) =0$. 
We will also assume $D^{(j)}(m_*) > 0$ whenever
$j\neq s$, so that $m_*$ does not simultaneously lie on another facet.

We will now present a physically motivated, and admittedly not completely
rigorous derivation of the generalized BEC force. One subtlety (among many)
that we sweep under the rug here is that the following argument is valid only
when $m_*$ is not a critical value of the density map of the \textit{restricted
functional theory on the facet}, a notion which will be defined precisely in
\cref{chap:abelian_functional_theory}. Since the set of critical values has
measure zero in the facet, the mistake we are about to make is not too serious.
In \cref{chap:abelian_functional_theory}, when we discuss abelian functional
theories, which generalize bosonic momentum space functional theories, we will
give a more general and strictly rigorous proof based on the constrained
search.

We will adopt the notation that $I_j$ denotes the set of all indices $\alpha$
for which the occupation number vector $\nalpha$ satisfies $D^{(j)}(\nalpha) =
0$. In other words, $I_j$ is the set of indices of permanents lying on the
facet labeled by $j$. 


Let $\ket{\Phi(\epsilon)}$, $\epsilon\in [0, \epsilon')$ be a curve such that
$$
  \iota^*(\ketbrap{\Phi(\epsilon)}) = m_* + \epsilon S^{(s)}
$$ 
for all $\epsilon$ and assume that $\ket{\Phi(\epsilon)}$ is a minimizer of the
constrained search at $m_*+\epsilon S^{(s)}$ for each $\epsilon$. That is,
\begin{equation}
  \label{eq:functional_near_facet}
  \Fp(m_* + \epsilon S^{(s)}) = 
  \braket{\Phi(\epsilon)|W|\Phi(\epsilon)}.
\end{equation}
Write $\ket{\Phi(\epsilon)} = \sum_{\alpha}c_\alpha(\epsilon)\ket{\nalpha}$. We
will assume that the coefficients $c_\alpha(\epsilon)$ take the following form:
\begin{equation}
  \label{eq:bec_force_coeff_ansatz}
  \begin{cases}
    c_\alpha(\epsilon) = p_\alpha + \epsilon q_\alpha + O(\epsilon^2)& \alpha \in I_s\\
    c_\alpha(\epsilon) = \sqrt{\epsilon}r_\alpha + O(\epsilon)& \alpha \notin I_s.
  \end{cases}
\end{equation}
Eq.~\eqref{eq:bec_force_coeff_ansatz} needs some justification. Since
$\braket{\Phi(\epsilon)|n_k|\Phi(\epsilon)}$ depends linearly on each
$|c_\alpha(\epsilon)|^2$, and only the ``off-facet'' permanents, i.e.,
$\ket{\nalpha}$ with $\alpha \notin I_s$, can contribute to the deviation
$\epsilon S^{(s)}$ from the facet, we must have
$c_\alpha(\epsilon)\propto \sqrt{\epsilon}$ for $\alpha \notin I_s$, up to
higher-order terms in $\epsilon$. 
On the other hand, it is not immediately clear
that the ``on-facet'' coefficients, that is, $c_\alpha(\epsilon)$ with $\alpha\in
I_s$, should not contain terms proportional to $\sqrt{\epsilon}$. 
Indeed, if $c_\alpha(0)=0$, it is conceivable that $c_\alpha(\epsilon)\sim
\sqrt{\epsilon}$ even if $\alpha \in I_s$. However, the vanishing of
$c_\alpha(0)$ actually implies that the state $\ket{\nalpha}$ is decoupled from
other permanents on the facet (this claim is made more precise by the ``No-Mixing
Lemma'' (\cref{lem:minimizing_state_zero_coefficient}), which we prove in
\cref{chap:abelian_functional_theory}), so $c_\alpha$ does not contribute to the
generalized BEC force. 

Plugging Eq.~\eqref{eq:bec_force_coeff_ansatz} into Eq.~\eqref{eq:functional_near_facet} yields
\begin{equation}
  \begin{aligned}
    &\Fp(m_* + \epsilon S^{(s)}) = \sum_{\alpha,\beta \in I_s}
      \bar p_\alpha p_\beta  W_{\alpha\beta}
    + 2\sqrt{\epsilon}\Re\sum_{\alpha\in I_s, \beta\notin I_{s}}
    \bar p_\alpha r_\beta W_{\alpha\beta} + O(\epsilon).
  \end{aligned}
\end{equation}
The coefficients $p_\alpha$ have to be chosen so as to minimize the first term,
and the resulting minimum is clearly just $\Fp(m_*)$. Similarly, the $r_\beta$
are fixed by minimizing the second term. Observe that there is always an
obvious choice of phases of the $r_\beta$ that will minimize the expression.
Hence, we have
\begin{equation}
  \label{eq:bec_force_intermediate_form}
  \begin{aligned}
    &  \frac{\Fp(m_* + \epsilon S^{(s)}) - \Fp(m_*)}{\sqrt{\epsilon}}
    = - 2 \sum_{\beta \notin I_s} \left|r_\beta\right |\cdot 
    \left|\sum_{\alpha\in I_s} \bar p_\alpha W_{\alpha\beta}\right|
     = -2\sum_{\beta \notin I_s} \left|r_\beta\right| \cdot 
     \left|\braket{\nbeta| W | \Phi}\right|,
  \end{aligned}
\end{equation}
where we have defined $\ket{\Phi}:=\ket{\Phi(0)} = \sum_{\alpha\in I_{s}}
p_\alpha\ket{\nalpha}$. The $\left|r_\beta\right|$ cannot be chosen arbitrarily: the
condition $\iota^*(\ketbrap{\Phi(\epsilon)}) = m_* + \epsilon S^{(s)}$ implies
$$
\sum_{\alpha\in I_s} \nalpha 
\left[\left|p_\alpha\right|^2 + 2\epsilon \Re(\bar p_\alpha q_\alpha)\right]
  + \epsilon\sum_{\alpha\notin I_s}\nalpha \left|r_\alpha\right|^2
  = m_* + \epsilon S^{(s)} 
$$
$$
\Rightarrow 2\Re\sum_{\alpha\in I_s} \nalpha \bar p_\alpha q_\alpha
+ \sum_{\alpha\notin I_s} \nalpha \left|r_\alpha\right|^2 = S^{(s)}.
$$
Take the inner product with $S^{(s)}$ and use $\braket{\nalpha, S^{(s)}}= \nu^{(s)}$ 
for $\alpha\in I_{s}$ to get
$$
2\nu^{(s)}\Re \sum_{\alpha\in I_s} \bar p_\alpha q_\alpha 
+  \sum_{\alpha\notin I_s} \left|r_\alpha\right|^2 
\braket{\nalpha, S^{(s)}} = \braket{S^{(s)}, S^{(s)}} = 1.
$$
But $\braket{\Phi(\epsilon)|\Phi(\epsilon)} = 1$ implies
$2\Re\sum_{\alpha\in I_s}\bar p_\alpha q_\alpha 
= -\sum_{\alpha\notin I_s}\left|r_\alpha\right|^2$, so we have
\begin{equation}
  \label{eq:bec_force_coeff_constraint}
  \sum_{\alpha\notin I_{s}}\left|r_\alpha\right|^2 D^{(s)}(\nalpha) = 1.
\end{equation}
That is, the last line of Eq.~\eqref{eq:bec_force_intermediate_form} should be
minimized subject to the constraint \eqref{eq:bec_force_coeff_constraint}. This
constrained minimization can be easily carried out and the result is
\begin{equation}
  \label{eq:amazing_exforce_formula}
  \doverd{\sqrt{\epsilon}}\Big|_{\epsilon=0}\Fp(m_*+\epsilon S^{(s)}) 
  = -2 \left[\sum_{\alpha\notin I_s}
  \frac{\left|\braket{\nalpha|W|\Phi}\right|^2}{D^{(s)}(\nalpha)}\right]^{\frac{1}{2}},
\end{equation}



Eq.~\eqref{eq:amazing_exforce_formula}, which presents an exact analytical
relation satisfied by the functional $\Fp$ near any facet, is the main
result of this section. It directly implies that the derivative $\dstraight_m
\Fp$ diverges as $1/\sqrt{\epsilon}$ when the density $m$ is at distance
$\epsilon\rightarrow 0$ to a facet, provided that the sum in
Eq.~\eqref{eq:amazing_exforce_formula} does not vanish. 
We would like to
highlight a number of features exhibited by this formula: 

\begin{enumerate}[label={(\arabic*)}]
  \item The sign is negative, indicating a repulsive force from the boundary as
    in Section~\ref{sec:BECforce_simlex}.
  \item The contribution of each permanent $\ket{\nalpha}$ is
    inversely weighted by the geometric factor $D^{(s)}(\nalpha)$ (see
    Fig.~\ref{fig:facet_diagram} for illustration), which is the distance from
    $\nalpha$ to the facet. In other words, distant permanents have smaller
    effects on the strength of the repulsive force. 
  \item The radicand
    is a sum of nonnegative terms. This means that the only
    way the repulsion strength $\dstraight\Fp(m_* + \epsilon S^{(j)}) /
    \dstraight\sqrt{\epsilon}|_{\epsilon = 0}$ can vanish is when
    $\braket{\nalpha| W |\Phi}=0$ for all $\alpha$ such that $D^{(s)}(\nalpha)>
    0$.
  \item To see how Eq.~\eqref{eq:amazing_exforce_formula} reduces to
    Eq.~\eqref{eq:BECforce_simplex_2} in the simplex setting, note that, in the
    latter situation, there is only one permanent $\ket{\nalpha}$ that
    is not on the facet, so the sum contains only one term and
    we recover Eq.~\eqref{eq:BECforce_simplex_2}. 
\end{enumerate}


\begin{ex}
  Let us illustrate Eq.~\eqref{eq:amazing_exforce_formula} by considering the
  Bose-Hubbard model (Eq.~\eqref{eq:hubbard_W}) with $(d,N,P) = (3, N, 0)$, where $N$ is a multiple of $3$
  (see \cref{fig:gradient_F_plots} for $N=6$ and $N=12$). The domain is
  a $2$-simplex (triangle), but this does not belong to what we call the
  ``simplex setting'' since there are occupation number vectors $m^{(\alpha)}$
  which are not extreme points. We will choose $s$ to be (the label of) the
  facet corresponding to the bottom edge (see
  \cref{fig:domains_dNP,fig:gradient_F_plots}), for which the inequality
  constraint is given by
\begin{equation}
  D^{(s)}(m) =  \frac{1}{\sqrt{6}}(2m_0-m_1-m_2)+\frac{N}{\sqrt{6}} \ge 0,
\end{equation}
so $S^{(s)} = \frac{1}{\sqrt{6}}(2,-1,-1)$ and $\nu^{(s)} = -N/\sqrt{6}$. 
Take $m_* = (0, Nt, N(1-t))$ for some $t\in (0,1)$, then
$D^{(s)}(m_*)=0$, so $m_*$ does lie on the facet labeled by $s$. 
To apply Eq.~\eqref{eq:amazing_exforce_formula}, first note that the permanents
on the facet are
$$
\ket{0,N,0}, \ket{0,N-3, 3}, \dotsb,  \ket{0,0,N}.
$$
Since $W$ does not couple the states on the facet (true for any pair
interaction) and $\ket{0,N,0}$ and $\ket{0,0,N}$ have the lowest energy among
all permanents on the facet (special to the Hubbard model), the minimizer states 
of the constrained search at $m_*$ are
\begin{equation}
  \ket{\Phi^*(\theta)} = \sqrt{t}\ket{0,N,0} + \sqrt{1-t}e^{i\theta}\ket{0,0,N}.
\end{equation}
The states $\ket{0,N,0}$ and $\ket{0,0,N}$ only couple to $\ket{1,N-2,1}$ and
$\ket{1,1,N-2}$ respectively, with matrix elements 
$$
    \braket{1,N-2,1|W|0,N,0} = \braket{1,1,N-2|W|0,0,N} =\frac{2}{3}\sqrt{N(N-1)}.
$$
Moreover, their distances to the facet are equal and given by 
$$
    D^{(s)}(1,N-2,1) = D^{(s)}(1,1,N-2) = \sqrt{6}/2.
$$
Assuming $N > 3$ so that $\ket{1,N-2,1}$ and $\ket{1,1,N-2}$ do not coincide,
Eq.~\eqref{eq:amazing_exforce_formula} becomes
\begin{equation}
  \label{eq:hubbard_exforce}
  \begin{aligned}
    \doverd{\sqrt{\epsilon}}\Big|_{\epsilon=0}\Fp(m_*+\epsilon S^{(j)}) 
    &= -2\left[\frac{ 
    \left(\frac{2}{3}\cdot \sqrt{N(N-1)}\right)^2(\sqrt{t}^2 + \sqrt{1-t}^2)
    }{\frac{\sqrt{6}}{2}}\right]^{\frac{1}{2}}\\ 
    &=-\frac{4\cdot 2^{\frac{1}{4}}\cdot 3^{\frac{3}{4}}}{9}\sqrt{N(N-1)}
  \end{aligned}
\end{equation}
The numerical value of the prefactor is about $1.205$. Interestingly, the
repulsion strength does not depend on the position $t$ on the facet in this
case.

In order to verify that Eq.~\eqref{eq:hubbard_exforce} gives the correct generalized BEC force,
we approximate the functional by 
$$
  \Fp(m_* + \epsilon S^{(s)}) 
  \approx \Fp(m_*) + \sqrt{\epsilon}\doverd{\sqrt{\epsilon}}\Big\vert_{\epsilon=0}
  \Fp(m_* + \epsilon S^{(s)})
$$
and compare the approximation to numerical values of $\Fp$. We will choose $m_*
= (0, \frac{N}{2}, \frac{N}{2})$, for which $\Fp(m_*) = N(N-1)/3$. It follows
that
\begin{equation}
  \label{eq:hubbard_approximate_functional}
  \frac{\Fp\left(m_* + \epsilon S^{(s)}\right)}{N^2}
  \approx \frac{N-1}{3N} - 1.205\sqrt{\epsilon} \frac{1}{N}\sqrt{\frac{N-1}{N}}.
\end{equation}
In Fig.~\ref{fig:exforce_numer_d3NP0}, this is compared to the exact functional
for $N=6,18,30$. Clearly, Eq.~\eqref{eq:hubbard_approximate_functional} yields
the correct generalized BEC force at $m_* = (0, \frac{N}{2}, \frac{N}{2})$.

\begin{figure}
  \centering
  \includegraphics[width=.64\textwidth]{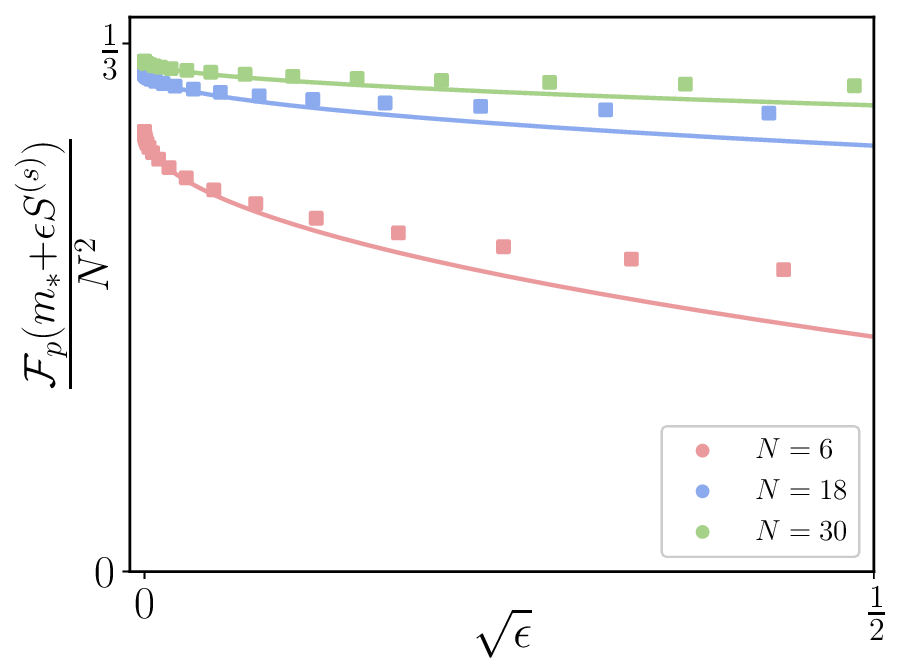}
  \caption{Comparison of the approximate functional given by
  Eq.~\eqref{eq:hubbard_approximate_functional} (solid lines) for the Hubbard model with
  exact numerical values (square dots).}
  \label{fig:exforce_numer_d3NP0}
\end{figure}

Finally, we would like to remark that Eq.~\eqref{eq:amazing_exforce_formula}
also accounts for the larger repulsion strength for $P=1$ compared to $P=0$ at
fixed $(d=3,N)$ when $N$ is a multiple of $3$, as can be observed in
Fig.~\ref{fig:gradient_F_plots}. When $P=1$, it is still true that a minimizing
state on the facet is always a superposition of the two extremal permanents (see
Fig.~\ref{fig:d12_hubbard_coupling}). However, these couple to five off-facet
permanents in total, as opposed to two in the case of $P=0$. As a result, the sum
in Eq.~\eqref{eq:amazing_exforce_formula} contains five terms when $P=1$
instead of two, resulting in a larger repulsion strength. This difference is
illustrated in Fig.~\ref{fig:d12_hubbard_coupling} for $N=12$.
\begin{figure}
  \centering
  \includegraphics[width=.7\textwidth]{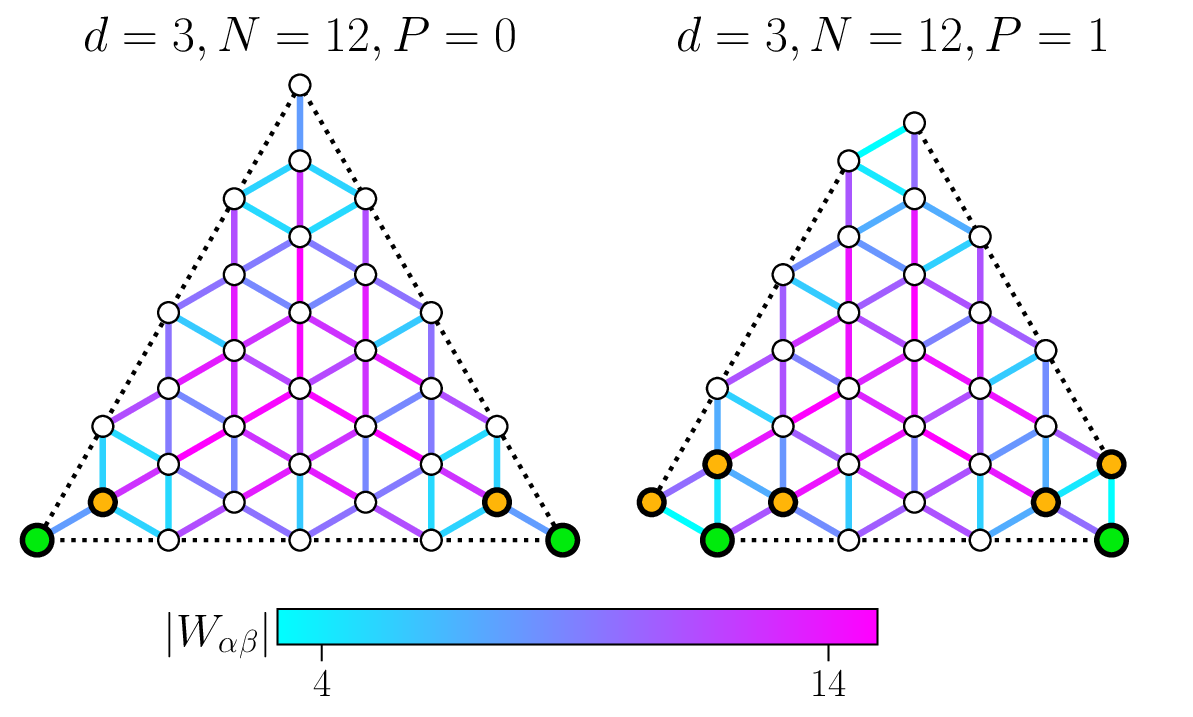}
  \caption{Graphical depiction of the underlying cause of the stronger
  generalized BEC force for $P=1$ (see the two bottom-left plots in
  Fig.~\ref{fig:gradient_F_plots}). In each diagram, two occupation number
  vectors $\nalpha, \nbeta$ are connected by an edge if $W_{\alpha\beta} \neq
  0$, with the value of $|W_{\alpha\beta}|$ encoded by the color. On a given
  facet, the extremal permanents (green) couple to two off-facet states (yellow)
  for $P=0$ and to five off-facet states for $P=1$.}
  \label{fig:d12_hubbard_coupling}
\end{figure}
\popQED
\end{ex}

\chapter{Abelian Functional Theory}
\label{chap:abelian_functional_theory}

The bosonic momentum space functional theory presented in
\cref{chap:momentum_rdmft} is a special case of an \textit{abelian} functional
theory. There, we consider the family of one-particle operators $h$ which are
diagonal in the momentum basis, which all commute and consequently allow us to
determine the domain of the functional with relative ease. In this chapter, we
single out a class of generalized functional theories which have this
commutativity property and investigate its consequences. An important result of
this chapter is the \textit{boundary force formula}
(\cref{thm:boundary_force_formula}), which we state and prove in
\cref{sec:abelian_boundary_force}.

\section{Definitions}
\begin{defn}
A generalized functional theory $(V, \mathcal{H}, \iota, W)$ is
\textbf{abelian} if the image of $\iota$ is commutative. In other words,
$[\iota(v), \iota(v')]=0$ for any $v,v'\in V$.
\end{defn}

Note that this definition is independent of the interaction $W$. That is, we do
\textit{not} require $W$ to commute with the external potentials.

\begin{ex}
  The discrete DFT in Example~\ref{ex:discrete_DFT} is abelian because the
  number operators $a_i^\dagger a_i^{\phantom{\dagger}}$ ($i=1, \dotsb, d$)
  commute. RDMFT (Example~\ref{ex:rdmft}), on the other hand, is not abelian
  (unless $\dim\mathcal{H}_1=1$) because the set of all one-particle operators
  $i\mathfrak{u}(\mathcal{H}_1)$, or more precisely its image in
  $i\mathfrak{u}(\mathcal{H})$, where $\mathcal{H}$ is the $N$-particle Hilbert
  space, is not commutative.
\end{ex}

\begin{ex}
  \label{ex:N_qubits_abelian}
  Consider a system of $N$ two-level spins, whose Hilbert space is $\mathcal{H}
  = (\mathbb{C}^2)^{\otimes N}$. As the space of external potentials, we take
  $V = \mathbb{R}^N$ together with potential map $\iota(v_1, \dotsb, v_N) =
  \sum_{i=1}^N v_i Z_i$, where $Z_i$ is the Pauli Z operator of the $i$-th
  spin. In contrast to Example~\ref{ex:nonabelian_spin_chain}, the functional
  theory defined by $(V, \mathcal{H}, \iota, W)$ is abelian for any interaction
  $W$ because the $Z_i$ commute.
\end{ex}

An (if not \textit{the most}) important feature of abelian functional theories
is that all external potentials can be simultaneously diagonalized, since they
all commute by assumption.

\begin{defn}
  \label{defn:abelian_weights}
  A density $\alpha\in V^*$ is a \textbf{weight} if there exists a nonzero
  vector $\ket{\Psi}\in \mathcal{H}$ such that
  $$
  \forall v\in V: \iota(v)\ket{\Psi} = \braket{\alpha, v}\ket{\Psi}.
  $$
  Given any $\alpha\in V^*$, the corresponding \textbf{weight space} is
  $$
  \mathcal{H}_\alpha:= \{\ket{\Psi}\in \mathcal{H}\mid \forall v\in V: \iota(v)\ket{\Psi} = \braket{\alpha,v}\ket{\Psi}\}.
  $$
  If $\alpha$ is a weight, its \textbf{multiplicity} is defined as the
  dimension of $\mathcal{H}_\alpha$. A nonzero vector in $\mathcal{H}_\alpha$
  is called a \textbf{weight vector}.
\end{defn}
Hence, a density $\alpha$ is a weight if and only if $\mathcal{H}_\alpha\neq
0$. It is an elementary fact that the Hilbert space decomposes into an
orthogonal direct sum of the weight spaces:
\begin{equation}
  \label{eq:weight_space_decomp}
  \mathcal{H} = \bigoplus_{\alpha\in \Omega} \mathcal{H}_\alpha,
\end{equation}
where $\Omega$ denotes the set of weights.

The importance of the weights is that they determine the domains of the pure
and ensemble functionals according to the following theorem, which is the
generalization of \cref{thm:domain_of_F}:

\begin{thm}
  \label{thm:abelian_nrep_set}
  The set of pure state representable densities coincides with the set of
  ensemble state representable densities, and both are given by the convex hull
  of the set of weights. In other words,
  \begin{equation}
    \iota^*(\mathcal{P}) = \iota^*(\mathcal{E}) = \conv(\Omega).
  \end{equation}
  \begin{proof}
    We will show
    $$
    \iota^*(\mathcal{P}) \overset{\text{(1)}}{\subset} \iota^*(\mathcal{E}) 
    \overset{\text{(2)}}\subset \conv(\Omega)
    \overset{\text{(3)}}\subset \iota^*(\mathcal{P}).
    $$

    Inclusion (1) follows directly from $\mathcal{P}\subset \mathcal{E}$ (all
    pure states are ensemble states).\\

    Let $(\ket{E_i})_{i\in I}$ be an orthonormal basis compatible with the
    weight space decomposition~\eqref{eq:weight_space_decomp}, where $I$ is some
    index set. 
    That is, for each $i\in I$, there exists a weight $\omega_i\in \Omega$ such that
    the basis vector $\ket{E_i}$ is in the weight space
    $\mathcal{H}_{\omega_i}$. 
    Take any ensemble state representable density
    $\rho \in \iota^*(\mathcal{E})$. 
    Then, by definition, 
    $$
      \rho =
      \iota^*\left(\sum_{i,j\in I}c_{ij}\ketbra{E_i}{E_j}\right),
    $$
    where the matrix $(c_{ij})_{ij}$ is positive semidefinite with unit trace. Hence, for
    any $v\in V$,
    $$
    \begin{aligned}
      &\braket{\rho,v} = \sum_{i,j\in I}c_{ij}\braket{E_j|\iota(v)|E_i}
      = \sum_{i,j\in I}c_{ij}\braket{\omega_i, v}\braket{E_j|E_i}\\
      &= \sum_{i,j\in I} c_{ij}\braket{\omega_i, v} \delta_{ij}
      = \sum_{i\in I} c_{ii} \braket{\omega_i, v}\\
      &= \left\langle \sum_{i\in I} c_{ii}\omega_i, v\right\rangle,
    \end{aligned}
    $$
    implying $\rho = \sum_{i\in I} c_{ii}\omega_i$. Since $c_{ii}\ge 0$ and $\sum_i
    c_{ii} = 1$, we conclude that $\rho$ is a convex combination of the
    weights. This shows inclusion (2).

    Take any $\rho = \sum_{\omega\in \Omega} t_\omega \omega$ with
    $\sum_{\omega\in\Omega} t_\omega = 1$ and $t_\omega \ge 0$. For each weight
    $\omega \in \Omega$, pick any weight vector $\ket{\Phi_\omega}\in
    \mathcal{H}_\omega$ with $\lVert \Phi_\omega\rVert = 1$. 
    Define $\ket{\Psi} = \sum_{\omega\in \Omega}
    \sqrt{t_\omega} \ket{\Phi_\omega}$. Then, for any $v\in V$,
    $$
    \begin{aligned}
      &\left\langle\iota^*(\ket{\Psi}\!\bra{\Psi}), v\right\rangle
     = \braket{\Psi|\iota(v)|\Psi} 
      =\sum_{\omega,\omega' \in \Omega} \sqrt{t_{\omega'}t_{\omega}} 
      \braket{\Psi_{\omega'}|\iota(v)|\Psi_\omega}\\
      &= \sum_{\omega,\omega'\in \Omega}\sqrt{t_{\omega'}t_\omega}\braket{\omega, v}\delta_{\omega'\omega}
      = \sum_{\omega\in \Omega} t_\omega \braket{\omega, v}\\
      &= \braket{\rho, v},
    \end{aligned}
    $$
    so $\iota^*(\ket{\Psi}\!\bra{\Psi}) = \rho$. This shows inclusion (3).
  \end{proof}
\end{thm}

\begin{rem}
  At least when the connected Lie group corresponding to $\iota(V)\subset
  i\mathfrak{u}(\mathcal{H})$ is compact, i.e., a torus,  
  Theorem~\ref{thm:abelian_nrep_set} is a corollary of \textit{Atiyah's
  theorem} \cite{atiyahConvexityCommutingHamiltonians1982} (also due to
  Guillemin and Sternberg \cite{guilleminConvexityPropertiesMoment1982}), which
  will be introduced in Chapter~\ref{chap:interlude}. What Atiyah, Guillemin,
  and Sternberg proved is a much more general statement about Hamiltonian
  torus actions on symplectic manifolds. In our case, the symplectic manifold
  is simply the projective Hilbert space (which can be identified with
  $\mathcal{P}$), for which a simple proof like the one presented above works.
  We will never need the full power of Atiyah's theorem in this thesis.
\end{rem}

In light of Theorem~\ref{thm:abelian_nrep_set}, we will simply call a density
$\rho \in V^*$ \textit{representable} if it is pure state representable or
ensemble state representable, since
$\iota^*(\mathcal{P})=\iota^*(\mathcal{E})$. Equivalently, a density $\rho\in
V^*$ is representable if and only if it belongs to $\conv(\Omega)$.
The set $\conv(\Omega)$, being the convex hull of finitely many points, is a
convex polytope. Clearly, the dimension of $\conv(\Omega)$ is equal to the
dimension of $\aff(\iota^*(\mathcal{P}))$, which is in turn equal to $\dim V -
\dim\iota^{-1}(\mathrm{span}\{\mathbbm{1}\})$ by
Proposition~\ref{thm:dimension_of_ens_rep}. In particular, if the potential map
$\iota$ is injective and its image does not contain $\mathbbm{1}$, then the
convex polytope $\conv(\Omega)$ will have full dimension in $V^*$. In general,
$\conv(\Omega)$ lies in a hyperplane of codimension
$\dim(\iota^{-1}(\mathrm{span}\{\mathbbm{1}\}))$. We summarize this discussion
in the following proposition.

\begin{propo}
  \label{thm:dimension_of_conv_Omega}
  The set of representable densities $\iota^*(\mathcal{P})=
  \iota^*(\mathcal{E})=\conv(\Omega)$ is a convex polytope of dimension $\dim V
  - \dim\iota^{-1}(\mathrm{span}\{\mathbbm{1}\})$.
\end{propo}

\begin{ex}
  \label{ex:abelian_qubit_functional_theory}
  Consider again the abelian functional theory of $N$ two-level spins in
  Example~\ref{ex:N_qubits_abelian}. Since $\iota(v) = \sum_{i=1}^N v_i
  Z_i$, 
  the weight vectors are exactly the simultaneous eigenvectors of
  the Pauli Z operators, which are product states of $\ket{\uparrow}$ and
  $\ket{\downarrow}$. 
  The corresponding weights are then $\Omega = \{-1,
  1\}^N\subset \mathbb{R}^N = (\mathbb{R}^N)^*$. 
  For each weight
  $\alpha=(\alpha_1, \dotsb, \alpha_N)$, $\alpha_i=\pm1$, the weight space
  $\mathcal{H}_\alpha$ is one-dimensional, so there is, up to scalars, a unique
  weight vector for $\alpha$. The set of representable densities is then
  $\iota^*(\mathcal{P}) = \iota^*(\mathcal{E}) = \conv(\Omega) = \{(\rho_1,
  \rho_2, \dotsb, \rho_N) \mid \rho_i\in [-1,1]\} = [-1,1]^N$, which is an
  $N$-dimensional hypercube (see \cref{fig:spin_hypercube} for $N=3$
  spins).
  \begin{figure}
    \centering
    \includegraphics[width=.35\textwidth]{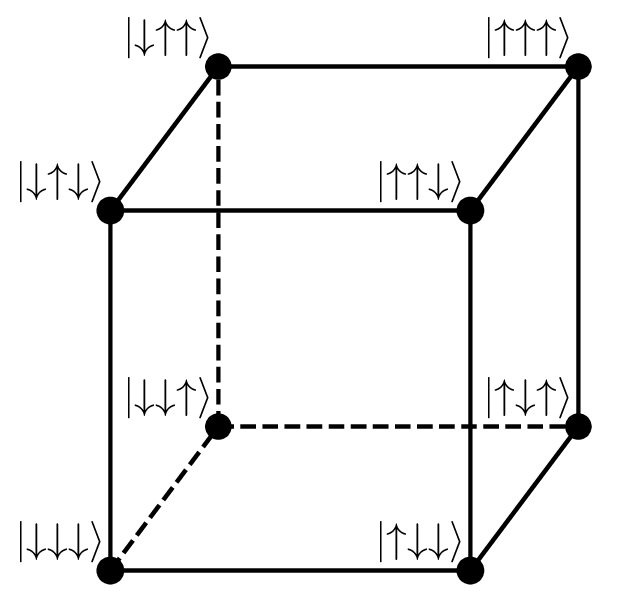}
    \caption{The weight structure for the abelian spin chain functional theory
    described in Example~\ref{ex:abelian_qubit_functional_theory} with $N=3$
    two-level spins. For three spins, there are $8$ possible product states which are
    simultaneous eigenstates of the Pauli Z operators, and they correspond to
    eight weights, given by $(\pm1,\pm1,\pm1)$ in $(\mathbb{R}^3)^* =
    \mathbb{R}^3$, forming vertices of a cube of side length $2$. According to
    Theorem~\ref{thm:abelian_nrep_set}, we have $\dom \Fp = \iota^*(\mathcal{P}) =
    \dom \Fe=\iota^*(\mathcal{E}) =[-1,1]^3$. 
    }
    \label{fig:spin_hypercube}
  \end{figure}
\end{ex}

\begin{ex}
  For the bosonic momentum space functional theory introduced in
  \cref{chap:momentum_rdmft} (see
  \cref{defn:bosonic_momentum_space_functional_theory}), the weight vectors are
  the permanents $\ket{m^{(\alpha)}}$ with momentum $P$, and the weights are
  the occupation number vectors $m^{(\alpha)}$ of the permanents.
\end{ex}

Now that we have established that $\iota^*(\mathcal{P})=\iota^*(\mathcal{E})=
\conv(\Omega)$ is a convex polytope, we can talk about its facets, which are by
definition the codimension-$1$ faces of $\conv(\Omega)$. In particular, we will
define the notion of the \textit{restricted functional theory} corresponding to
each facet of $\conv(\Omega)$. This will turn out to be useful when we discuss
boundary forces in subsequent sections, where properties of the restricted
functional theory will play an important role in the derivation of the boundary
force formula.

\begin{defn}
  \label{def:facet_functional_theory}
  Let $(V, \mathcal{H}, \iota, W)$ be an abelian functional theory and let $F$
  be a facet of the convex polytope $\conv(\Omega)$. Let $\Omega_F:= \Omega \cap
  F\subset \Omega$ be the subset of weights lying on $F$. Consider
  the subspace $\mathcal{H}_{F} = \bigoplus_{\omega\in \Omega_F}
  \mathcal{H}_\omega \subset \mathcal{H}$ and the corresponding orthogonal
  projection $\Pi_{F}: \mathcal{H}\rightarrow \mathcal{H}_{F}$. The
  \textbf{restricted functional theory on the facet $F$} or simply
  the \textbf{facet functional theory on $F$} is the generalized functional theory
  given by $(V, \mathcal{H}_F, \iota_{F},
  \Pi_{F}^{\phantom{\dagger}}W\Pi_{F}^\dagger)$, where $\iota_{F}(v) =
  \Pi_{F}^{\phantom{\dagger}}\iota(v)\Pi_{F}^\dagger$. 
\end{defn}

Note that we do not replace the vector space $V$ that parametrizes the space of
external potentials. This is mostly a matter of convention: intuitively,
elements in $V$ which are perpendicular to $F$ will be mapped to something
proportional to $\mathbbm{1}_F$ under $\iota_F$, where $\mathbbm{1}_F$ denotes
the identity on $\mathcal{H}_F$, which is fine, since we have always allowed
the possibility for the image of the potential map to contain the identity. In
fact, this is consistent with Proposition~\ref{thm:dimension_of_conv_Omega} as
the dimension of the facet $F$ is precisely one lower than that of $\conv(\Omega)$.

\begin{propo}[Basic properties of restricted functional theories]
  Let $F\subset \conv(\Omega)$ be any facet. Then:
  \begin{enumerate}[label={(\arabic*)}]
    \item The restricted functional theory on $F$ is abelian.
    \item $\Omega_F := \Omega \cap F$ is the set of weights of the restricted functional theory
    \item $\dim(\iota_F^{-1}\mathrm{span}\{\mathbbm{1}_F\}) =
      \dim(\iota^{-1}\mathrm{span}\{\mathbbm{1}\}) + 1$.
    \item For all $\Gamma \in \mathcal{P}$, $\iota^*(\Gamma)\in F$ if and only
      if $\Gamma \in \mathcal{P}_F$, where $\mathcal{P}_F$ denotes the set of pure states on $\mathcal{H}_F$.
    \item For all $\Gamma \in \mathcal{E}$, $\iota^*(\Gamma)\in F$ if and only
      if $\Gamma \in \mathcal{E}_F$, where $\mathcal{E}_F$ denotes the set of ensemble states on $\mathcal{H}_F$.
    \item $\Fp^{(F)} = \Fp|_F$, where $\Fp^{(F)}$ denotes the pure functional of the restricted functional theory.
    \item $\Fe^{(F)} = \Fe|_F$, where $\Fe^{(F)}$ denotes the ensemble functional of the restricted functional theory.
  \end{enumerate}
  \begin{proof}
    (1) and (2) are obvious. (3) follows from
    Proposition~\ref{thm:dimension_of_conv_Omega} and the fact that $\dim
    \conv(\Omega_F)  = \dim F = \dim \conv(\Omega) - 1$.

    Take a pure state $\ketbrap{\Psi}\in \mathcal{P}$ such that
    $\iota^*(\ketbrap{\Psi})\in F$. Let $\ket{\Psi} = \sum_{\omega\in \Omega}
    \ket{\Psi_\omega}$ be the weight space decomposition of $\ket{\Psi}$, with
    $\ket{\Psi_\omega}\in \mathcal{H}_\omega$. Then 
    $$
      \braket{\iota^*(\ket{\Psi}\!\bra{\Psi}), v}
       = \sum_{\omega\in \Omega} \braket{\Psi_\omega|\iota(v)|\Psi_\omega}
        = \sum_{\omega \in \Omega} \braket{\omega, v} \lVert\Psi_\omega\rVert^2
    $$
    for all $v\in V$. Hence, $\iota^*(\ketbrap{\Psi}) = \sum_{\omega \in
    \Omega}\lVert \Psi_\omega\rVert^2 \omega$. But this convex combination has
    to lie on the facet $F$, so $\ket{\Psi_\omega}$ must vanish if $\omega
    \notin F$, which implies $\ket{\Psi}\in \mathcal{H}_F$. This shows (4).

    Let $\Gamma = \sum_i t_i \ketbrap{\Psi_i}\in \mathcal{E}$ be any ensemble
    state such that $\iota^*(\Gamma)\in F$. Then by linearity of $\iota^*$, we
    have $\sum_i t_i \iota^*(\ketbrap{\Psi_i})\in F$, so
    $\iota^*(\ketbrap{\Psi_i})\in F$ for each $i$. (4) then implies
    $\ket{\Psi_i}\in \mathcal{H}_F$, which shows (5).

    By definition, for $\rho\in F$,
    $$
    \begin{aligned}
      &   \Fp^{(F)}(\rho) = \min\Big\{\braket{\Psi|W|\Psi}\mid \ket{\Psi}\in
    \mathcal{H}_F, \lVert \Psi\rVert=1, \iota^*(\ketbrap{\Psi})=\rho\Big\}\\
      &   \Fp|_F(\rho) = \min\Big\{\braket{\Psi|W|\Psi}\mid \ket{\Psi}\in
    \mathcal{H}, \lVert \Psi\rVert=1, \iota^*(\ketbrap{\Psi})=\rho\Big\}.
    \end{aligned}
    $$
    By (4), we can replace the condition $\ket{\Psi}\in \mathcal{H}$ in the
    second line with $\ket{\Psi}\in \mathcal{H}_F$. This shows (6).
    Similarly, (7) follows from writing down the definitions of $\Fe^{(F)}$ and $\Fe|_F$ and
    applying (5).
  \end{proof}
\end{propo}

\begin{ex}
  Consider again the abelian spin functional theory constructed in
  \cref{ex:abelian_qubit_functional_theory}. For each $j\in
  \{1,2,\dotsb, N\}$, a pair of facets of $\conv(\Omega) = [-1,1]^N$ are given
  by $F_j^\pm := \{\rho \in [-1,1]^N\mid \rho_j = \pm 1\}$, and every 
  facet of $\conv(\Omega)$ is of this form. For each $(j, q)$, where $q=\pm 1$
  is a sign, the facet Hilbert space is $\mathcal{H}_{F_j^q} = \{\ket{\Psi}\in
  \mathcal{H}\mid Z_j\ket{\Psi} = q\ket{\Psi}\}$, the $q$-eigenspace of
  $Z_j$. It is easy to check that $e_j \in V$, the vector whose $j$-th
  entry is one and all other entries are zero, satisfies $\iota_{F_j^q}(e_j) =
  q\mathbbm{1}_{F_j^q}$.
\end{ex}

Finally, we introduce the notion of \textit{strong orthogonality} of two
vectors in $\mathcal{H}$, which will be relevant for the proof of the boundary
force in \cref{sec:abelian_boundary_force}. For any vector $\ket{\Psi}\in
\mathcal{H}$, we will denote its component in a weight space
$\mathcal{H}_\omega$ by $\ket{\Psi_\omega}$.

\begin{defn}
  Two vectors $\ket{\Phi}, \ket{\Phi'}$ are \textbf{strongly orthogonal} if their
  components are orthogonal in each weight space. That is,
  $\braket{\Phi_\omega|\Phi_\omega'}=0$ for all $\omega \in \Omega$.
\end{defn}


\begin{lem}
  \label{lem:strong_ortho_density_addition}
  For normalized and strongly orthogonal states $\ket{\Phi}$ and $\ket{\Phi'}$, define
  $\ket{\Psi} =x\ket{\Phi}+y\ket{\Phi'}$, where $|x|^2+|y|^2=1$. Then
  \begin{equation}
    \iota^*(\ketbrap{\Psi}) = |x|^2\iota^*(\ketbrap{\Phi})
      + |y|^2\iota^*(\ketbrap{\Phi'}).
  \end{equation}
  \begin{proof}
    $$
    \begin{aligned}
      &\left\langle \iota^*(\ketbrap{\Psi}), v\right\rangle
    = \sum_{\alpha,\beta\in \Omega}(\bar x\bra{\Phi_\alpha} + \bar y \bra{\Phi'_\alpha})
    \iota(v)(x\ket{\Phi_\beta}+y\ket{\Phi'_\beta})\\
    &= |x|^2\braket{\Phi|\iota(v)|\Phi}  + |y|^2 \braket{\Phi'|\iota(v)|\Phi'}
      + 2\Re\Big[\bar x y\sum_{\alpha,\beta\in\Omega}\braket{\beta,
      v}\underbrace{\braket{\Phi_\alpha|\Phi'_\beta}}_{=0}
      \Big]\\
      &= \left\langle|x|^2\iota^*(\ketbrap{\Phi}) + |y|^2\iota^*(\ketbrap{\Phi'}), v\right\rangle
    \end{aligned}
    $$
  \end{proof}
\end{lem}

\section{Unique $v$-representability}
In the abelian setting, slightly more can be said about unique
$v$-representability than what we proved in Section~\ref{sec:hk_theorem} for
any general functional theory. Recall that a density $\rho\in V^*$ is said to
be pure state $v$-representable if there exists an external potential $v\in V$ such that
$\iota(v)+W$ has $\rho$ as the density of a pure ground state
(Definition~\ref{def:representability}). It is uniquely $v$-representable if
any two such external potentials $v,v'$ satisfy $\iota(v)-\iota(v')\propto
\mathbbm{1}$ (Definition~\ref{def:unique_vrep}). In
Section~\ref{sec:hk_theorem}, after proving the weak Hohenberg-Kohn theorem, we
showed that its original version, which is stronger, is equivalent to the
statement that all pure state $v$-representable densities are uniquely so. 
We then gave a simple counterexample (Example~\ref{ex:HK_counterexample}),
which led to the conclusion that the strong HK theorem is false in
general. However, we were able to provide a partial remedy to this failure.
Namely, Theorem~\ref{thm:unique_vrep_regular_value} states that the strong HK
theorem can only fail at the critical values of the map
$\iota^*|_{\mathcal{P}}: \mathcal{P}\rightarrow \aff(\iota^*(\mathcal{P}))$,
which implies, through Sard's theorem, that the failure of the strong HK
theorem only occurs on a set of measure zero (relative to
$\aff(\iota^*(\mathcal{P}))$). When a functional theory is abelian, there is a
satisfactory characterization of the set of critical values, which we will give
now.

For any vector $\ket{\Psi}\in \mathcal{H}$, we denote by
$\Omega_{\Psi}\subset \Omega$ the \textit{support} of $\ket{\Psi}$, which
is by definition the set of weights $\omega$ such that $\ket{\Psi}$ has nonzero
component in the weight space $\mathcal{H}_\omega$. 
In other words, if
$\ket{\Psi} = \sum_{\omega \in \Omega} \ket{\Psi_\omega}$ is the weight space
decomposition, then $\Omega_{\Psi} = \{\omega\in \Omega \mid
\ket{\Psi_\omega}\neq 0\}$. Clearly, $\iota^*(\ket{\Psi}\!\bra{\Psi})\in
\conv(\Omega_{\Psi})$.

\begin{lem}
  \label{lem:derivative_of_density_map_abelian}
  For any pure state $\ketbrap{\Psi}\in \mathcal{P}$,
  $$
  \im(D_{\ket{\Psi}\!\bra{\Psi}}\iota^*|_{\mathcal{P}}) = 
  \overrightarrow{\aff(\Omega_{\Psi})}.
  $$
  \begin{proof}
    Let $\ket{\Psi} = \sum_{\omega \in \Omega_{\Psi}}\ket{\Psi_\omega}$
    be the weight space decomposition. Let $\ket{\phi}\in
    T_{\ketbrap{\Psi}}\mathcal{P}\cong \ket{\Psi}^\perp$ be any tangent vector.
    Then by Lemma~\ref{lem:derivative_of_density_map}, we have 
    $$
    \left\langle D_{\ketbrap{\Psi}}\iota^*|_{\mathcal{P}}(\ket{\phi}), v\right\rangle
    = \sum_{\omega\in \Omega_{\Psi}}
    \braket{\omega, v}2\Re\braket{\phi|\Psi_\omega}
    $$
    for all $v\in V$. It follows that
    $$
    D_{\ket{\Psi}\!\bra{\Psi}}\iota^*|_{\mathcal{P}}(\ket{\phi})
    =2\sum_{\omega \in \Omega_{\Psi}} \mathrm{Re}\braket{\phi|\Psi_\omega}\omega.
    $$
    This lies in $\overrightarrow{\aff(\Omega_\Psi)}$ because
    $\sum_{\omega\in \Omega_{\Psi}}\braket{\phi|\Psi_\omega} = \braket{\phi|\Psi} = 0$.
    Since $\ket{\phi}$ was an arbitrary tangent vector, this shows ``$\subset$''.

    Conversely, any $\alpha\in \overrightarrow{\aff(\Omega_{\Psi})}$ can
    be written as a linear combination of vectors of the form $\omega-\omega'$,
    where $\omega,\omega'\in \Omega_{\Psi}$. For any such pair, take
    $$
    \ket{\phi} = \frac{\ket{\Psi_\omega}}{\braket{\Psi_\omega|\Psi_\omega}} -
    \frac{\ket{\Psi_{\omega'}}}{\braket{\Psi_{\omega'}|\Psi_{\omega'}}}.
    $$
    It is straightforward to verify that $\braket{\phi|\Psi} =0$, 
    so $\ket{\phi}$ is a tangent vector to
    $\mathcal{P}$ at $\ketbrap{\Psi}$. We have
    $$
    D_{\ket{\Psi}\!\bra{\Psi}}\iota^*|_{\mathcal{P}}(\ket{\phi})
    = 2 (\omega - \omega').
    $$
    Since the derivative is linear, this shows ``$\supset$''.
  \end{proof}
\end{lem}

\begin{figure}[htb]
  \centering
  \subfloat[$4$ bosons on $3$ sites, $P=1$]{
  \includegraphics[width=.25\textwidth]{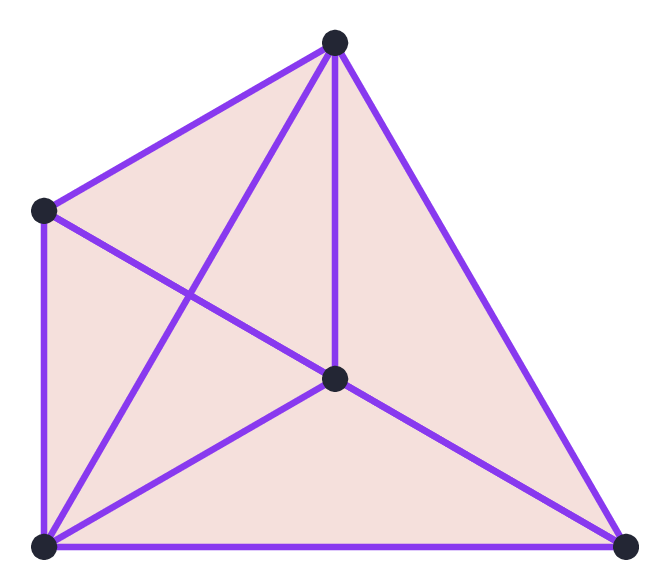}
  }
  \hspace{.5em}
  \subfloat[$6$ bosons on $3$ sites, $P=0$]{
  \includegraphics[width=.25\textwidth]{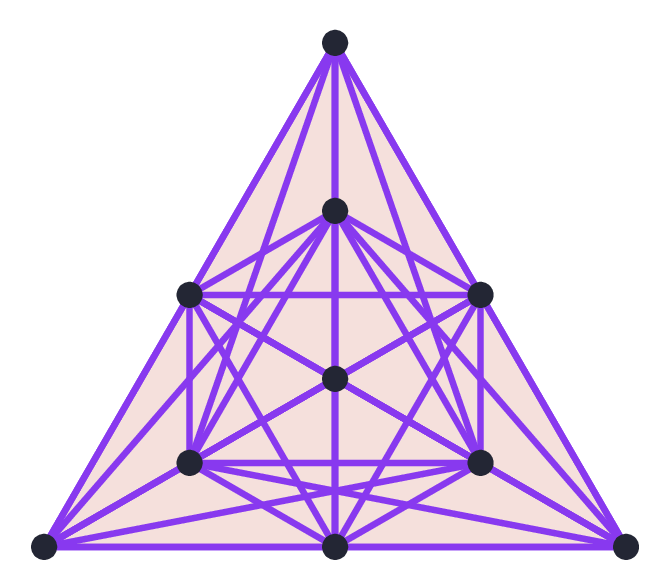}
  }
  \hspace{.5em}
  \subfloat[$6$ bosons on $3$ sites, $P=1$]{
  \includegraphics[width=.25\textwidth]{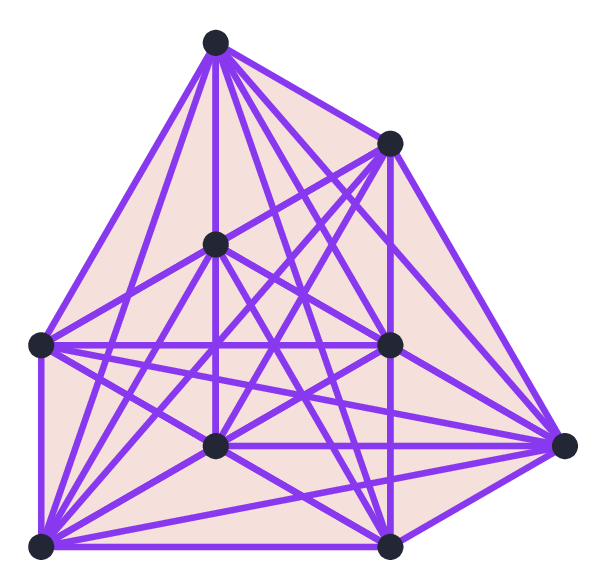}
  }

  \subfloat[$2$ fermions on $4$ sites]{
    \begin{minipage}{.27\textwidth}
      \centering
    \begingroup
\definecolor{critcolorA}{HTML}{8839ef}
\definecolor{critcolorB}{HTML}{e64553}
\definecolor{critcolorC}{HTML}{a0ed9f}

\definecolor{crust}{HTML}{11111b}

\newcommand{\opacityA}{0.66} 
\newcommand{\opacityB}{0.6} 

\begin{tikzpicture}
[x={(-0.80505575cm,-0.49347504cm)},y={(1.68783448cm,-0.23537552cm)},z={(0.00000000cm,1.78828989cm)}]


  \coordinate (F) at (1,0,0);
  \coordinate (B) at (-1,0,0);
  \coordinate (R) at (0,1,0);
  \coordinate (L) at (0,-1,0);
  \coordinate (U) at (0,0,1);
  \coordinate (D) at (0,0,-1);

  \coordinate (auxFLD) at (0.65,-0.8,-.97);
  \coordinate (auxBRU) at (-0.65,0.8,1.2);

  \coordinate (O) at (0,0,0);

  \path (auxFLD)--(auxBRU);

  \fill[crust] (B) circle (.2em);

  \draw[crust,line width=1.2pt] (B)--(U);
  \draw[crust,line width=1.2pt] (B)--(D);
  \draw[crust,line width=1.2pt] (B)--(L);
  \draw[crust,line width=1.2pt] (B)--(R);

  \fill[critcolorA,opacity=\opacityA] (B)--(D)--(O)--cycle;
  \fill[critcolorC,opacity=\opacityA] (B)--(L)--(O)--cycle;
  \fill[critcolorC,opacity=\opacityA] (B)--(R)--(O)--cycle;
  \fill[critcolorA,opacity=\opacityA] (B)--(U)--(O)--cycle;

  \draw[gray, line width=.7pt] (B)--(O);

  \fill[critcolorB,opacity=\opacityA] (U)--(L)--(O)--cycle;
  \fill[critcolorB,opacity=\opacityA] (D)--(R)--(O)--cycle;
  \fill[critcolorB,opacity=\opacityA] (D)--(L)--(O)--cycle;
  \fill[critcolorB,opacity=\opacityA] (U)--(R)--(O)--cycle;

  \fill[critcolorC,opacity=\opacityA] (F)--(L)--(O)--cycle;

  \draw[gray, line width=.7pt] (L)--(O);

  \fill[critcolorA,opacity=\opacityA] (F)--(D)--(O)--cycle;
  \fill[critcolorA,opacity=\opacityA] (F)--(U)--(O)--cycle;

  \draw[gray, line width=.7pt] (D)--(O);

  \fill[critcolorC,opacity=\opacityA] (F)--(R)--(O)--cycle;

  \draw[gray, line width=.7pt] (U)--(O);
  \draw[gray, line width=.7pt] (R)--(O);
  \draw[gray, line width=.7pt] (F)--(O);

  \draw[crust,line width=1.2pt] (U)--(R)--(D)--(L)--cycle;
  \draw[crust,line width=1.2pt] (F)--(U);
  \draw[crust,line width=1.2pt] (F)--(D);
  \draw[crust,line width=1.2pt] (F)--(L);
  \draw[crust,line width=1.2pt] (F)--(R);

  \fill[crust] (F) circle (.2em);
  \fill[crust] (L) circle (.2em);
  \fill[crust] (R) circle (.2em);
  \fill[crust] (U) circle (.2em);
  \fill[crust] (D) circle (.2em);


\end{tikzpicture}
\endgroup
    \end{minipage}
  }
  \hspace{.7em}
  \subfloat[$3$ qubits]{
    \begin{minipage}{.27\textwidth}
    \centering
    \begingroup

\definecolor{mauve}{HTML}{8839ef}
\definecolor{mauvedark}{HTML}{4b2978}
\definecolor{peach}{HTML}{e64553}
\definecolor{lightgreen}{HTML}{a0ed9f}
\definecolor{crust}{HTML}{11111b}

\newcommand{\opacityA}{0.85} 
\newcommand{\opacityB}{0.65} 

\begin{tikzpicture}[x={(-0.53383376cm,-0.32722409cm)},y={(1.11920575cm,-0.15607789cm)},z={(0.00000000cm,1.18581790cm)}]
  \coordinate (FRU) at (1,1,1);
  \coordinate (FRD) at (1,1,-1);
  \coordinate (FLU) at (1,-1,1);
  \coordinate (FLD) at (1,-1,-1);

  \coordinate (BRU) at (-1,1,1);
  \coordinate (BRD) at (-1,1,-1);
  \coordinate (BLU) at (-1,-1,1);
  \coordinate (BLD) at (-1,-1,-1);

  \coordinate (F) at (1,0,0);
  \coordinate (B) at (-1,0,0);
  \coordinate (R) at (0,1,0);
  \coordinate (L) at (0,-1,0);
  \coordinate (U) at (0,0,1);
  \coordinate (D) at (0,0,-1);

  \coordinate (FU) at (0.5,0,0.5);
  \coordinate (FD) at (0.5,0,-0.5);
  \coordinate (FR) at (0.5,0.5,0);
  \coordinate (FL) at (0.5,-0.5,0);

  \coordinate (BU) at (-0.5,0,0.5);
  \coordinate (BR) at (-0.5,0.5,0);

  \coordinate (BD) at (-0.5,0,-0.5);
  \coordinate (BL) at (-0.5,-0.5,0);

  \coordinate (RD) at (0,0.5,-0.5);
  \coordinate (LU) at (0,-0.5,0.5);

  \coordinate (RU) at (0,0.5,0.5);

  %
  %
  %

  \draw[gray,line width=.7pt] (BLU)--(BU)--(BRU);

  \draw[crust,line width=1.2pt] (BLD)--(BLU);
  \draw[crust,line width=1.2pt] (BLD)--(FLD);
  \draw[crust,line width=1.2pt] (BLD)--(BRD);

  \draw[gray,line width=.7pt] (U)--(B)--(D);
  \draw[gray,line width=.7pt] (L)--(B)--(R);

  \draw[crust,thick] (FLU)--(BLD)--(FRD)--cycle;

  \draw[crust,thick] (BRU)--(BLD);

  \draw[crust,thick] (BLU)--(FLD)--(BRD)--cycle;

  \fill[crust] (BRU) circle (.2em);
  \fill[crust] (BLU) circle (.2em);
  \fill[crust] (BLD) circle (.2em);

  \fill[peach!15,opacity=\opacityA] (BRD)--(FLD)--(BLU)--cycle;  

  \fill[mauve!5,opacity=\opacityA] (FRD)--(BRU)--(BLD)--cycle;  

  \fill[mauve!25,opacity=\opacityA] (BRU)--(FLU)--(BLD)--cycle;  

  \fill[mauvedark,opacity=\opacityA] (BLD)--(L)--(D)--cycle;  

  \fill[lightgreen,opacity=\opacityB] (BRU)--(BU)--(BLU)--cycle;


  \draw[gray,line width=.7pt] (L)--(U)--(R)--(D)--cycle;


  \draw[gray,line width=.7pt] (FRD)--(RD)--(BRD);
  \draw[gray,line width=.7pt] (FLU)--(LU)--(BLU);

  \draw[gray,line width=.7pt] (BRU)--(BR)--(BRD);


  \fill[lightgreen,opacity=\opacityB] (BRU)--(BR)--(BRD)--cycle;

  \fill[lightgreen,opacity=\opacityB] (BLD)--(BD)--(BRD)--cycle;

  \fill[peach!5,opacity=\opacityA] (BLU)--(U)--(B)--cycle;  
  \fill[peach!5,opacity=\opacityA] (FRU)--(R)--(U)--cycle;  
  \fill[peach!5,opacity=\opacityA] (BRD)--(R)--(B)--cycle;  

  \fill[peach,opacity=\opacityA] (BRD)--(R)--(D)--cycle;  

  \fill[lightgreen,opacity=\opacityB] (BRD)--(RD)--(FRD)--cycle;

  \fill[mauvedark,opacity=\opacityA] (FRD)--(F)--(D)--cycle;  

  \fill[mauve,opacity=\opacityA] (FRD)--(F)--(R)--cycle;  
  \fill[mauve,opacity=\opacityA] (BRU)--(R)--(U)--cycle;  

  \fill[peach!10,opacity=\opacityA] (BLU)--(L)--(U)--cycle;  
  \fill[lightgreen,opacity=\opacityB] (FLU)--(LU)--(BLU)--cycle;
  \fill[mauve,opacity=\opacityA] (FLU)--(F)--(U)--cycle;  
  \fill[mauvedark,opacity=\opacityA] (FLU)--(F)--(L)--cycle;  

  \fill[lightgreen,opacity=\opacityB] (FRU)--(FU)--(FLU)--cycle;

  \fill[peach!10,opacity=\opacityA] (FRU)--(F)--(U)--cycle;  
  \fill[peach!10,opacity=\opacityA] (FLD)--(F)--(L)--cycle;  


  \fill[lightgreen,opacity=\opacityB] (FRU)--(RU)--(BRU)--cycle;
  \draw[gray,line width=.7pt] (FRU)--(RU)--(BRU);

  \fill[peach,opacity=\opacityA] (FRU)--(F)--(R)--cycle;  
  \fill[peach,opacity=\opacityA] (FLD)--(F)--(D)--cycle;  

  \draw[gray, line width=.7pt] (U)--(F)--(D);
  \draw[gray, line width=.7pt] (L)--(F)--(R);

  \draw[crust,thick] (FLU)--(BRU)--(FRD)--cycle;

  \draw[crust,thick] (FRU)--(BLU);
  \draw[crust,thick] (FRU)--(BRD);
  \draw[crust,thick] (FRU)--(FLD);

  \draw[gray, line width=.7pt] (FRU)--(FR)--(FRD);
  \draw[gray, line width=.7pt] (FLU)--(FL)--(FLD);

  \draw[gray, line width=.7pt] (FLU)--(FU)--(FRU);
  \draw[gray, line width=.7pt] (FLD)--(FD)--(FRD);

  \fill[lightgreen,opacity=\opacityB] (FRU)--(FR)--(FRD)--cycle;
  \fill[lightgreen,opacity=\opacityB] (FLD)--(FD)--(FRD)--cycle;
  \fill[lightgreen,opacity=\opacityB] (FLU)--(FL)--(FLD)--cycle;
  \fill[lightgreen,opacity=\opacityB] (FLU)--(FU)--(FRU)--cycle;

  \draw[crust,line width=1.2pt] (FLU)--(FRU)--(FRD)--(FLD)--cycle;
  \draw[crust,line width=1.2pt] (FLU)--(BLU)--(BRU)--(FRU);
  \draw[crust,line width=1.2pt] (BRU)--(BRD)--(FRD);

  \fill[crust] (BRD) circle (.2em);
  \fill[crust] (FLD) circle (.2em);
  \fill[crust] (FRD) circle (.2em);
  \fill[crust] (FRU) circle (.2em);
  \fill[crust] (FLU) circle (.2em);

\end{tikzpicture}

\endgroup
    \end{minipage}
  }
  \caption{Critical values of several abelian functional theories. (a)--(c):
  translation invariant bosonic functional theories discussed in
  \cref{chap:momentum_rdmft}, with $d=3$ sites. The critical values are
  the points lying in the convex hull of two weights, indicated in purple.
  The weight structure of (d) is that of two (spinless) fermions placed on four
  lattice sites with local external potentials. The six vertices correspond to
  the six Slater determinants $\ket{1,1,0,0}, \ket{1,0,1,0}, \dotsb,
  \ket{0,0,1,1}$. In (e), we consider the abelian functional theory of three two-level spins
  (\cref{ex:N_qubits_abelian}). The set of weights is $\{-1,1\}^3$,
  which forms the vertices of a cube (see also \cref{fig:spin_hypercube}).
  Since $\dim \Omega=3$ for both (d) and (e), the critical values
  are the points lying in planes spanned by the weights. In both cases, the
  boundary $\partial \conv(\Omega)$ (points on which are always critical
  values) is omitted for ease of visualization. The set of critical values of
  (d) (minus the boundary) is the union of three squares, shown in purple, red, and green.
  For (e), it is the union of two hollow tetrahedra (red and purple) and 
  six rectangles (green).
  }
  \label{fig:abelian_critical_values_example}
\end{figure}
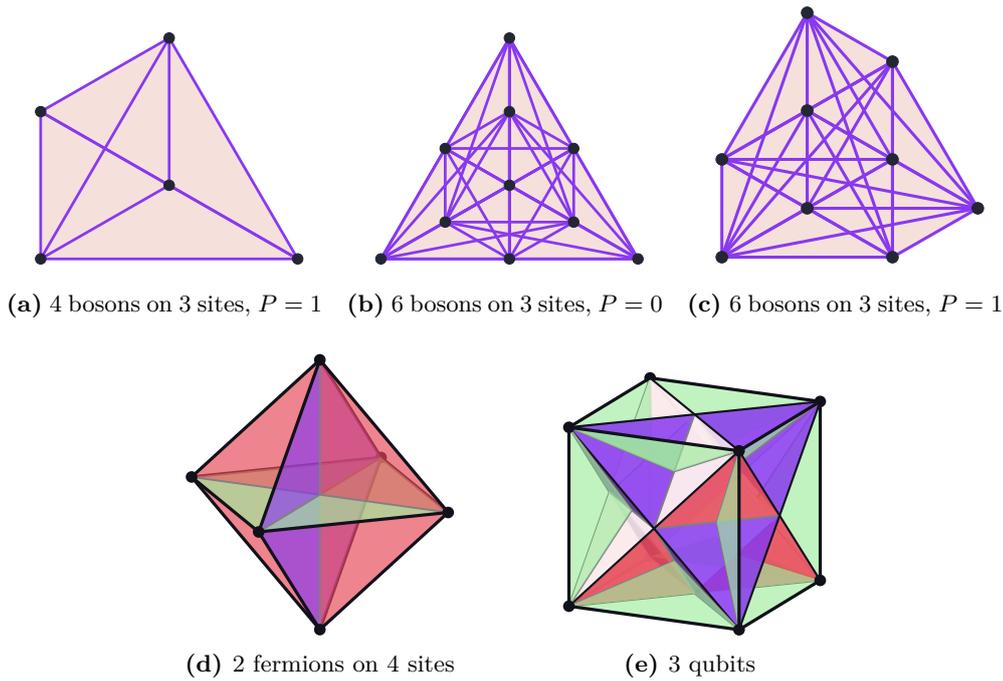

\begin{propo}
  \label{thm:abelian_critical_value_charcterization}
  Let $(V, \mathcal{H}, \iota, W)$ be an abelian functional theory. Let $\rho
  \in \conv(\Omega)$ be a representable density. Then $\rho$ is a critical value
  of the map $\iota^*|_{\mathcal{P}}: \mathcal{P}\rightarrow \aff(\Omega)$
  if and only if $\rho$ lies in the convex hull of $\dim\conv(\Omega) = \dim V - \dim
  \iota^{-1}(\mathrm{span}\{\mathbbm{1}\})$ weights. (See also Lemma~3.3 of Ref.~\cite{walterMultipartiteQuantumStates2014}.)
  \begin{proof}
    If $\rho$ is a critical value, then there is a pure state $\ketbrap{\Psi}$
    such that $D_{\ketbrap{\Psi}}\iota^*|_{\mathcal{P}}$ is not surjective
    onto $\overrightarrow{\aff(\Omega)}$. By
    Lemma~\ref{lem:derivative_of_density_map_abelian}, we have
    $\overrightarrow{\aff(\Omega_{\Psi})}
    \subsetneq\overrightarrow{\aff(\Omega)}$. It follows that
    $\dim\conv(\Omega_{\Psi}) < \dim \conv(\Omega)$. Since $\rho \in
    \conv(\Omega_{\Psi})$, it is a convex combination of $\dim
    \conv(\Omega_{\Psi}) + 1 \le \dim \conv(\Omega)$ weights by
    Carath\'eodory's theorem.

    Conversely, suppose $\rho$ lies in the convex hull of $\dim \conv(\Omega)$
    weights. In other words, there is a subset $\Omega'\subset \Omega$ and
    coefficients $t_\omega\ge 0$ such that $\sum_{\omega\in \Omega'}t_\omega \omega
    = \rho$, $\sum_{\omega\in \Omega'}t_{\omega} = 1$, and $|\Omega'|\le \dim \conv(\Omega)$.
    Consider the state $\ket{\Psi}:= \sum_{\omega\in
    \Omega'}\ket{\Psi_\omega}$, where the $\ket{\Psi_\omega}$ satisfy
    $\ket{\Psi_\omega}\in\mathcal{H}_\omega$ and $\lVert \Psi_\omega\rVert^2 =
    t_\omega$ for all $\omega \in \Omega'$. Then again by
    Lemma~\ref{lem:derivative_of_density_map_abelian}
    we have $\im(D_{\ket{\Psi}\!\bra{\Psi}}\iota^*|_{\mathcal{P}}) =
    \overrightarrow{\aff(\Omega_{\Psi})} =
    \overrightarrow{\aff(\Omega')}$. But $\dim
    \overrightarrow{\aff(\Omega')}\le |\Omega'|-1 \le \dim \conv(\Omega)-1$, so
    $D_{\ket{\Psi}\!\bra{\Psi}}\iota^*|_{\mathcal{P}}$ cannot be surjective
    onto $\overrightarrow{\aff(\Omega)}$. This shows that $\rho$ is a critical value.
  \end{proof}
\end{propo}

\begin{rem}
  By Carath\'eodory's theorem, any representable density $\rho$ can be written
  as a convex combination of $\dim \conv(\Omega) + 1$  weights. However, almost none
  can be written as that of $\dim \conv(\Omega)$ weights, since those which can be
  expressed this way all lie in a finite union of hyperplanes of dimension
  $\dim \conv(\Omega)-1$, which intersects $\conv(\Omega)$ in a set of measure zero. 
  Alternatively, we can arrive at the same conclusion by applying Sard's
  theorem and Proposition~\ref{thm:abelian_critical_value_charcterization}.
\end{rem}
\begin{rem}
  A special case of
  Proposition~\ref{thm:abelian_critical_value_charcterization}
  is when $\rho$ lies on a facet $F$ of $\conv(\Omega)$. In this case, $\rho$
  can be written as a convex combination of $\dim F+1 = \dim \conv(\Omega)$ weights,
  hence $\rho$ must be a critical value. On the other hand, we know this quite
  intuitively already from the fact that $\rho$ lies on a facet of
  $\conv(\Omega)$: the derivative of $\iota^*|_{\mathcal{P}}$ at any pure
  state in the preimage of $\rho$ cannot be surjective onto
  $\overrightarrow{\aff(\Omega)}$, because otherwise we would be able to walk
  outside of $\conv(\Omega)$ by perturbing the pure state in the right direction!
\end{rem}

Combining Proposition~\ref{thm:abelian_critical_value_charcterization} and
Proposition~\ref{thm:unique_vrep_regular_value}, we get:
\begin{cor}
  Let $\rho\in \conv(\Omega)$ be a representable density. If $\rho$ is pure
  state $v$-representable and does not lie in the convex hull of $\dim \Omega$
  weights, then $\rho$ is uniquely $v$-representable.
\end{cor}

In \cref{fig:abelian_critical_values_example}, we show the sets of critical
values in various abelian functional theories with $\dim \conv(\Omega) = 2$ or
$\dim \conv(\Omega)=3$. By \cref{thm:unique_vrep_regular_value}, these lower
dimensional polytopes are where the strong Hohenberg-Kohn theorem
(\cref{thm:strong_HK}) could potentially fail.

\section{Boundary Forces}
\label{sec:abelian_boundary_force}

Finally, we turn our attention to boundary forces, with the main objective
being to generalize and formalize the discussion in
\cref{sec:generalized_bec_force}.
The following theorem is the main result of this section:
\begin{thm}
  \label{thm:boundary_force_formula}
  Let $(V, \mathcal{H}, \iota, W)$ be an abelian functional theory. Let
  $F\subset \conv(\Omega)$ be a facet and $\rho_* \in F$ a density which is a
  regular value of the facet density map. In other words, $\rho_*$ does not
  belong to the convex hull of $\dim F$ weights
  (Proposition~\ref{thm:abelian_critical_value_charcterization}). Take an
  inward-pointing vector $\eta \in \overrightarrow{\aff(\Omega)}$, and take
  $S\in V$ perpendicular to $F$ such that $\braket{\eta, S}=1$. Then
  \begin{equation}
    \label{eq:boundary_force_formula}
    \lim_{\epsilon\rightarrow 0^+}\frac{\Fp(\rho_* + \epsilon\eta) -
    \Fp(\rho_*)}{\sqrt{\epsilon}} = -2\left[\sum_{\omega \in \Omega \setminus
    \Omega_F}\frac{\lVert \Pi_{\omega}W\ket{\Phi}\rVert^2}{\braket{\omega, S} -
    \braket{\gamma, S}}\right]^{\frac{1}{2}}
  \end{equation}
  for any $\gamma \in \aff(F)$, where $\ket{\Phi}$ is a minimizer of the
  constrained search \eqref{eq:def_pure_functional} at $\rho_*$. If there are
  multiple minimizers, $\ket{\Phi}$ is taken to minimize
  Eq.~\eqref{eq:boundary_force_formula}.
\end{thm}

\begin{rem}
	With the notation of \cref{chap:momentum_rdmft}, the facet $F$ corresponds to
	an inequality constraint $D^F(\rho) = \braket{\rho, S} - \nu^F \ge 0$.
	It is easy to verify that $\nu^F= \braket{\gamma, S}$,
	so the denominator in Eq.~\eqref{eq:boundary_force_formula}
	is nothing but $D^F(\omega)$.
\end{rem}

Equation~\eqref{eq:boundary_force_formula} implies $\Fp(\rho_* +
\epsilon\eta)\approx\Fp(\rho_*) - \sqrt{\epsilon}\mathcal{G}$ for small
$\epsilon$, where $-\mathcal{G}$ is the right hand side of
Eq.~\eqref{eq:boundary_force_formula}, and is
the \textit{repulsion strength} in \cref{sec:generalized_bec_force}.

In the rest of this section, we will present two derivations of this formula.
The first one applies perturbation theory to ground states of Hamiltonians
$H(v)$, where $v$ is taken to be large in the direction of $S$, to deduce
properties of ground states mapping to regions near the facet $F$. This approach
has the advantage of being more or less intuitive and transparent. However, it
has a crucial flaw: in order to use perturbation theory at all, we must assume,
among other things, pure state $v$-representability. What this means is that we
are actually making a statement about the Hohenberg-Kohn functional $\FHK$, and
not about the pure functional $\Fp$ as in the statement of
Theorem~\ref{thm:boundary_force_formula}. Thanks to
Proposition~\ref{thm:functionals_agree_on_v_rep}, we know that $\FHK$ and $\Fp$
at least agree on their common domain, so this approach is still valid for
$\Fp$ as long as we stay within the set of pure state $v$-representable
densities, which is in general a proper subset of $\conv(\Omega)$.

The second approach, which is much more involved, proves
Theorem~\ref{thm:boundary_force_formula} directly from the definition of the
pure functional $\Fp$ (Definition~\ref{def:pure_ensemble_functional}). In other
words, the proof will be completely agnostic to $v$-representability issues.
Despite the considerable number of technicalities one faces,
the intuition is very simple: whenever a density $\rho$ is close to the fixed
density $\rho_*$ on the facet $F$, say at ``distance'' $\epsilon$ (of course,
we do not have an inner product on $V^*$, so we will have to be more precise),
any state $\ket{\Psi}$ such that $\iota^*(\ketbrap{\Psi})=\rho$ must assume
such a form where its overlap with $\mathcal{H}_\omega$ also approaches zero at
a rate proportional to $\sqrt{\epsilon}$, where $\omega$ is any weight not on
the facet. In that regime, the zeroth order contribution to the functional must
be from the facet Hilbert space $\mathcal{H}_F$ (see
Definition~\ref{def:facet_functional_theory}), to which a correction of order
$\sqrt{\epsilon}$ from facet--off-facet interactions is to be added. This
latter term is then the source of the boundary force.

\subsection{Heuristic Derivation by Perturbation Theory}
\label{sec:abelian_boundary_force_perturbation}
		Here, we provide a quick derivation of
		Eq.~\eqref{eq:boundary_force_formula} based on perturbation theory. As a
		precursor to the much more rigorous proof in
		Section~\ref{sec:boundary_force_rigorous_proof}, the present section will
		not dwell on precise language. The aim is to gain some intuition for why
		the boundary force should exist, and why the derivative of the functional
		should diverge as $1/\sqrt{\epsilon}$ as we get close to a facet. 

		Throughout this section, we will assume harmlessly that the functional domain
		$\iota^*(\mathcal{P})=\iota^*(\mathcal{E})$ has full dimension in $V^*$.
		That is, we will assume $\dim \conv(\Omega) = \dim V^*$, or equivalently
		(Proposition~\ref{thm:dimension_of_conv_Omega}) that $\iota: V\rightarrow
		i\mathfrak{u}(\mathcal{H})$ is injective and $\mathbbm{1}\notin
		\iota(V)$. 
		We will make the following further (extremely strong) assumptions: 
  	\begin{enumerate}[label={(\arabic*)}]
			\item The densities in $\relint(\conv(\Omega))$ (which is just the usual
				interior) are pure state $v$-representable.
			\item $\Fp$ is differentiable on $\relint(\conv(\Omega))$.
			\item The ground state energy function $E(v) :=
						\min_{\Gamma\in \mathcal{E}}\braket{\Gamma, \iota(v)+W}$ is differentiable.
			\item The densities in $\relint(F)$ are all uniquely pure state $v$-representable 
				in the restricted functional theory on $F$.
		\end{enumerate}
 		By Proposition~\ref{thm:functionals_agree_on_v_rep}, (1) implies that the
 		three functionals agree on the relative interior:
		$$
			\Fp|_{\relint\conv(\Omega)} \equiv \Fe|_{\relint\conv(\Omega)}
			\equiv \FHK|_{\relint\conv(\Omega)}
		$$
		Hence, in this section, we will ignore the distinction between the
		three functionals, and simply write $\mathcal{F}$ for all three.
		Assumptions (2) and (3) together allow us to use the following: if 
		$\rho = \dstraight_v E$, then  $-v = \dstraight_\rho \mathcal{F}$.
		(If differentiability is not assumed, we would
		have to replace the derivatives with the super/sub-differential instead.)

		Set $U:= \ker(\eta) \subset V$ and $\tilde U:=S^\circ =
		\{\beta\in V^*| \braket{\beta, S} = 0\} \subset V^*$. We then get decompositions
		$$
			\begin{aligned}
				& V = \mathbb{R} S\oplus U \\
				 & V^* = \mathbb{R}\eta \oplus \tilde U.
			\end{aligned}
		$$
		Note that $\tilde U$ is just $\overrightarrow{\aff(F)}$. 
		Let $\pi_S: V\rightarrow \mathbb{R}S$ and $\pi_U: V\rightarrow U$ be the
		projections with respect to the given decomposition of $V$ (i.e.,
		$\pi_S|_{\mathbb{R}S} = \mathrm{Id}_{\mathbb{R}S}$ and $\ker \pi_S = U$,
		similarly for $\pi_U$). It is easy to show that $\pi_S(v) = \braket{\eta, v}S$.
		We are guided by the following intuition: for $v \in V$, we can write $v =
		v_S + v_U$ with respect to $V = \mathbb{R}S \oplus U$. If $v_S$ is sent to
		infinity in the correct direction, the corresponding density will tend to
		the facet $F\subset \conv(\Omega)$.

		For $v\in V$, we have $v = \pi_S(v) + \pi_U(v)$. The Hamiltonian can then
		be written as
		\begin{equation}
			\label{eq:facet_hamiltonian}
			H(v) =\iota(v) + W = \iota(\pi_S(v)) + \iota(\pi_U(v)) + W.
		\end{equation}
		The subspace of ground states of such a Hamiltonian in the limit
		$\braket{\eta, \pi_S(v)} = \braket{\eta, v}\rightarrow +\infty$ is exactly the facet Hilbert
		space $\mathcal{H}_F\subset\mathcal{H}$, with energy
		\begin{equation}
			\label{eq:abelian_perturb_zeroth_order}
			E_0 = \braket{\Phi|\iota(\pi_S(v))|\Phi} = \braket{\alpha,\pi_S(v)} =
			\braket{\gamma, \pi_S(v)},
		\end{equation}
		where $\ket{\Phi}$ is any normalized state in $\mathcal{H}_F$ and
		$\alpha\in \Omega_F$ is any weight on the facet.		
		The last equality holds because $\gamma - \alpha\in \overrightarrow{\aff(F)}$, and
		$S$ is, by assumption, perpendicular to the facet $F$.
		We now apply perturbation theory to the Hamiltonian
		\eqref{eq:facet_hamiltonian}, treating $\iota(\pi_U(v)) + W$ as a small
		perturbation. Since the ground states are degenerate, the first order
		correction is the lowest eigenvalue of $\Pi_F^{\phantom{\dagger}}[\iota(\pi_U(v)) +W]
		\Pi_F^\dagger$ (i.e., the perturbing Hamiltonian restricted/compressed to
		the ground state subspace $\mathcal{H}_F$), which is nothing but the ground
		state energy of the external potential $\pi_U(v) \in V$ in the restricted
		functional theory on $F$, corresponding to a unique ground state
		$\ket{\Psi(v)}\in \mathcal{H}_F$ by assumption (4). That is, we have the
		first order energy correction
		\begin{equation}
			\label{eq:abelian_perturb_first_order}
			E_1 = E_F(\pi_U(v)),
		\end{equation}
		where $E_F: V\rightarrow\mathbb{R}$ is the ground state energy function of
		the facet functional theory. It is now straightforward to obtain the second
		order correction to the ground state energy: the excited states are the weight states
		not lying on $F$, with each weight space $\mathcal{H}_\omega$ having energy
		$\braket{\omega, \pi_S(v)}$. Combining the second order correction with
		Eq.~\eqref{eq:abelian_perturb_zeroth_order} and
		Eq.~\eqref{eq:abelian_perturb_first_order}, we get the ground state energy
		up to second order:
		$$
			E(v) \approx \braket{\gamma, \pi_S(v)} + E_F(\pi_U(v)) + 
					\sum_{\omega\in \Omega \setminus \Omega_F}
					\frac{\overbrace{\lVert \Pi_\omega W\ket{\Phi(v)}\rVert^2}^{=:r_\omega(v)}}
					{\braket{\gamma, \pi_S(v)}-\braket{\omega,\pi_S(v)}}.
			$$
		Note that $\braket{\gamma, \pi_S(v)} + E_F(\pi_U(v))$ is simply $E_F(v)$,
		due to the fact that $\iota_F(\pi_S(v)) = \braket{\gamma,\pi_S(v)}
		\mathbbm{1}_{\mathcal{H}_F}$.
		It follows that
		\begin{equation}
			E(v) \approx E_F(v) + \sum_{\omega\in \Omega\setminus \Omega_F}
			\frac{r_\omega(v)}{{\braket{\gamma, \pi_S(v)}-\braket{\omega,\pi_S(v)}}}.
		\end{equation}
		Taking the derivative at $v\in V$, we get
		\begin{equation}
			\label{eq:facet_derivative_approx}
			\begin{aligned}
				\dstraight_v E  \approx 
				\underbrace{\dstraight_v E_F}_{\in \iota_F^*(\mathcal{E}_F) = F}
				-\underbrace{\sum_{\omega\in \Omega\setminus \Omega_F}
					\frac{r_\omega(v)}
					{\braket{\gamma-\omega, \pi_S(v)}^2}
					(\gamma\circ\pi_S - \omega\circ \pi_S)
				}_{\in \mathbb{R}\eta}
				+ \underbrace{\sum_{\omega\in \Omega\setminus \Omega_F} 
				\frac{\dstraight_v r_\omega}{\braket{\gamma-\omega,\pi_S(v)}}}_{\in \tilde U},
			\end{aligned}
		\end{equation}
		which is valid only for $\braket{\eta, v} \rightarrow+\infty$.
		The second term is proportional to $\eta$ because it vanishes on $\ker \eta
		= U$. Similarly, the third term belongs to $\tilde U$ because
		$$
			\braket{\dstraight_{v}r_\omega, S} = \doverd{t}\Big|_{t=0}
			r_{\omega}(v+tS) = 0,
		$$
		where the last equality follows from $\ket{\Phi(v)} = \ket{\Phi(v+tS)}$
		since $\iota_F(e)$ is proportional to the identity in $\mathcal{H}_F$.
		Equation~\eqref{eq:facet_derivative_approx} has an intuitive
		interpretation. As $\braket{\eta, v}\rightarrow +\infty$, we are driving the ground
		state density (left hand side) close to the facet $F$, and this equation
		tells us how the ground state density $\rho$ approaches $F$: the
		first term on the right hand side, which is the derivative of the energy
		function of the facet functional theory, can be interpreted as a density on
		$F$, and the last two terms represent how $\rho$ deviates from the density
		on the facet, with the deviation decomposed into a ``longitudinal''
		component (in $\mathbb{R}\eta$) and a ``transverse'' one (in $\tilde U=
		\overrightarrow{\aff(F)}$). Since $\braket{\gamma-\omega, \pi_S(v)} =
		\braket{\gamma-\omega, S}\cdot\braket{\eta, v}$, we immediately see that the
		longitudinal component is inversely proportional to $\braket{\eta, v}^2$.
		In other words, in order for $\rho$ to be $\epsilon$-close to the facet
		$F$, we must have $\braket{\eta, v}\propto 1/\sqrt{\epsilon}$. But the
		external potential $v$ is just the negative of the derivative of the
		functional $\mathcal{F}$ at $\rho$, so $\dstraight_\rho \mathcal{F}$ must
		behave like $1/\sqrt{\epsilon}$. We will now try to make this argument more precise.

		For $b$ small enough, define a path $c: (0, b)\rightarrow V$ by
		$\dstraight_{c(\epsilon)}E = \rho_* + \epsilon \eta$. We know this
		characterizes $c$ uniquely because $v\mapsto \dstraight_v E$ and $\rho
		\mapsto -\dstraight_\rho \mathcal{F}$ are inverses of each other.
		Applying Eq.~\eqref{eq:facet_derivative_approx} to
		$\dstraight_{c(\epsilon)}E$ and rearranging, we get
		$$
			\begin{aligned}
				&\underbrace{\dstraight_{c(\epsilon)}E_F - \rho_*
					+ \sum_{\omega\in \Omega\setminus \Omega_F}
					\frac{\dstraight_{c(\epsilon)}r_\omega}
					{\braket{\gamma-\omega,\pi_S(c(\epsilon))}}
					}_{\in \tilde U}\\
				&\hspace{3em}-  \underbrace{\sum_{\omega\in \Omega\setminus \Omega_F}
					\frac{r_\omega(c(\epsilon))}
					{\braket{\gamma-\omega,\pi_S(c(\epsilon))}^2}
					((\gamma-\omega)\circ\pi_S) - \epsilon \eta}_{\in \mathbb{R}\eta} 
				   \approx 0 \in V^*.
			\end{aligned}
		$$
		Since $(\mathbb{R}\eta) \cap \tilde U = 0$, the
		two underlined groups of terms must vanish separately. For the second group
		of terms, we evaluate everything on $S$, keeping in mind that
		$\braket{\eta,S} = 1$ and $\pi_S(S) = S$:
		\begin{equation}
			\label{eq: facet_pot_curve_components}
			\left\{
				\begin{aligned}
				 	&0 \approx \dstraight_{c(\epsilon)}E_F - \rho_* 
				 		+ \sum_{\omega\in \Omega\setminus\Omega_F} 
					\frac{\dstraight_{c(\epsilon)}r_\omega}
					{\braket{\gamma-\omega,\pi_S(c(\epsilon))}}\\
				 	&\epsilon \approx -\sum_{\omega\in \Omega\setminus\Omega_F}
							\frac{r_\omega(c(\epsilon))}
							{\braket{\gamma-\omega,\pi_S(c(\epsilon))}^2} 
							\braket{\gamma-\omega,S}				
			  \end{aligned} 
			\right.
		\end{equation}
		Write $c(\epsilon) = \tau(\epsilon)S + c_U(\epsilon)$ where $\tau(\epsilon)
			\in \mathbb{R}$ and $c_U(\epsilon)\in U$. Then the second equation of
		Eq.~\eqref{eq: facet_pot_curve_components} becomes
		$$
			\begin{aligned}
				  \epsilon \approx \frac{1}{\tau(\epsilon)^2}\sum_{\omega\in \Omega\setminus \Omega_F}
				\frac{r_\omega(c(\epsilon))}{\braket{\omega-\gamma, S}}
				\Rightarrow \tau(\epsilon)\approx \frac{1}{\sqrt{\epsilon}} 
				\left(\sum_{\omega\in \Omega\setminus \Omega_F} \frac{r_\omega(c(\epsilon))}{\braket{\omega-\gamma, S}}\right)^{\frac{1}{2}}.
			\end{aligned}
		$$
		It follows that
		$$
		\begin{aligned}
			&\doverd{\epsilon}\mathcal{F}(\rho_*+\epsilon\eta) = \braket{\eta, \dstraight_{\rho_* + \epsilon \eta} \mathcal{F}}
			= \braket{\eta, \dstraight_{\dstraight_{c(\epsilon)}E}\mathcal{F}} = -\braket{\eta, c(\epsilon)}
			= - \tau(\epsilon) \\
			&\approx -\frac{1}{\sqrt{\epsilon}}\left(\sum_{\omega\in \Omega\setminus
			\Omega_F}\frac{\lVert \Pi_\omega W\ket{\Phi(c(\epsilon))}\rVert^2}{\braket{\omega-\gamma, S}}\right)^{\frac{1}{2}}.
		\end{aligned}
		$$
		We may replace $\ket{\Phi(c(\epsilon))}$ with its limit as $\epsilon
		\rightarrow 0$, potentially making an error of higher order in $\epsilon$.
		The limit is nothing but $\ket{\Phi}$, the unique state corresponding to
		the density $\rho_*\in F$ in the facet functional theory. Therefore,
		for $\epsilon$ small,
		\begin{equation}
			\doverd{\sqrt{\epsilon}}\mathcal{F}(\rho_*+\epsilon\eta)
			= 2\sqrt{\epsilon}\doverd{\epsilon}\mathcal{F}
			\approx -2
				\left(\sum_{\omega\in \Omega\setminus \Omega_F}\frac{\lVert \Pi_\omega
				W\ket{\Phi}\rVert^2}{\braket{\omega-\gamma, S}}\right)^{\frac{1}{2}},
		\end{equation}
		which is the claimed boundary force formula
		(Eq.~\eqref{eq:boundary_force_formula}).
 

\subsection{Proof by Constrained Search}
\label{sec:boundary_force_rigorous_proof}
Here, we present a rigorous proof of Theorem~\ref{thm:boundary_force_formula}
by constrained search. Since the proof is long, it will bit split into sections
for the sake of comprehensibility. 


\subsubsection{Step I: The ``No Mixing Lemma''}
The first step is to have good control over minimizers $\ket{\Phi}$ of the
constrained search on the facet. Informally speaking, we will show that ``a
pure state which is a minimizer of the constrained search does not mix with a
state which is strongly orthogonal to it under the interaction $W$''. Since
this statement is entirely general, in the sense that it applies to any density
lying in the relative interior of the functional domain (convex hull of all
weights) of any abelian functional theory, we will not consider the facet
functional theory here. In Step III, we will apply this result to the facet
functional theory.

\begin{lem}[No Mixing Lemma]
  \label{lem:minimizing_state_zero_coefficient}
  Let $(V,\mathcal{H},\iota,W)$ be any abelian functional theory with weights $\Omega$.
  Suppose $\rho_*$ lies in the relative interior of $\conv(\Omega)$. Let
  $\ket{\Phi}$ be a minimizer of the pure state constrained search at $\rho_*$.
  In other words, $\iota^*(\ketbrap{\Phi})=\rho_*$ and
  $\braket{\Psi|W|\Psi} \ge \braket{\Phi |W|\Phi}$ for all states $\ket{\Psi}$
  such that $\iota^*(\ketbrap{\Psi})=\rho_*$. 
  Let $\ket{\delta}$ be a weight vector. If $\braket{\delta|\Phi}= 0$, then
  $\braket{\delta|W|\Phi} = 0$.
  \begin{proof}
    We argue by contradiction. The strategy is as follows: suppose
    $\braket{\delta|W |\Phi}\neq 0$. We will find a state $\ket{\Psi}$ such
    that $\iota^*(\ketbrap{\Psi})=\rho_*$ and $\braket{\Psi|W|\Psi} <
    \braket{\Phi|W|\Phi}$, which contradicts the assumption that $\ket{\Phi}$
    is a minimizer of the constrained search.
    Finding such a state will be achieved by constructing a one-parameter
    family of states $\ket{\Psi(t)}$ depending smoothly on $t$ with
    $\ket{\Psi(0)} = \ket{\Phi}$, and showing that
    $\iota^*(\ketbrap{\Psi(t)}) = \rho_*$ for all $t$ and that
    $\doverd{t}|_{t=0}\braket{\Psi(t)|W|\Psi(t)} \neq 0$.

    Let $\omega\in \Omega$ denote the weight of $\ket{\delta}$. In other words,
    $\iota(v)\ket{\delta} = \braket{\omega,v}\ket{\delta}$ for all external
    potentials $v\in V$. We will distinguish between two cases depending on
    whether $\omega$ coincides with $\rho_*$ or not. This is technically not
    necessary as the argument in the latter case also applies to the former,
    but we nevertheless choose to present the proof this way for more
    clarity. Without loss of generality, we assume $\lVert\delta\rVert=1$. \\

    \noindent
    \textit{Case (a)}: $\omega = \rho_*$\\
    For any angle $\phi$, define $\ket{\Psi(t)}:= te^{i\phi}\ket{\delta} +
    \sqrt{1-t^2}\ket{\Phi}$. Then $\iota^*(\ket{\Psi(t)}\!\bra{\Psi(t)}) =
    t^2\omega + (1-t^2)\omega = \omega = \rho_*$. The derivative of the
    expectation value $\braket{\Psi(t)|W|\Psi(t)}$ at $t=0$ is given by
    \begin{equation}
      \doverd{t}\Big|_{t=0}\braket{\Psi(t)|W|\Psi(t)} =
      2\Re\left[e^{-i\phi}\braket{\delta|W | \Phi}\right].
    \end{equation}
    Since $\braket{\delta|W|\Phi}\neq 0$, there is a suitable choice of angle
    $\phi$ such that the expression above does not vanish.\\

    \noindent
    \textit{Case (b)}: $\omega \neq \rho_*$\\
    If $\omega\neq \rho_*$, consider the ray starting from $\omega$ extending
    in the direction of $\rho_*$. This ray intersects the boundary of
    $\conv(\Omega)$ in some point $\mu$. Since the boundary of $\conv(\Omega)$
    is exactly the union of all its facets, $\mu$ belongs to at least one
    facet, which we will call $H$ (see Figure~\ref{fig:weight_ray_argument}). 

    Let $(\ket{E_i})_{i=1}^{\dim \mathcal{H}}$ be an orthonormal basis for $\mathcal{H}$
    which is compatible with the weight space decomposition. That
    is, each $\ket{E_i}$ is a weight vector with weight $\omega_i\in \Omega$.
    Without loss of generality, way may take $\ket{E_1} = \ket{\delta}$. Let
    $I_\Phi:= \{i\mid \braket{E_i|\Phi}\neq 0\}$ and $I_H:= \{i\mid \omega_i
    \in H\}$.

    Since $\rho_*$ is assumed to be in the relative interior of
    $\conv(\Omega)$, we have $\rho_*\neq \mu$. It follows that $\omega,
    \rho_*, \mu$ are distinct and collinear, and hence $\rho_*$ is a strict
    convex combination of the other two. In other words, $\rho_* = u_1 \omega +
    u_2 \mu$ with $u_1+u_2 = 1$ and $u_1, u_2 \in (0,1)$. Since $\mu$ lies
    on the facet $H$, we can express $\mu$ as a convex combination $\mu =
    \sum_{i\in I_H} q_i \omega_i$.

    Write $\ket{\Phi} = \sum_{i\in I_\Phi}c_ie^{i\theta_i}\ket{E_i}$, where the
    $c_i$ are positive. For arbitrary angles $(\phi_i)_{i\in
    I_H\setminus I_\Phi}$ and $\phi$, consider the smooth curve
    \begin{equation}
      \label{eq:state_curve_orthogonal}
      \begin{aligned}
        \ket{\Psi(t)} := 
        &\sum_{i\in I_\Phi\setminus I_H} \sqrt{c_i^2(1-t^2)}e^{i\theta_i}\ket{E_i}
        + \sum_{i\in I_\Phi\cap I_H} \sqrt{c_i^2(1-t^2)+u_2t^2q_i}e^{i\theta_i}\ket{E_i}\\
        &+t\sqrt{u_2}\sum_{i \in I_H \setminus I_\Phi} \sqrt{q_i}e^{i\phi_i} \ket{E_i}
        +t\sqrt{u_1}e^{i\phi}\ket{\delta},
      \end{aligned}
    \end{equation}
    which is defined on a sufficiently small open neighborhood of $0$. We claim
    that the four terms in Eq.~\eqref{eq:state_curve_orthogonal} are mutually
    strongly orthogonal. This is clear for the first three terms: they are
    spanned by disjoint sets of weight vectors. To see that
    $\ket{E_1}=\ket{\delta}$ is strongly orthogonal to the rest, note that
    $1\notin I_\Phi$ because $\braket{E_1|\Phi}=0$ by assumption. Moreover,
    $1\notin I_H$ because $\omega$ clearly cannot lie on the facet $H$: indeed,
    if $\omega$ belonged to $H$, then $\rho_* = u_1\omega+u_2\mu$ would also
    lie on $H$, contradicting the assumption that $\rho_*\in\relint(\conv(\Omega))$.
    Thus $1\notin I_\Phi\cup I_H$ and we obtain the desired conclusion.

    The mutual orthogonality implies
    $$
    \begin{aligned}
      \lVert \Psi(t)\rVert^2
      &= \sum_{i\in I_\Phi\setminus I_H} c_i^2(1-t^2)
        + \sum_{i\in I_\Phi\cap I_H} [c_i^2(1-t^2)+u_2t^2q_i)]
        +t^2u_2\sum_{i \in I_H \setminus I_\Phi}q_i 
        +t^2u_1\\
        &= (1-t^2)\sum_{i\in I_\Phi}c_i^2 + t^2u_2\sum_{i\in I_H}q_i + t^2u_1\\
        &= 1-t^2 + t^2u_2 + t^2u_1 = 1.
    \end{aligned}
    $$
    So $\ket{\Psi(t)}$ is normalized for all $t$. The mutual strong
    orthogonality implies that the density of $\ketbrap{\Psi(t)}$ is the sum of 
    the densities of its constituents (\cref{lem:strong_ortho_density_addition}):
    $$
    \begin{aligned}
      &\iota^*(\ketbrap{\Psi(t)})\\
      &= \sum_{i\in I_\Phi\setminus I_H} c_i^2(1-t^2)\omega_i  
      + \sum_{i\in I_\Phi\cap I_H} [c_i^2(1-t^2)+ u_2t^2q_i]\omega_i
      + t^2u_2\sum_{i\in I_H\setminus I_\Phi}q_i\omega_i + t^2u_1\omega\\
      &=(1-t^2)\sum_{i\in I_\Phi}c_i^2\omega_i + t^2u_2\sum_{i\in I_H}q_i\omega_i
      + t^2u_1\omega\\
      &= (1-t^2) \rho_* + t^2 u_2 \mu + t^2u_1 \omega
      = (1-t^2)\rho_* + t^2 \rho_* \\
      &= \rho_*
    \end{aligned}
    $$
    So we see that $\ketbrap{\Psi(t)}$ and $\ket{\Phi}$ share the same density.
    Finally, it is easy to verify from Eq.~\eqref{eq:state_curve_orthogonal} that 
    $\ket{\Psi(0)}=\ket{\Phi}$.

    Now we compute the derivative of the expectation value
    $\braket{\Psi(t)|W|\Psi(t)}$ at $t=0$. The only contributions we get are
    from differentiating the last two terms of
    Eq.~\eqref{eq:state_curve_orthogonal}, since the other two are smooth in
    $t^2$. Therefore,
    \begin{equation}
      \begin{aligned}
        &\doverd{t}\Big|_{t=0}\braket{\Psi(t)|W|\Psi(t)}\\
        &=2\Re\left[\sqrt{u_2}\sum_{i\in I_H\setminus I_\Phi}\sqrt{q_i}
        e^{-i\phi_i}\braket{E_i|W |\Phi}\right]
        + 2\Re \left[\sqrt{u_1} e^{-i\phi} \braket{\delta|W|\Phi}\right].
      \end{aligned}
    \end{equation}
    Remember that the angles $(\phi_i)_{i\in I_H\setminus I_\Phi}$ and $\phi$
    can be chosen arbitrarily. We then choose the $\phi_i$ so that the first
    term is nonnegative. Because $\braket{\delta|W|\Phi}\neq 0$ and $u_1> 0$,
    we can choose $\phi$ so that the second term is positive. But this implies
    that there is a $t_0$ such that $\braket{\Psi(t_0)|W|\Psi(t_0)} <
    \braket{\Phi|W|\Phi}$.
  \end{proof}
  \begin{figure}
    \centering
    \includegraphics[width=.5\textwidth]{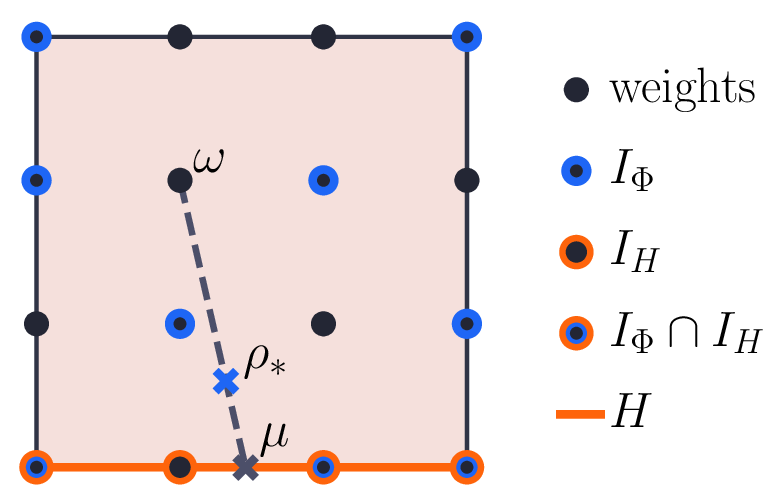}
    \caption{Illustration of the construction of the curve $\ket{\Psi(t)}$. In
    this diagram, we have assumed that each weight space has multiplicity one
    for simplicity.
    }
    \label{fig:weight_ray_argument}
  \end{figure}
\end{lem}

Lemma~\ref{lem:minimizing_state_zero_coefficient} can be made slightly
stronger:
\begin{cor}
\label{lem:minimizing_state_zero_coefficient_strong}
  Let $\rho_*, \ket{\Phi}$ be as above. Then $\braket{\Gamma|W|\Phi}=0$ for any
  vector $\ket{\Gamma}$ that is strongly orthogonal to $\ket{\Phi}$.
  \begin{proof}
    Write $\ket{\Gamma} = \sum_{\omega \in \Omega_\Gamma} \ket{\Gamma_\omega}$.
    We have $\braket{\Gamma_\omega|\Phi}=0$ by strong orthogonality. Applying
    Lemma~\ref{lem:minimizing_state_zero_coefficient} to each
    $\ket{\Gamma_\omega}$ then yields
    $\braket{\Gamma_\omega|W|\Phi}=0$, which implies
    $\braket{\Gamma|W|\Phi}=0$.
  \end{proof}
\end{cor}

To see how Lemma~\ref{lem:minimizing_state_zero_coefficient} can fail if we do
not assume $\rho_* \in \relint(\conv(\Omega))$, let $\rho_*$ be any density on
a facet $F\subset \conv(\Omega)$. Let $\ket{\Phi}$ be a minimizer of the
constrained search at $\rho_*$. Then $\ket{\Phi}$ is orthogonal to $\ket{E_i}$
for any $i$ such that $\omega_i\notin F$. 
But there always exists an interaction $W$ such that $\braket{E_i|W|\Phi}\neq
0$.

%
%
%
%

\subsubsection*{Step II: Facet -- Off-Facet Decomposition}
\label{section:proof_step2}
Next, we will decompose the pure functional $\Fp$ into three pieces. In essence,
the strategy is to split each state $\ket{\Phi}$ whose density
$\iota^*(\ketbrap{\Phi})$ is close to a facet $F\subset\conv(\Omega)$
into $\ket{\Phi_F} + \ket{\Phi'}$, where $\ket{\Phi_F}$ has support only on the
facet, and $\ket{\Phi'}$ has support only off-facet. The expectation value
$\braket{\Phi|W|\Phi}$ then splits into three terms, called ``facet-facet'',
``facet--off-facet'', and ``off-facet--off-facet'' contributions, which is the
content of Theorem~\ref{thm:pure_functional_facet_decomposition}. The punchline
is then that for the boundary force, only the facet--off-facet term will
matter, which is shown in Step~III.

For the actual boundary force computation, which we carry out in Step~IV, it is
necessary to have a manageable characterization of the fiber
$(\iota^*)^{-1}(\rho)$ for a density $\rho$ close to the facet $F$. To study
the fiber, we first restrict our attention to the squared moduli of
the coefficients $(y_i)_i$ (with respect to some chosen basis) of a state $\ketbrap{\Psi}\in
(\iota^*)^{-1}(\rho)$, which leads us naturally to the definition of the
\textit{classical density map}, which is an affine map. We will show
(Proposition~\ref{thm:classical_density_fiber_splitting}) that whenever a
density $\rho$ is close to some density $\rho_*$ on the facet, the fiber of the
classical density map over $\rho$ is approximately the product of the fiber
over $\rho_*$ and another convex polytope, with the latter parametrizing,
roughly speaking, the different directions in which a state can deviate from
one mapping to the facet. We then argue
(Lemma~\ref{lem:comb_equiv_polytope_surj_map},
Corollary~\ref{thm:fiber_polytope_parametrization}) that the fiber can be
replaced by an actual, honest product of two convex polytopes due to
topological reasons, which allows us to obtain the so-called facet--off-facet
decomposition of the pure functional
(Theorem~\ref{thm:pure_functional_facet_decomposition}).

\begin{defn}
  \label{defn:classical_density}
  Let $(V, \mathcal{H}, \iota, W)$ be an abelian functional theory together
  with an orthonormal basis $(\ket{E_i})_{i\in I}$, $|I| = \dim \mathcal{H}$,
  for $\mathcal{H}$ which is compatible with the weight space decomposition
  $\mathcal{H} = \bigoplus_{\omega \in \Omega}\mathcal{H}_\omega$. For each
  $i\in I$, let $\omega_i\in \Omega$ be the weight of $\ket{E_i}$. The
  \textbf{space of classical states} is the $(\dim\mathcal{H}-1)$-simplex
  $\Delta_I:= \{y\in \mathbb{R}^I\mid \sum_{i\in I} y_i = 1,y_i\ge 0\}$. The
  \textbf{classical density map} is
  \begin{equation}
    \begin{aligned}
      \cldmap: &\;\Delta_I \rightarrow \conv(\Omega)\\
      & y \mapsto \sum_{i\in I} y_i \omega_i.
    \end{aligned}
  \end{equation}
  $\cldmap$ can be extended to an affine map $\bar\cldmap:\aff(\Delta_I) =
  \{y\in \mathbb{R}^I\mid \sum_i y_i =1\} \rightarrow \aff(\Omega)$. We will
  denote the translation space of the space of classical states by
  $\mathfrak{D}:=\overrightarrow{\aff(\Delta_I)} = \{t\in \mathbb{R}^I \mid
  \sum_i t_i = 0\}$. Finally, we write $\mathfrak{N}$ for the kernel of
  $D\cldmap: \mathfrak{D}\rightarrow \overrightarrow{\aff(\Omega)}$.
\end{defn}

\begin{propo}
  \label{thm:fiber_full_dimension}
  If $\rho$ lies in the relative interior of $\conv(\Omega)$, then
  $\cldmap^{-1}(\rho)$ is a convex polytope of dimension $\dim \mathfrak{N}$.
  \begin{proof}
    $\cldmap^{-1}(\rho) = {\bar\cldmap}^{-1}(\rho)\cap \Delta_I$ is a convex
    polytope because it is the intersection of an affine space with a simplex.
    Since ${\bar\cldmap}^{-1}(\rho)\subset\aff(\Delta_I)$, such an intersection
    will have the same dimension as $\dim {\bar\cldmap}^{-1}(\rho) = \dim
    \mathfrak{N}$ as long as ${\bar\cldmap}^{-1}(\rho)\cap \relint(\Delta_I)
    \neq \varnothing$. Showing the latter condition is equivalent to finding
    $t\in \Delta_I$, $t_i >0$, such that $\cldmap(t) = \rho$. 

    Take any $i\in I$. If $\omega_i = \rho$, set $t^{(i)}_j = \delta_{ij}$.
    Otherwise, consider the ray from $\omega_i$ through $\rho$. This ray
    intersects the convex polytope $\conv(\Omega)$ in a point $\mu$, which lies
    on at least one facet $F$. Expressing $\rho$ as a strict (because $\rho$ is
    in the relative interior) convex combination of $\omega_i$ and $\mu$, and
    writing $\mu$ as a convex combination of $\omega_j, j\in I_F$, we obtain a
    convex combination $\rho = \sum_{j\in I} t^{(i)}_j \omega_j$, where
    $t^{(i)}_i > 0$. (Note that this argument is very similar to the one used
    in the proof of Lemma~\ref{lem:minimizing_state_zero_coefficient})

    Finally, we have 
    $$
      \rho = \frac{1}{|I|}\sum_{i\in I}\sum_{j\in I}t^{(i)}_j\omega_j 
      = \sum_{j\in I}\underbrace{\left(\sum_{i\in
      I}\frac{t^{(i)}_j}{|I|}\right)}_{=:t_j}\omega_j = \sum_{j\in I}
      t_j\omega_j.
    $$
    For each $j\in I$, the coefficient $t_j$ is a sum of nonnegative terms,
    with at least one being positive, namely $t_j^{(j)} >0$. It follows that
    $t_j>0$ and $t\in \relint(\Delta_I)$.
  \end{proof}
\end{propo}

The motivation for studying the classical density map $\cldmap$ is that
understanding its fibers allows us to understand the fibers of the (pure state)
density map $\iota^*|_{\mathcal{P}}$ up to phases. More precisely:
\begin{propo}
  For any $\rho\in \conv(\Omega)$,
  \begin{equation}
    (\iota^*)^{-1}(\rho) \cap \mathcal{P}
     = \left\{\ketbrap{\Psi}\Big\vert\ket{\Psi} = \sum_{i\in I}\sqrt{y_i}\xi_i
     \ket{E_i},  \; y\in \cldmap^{-1}(\rho), \xi_i\in S^1\right\},
  \end{equation}
  where $S^1 = \{\xi \in \mathbb{C} \mid |\xi| = 1\}$ is the set of phases.
  \begin{proof}
    This is easily shown by direct calculation.
  \end{proof}
\end{propo}

Hence, we immediately obtain a formula of the pure functional expressed as a
minimization over both the phases and the fiber of $\cldmap$:

\begin{cor}
  \label{thm:functional_classical_fiber}
  For any $\rho \in \conv(\Omega)$, the value of the pure functional at $\rho$ is
  \begin{equation}
    \Fp(\rho) = \min_{\xi \in (S^1)^I} \min_{y\in \cldmap^{-1}(\rho)}
    \sum_{i,j\in I} \sqrt{y_i y_j}\bar \xi_i \xi_j W_{ij},
  \end{equation}
  where $W_{ij}:= \braket{E_i|W|E_j}$.
\end{cor}

The rest of this section will be dedicated to understanding the fiber
$\cldmap^{-1}(\rho)$ when $\rho$ is close to a facet $F\subset \conv(\Omega)$.

If $F\subset \conv(\Omega)$ is a facet, there is a natural choice of
orthonormal basis for $\mathcal{H}_F$, namely $(\ket{E_i})_{i\in I_F}$, where
$I_F:= \{i\in I\mid \omega_i \in F\}$. Moreover, this basis is easily seen to
be compatible with the weight space decomposition $\mathcal{H}_F =
\bigoplus_{\omega \in \Omega_F}\mathcal{H}_\omega$. Hence, there corresponds a
classical density map for the facet functional theory, which we will denote by
$\cldmap_F: \Delta_{I_F} \rightarrow \conv(\Omega_F)$. We will think of
$\Delta_{I_F}$ as a subset of $\Delta_{I}$ by mapping $y\in \Delta_{I_F}$ to
$\tilde y\in \Delta_I$, where $\tilde y_i =0$ if $i\notin I_F$ and $\tilde y_i
= y_i$ if $i\in I_F$. Similarly, we think of $\mathfrak{D}_F :=
\overrightarrow{\aff(\Delta_{I_F})}$ as a subspace of $\mathfrak{D}$ in the same
way.

\begin{propo} 
  \label{thm:classical_density_map_facet_commute}
  $\cldmap_F$ and $D\cldmap_F$ are the restrictions of $\cldmap$ and $D\cldmap$
  respectively:
  \begin{equation}
  \begin{tikzcd}
    \Delta_F \arrow{r}{\cldmap_F} \arrow[hookrightarrow]{d}& \conv(\Omega_F)\arrow[hookrightarrow]{d}\\
    \Delta \arrow[swap]{r}{\cldmap}& \conv(\Omega)\\
  \end{tikzcd}
    \hspace{4em}
  \begin{tikzcd}
    \mathfrak{D}_F \arrow{r}{D\cldmap_F} \arrow[hookrightarrow]{d}& \overrightarrow{\aff(\Omega_F)}\arrow[hookrightarrow]{d}\\
    \mathfrak{D} \arrow[swap]{r}{D\cldmap}& \overrightarrow{\aff(\Omega)}\\
  \end{tikzcd}
  \end{equation}
  \begin{proof}
    This is obvious from the definitions of $\cldmap$ and $\cldmap_F$.
  \end{proof}
\end{propo}


\begin{cor}
  \label{thm:parallel_null_space}
  $\mathfrak{N}_F = \mathfrak{N}\cap \mathfrak{D}_F$.
\end{cor}

\begin{propo}
  \label{thm:dim_of_facet_offfacet_null_spaces}
  \;\\
  \vspace{-2em}
  \begin{enumerate}[label={(\arabic*)}]
    \item $\dim \mathfrak{N} = \dim \mathcal{H} - \dim \conv(\Omega) - 1$
    \item $\dim \mathfrak{N}_F = \dim \mathcal{H}_F - \dim F - 1$
  \end{enumerate}
  \begin{proof}
    (1) follows from applying the rank-nullity theorem to $D\cldmap$. (2)
    follows from (1) applied to the facet functional theory.
  \end{proof}
\end{propo}

\begin{figure}
  \centering
  \includegraphics[width=.35\textwidth]{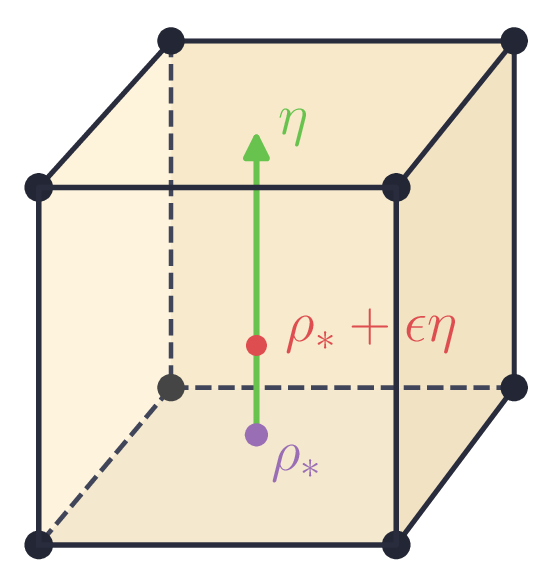}
  \caption{Starting with a density $\rho_*$ on a facet (in this case the bottom
    face of the cube), we pick an inward direction $\eta$ and study the fiber of $\cldmap$
    over $\rho_* +\epsilon\eta$.
    }
  \label{fig:wander_offfacet}
\end{figure}

Pick a complement $\mathfrak{N}'$ of $\mathfrak{N}_F$ in $\mathfrak{N}$. For
example, we could take $\mathfrak{N}'$ to be the orthogonal complement of
$\mathfrak{N}_F$. Roughly speaking, we want to use $\mathfrak{N}'$ to
parametrize the ways a classical state can distribute its coefficients
(interpreted as probabilities) in the off-facet weights, while $\mathfrak{N}_F$
parametrizes how the coefficients within the facet $F$ are distributed. 
This idea is made precise in the following proposition:

\begin{propo}
  \label{thm:classical_density_fiber_splitting}
  Let $\rho_*$ be in the relative interior of a facet $F\subset \conv(\Omega)$,
  let $\eta\in \overrightarrow{\aff(\Omega)}$ be an inward-pointing vector
  (relative to $\conv(\Omega)$, see Fig.~\ref{fig:wander_offfacet}), and fix an
  $l\in \mathfrak{D}$ that solves $D\cldmap(l) = \eta$. Then for any $\epsilon
  \ge 0$ such that $\rho_* + \epsilon \eta \in \conv(\Omega)$, the fiber of the
  classical density map $\cldmap$ above $\rho_*+\epsilon\eta$, i.e., the set of
  all $y\in \Delta_I$ that solve $\cldmap(y) = \rho_* + \epsilon \eta$, is
  given by
  \begin{equation}
    \label{eq:fiber_facet_decomposition}
    \begin{aligned}
    \cldmap^{-1}(\rho_* + \epsilon \eta)
     = \Big\{
       y + \epsilon q \Big\vert 
       q\in \mathbf{P}', y\in \mathbf{P}^\parallel(\epsilon q)
       \Big\},
    \end{aligned}
  \end{equation}
  where
  \begin{equation}
    \label{eq:decomposing_polytopes}
    \begin{aligned}
      &\mathbf{P}':= \{q\in l+\mathfrak{N}'\mid  \forall i\in I\setminus I_F: q_i\ge 0\}\\
      &\mathbf{P}^\parallel(\epsilon q) := 
      \{y\in \aff(\Delta_{I_F})\mid \bar\cldmap(y) = \rho_*, \forall i \in I_F: y_i \ge -\epsilon q_i\}.
    \end{aligned}
  \end{equation}
  Additionally, we have $\mathbf{P}^\parallel(0) = \cldmap^{-1}(\rho_*)$.
  \begin{proof}
    $\mathbf{P}^\parallel(0) = \{y\in \aff(\Delta_{I_F})\mid \bar\cldmap(y) =
    \rho_*\}\cap \Delta_{I_F} = \cldmap^{-1}(\rho_*)$. This proves the last assertion.

    If $\epsilon = 0$, the right hand side of
    Eq.~\eqref{eq:fiber_facet_decomposition} reduces to
    $\mathbf{P}^\parallel(0) =\cldmap^{-1}(\rho_*)$, which is the same as
    the left hand side. Thus we assume $\epsilon > 0$ in the following.

    Suppose $z\in \Delta_I$ is a classical state which solves $\cldmap(z) =
    \rho_* + \epsilon \eta$. Take any $y'\in \Delta_{I_F}$ such that $\cldmap(y')
    = \rho_*$. Then $D\cldmap(z - y' - \epsilon l) = 0$, so $z =y' + \epsilon l
    + s + t$, where $s\in \mathfrak{N}_F$ and $t\in
    \mathfrak{N}'$. Thus $z = y + \epsilon q$ with $y = y'+s$ and $q =
    l+\epsilon^{-1}q$. Because $z\in \Delta_I$, we have $y_i + \epsilon
    q_i\ge0$ for all $i\in I$. If $i\in I\setminus I_F$, this implies $q_i \ge
    0$. This shows ``$\subset$'' in Eq.~\eqref{eq:fiber_facet_decomposition}.

    Conversely, take $z = y+\epsilon q$ with $q\in \mathbf{P}'$ and $y\in
    \mathbf{P}^\parallel(\epsilon q)$. Then 
    $$
      \cldmap(z) = \cldmap(y) + \epsilon D\cldmap(q) = \rho_* + \epsilon \eta.
    $$
    Clearly $z\in \Delta_I$ (the inequalities in
    Eq.~\eqref{eq:decomposing_polytopes} ensure that $z_i \ge 0$ for all $i\in
    I$). It follows that $z \in \cldmap^{-1}(\rho_* + \epsilon \eta)$. This shows ``$\supset$''.
  \end{proof}
\end{propo}

\newcolumntype{M}[1]{>{\centering\arraybackslash}m{#1}}

\begin{table}
  \centering
  \begin{tabular}{M{3.4cm}ccM{3.4cm}M{3.4cm}}
    \hline
    $\Omega, \conv(\Omega), F$& $\dim\mathcal{H}$  & $\dim \mathcal{H}_F$&  $\mathfrak{N}_F$ & $\mathfrak{N}'$
    \\\hline
     \includegraphics[width=.16\textwidth]{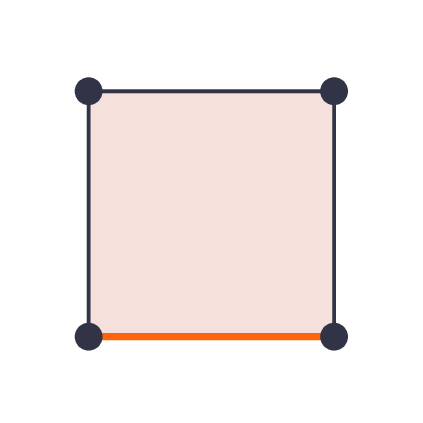} & 
     4& 2 & &
     \includegraphics[width=.16\textwidth]{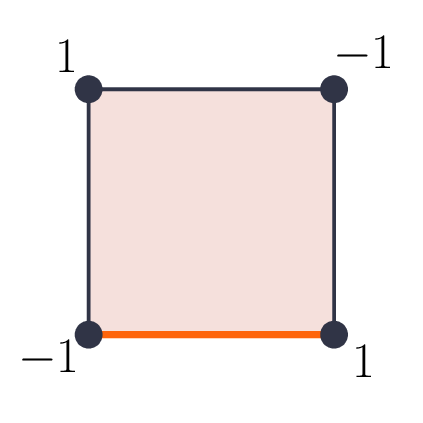}\\\hline
     \includegraphics[width=.18\textwidth]{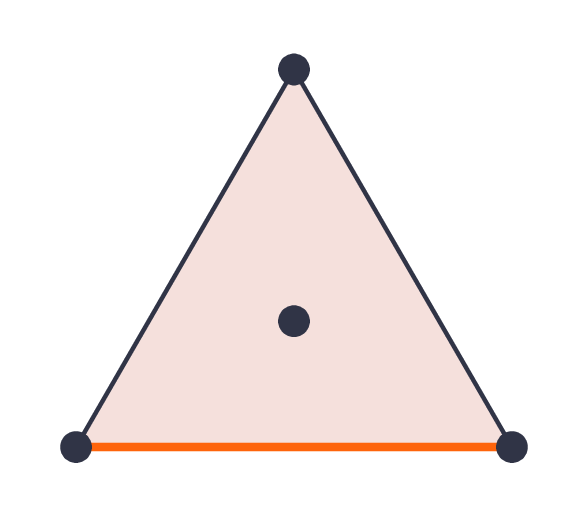} & 
     4& 2 & &
     \includegraphics[width=.18\textwidth]{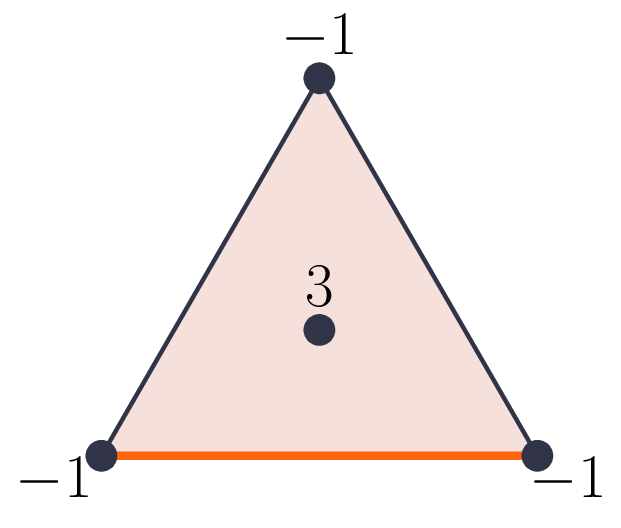}\\\hline
     \includegraphics[width=.2\textwidth]{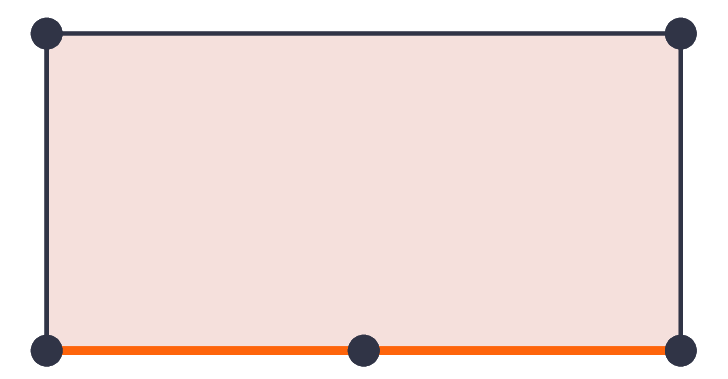} & 
     5& 3 &
     \includegraphics[width=.21\textwidth]{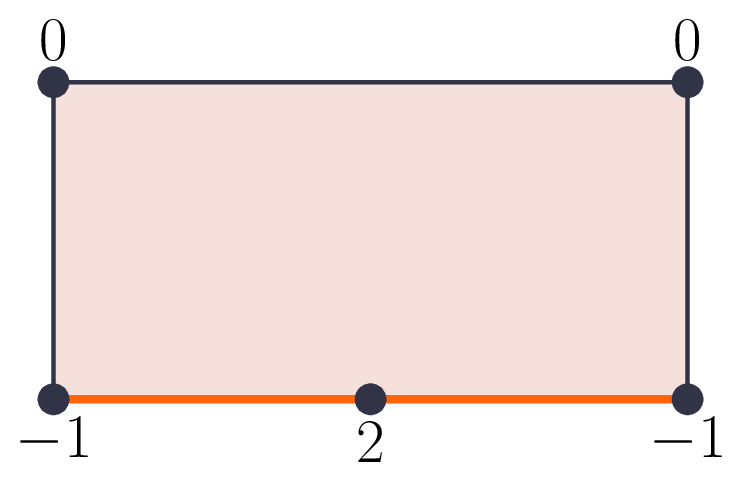}&
     \includegraphics[width=.21\textwidth]{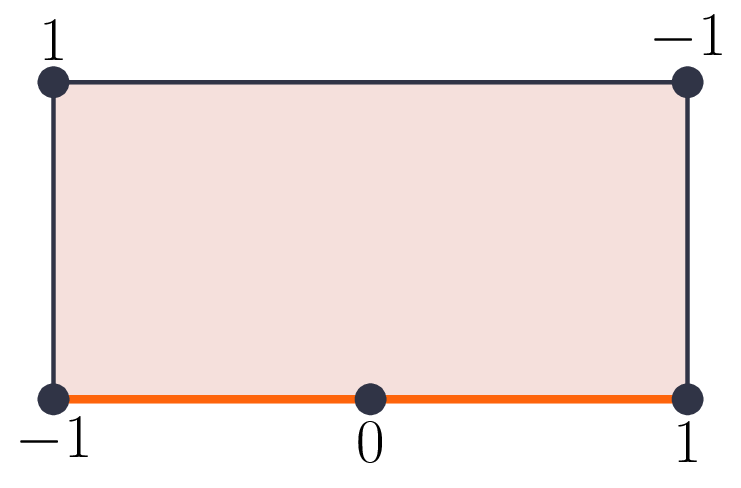}
     \\\hline
     \includegraphics[width=.21\textwidth]{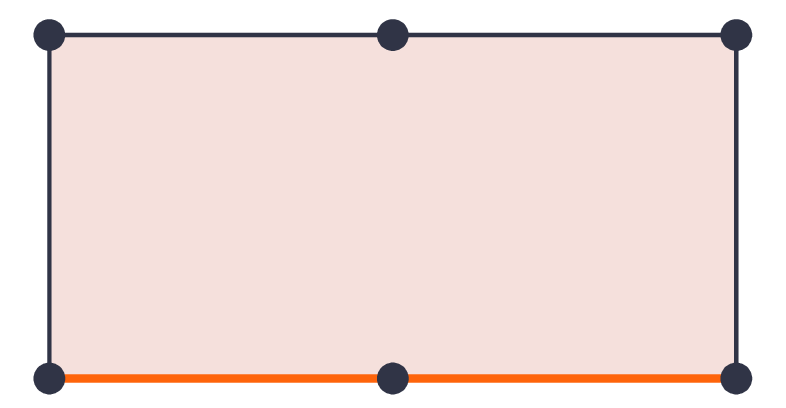} & 
     6& 3 &
     \includegraphics[width=.21\textwidth]{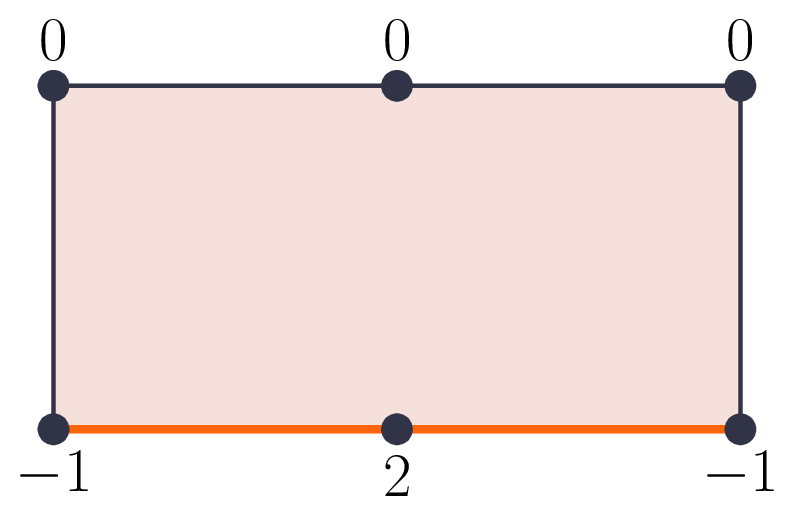} &
     \includegraphics[width=.19\textwidth]{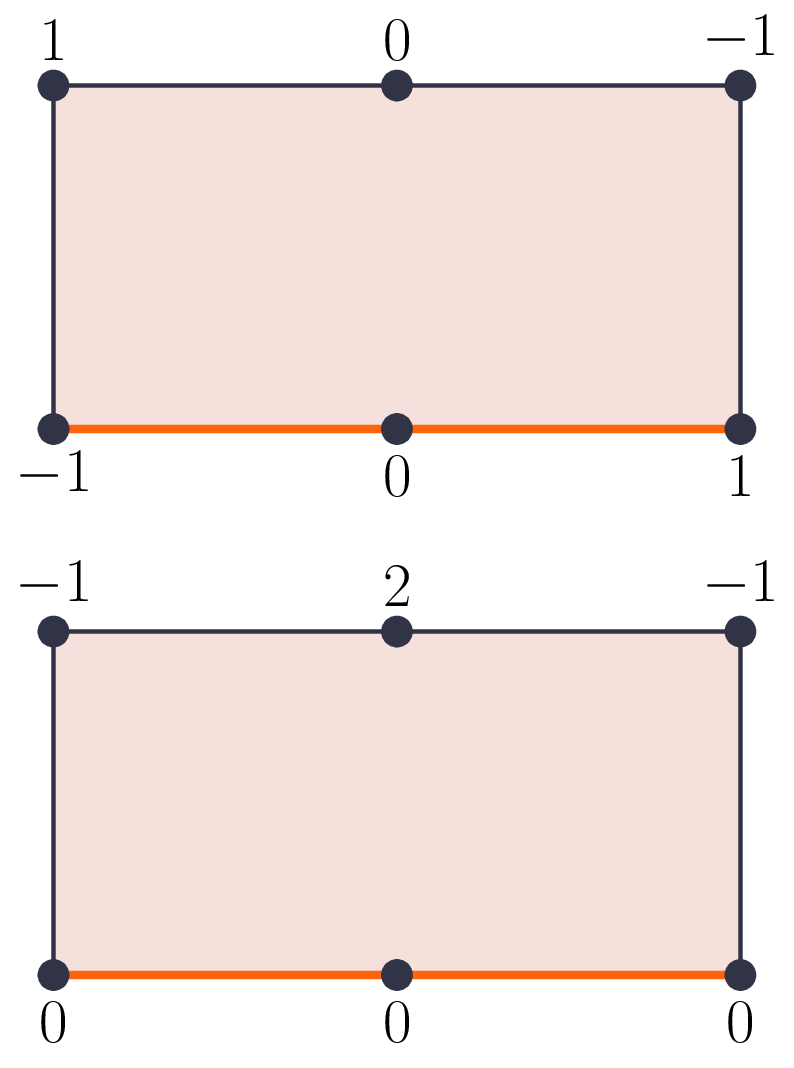}\\
     \hline
     \includegraphics[width=.19\textwidth]{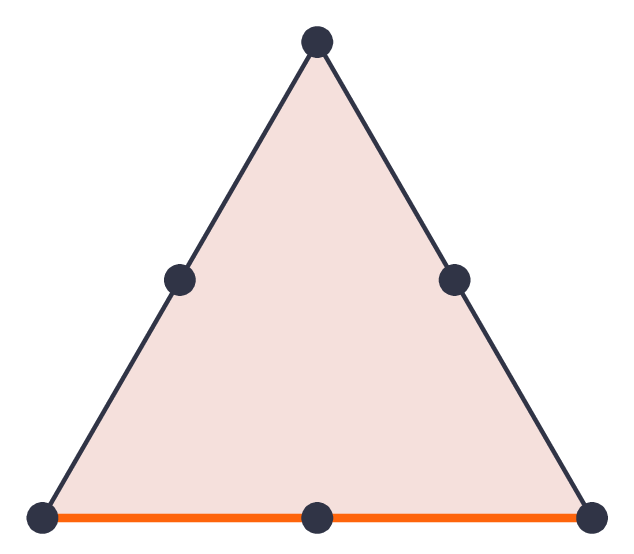} & 
     6& 3 &
     \includegraphics[width=.19\textwidth]{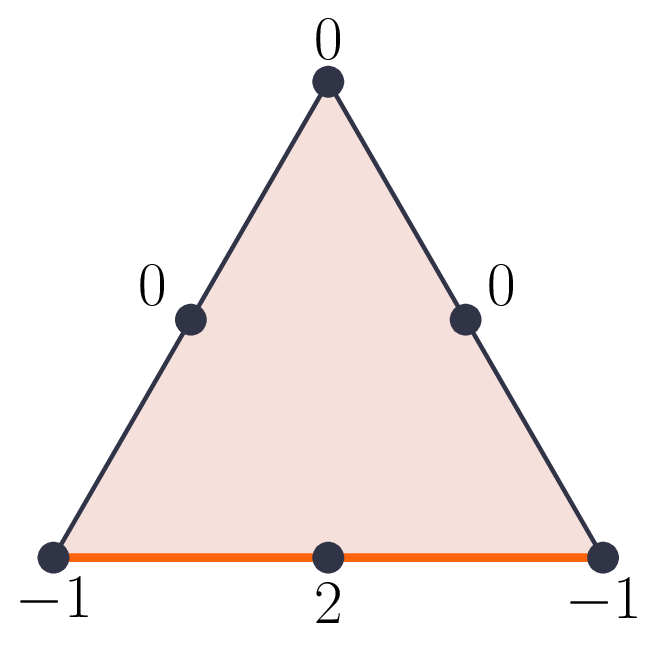} &
     \includegraphics[width=.15\textwidth]{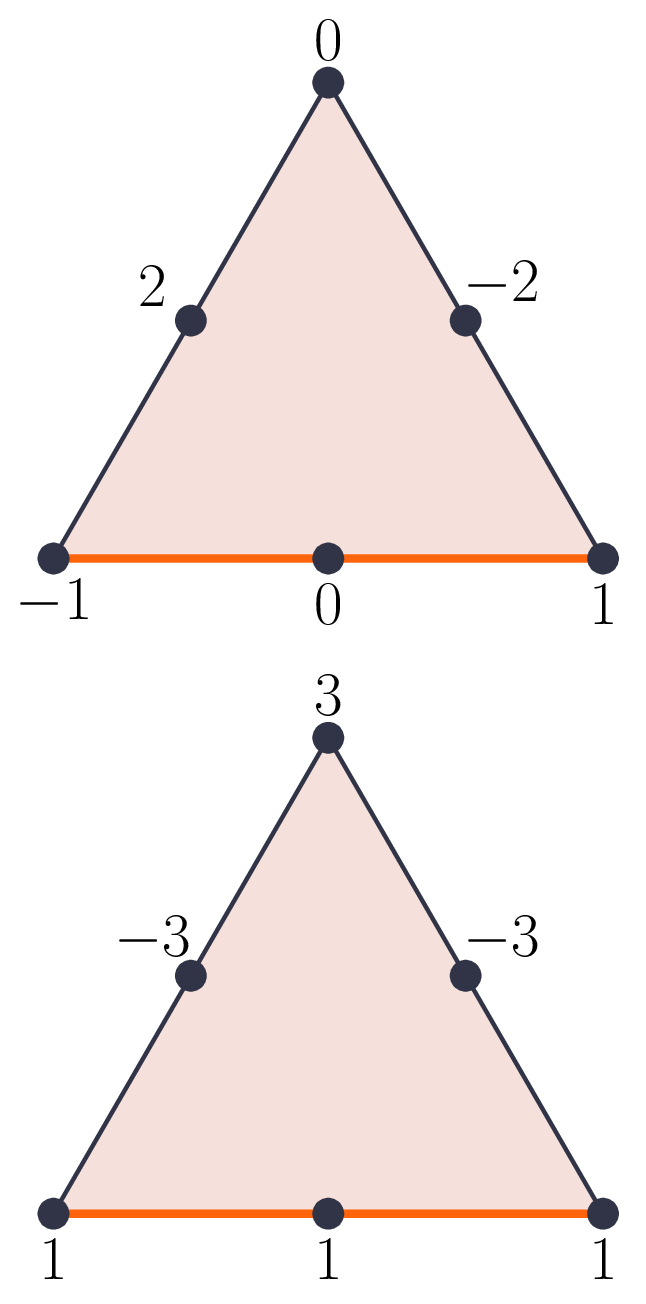}\\
     \hline
     \includegraphics[width=.21\textwidth]{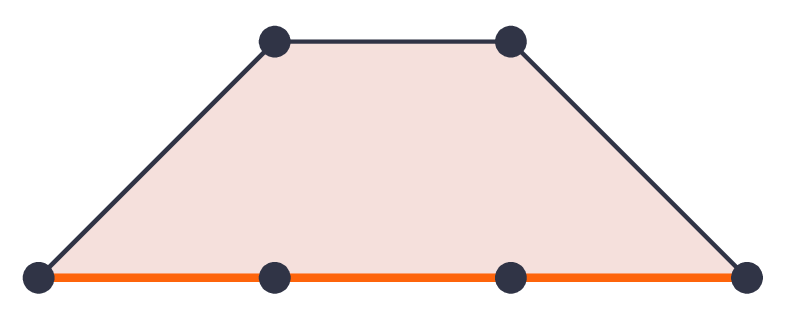} & 
     6& 4 &
     \includegraphics[width=.18\textwidth]{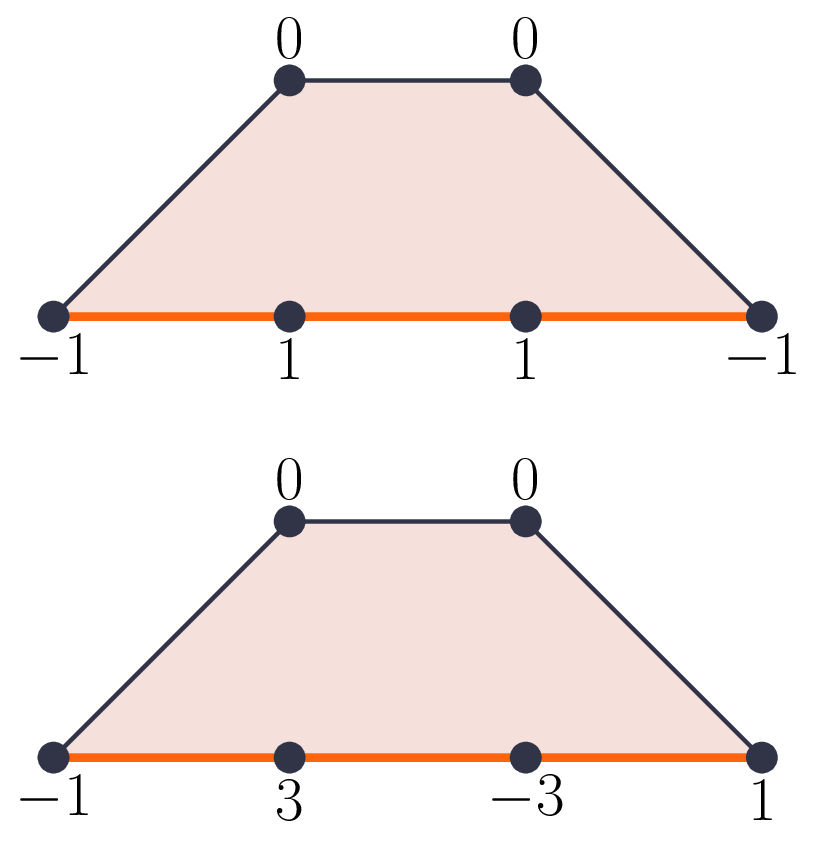} &
     \includegraphics[width=.21\textwidth]{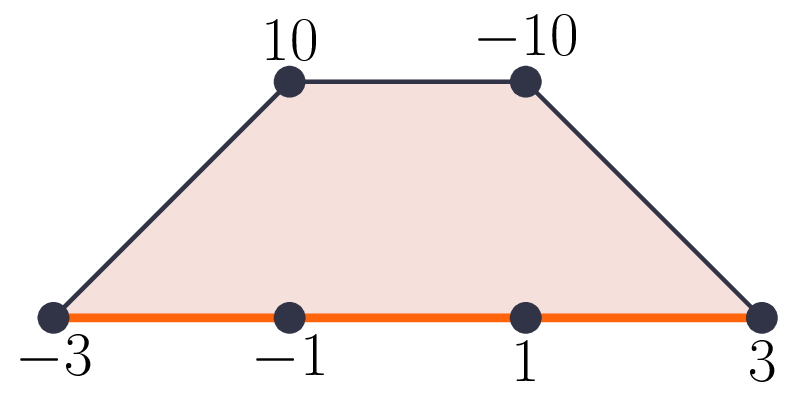}\\
     \hline
  \end{tabular}
  \caption{The subspaces $\mathfrak{N}_F$ and $\mathfrak{N}'$ for several sets
  of weights $\Omega$ and choices of facet $F\subset \conv(\Omega)$. For
  simplicity, the multiplicity of each weight space is always one, and we take
  $\mathfrak{N}' = \mathfrak{N}_F^\perp$. We present $\mathfrak{N}_F$ and
  $\mathfrak{N}'$ by listing a set of spanning vectors, which all belong to
  $\mathfrak{N}\subset\mathfrak{D}:= \{t\in \mathbb{R}^I\mid \sum_{i\in I} t_i = 0\}$.}
  \label{tab:kernel_facet_offfacet_decomposition_examples}
\end{table}

\begin{propo}
  \label{thm:parametrizing_spaces_are_convex_polytopes}
  \;\\
  \vspace{-2em}
  \begin{enumerate}[label={(\arabic*)}]
    \item $\mathbf{P}'$ is a convex polytope.
    \item $\mathbf{P}^\parallel(\epsilon q)$ is a convex polytope
      for all $\epsilon\ge 0$ and $q\in \mathbf{P}'$.
    \item If $\rho_*\in \relint(F)$, then $\dim \mathbf{P}^\parallel(0) = \dim
      \mathfrak{N}_F$.
  \end{enumerate}
  \begin{proof}
    Since $\mathbf{P}'$ is a subset of $l + \mathfrak{N}'$ defined as an
    intersection of half spaces, it suffices to show that $\mathbf{P}'$
    does not contain a ray. For the sake of contradiction, suppose
    $x + \mathbb{R}_{\ge 0} q$, $q\in \mathfrak{N}'\setminus \{0\}$, were such
    a ray contained in $\mathbf{P}'$. Let $v\in V$ be a vector (external
    potential) such that $\braket{\omega_i, v} \ge c$ for all $i\in I$, with
    equality if and only if $i \in I_F$. Then
    $$
    \begin{aligned}
      &0 = \left\langle D\cldmap(q), v\right\rangle
    = \left\langle\sum_{i\in I} q_i\omega_i, v\right\rangle
      = \sum_{i\in I_F}q_i \braket{\omega_i, v} + \sum_{i\in I\setminus I_F} q_i \braket{\omega_i, v}\\
      &= \sum_{i\in I_F} q_i (\braket{\omega_i, v}-c) + \sum_{i\in I\setminus I_F} q_i (\braket{\omega_i ,v}-c)\\
      &=\sum_{i\in I\setminus I_F} q_i \underbrace{(\braket{\omega_i, v}-c)}_{>0},
    \end{aligned}
    $$
    where we have used $\sum_i q_i =0$. Since $q\in \mathfrak{N}'\setminus
    \{0\}$, $q_j \neq 0$ for at least one $j\in I$.
    It follows from the above equation that $q_i < 0$ for at least one $i\in
    I\setminus I_F$. We can then find $x + \lambda q$ on the ray satisfying
    $(x+\lambda q)_i < 0$, contradicting the defining inequalities of
    $\mathbf{P}'$.

    The same argument applies to $\mathbf{P}^\parallel(\epsilon q)$: if $y
    +\lambda p\in \mathbf{P}^\parallel(\epsilon q)$ for all $\lambda \ge 0$,
    where $p\in \ker D\cldmap \cap \mathfrak{D}_F = \mathfrak{N}_F$ and $p\neq
    0$, then $p_i < 0$ for some $i\in I_F$, so for $\lambda$ large enough we 
    get $y_i + \lambda p_i + \epsilon q_i < 0$.

    (Alternatively, if $\mathbf{P}'$ or $\mathbf{P}^\parallel(\epsilon q)$
    contained a ray, then so would $\cldmap^{-1}(\rho_*+\epsilon\eta)$, which
    is a convex polytope.)

    (3) follows from applying Proposition~\ref{thm:fiber_full_dimension} to
    $\rho_* \in \relint(F)$.
  \end{proof}
\end{propo}

Let us recapitulate what we have done so far. The objective is to study the
behavior of the pure functional $\Fp$ near one of the facets of its domain, which is
$\conv(\Omega)$. To do so, we pick a density $\rho_*$ in the relative interior
of a facet $F\subset \conv(\Omega)$, and walk inwards, in the direction of the
fixed vector $\eta \in \overrightarrow{\aff(\Omega)}$. The value of the pure
functional at $\rho_* + \epsilon \eta$ is defined as a minimization of the
expectation value $\braket{\Psi|W|\Psi}$ over $\ketbrap{\Psi}\in
(\iota^*)^{-1}(\rho_* +\epsilon \eta)$. To understand the fiber, we observe
that, with respect to the suitably chosen basis $(\ket{E_i})_{i\in I}$ for the
Hilbert space $\mathcal{H}$, the density map is, up to phases, an affine map
$\cldmap$ sending the squared moduli of the coefficients, which we
interpret as a ``classical state'' or probability distribution, to a density in
$\conv(\Omega)$. Thus, the fibers of $\cldmap$ contain almost all the information
needed to reconstruct the fibers of
$\iota^*|_{\mathcal{P}}$. 
We then find out, in
Propositions~\ref{thm:classical_density_fiber_splitting} and
\ref{thm:parametrizing_spaces_are_convex_polytopes}, that each element in fiber
$\cldmap^{-1}(\rho_*+\epsilon \eta)$ can be written as $p+ \epsilon q$, where
$q$ is parametrized by a fixed convex polytope, namely $\mathbf{P}'$, and $p$
is parametrized by a convex polytope which depends on $\epsilon q$, namely
$\mathbf{P}^\parallel(\epsilon q)$. The intuition is that $\epsilon q\in
\mathbf{P}'$ parametrizes ``how we deviate from a (classical) state mapping to
the facet'', whereas $p\in \mathbf{P}^\parallel(\epsilon q)$ is supposed to be
a ``reference state'' mapping to $\rho_*$. We see that this splitting works
exactly because any state whose density lies close to the facet must be itself
also close to a state which maps to the facet.

However, this parametrization is imperfect: it would be nice if
$\mathbf{P}^\parallel(\epsilon q)$ did not depend on $\epsilon q$. On the other
hand, $\mathbf{P}^\parallel(\epsilon q)$ is ``almost'' the same as
$\mathbf{P}(0)$ as long as $\epsilon$ is small, which is clear from
Eq.~\ref{eq:decomposing_polytopes}. 
Hence, we might hope to replace $\mathbf{P}^\parallel(\epsilon q)$ by
$\mathbf{P}(0)$ as the parameter space for $p$, which can be done if we can find
a surjective map $\mathbf{P}(0)\rightarrow \mathbf{P}^\parallel(\epsilon q)$
for all $\epsilon$ (small enough) and $q\in \mathbf{P}'$.
In order to accomplish this, we need the following two lemmas:

\begin{lem}
  \label{lem:existence_of_gen_bary_coord}
  Let $P$ be any convex polytope and let $v^{(1)}, \dotsb, v^{(n)}$ be its
  vertices. There exists a continuous map $\sigma: P \rightarrow \Delta_{n-1}$
  (where $\Delta_{n-1}$ is the standard $(n-1)$-simplex) such that $p =
  \sum_{k=1}^n \sigma(p)_kv^{(k)}$ for all $p\in P$. Furthermore, any such map has
  the following property: if $f\in P$ is a face and $p\in f$, then
  $\sigma(p)_k=0$ whenever $v^{(k)}\notin f$.
  \begin{proof}
    Triangulate $P$ by simplices whose vertices are vertices of $P$ and glue
    together the barycentric coordinates of each simplex. This is well-defined
    on regions where the simplices overlap because barycentric coordinates for
    a simplex are unique.

    The second assertion is a direct consequence of the defining property of
    $\sigma$.
  \end{proof}
\end{lem}

\begin{rem}
  More succinctly, we can summarize the statement of the above lemma  by saying
  that the ``weighted average'' map $\Delta_{n-1}\rightarrow P$, $t \mapsto
  \sum_{k=1}^nt_k v^{(k)}$, admits a continuous section. Such a map is called a
  system of \textit{generalized barycentric coordinates}. For our purpose
  below, continuity is enough.
  Alternatively, Lemma~\ref{lem:existence_of_gen_bary_coord} can be shown using
  Michael continuous selection theorem \cite{michaelContinuousSelections1956},
  making use of the fact that the surjection $\Delta_{n-1}\rightarrow P$ is lower
  hemicontinuous.
  \end{rem}

Recall that two convex polytopes $P,Q$ are said to be \textit{combinatorially
equivalent} if their face lattices are isomorphic. In other words, there exists
a bijection $\phi: \mathrm{Faces}(P)\rightarrow \mathrm{Faces}(Q)$ preserving
inclusion.

\begin{lem}
  \label{lem:comb_equiv_polytope_surj_map}
  Let $P, Q$ be combinatorially equivalent convex polytopes and let $\phi:
  \mathrm{Faces}(P)\rightarrow \mathrm{Faces}(Q)$ be a lattice
  isomorphism. Let $v^{(1)}, \dotsb, v^{(n)}\in P$ be the vertices of $P$. Let $\sigma:
  P \rightarrow \Delta_{n-1}$ be a system of generalized barycentric
  coordinates. Define a map
  $$
  \begin{aligned}
    R: &\;P \rightarrow Q\\ 
    & p \mapsto \sum_{k=1}^n \sigma(p)_k\phi(v^{(k)}) .
  \end{aligned}
  $$
  Then $R$ maps each face $f\subset P$ surjectively onto $\phi(f)$. In
  particular, $R$ is surjective.

  \begin{proof}
    Let $f\subset P$ be any face. Take any point $p\in f$, then $R(p) =
    \sum_{k=1}^n \sigma(p)_k \phi(v^{(k)}) \in \conv(\{\phi(v^{(k)}) \mid
    v^{(k)} \in f\}) = \phi(f)$
    due to the second part of Lemma~\ref{lem:existence_of_gen_bary_coord}. This
    shows $R(f) \subset \phi(f)$. Applying the same argument to the facets
    of $f$, we get $R(\partial f) \subset \partial \phi(f)$. In the
    following, we will prove that the map $R|_{\partial f}: \partial f
    \rightarrow \partial \phi(f)$ has degree $1$ over $\mathbb{Z}/2\mathbb{Z}$
    by induction on the dimension of the faces of $P$. Once this is shown, we
    can prove surjectivity of $R|_f$ onto $\phi(f)$ for any $k$-face by
    contradiction: suppose $y\in \phi(f)$ is not in the image of $R|_f$,
    then we can construct a continuous map $b: f \rightarrow \partial \phi(f)$
    which sends $x \in f$ to the intersection of $\partial \phi(f)$ with the
    ray from $y$ through $f(x)$. Clearly, $b|_{\partial f} \equiv
    R|_{\partial f}$. Hence, $R|_{\partial f}$ is equal to the
    composition 
    $$
      \partial f \hookrightarrow f \overset{b}{\rightarrow} \partial \phi(f).
    $$
    But $f$ is contractible, so the induced map $H_{k-1}(\partial f)\rightarrow
    H_{k-1}(\partial\phi(f))$ must be trivial, a contradiction.\\

    \noindent
    \textit{Base case:}\\
    First, note that $R(v^{(k)}) = \sum_{{k'}=1}^n \phi(v^{(k')})\delta_{kk'} =
    \phi(v^{(k)})$, where again we have used the second assertion of
    Lemma~\ref{lem:existence_of_gen_bary_coord}.
    Let $f\subset P$ be any $1$-face. Its boundary is $\{v^{(k)}, v^{(l)}\}\cong S^0$
    with $k\neq l$. Similarly, $\partial\phi(f) = \{\phi(v^{(k)}),
    \phi(v^{(l)})\}\cong S^0$. Since $R(v^{(k)}) = \phi(v^{(k)})$ and
    $R(v^{(l)}) = \phi(v^{(l)})$, the map $R|_{\partial f}: \partial
    f\rightarrow \partial \phi(f)$ has degree $1$ over
    $\mathbb{Z}/2\mathbb{Z}$. (We choose to work over $\mathbb{Z}/2\mathbb{Z}$
    so we can ignore orientations. Over $\mathbb{Z}$, we would get $\pm 1$
    depending on the chosen orientations. Hereafter in this proof, we shall omit ``over
    $\mathbb{Z}/2\mathbb{Z}$''.)\\

    \noindent
    \textit{Inductive step:}\\
    Suppose $R|_{\partial s}: \partial s\rightarrow \partial \phi(s)$ has
    degree $1$ for all faces $s\subset P$ of dimension at most $m$. Take any
    $(m+1)$-face $f\subset P$.

    Let $X_\bullet$, resp. $Y_\bullet$, denote the regular CW complexes
    constructed from the proper faces of $f$, resp. $\phi(f)$. That is, $X_0$
    is the vertex set of $f$ and $X_k$ is obtained from $X_{k-1}$ by gluing one
    open $k$-disk to $X_{k-1}$ for each $k$-face of $f$. For each face
    $s\subset f$, we will denote the corresponding $k$-cell also by $s$. We can
    identify $X_k$ with the union of all faces of $f$ of dimension at most $k$.

    Since $R(s)\subset \phi(s)$ for all faces $s$, the map $R: P
    \rightarrow Q$ is a cellular map, i.e., it satisfies $R(X_k) \subset
    Y_k$ for all $k=0, \dotsb, m$. Indeed, any point $x\in X_k$ is contained in
    a $k$-face $s\subset f$. So $R(x)\in R(s) \subset \phi(s)\subset Y_k$
    because $\phi(s)$ is a $k$-face of $\phi(f)$. Therefore, $R$ induces a
    map between the cellular homology groups. We will now show that $R_*:
    H_\bullet^{\text{cell}}(X_\bullet)\rightarrow
    H_\bullet^{\text{cell}}(Y_\bullet)$ is an isomorphism (remember that we
    work over $\mathbb{Z}/2\mathbb{Z}$).
    
    Let $R_\sharp: H_k(X_k, X_{k-1})\rightarrow H_k(Y_k, Y_{k-1})$ denote
    the induced map between the cellular chain complexes. The group $H_k(X_k,
    X_{k-1})$, resp. $H_k(Y_k, Y_{k-1})$, is free abelian over the set of
    $k$-faces of $f$, resp. $\phi(f)$. Take any $k$-face $s\subset f$. Let
    $\tilde s$ be a singular $k$-chain which represents $s$, with the latter viewed as an
    element of $H_k(X_k, X_{k-1})$. That is, $\tilde s$ is a relative cycle in
    $(X_k, X_{k-1})$ which is not a relative boundary. Then $R(\tilde s)$ is
    a singular $k$-chain which is a relative cycle in $(Y_k, Y_{k-1})$ since
    $\partial R(\tilde s) = R(\partial \tilde s) \subset \partial \phi(s)$.
    Suppose $R(\tilde s)$ were a relative boundary, then $R(\tilde s) =
    \partial c$ for some $(k+1)$-chain $c$ in $Y_k$. But then $\partial
    R(\tilde s) = R(\partial \tilde s) = 0$, which contradicts the
    induction hypothesis that $R|_{\partial s}: \partial s \rightarrow
    \phi(\partial s)$ has $\mathbb{Z}/2\mathbb{Z}$-degree $1$ because $\partial\tilde s$
    represents the generator of $H_{k-1}(\partial s)\cong \mathbb{Z}_2$. It
    follows that $R_\sharp(s) = \phi(s)\in H_k(Y_k, Y_{k-1})$. Since $s$ was
    an arbitrary $k$-face of $f$, and the set of all $k$-faces of $f$ (resp.
    $\phi(f)$) generates $H_k(X_k, X_{k-1})$ (resp. $H_k(Y_k, Y_{k-1})$)
    freely, we have shown that $R_\sharp: H_k(X_k, X_{k-1})\rightarrow
    H_{k-1}(Y_k, Y_{k-1})$ is an isomorphism.

    Therefore, $R_*: H_\bullet^{\text{cell}}(X_\bullet)\rightarrow
    H_\bullet^{\text{cell}}(Y_\bullet)$ is an isomorphism. In particular,
    $$
      R_*: H_m^{\text{cell}}(X_\bullet)\cong\mathbb{Z}/2\mathbb{Z}\rightarrow
    H_m^{\text{cell}}(Y_\bullet) \cong \mathbb{Z}/2\mathbb{Z}$$ 
    is an isomorphism.
  \end{proof}
\end{lem}

Now we will apply Lemma~\ref{lem:existence_of_gen_bary_coord} and
Lemma~\ref{lem:comb_equiv_polytope_surj_map} with $P = \mathbf{P}^\parallel(0)$
and $Q = \mathbf{P}'(\epsilon q)$.
Before continuing, it is important to remind ourselves that many objects here
depend on various previous choices, and the dependence is often suppressed. For
example, the classical density map $\cldmap$ is constructed from a chosen
orthonormal basis $(\ket{E_i})_{i\in I}$, and the vector spaces
$\mathfrak{N}^{\parallel}, \mathfrak{N}'$ depend on the facet $F$. The convex
polytopes $\mathbf{P}', \mathbf{P}^\parallel(\epsilon q)$ depend on not
only $\rho_*, \eta$, but also on the choice of a particular solution $l\in
(D\cldmap)^{-1}(\eta)$. 

\begin{propo}
  \label{thm:comb_equiv_of_polytopes}
  Suppose $\rho_*\in F$ is a regular value of the facet density map. Let $v^{(1)},
  \dotsb, v^{(n)}$ be the vertices of $\mathbf{P}^\parallel(0)$. Then there exists
  an $\bar \epsilon>0$ and continuous maps $\Pi^{(k)}: \mathbf{P}'\rightarrow
  \mathfrak{N}^{\parallel}$, $k=1,\dotsb, n$, such that $v^{(k)}\mapsto v^{(k)}
  + \epsilon\Pi^{(k)}(q)$ gives a face lattice isomorphism between
  $\mathbf{P}^\parallel(0)$ and $\mathbf{P}^\parallel(\epsilon q)$ for all
  $\epsilon \in [0,\bar\epsilon]$ and $q\in \mathbf{P}'$ (see
  Fig.~\ref{fig:sliding_vertices}). 
  \begin{figure}
    \centering
    \includegraphics[width=.48\textwidth]{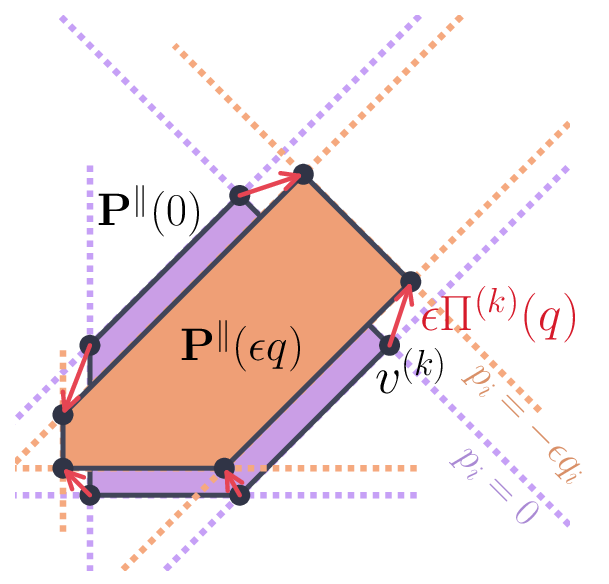}
    \caption{Schematic illustration of the relation between
    $\mathbf{P}^\parallel(0)$ (purple) and $\mathbf{P}^\parallel(\epsilon q)$
    (orange) for a fixed $q$ when $\epsilon$ is small.
    The dashed lines represent bounding hyperplanes corresponding to the
    inequalities $p_i \ge \epsilon q_i$. If $\rho_*$ is a regular value (for
    the facet density map), each vertex of the polytope
    $\mathbf{P}^\parallel(0)$ lies in exactly $\dim\mathfrak{N}_F$, in this
    case two, hyperplanes. As long as $\epsilon$ is small enough, the polytope
    $\mathbf{P}^\parallel(\epsilon q)$ will be bounded by the translations of
    the same set of hyperplanes, and there will be no ``new vertices''.
    }
    \label{fig:sliding_vertices}
  \end{figure}
  \begin{proof}
    The intuition is very simple: the convex polytope
    $\mathbf{P}^\parallel(\epsilon q)$ is defined to be the subset of
    $\bar\cldmap_F^{-1}(\rho_*)$ bounded by the $|I_F|$ inequalities given by
    $y_i \ge -\epsilon q_i$, where $i\in I_F$. When we perturb this system of
    inequalities by slowly increasing $\epsilon$ from zero, we are effectively
    translating the bounding hyperplanes of the polytope. If the initial
    ($\epsilon=0$) arrangement of the hyperplanes is not too ``degenerate'',
    then we expect the vertices of $\mathbf{P}^\parallel(\epsilon q)$ to be
    ``stable'' for small $\epsilon$. We will make this more precise in the
    following.

    For $i\in I_F$, define 
    $$
      H_i(\epsilon q) :=
      \{y\in\bar\cldmap_F^{-1}(\rho_*)\mid y_i = -\epsilon
      q_i\}\subset\bar\cldmap_F^{-1}(\rho_*).
    $$
    Intuitively, each $H_i(\epsilon
    q)$ should be a hyperplane in $\bar\cldmap_F^{-1}(\rho_*)$, but a priori it
    can happen that $H_i (\epsilon q) = \varnothing$ or $H_i(\epsilon q) =
    \bar\cldmap^{-1}_F(\rho_*)$. Nevertheless, the latter is impossible at
    $\epsilon = 0$: if $H_i(0)$ were all of $\bar\cldmap_F^{-1}(\rho_*)$ for
    some $i\in I_F$, then $\bar\cldmap_F^{-1}(\rho_*)$ is disjoint from
    $\relint(\Delta_{I_F})$, which is ruled out since $\rho_* \in \relint(F)$
    (see the proof of
    Proposition~\ref{thm:parametrizing_spaces_are_convex_polytopes}). Thus,
    each $H_i(0)$ is either empty or a hyperplane in
    $\bar\cldmap_F^{-1}(\rho_*)$. By compactness of $\mathbf{P}'$
    (Proposition~\ref{thm:parametrizing_spaces_are_convex_polytopes}), this
    continues to be true for $H_i(\epsilon q)$ for all $q\in\mathbf{P}'$ as
    long as $\epsilon$ is small enough, say $\epsilon \in [0,\bar\epsilon_1]$
    (where $\bar\epsilon_1 > 0$).

    For each $k\in \{1, \dotsb, n\}$, let $I_k\subset I_F$ be the index set of
    hyperplanes $H_i(0)$ that contain $v^{(k)}$. That is, $I_k:= \{i\in I\mid
    v^{(k)}_i = 0\}$. We claim that $|I_k| = \dim \mathfrak{N}_F$ for
    all $k$. Indeed, suppose $|I_k| > \dim \mathfrak{N}_F$, then
    $\rho_*=\sum_{i\in I_F}v^{(k)}_i \omega_i$ lies in the convex hull of $|I_F| -
    |I_k| < \dim \mathcal{H}_F - \dim \mathfrak{N}_F= \dim
    \conv(\Omega_F)+1$ weights, which implies that $\rho_*$ is a critical
    value (Proposition~\ref{thm:abelian_critical_value_charcterization}),
    contradicting our assumption. On the other hand, suppose $|I_k| < \dim
    \mathfrak{N}_F$.  We can then find a nonzero $\delta
    \in\mathfrak{N}_F$ such that $v^{(k)} + \mathbb{R}\delta\subset
    \bigcap_{i\in I_k}H_i(0)$. But $v^{(k)}_i > 0$ for all $i\in I_F\setminus
    I_k$, so we can find an open neighborhood $U$ of $0 \in \mathbb{R}$ such that $v^{(k)} +
    U\delta\subset \mathbf{P}^\parallel(0)$, which implies that $v^{(k)}$ is
    not extremal in $\mathbf{P}^\parallel(0)$ and hence not a vertex. So each
    vertex of $\mathbf{P}^\parallel(0)$ is contained in exactly
    $\dim\mathfrak{N}_F$ bounding hyperplanes. In technical terms, the
    convex polytope $\mathbf{P}^\parallel(0)$ is \textit{simple}.

    Let $I_{\text{rel}}:= \bigcup_{k=1}^nI_k$ be the set of indices of the
    ``relevant'' hyperplanes, for these are the ones that actually bound the
    polytope $\mathbf{P}^\parallel(0)$, and write $I_{\text{irrel}}:= I_F\setminus
    I_{\text{rel}}$.
    Again by compactness of $\mathbf{P}'$, we can find an $\bar \epsilon_2 \in
    (0, \bar\epsilon_1]$ such that for all $\epsilon \in [0, \bar\epsilon_2]$,
    for all $q\in \mathbf{P}'$, for all $k\in \{1, \dotsb, n\}$, and for all $j
    \in I_{\text{irrel}}$, it holds that $H_j(\epsilon q)\cap \bigcap_{i\in
    I_k}H_i(\epsilon q) = 0$. (In words: as long as $\epsilon$ is small enough,
    the irrelevant hyperplanes will stay irrelevant.)

    Now, for each $k\in \{1, \dotsb, n\}$ and $q\in \mathbf{P}'$, there is a
    unique $\Pi^{(k)}(q) \in \mathfrak{N}_F$ that solves
    $\Pi^{(k)}(q)_i =q_i$ for all $i \in I_k$. It is not hard to see that
    $\Pi^{(k)}(q)$ depends linearly on $q$. Then $(v^{(k)}+ \epsilon
    \Pi^{(k)}(q))_{k=1}^n$ are exactly the vertices of the convex polytope
    $\mathbf{P}^\parallel(\epsilon q)$ whenever $\epsilon \in [0,\bar\epsilon_3]$
    for some $\bar\epsilon_3 \in (0,\bar\epsilon_2]$.

    Finally, we show that the assignment $v^{(k)} \mapsto v^{(k)} + \epsilon
    \Pi^{(k)}(q)$ gives a lattice isomorphism between
    $\mathbf{P}^\parallel(0)$ and $\mathbf{P}^\parallel (\epsilon q)$ whenever
    $\epsilon \in [0, \bar\epsilon)$ for some $\bar\epsilon\in (0,\bar\epsilon_3]$.
    To be more precise, we define a map $\phi:
    \mathrm{Faces}(\mathbf{P}^\parallel(0)) \rightarrow
    \mathrm{Faces}(\mathbf{P}^\parallel(\epsilon q))$ in the following way:
    if $f = \conv(\{v^{(k_1)}, \dotsb v^{(k_m)}\})$ is a face of
    $\mathbf{P}^\parallel(\epsilon q)$, we set 
    $$
      \phi(f) = \conv(\{v^{(k_1)}+\epsilon \Pi^{(k_1)}(q), \dotsb, v^{(k_m)} +
      \epsilon \Pi^{(k_m)}(q)\}).
    $$
    It should be clear that $\phi(f)\subset \mathbf{P}^\parallel(\epsilon q)$
    is a face. Indeed, take $\alpha\in \mathbb{R}^{I_F}$ and $c\in \mathbb{R}$ such that
    $$
    \forall 1\le k \le n: 
    \begin{cases}
       v^{(k)} \in f \Rightarrow \braket{\alpha, v^{(k)}} = c\\
      v^{(k)} \notin f \Rightarrow \braket{\alpha, v^{(k)}} > c.
    \end{cases}
    $$
    Then
    $$
    \braket{\alpha, v^{(k)} + \epsilon \Pi^{(k)}(q)}
     = \braket{\alpha, v^{(k)}} + \epsilon \braket{\alpha, \Pi^{(k)}(q)}.
    $$
    The second term does not depend on $k$ as long as $v^{(k)}\in f$. Therefore, we have
    $$
    \forall 1\le k \le n: 
    \begin{cases}
      v^{(k)} \in f \Rightarrow \braket{\alpha, v^{(k)} + \epsilon \Pi^{(k)}(q)} = c'\\
      v^{(k)} \notin f \Rightarrow \braket{\alpha, v^{(k)} + \epsilon \Pi^{(k)}(q)} > c'
    \end{cases}
    $$
    for some $c'$ as long as $\epsilon$ is small enough, showing that
    $\phi(f)\subset \mathbf{P}^\parallel(\epsilon q)$ is a face.
    Since $\mathbf{P}'$ is compact and the number of faces is finite, there
    exists an $\bar\epsilon\in (0,\bar\epsilon_3]$ such that $\phi:
    \mathrm{Faces}(\mathbf{P}^\parallel(0)) \rightarrow
    \mathrm{Faces}(\mathbf{P}^\parallel(\epsilon q))$ is a lattice isomorphism
    for all $q\in \mathbf{P}'$ and $\epsilon \in [0, \bar\epsilon]$.
  \end{proof}
\end{propo}

\begin{cor}
  \label{thm:fiber_polytope_parametrization}
  Let $\sigma: \mathbf{P}^\parallel(0) \rightarrow \Delta_{n-1}$ be any 
  continuous system of generalized barycentric coordinates.
  Consider the ``parametrization map''
  \begin{equation}
    \begin{aligned}
      \Upsilon:\;& [0,\bar\epsilon]\times \mathbf{P}'\times \mathbf{P}^\parallel(0)
        \rightarrow \aff(\Delta_{I_F}) \\
    & (\epsilon, q, p) \mapsto \sum_{k=1}^n \sigma(p)_k
      (v^{(k)}+\epsilon\Pi^{(k)}(q))
      = p + \epsilon\sum_{k=1}^n\sigma(p)_k\Pi^{(k)}(q)
      .
    \end{aligned}
  \end{equation}
  For any $(\epsilon, q)$, the map $\Upsilon(\epsilon, q, \cdot)$ maps
  $\mathbf{P}^\parallel(0)$ surjectively onto $\mathbf{P}^\parallel(\epsilon
  q)$. Moreover, $\Upsilon$ depends continuously on $p$ and $q$, and
  affinely on $\epsilon$.
  \begin{proof}
    Apply Lemma~\ref{lem:comb_equiv_polytope_surj_map} to the lattice
    isomorphism given by Proposition~\ref{thm:comb_equiv_of_polytopes}.
  \end{proof}
\end{cor}

Using Corollary~\ref{thm:functional_classical_fiber}, we obtain the main result
of this section:
\begin{thm}[facet--off-facet decomposition of the pure functional]
  \label{thm:pure_functional_facet_decomposition}
  Let $\rho_*\in \relint(F)$ be a density on the facet which is a regular value and let $\eta \in
  \overrightarrow{\aff(\Omega)}$ be an inward vector. Choose some $l\in
  D\cldmap^{-1}(\eta)$, and let $\mathbf{P}'$ and $\mathbf{P}^\parallel$ be
  defined as in Proposition~\ref{thm:classical_density_fiber_splitting}
  according to the choice of $l$. Let $\bar \epsilon$ and  $\Pi^{(k)}$ be given
  by Proposition~\ref{thm:comb_equiv_of_polytopes}. Then
\begin{equation}
  \label{eq:F_facet_offfacet_decomposition_main}
    \Fp(\rho_* + \epsilon\eta)
    =
     \min_{\xi\in (S^1)^{I}}
     \min_{q\in \mathbf{P}'}
     \min_{p \in \mathbf{P}^\parallel(0)}
    \Big(
      Q_{\!f\!f}(\epsilon; \xi, p, q)
      + Q_{\!oo} (\epsilon; \xi, p, q)
      + Q_{\!f\!o} (\epsilon; \xi, p, q)
      \Big)
\end{equation}
  for $\epsilon \in [0, \bar\epsilon]$, where the subscripts stand for
  ``facet--facet'', ``off-facet--off-facet'' and ``facet--off-facet''
  respectively. Individually, these terms are (suppressing the dependence on
  $\epsilon, \xi,p,q$)
\begin{equation}
  \label{eq:F_facet_offfacet_decomposition}
  \begin{aligned}
    &Q_{\!f\!f} := \sum_{i,j\in I_F} 
  \bar\xi_i \xi_j W_{ij}
    \sqrt{p_i+ \epsilon q_i+\epsilon\sum_{k=1}^n \sigma(p)_k \Pi^{(k)}(q)_i}
    \sqrt{p_j+ \epsilon q_j+\epsilon\sum_{k=1}^n \sigma(p)_k \Pi^{(k)}(q)_j}
\\[.5em]
    &Q_{\!f\!o} := 2\mathrm{Re}\sum_{i\in I_F} \sum_{j\in I \setminus I_F}
    \bar\xi_i\xi_jW_{ij}\sqrt{p_i+ \epsilon q_i+\epsilon\sum_{k=1}^n \sigma(p)_k \Pi^{(k)}(q)_i}
      \cdot \sqrt{\epsilon} \sqrt{q_j}\\[.5em]
    &Q_{\!oo} := 
    \sum_{i,j \in I \setminus I_F} 
      \bar\xi_i\xi_j W_{ij}
      \epsilon \sqrt{q_iq_j}.
  \end{aligned}
\end{equation}
We will write $Q(\epsilon;\xi, p, q)$ for the sum of the three, so that
\begin{equation}
  \label{eq:pure_functional_minim_decomposed}
  \Fp(\rho_*+ \epsilon\eta) = 
  \min_{\xi\in (S^1)^{I}}
     \min_{q\in \mathbf{P}'}
     \min_{p \in \mathbf{P}^\parallel(0)}
  Q(\epsilon; \eta, p, q). 
\end{equation}
\end{thm}
Looking at Eqs.~\eqref{eq:F_facet_offfacet_decomposition_main} and
\eqref{eq:F_facet_offfacet_decomposition}, we can already anticipate where the
boundary force might stem from: the facet--off-facet term $Q_{fo}(\epsilon;
\xi,p,q)$ contains a factor of $\sqrt{\epsilon}$. If there were no
minimizations in Eq.~\eqref{eq:F_facet_offfacet_decomposition_main}, we would
know immediately at least one source of the boundary force (it is not clear
whether the rest of the terms contribute). It then remains to carry out the
minimizations and investigate how the $\sqrt{\epsilon}$ behavior survives after
doing so.

\subsubsection*{Step III: Exchanging $\doverd{\sqrt{\epsilon}}$ and $\min$}
\label{section:proof_step3}
At first sight, it is not clear whether and how one can differentiate the pure
functional $\Fp$, which is defined as a constrained minimization
(Definition~\ref{def:pure_ensemble_functional}). Now that we have cast $\Fp$
into a slightly nicer form, namely
Eq.~\eqref{eq:pure_functional_minim_decomposed}, it is still not obvious how
taking the derivative interacts with minimization.

There is a convenient tool for dealing with such problems, which we have used once in 
\cref{sec:BECforce_simlex}. Namely,
Danskin's theorem
\cite{danskinTheoryMaxMinApplications1966} asserts that if $\tilde Q: [0,
\varepsilon)\times X\rightarrow \mathbb{R}$ is a continuous function, where $X$ is a
compact metric space, and $\mathcal{\tilde F}(\epsilon) := \min_{x\in X}\tilde
Q(\epsilon; x)$, then $\mathcal{\tilde F}'(0_+)$ exists and is equal to
$\min_{x\in X_0} \tilde Q'(0; x)$ provided that $\tilde Q$ is continuously differentiable
with respect to $\epsilon$, where $X_0:= \argmin_{x\in X}\tilde Q(0; x)$.
In our case, the objective function $Q= Q_{\!f\!f} + Q_{\!f\!o}+ Q_{\!o\!o}$
(defined in Eq.~\eqref{eq:F_facet_offfacet_decomposition}) is not necessarily
continuously differentiable with respect to $\sqrt{\epsilon}$. To see what
could go wrong, note that $Q(\epsilon; \xi,p,q)$ contains terms of the form
$\sqrt{A + B\epsilon}$, where $A,B$ depend continuously on $p$ and $q$. Such a
function is not continuously differentiable at points where $A=\epsilon =0$ and
$B > 0$. Fortunately, the No Mixing Lemma
(Lemma~\ref{lem:minimizing_state_zero_coefficient}) will help us conclude that
these discontinuities will not matter \textit{at minimizers of the constrained
search}. This is the content of Lemma~\ref{lem:derivative_of_F_approaches_fo},
which moreover confirms our intuition that only $Q_{\!f\!o}$ should matter for
the boundary force.

Hereafter, a prime ($'$) on $Q$, $Q_{\!f\!o}$, $Q_{\!f\!f}$, or $Q_{\!oo}$ will
indicate differentiation with respect to $\sqrt{\epsilon}$. To save space, we
will often denote $(\xi, p, q)$ collectively by the variable $x\in (S^1)^{I}\times
\mathbf{P}^\parallel(0) \times \mathbf{P}'=: X$. With this notation,
Eq.~\eqref{eq:F_facet_offfacet_decomposition_main} becomes
$$
\Fp(\rho_*+\epsilon\eta) = \min_{x\in X} Q(\epsilon;x).
$$
At each $\epsilon$, the set of minimizers will be denoted by $X_\epsilon\subset
X$. That is, $\Fp(\rho_*+\epsilon\eta) = Q(\epsilon; x)$ for all $\epsilon$ and
$x\in X_\epsilon$.

\begin{lem}
  \label{lem:derivative_of_F_is_fo}
  \begin{equation}
    \begin{aligned}
      Q'(0;x) = Q_{\!f\!o}'(0; x)
      = 2\Re\sum_{i\in I_F}\sum_{j\in I\setminus I_F}
      \bar\xi_i\xi_j W_{ij} \sqrt{p_i} \sqrt{q_j}
    \end{aligned}
  \end{equation}
  for any $x =(\xi, p, q) \in X_0$.
  \begin{proof}
    We will show $Q_{\!f\!f}'(0; x) =0$ and $Q_{\!oo}'(0; x) = 0$.

    The latter is clear (see Eq.~\eqref{eq:F_facet_offfacet_decomposition}):
    \begin{equation}
      \doverd{\sqrt{\epsilon}} Q_{\!oo}(0; x) =
      \doverd{z}\Big|_{z=0}\Big(z^2 \sum_{i,j\in I\setminus I_F}\dotsb\Big) = 0
    \end{equation}
    The former, on the other hand, requires slightly more work to establish.
    Observe, with $Q_{\!f\!f}$ given in
    Eq.~\eqref{eq:F_facet_offfacet_decomposition}, that the potentially nonzero
    contributions to $Q'_{\!f\!f}$ arise only when differentiating a radical
    which vanishes at $\epsilon =0$. Let $\ket{\Phi(\epsilon)}$ denote the
    state corresponding to $(\epsilon, \xi, p, q)$, that
    is,
    \begin{equation}
      \begin{aligned}
        \ket{\Phi(\epsilon)} &= \sum_{i \in I_F}\xi_i \sqrt{y_i + q_i +
        \epsilon \sum_{k=1}^n \sigma(p)_k\Pi^{(k)}(q)_i}.
      \end{aligned}
    \end{equation}
    Then 
    $$
      Q'_{\!f\!f}(0;\xi,p,q) = 2\Re \braket{\Phi(0)|W|\Phi'(0)},
    $$
    where $\ket{\Phi'(\epsilon)} := \doverd{\sqrt{\epsilon}}
    \ket{\Phi(\epsilon)}$. Now, for each $i\in I_F$, if $\braket{E_i|\Phi(0)}\neq
    0$, then $\braket{E_i|\Phi'(0)}=0$ because the function $\epsilon \mapsto
    \sqrt{A\epsilon + B}$ is smooth for $B > 0$, and consequently the derivative with
    respect to $\sqrt{\epsilon}$ must vanish. In other words, $\ket{\Phi(0)}$
    and $\ket{\Phi'(0)}$ are strongly orthogonal. Since $\ket{\Phi(0)}$ is a
    minimizer of the constrained search at $\epsilon = 0$ (i.e., at $\rho_*$),
    we have $\braket{\Phi(0)|W|\Phi'(0)} = 0$ by
    Lemma~\ref{lem:minimizing_state_zero_coefficient_strong}.
  \end{proof}
\end{lem}

\begin{lem}
  \label{lem:derivative_of_F_approaches_fo}
  Let $(\epsilon_\nu)_\nu$ be a sequence converging to $0$ with $\epsilon_\nu
  \in [0, \bar\epsilon)$. Let $(x_\nu)_\nu = (\xi^\nu, p^\nu, q^\nu)_\nu$ be
  any sequence in $X$ converging to $x^* = (\xi^*, p^*, q^*) \in X$. If $x^*
  \in X_0$ (i.e., $x^*$ is a minimizer at $\epsilon = 0$), then
  \begin{equation}
    \label{eq:F_derivative_limit}
    \begin{aligned}
      \lim_{\nu\rightarrow\infty} Q'(\epsilon_{\nu}; x_{\nu})
      = Q'_{\!f\!o}(0; x^*).
    \end{aligned}
  \end{equation}
  That is, $Q$ is continuously differentiable at $(0, x^*)$.

  \begin{proof}


    To show Eq.~\eqref{eq:F_derivative_limit}, note that the only terms that
    might ruin the equality are those arising from differentiating (with
    respect to $\sqrt{\epsilon}$) the radical
    \begin{equation}
      \label{eq:problematic_radical}
      \sqrt{p_j + \epsilon q_j +\epsilon \sum_{k=1}^n \sigma(p)_k\Pi^{(k)}(q)_j}
    \end{equation}
    for some $j\in I_F$ for which the radical itself vanishes in the limit
    $\nu\rightarrow \infty$, since everything else is continuous after
    differentiation at $\epsilon=0$ (see
    Eq.~\eqref{eq:F_facet_offfacet_decomposition}). In the rest of this proof,
    we will show that while the derivative of such a ``problematic radical''
    need not converge to zero in the limit $\nu\rightarrow \infty$, the
    derivative itself never contributes to the limit because it is bounded and
    it is always multiplied by something that converges to zero.

    Fix a $j\in I_F$ such that 
    \begin{equation}
      \label{eq:radical_goes_to_zero}
      p_j^\nu + \underbrace{\epsilon_\nu q_j^\nu+ \epsilon_\nu \sum_{k=1}^n
      \sigma(p^\nu)_k\Pi^{(k)}(q^\nu)_j}_{=:\epsilon_\nu G_j(p^\nu,q^\nu)}
      \xrightarrow{\nu\rightarrow\infty} 0.
    \end{equation}
    We want to study the derivative
    \begin{equation}
      \label{eq:problematic_radical_derivative}
      \mathcal{N}_j^\nu := \doverd{\sqrt{\epsilon}}\Big|_{\epsilon = \epsilon_\nu}\sqrt{p^\nu_j + \epsilon G_j(p^\nu, q^\nu)},
    \end{equation}
    where $\mathcal{N}$ stands for ``nasty''. In $Q_{\!f\!f}'(\epsilon_\nu;
    x_\nu)$, each of these factors always comes multiplied by 
    $$
      \sum_{i\in I_F}\bar\xi_i^\nu W_{ij}\sqrt{p_i^\nu+\epsilon_\nu q^\nu_i +
      \epsilon_\nu \sum_{k=1}^n\sigma(p^\nu)_k \Pi^{(k)}(q^\nu)_i     }
    $$
    (see Eq.~\eqref{eq:F_facet_offfacet_decomposition}), which converges to
    zero by the No Mixing Lemma (Lemma~\ref{lem:minimizing_state_zero_coefficient}). 
    In $Q_{\!f\!o}'$, on
    the other hand, it is always multiplied by $\sum_{i\in I\setminus I_F}\bar
    \xi_i^\nu W_{ij}\sqrt{\epsilon_\nu}\sqrt{q_i^{\nu}}$, which also converges
    to zero. Therefore, if we can show that the sequence
    $(\mathcal{N}^j_\nu)_\nu$ is bounded, all the anomalous terms will vanish.

    Now, there are several possibilities for the derivative
    \eqref{eq:problematic_radical_derivative}. First, suppose $p_j^\nu +
    \epsilon_\nu G_j(p^\nu, q^\nu) = 0$. If $\epsilon_\nu =0$, then $p_j^\nu =
    0$, which implies $G_j(p^\nu, q^\nu)\ge 0$ (otherwise we would be able make
    the radicand negative by slightly increasing $\epsilon_\nu$). Consequently, we
    have $\mathcal{N}_j^\nu = \sqrt{G_j(p^\nu, q^\nu)}$, which is bounded
    because $G_j$ is continuous and $p^\nu, q^\nu$ belong to
    $\mathbf{P}^\parallel(0)$ and $\mathbf{P}'$ respectively, which are both
    compact. If $\epsilon_\nu \in (0, \bar\epsilon)$, then $G_j(p^\nu, q^\nu) = 0$ because
    otherwise the radicand could be made negative by perturbing $\epsilon_\nu$.
    In this case, $\mathcal{N}_j^\nu = 0$. 

    Next, we deal with the case $p^\nu_j+\epsilon_\nu G_j(p^\nu, q^\nu)>0$, for which
    $$
    \mathcal{N}_j^\nu =\frac{\sqrt{\epsilon_{\nu}}G_j(p^\nu, q^\nu)}{\sqrt{p^\nu_j +
         \epsilon_\nu G_j(p^\nu,q^\nu)}}.
    $$
    If $p^\nu_j = 0$, then $\mathcal{N}_j^\nu$ is simply $\sqrt{G_j(p^\nu, q^\nu)}$.
    Otherwise, we separate two cases depending on the sign of $G_j(p^\nu,
    q^\nu)$. If $G_j(p^\nu, q^\nu) \ge 0$, then $0 \le \mathcal{N}_j^\nu \le
    \sqrt{G_j(p^\nu, q^\nu)}$, so $\mathcal{N}_j^\nu$ is bounded. Suppose
    $G_j(p^\nu, q^\nu) <0$.  Since $p^\nu_j + \bar\epsilon G_j(p^\nu, q^\nu)\ge
    0$, we have $G_j(p^\nu, q^\nu) \ge -p^\nu_j / \bar\epsilon$, which holds
    for all $\nu$. It follows that
    $$
    \begin{aligned}
      0 &\le -\frac{\sqrt{\epsilon_\nu}G_j(p^\nu, q^\nu)}{\sqrt{p^\nu_j +
      \epsilon_\nu G_j(p^\nu, q^\nu)}}
      = \sqrt{-G_j(p^\nu, q^\nu)}\sqrt{\frac{-\epsilon_\nu G_j(p^\nu,
          q^\nu)}{p_j^\nu + \epsilon_\nu G_j(p^\nu, q^\nu)}}\\
      &= \sqrt{-G_j(p^\nu, q^\nu)}\sqrt{-1+\frac{p_j^\nu}{p_j^\nu + \epsilon_\nu G_j(p^\nu, q^\nu)}}\\
      & \le \sqrt{-G_j(p^\nu, q^\nu)}\sqrt{-1+\frac{p_j^\nu}{p_j^\nu - \frac{\epsilon_\nu}{\bar\epsilon}p_j^\nu}} 
      =\sqrt{-G_j(p^\nu, q^\nu)}\sqrt{-1 + \frac{\bar\epsilon}{\bar\epsilon-\epsilon_\nu}}
      \rightarrow 0.
    \end{aligned}
    $$
    This concludes the proof that none of the anomalous terms survive the limit
    $\nu\rightarrow \infty$.
  \end{proof}
\end{lem}

The following is adapted from the proof of Danskin's theorem
\cite{danskinTheoryMaxMinApplications1966}.

\begin{thm}
  \label{thm:deriv_of_min_is_min_of_deriv}
  \begin{equation}
    \begin{aligned}
      &\doverd{\sqrt{\epsilon}}\Big|_{\epsilon = 0 }
      \Fp(\rho_*+\epsilon\eta)
    =\min_{x\in X_0} Q'_{\!f\!o}(0; x)\\
    \end{aligned}
  \end{equation}
  \begin{proof}
    Let $(\epsilon_\nu)_\nu$ be a sequence converging to $0$ such that
    $\epsilon_\nu \in (0, \bar\epsilon)$. For each $\nu$, pick a minimizer
    $x_\nu\in X_{\epsilon_\nu}$. Also pick $x\in X_0$. Then
  \begin{equation}
    \label{eq:diff_quotient_upper_bound}
    \begin{aligned}
      &\frac{\Fp(\rho_* + \epsilon_\nu\eta) - \Fp(\rho_*)}{\sqrt{\epsilon_\nu}}
      = \frac{Q(\epsilon_\nu;x_\nu) - Q(0;x)}{\sqrt{\epsilon_\nu}}\\
      &= \frac{1}{\sqrt{\epsilon_\nu}}\Big(Q(\epsilon_\nu; x_\nu) 
      - Q(\epsilon_\nu; x)
      + Q(\epsilon_\nu; x) - Q(0; x)\Big)\\
      &\le \frac{Q(\epsilon_\nu; x) - Q(0; x)}{\sqrt{\epsilon_\nu}} 
      = Q'(b_\nu \epsilon_\nu ; x),
    \end{aligned}
  \end{equation}
    where $b_\nu \in [0,1]$ and we used the Mean Value Theorem in the last
    equality. Taking $\limsup_{m\rightarrow \infty}$ of
    Eq.~\eqref{eq:diff_quotient_upper_bound}, we get
      \begin{equation}
    \label{eq:limsup_upperbound}
    \begin{aligned}
      \limsup_{\nu\rightarrow \infty}&
      \frac{\Fp(\rho_* + \epsilon_\nu
      \eta) - \Fp(\rho_*)}{\sqrt{\epsilon_\nu}}
      \le  Q'(0; x) 
      = Q'_{\!f\!o}(0; x),
    \end{aligned}
  \end{equation}
    where we used the continuity of $\epsilon \mapsto Q'(\epsilon; x)$
    and the last equality is due to Lemma~\ref{lem:derivative_of_F_is_fo}. 
    Note that Eq.~\eqref{eq:limsup_upperbound} holds for any $x\in X_0$.

    By compactness of $X$, we extract a subsequence $(x_{\nu_l})_l$ such that
    $x_{\nu_l}\rightarrow x^*$ for some $x^*\in X$. We claim that $x^*\in X_0$.
    Indeed, for any $y\in X$ it holds that 
    $$
        Q(0; y) = \lim_{l\rightarrow\infty} Q(\epsilon_{\nu_l}; y)
        \ge \lim_{l\rightarrow \infty} Q(\epsilon_{\nu_l}; x_{\nu_l})
        = Q(0; x^*).
    $$
    By further extracting a subsequence if necessary, we may assume that 
    $$
    \lim_{l\rightarrow\infty}\frac{Q(\epsilon_{\nu_l};x_{\nu_l}) - Q(0; x^*)}{\sqrt{\epsilon_{\nu_l}}}
    =
    \liminf_{\nu\rightarrow\infty} \frac{Q(\epsilon_{\nu}; x_{\nu}) - Q(0;x^*)}{\sqrt{\epsilon_\nu}}.
    $$
    We now bound the difference quotient from below. We have
  \begin{equation}
    \label{eq:diff_quotient_lower_bound}
    \begin{aligned}
      &\frac{Q(\epsilon_{\nu_l};x_{\nu_l}) - Q(0; x^*)}{\sqrt{\epsilon_{\nu_l}}}
      = \frac{1}{\sqrt{\epsilon_{\nu_l}}}\Big(
        Q(\epsilon_{\nu_l}; x_{\nu_l}) - Q(0; x_{\nu_l}) 
        + Q(0; x_{\nu_l}) - Q(0; x^*)
        \Big)\\
      &\ge \frac{Q(\epsilon_{\nu_l}; x_{\nu_l}) - Q(0; x_{\nu_l})}{\sqrt{\epsilon_{\nu_l}}}
      = Q'(\tilde b_{l}\epsilon_{\nu_l}; x_{\nu_l}),
    \end{aligned}
  \end{equation}
    for some $(\tilde b_{l})_l$ with $\tilde b_l\in [0,1]$, where again we have used
    the Mean Value Theorem. Now we apply
    Lemma~\ref{lem:derivative_of_F_approaches_fo} to the sequences
    $(\tilde b_l\epsilon_{\nu_l})_l$ and $(x_{\nu_l})_l$ to conclude
    \begin{equation}
      \label{eq:liminf_lowerbound}
      \liminf_{\nu\rightarrow \infty}\frac{\Fp(\rho_*+\epsilon_{\nu}\eta) -
      \Fp(\rho_*)}{\sqrt{\epsilon_{\nu}}}
      \ge \lim_{l\rightarrow\infty}Q'(\tilde b_l\epsilon_{\nu_l}; x_{\nu_l})
      = Q'_{\!f\!o}(0; x^*).
    \end{equation}
  This, together with Eq.~\eqref{eq:limsup_upperbound}, proves the theorem.
  \end{proof}
\end{thm}

\subsubsection*{Step IV: Relaxation of $\mathbf{P}'$; Completing the Proof}
\label{section:proof_step4}

By Theorem~\ref{thm:deriv_of_min_is_min_of_deriv} and
Lemma~\ref{lem:derivative_of_F_is_fo}, we now have 
\begin{equation}
  \label{eq:boundary_force_as_minimization}
  \begin{aligned}
    \doverd{\sqrt{\epsilon}}\Big|_{\epsilon=0}\Fp(\rho_*+\epsilon\eta)
  =  \min_{(\xi, p, q) \in X_0} \Big(
    2\Re
    \sum_{i\in I_F}
    \sum_{j\in I\setminus I_F} 
    \bar\xi_i \xi_j W_{ij}\sqrt{p_i}\sqrt{q_j}\Big),
  \end{aligned}
\end{equation}
where $X_0$ denotes the set of minimizers of $Q(0; x)$:
\begin{equation}
  X_0 := \mathop{\mathrm{argmin}}\limits_{(\xi,p,q)\in X} 
  \left(\sum_{i,j\in I_F}\bar\xi_i\xi_j W_{ij}
  \sqrt{p_i}\sqrt{p_j}
  \right) \subset (S^1)^{I}\times \mathbf{P}^\parallel(0) \times \mathbf{P}' =: X.
\end{equation}
Since 
$$
  Q(0;\xi,p,q) = Q_{\!f\!f}(0;\xi,p,q) = \sum_{i,j\in I_F}\bar\xi_i \xi_j 
  W_{ij}\sqrt{p_i} \sqrt{p_j}
$$ 
(see Eq.~\eqref{eq:F_facet_offfacet_decomposition}) does not depend on $q$ and
$\xi_i$ for $i\in I\setminus I_F$, the set $X_0$ decomposes as a product $X_0 =
\tilde X_0 \times \mathbf{P}'\times (S^1)^{I\setminus I_F}$, where $\tilde
X_0\subset \mathbf{P}^\parallel(0) \times (S^1)^{I_F}$ consists of all pairs
$(p, \zeta)$ that minimize $Q_{\!f\!f}$ at $\epsilon =0$. In other words, the
minimization in Eq.~\eqref{eq:boundary_force_as_minimization} can be split into
two steps:
\begin{equation}
  \label{eq:G_with_min_split}
  \begin{aligned}
    \doverd{\sqrt{\epsilon}}\Big|_{\epsilon=0}\Fp(\rho_*+\epsilon\eta) &= 
    \min_{(p, \zeta)\in \tilde X_0}\;\min_{q\in \mathbf{P}', \chi\in
    (S^1)^{I\setminus I_F}} 2\Re\left(
    \sum_{i \in I_F} \sum_{j\in I\setminus I_F} 
      \bar\zeta_i \chi_j W_{ij}\sqrt{p_i}\sqrt{q_j}\right)\\
      &=\min_{(p,\zeta)\in \tilde X_0} \min_{q\in \mathbf{P}', \chi\in (S^1)^{I\setminus I_F}}
      2\Re
      \sum_{j\in I\setminus I_F} \chi_j\sqrt{q_j} 
      \sum_{i\in I_F} \bar\zeta_i \sqrt{p_i} W_{ij},
  \end{aligned}
\end{equation}
where we have split the collection of phases $\xi=(\xi_i)_{i\in I}$ into the
pair $(\zeta, \chi)$ with $\zeta\in (S^1)^{I_F}$ and $\chi \in
(S^1)^{I\setminus I_F}$. We will deal with the second (inner) minimization in
Eq.~\eqref{eq:G_with_min_split}, that is, the one over $(q, \chi)\in
\mathbf{P}'\times (S^1)^{I\setminus I_F}$. 

First, note that the minimization over the phases $(\chi_j)_{j\in I\setminus
I_F}$ is trivial since they appear independently. Consequently, we have
\begin{equation}
  \label{eq:G_with_min_split_2}
  \begin{aligned}
    &\doverd{\sqrt{\epsilon}}\Big|_{\epsilon=0}\Fp(\rho_*+\epsilon\eta) 
      =-2\max_{(p,\zeta)\in \tilde X_0} \max_{q\in \mathbf{P}'}
      \sum_{j\in I\setminus I_F} \sqrt{q_j} 
      \left|\sum_{i\in I_F} \bar\zeta_i \sqrt{p_i} W_{ij}\right|\\
      &=-2 \max_{(p,\zeta)\in \tilde X_0} \max_{q\in \mathbf{P}'}
      \sum_{j\in I\setminus I_F}\sqrt{q_j}
      \left|
      \braket{\Phi(p, \zeta)|W|E_j}
      \right|,
  \end{aligned}
\end{equation}
where we have defined  $\ket{\Phi(p, \zeta)} := \sum_{i\in I_F}\zeta_i
\sqrt{p_i}\ket{E_i}$. Hence, we are left with the minimization over $q\in
\mathbf{P}'$. We noted before (see
Proposition~\ref{thm:parametrizing_spaces_are_convex_polytopes}) that
$\mathbf{P}'$ is a convex polytope. However, its shape could still be fairly
complicated in general, and it is a priori not clear whether it is easy to
carry out the minimization over $q\in \mathbf{P}'$ in
Eq.~\eqref{eq:G_with_min_split_2}. On the other hand, the objective function
depends only on $(q_j)_{j\in I\setminus I_F}$, so not all defining constraints of
the polytope $\mathbf{P}'$ are necessarily relevant. To make this more precise,  
let $\mathrm{pr}^F$ denote the projection that sets the entries $y_i$
of a vector $y\in \mathfrak{D}\subset \mathbb{R}^I$ to zero whenever $i\in I_F$. That is,
\begin{equation}
    \mathrm{pr}^F(y)_i = \begin{cases}
      0 & i \in I_F\\
      y_i & i \in I \setminus I_F.
    \end{cases}
\end{equation}
Then we may replace $\mathbf{P}'$ with $\mathrm{pr}^F(\mathbf{P}')$ in
Eq.~\eqref{eq:G_with_min_split_2}. The hope now is that
$\mathrm{pr}^F(\mathbf{P}')$ admits a simpler characterization, which is
fortunately true thanks to the following lemma:

\begin{lem}[Relaxation of the polytope $\mathbf{P}'$]
  \label{lem:replace_P_relax}
  Let $S\in V$ be an external potential such that $\braket{\eta, S}=1$ and
  $\braket{\overrightarrow{\aff(F)}, S} = 0$. Then
  $\mathrm{pr}^F(\mathbf{P}') =
  \mathrm{pr}^F(\mathbf{P}'_{\mathrm{relax}})$, where
  $$
  \mathbf{P}'_{\mathrm{relax}} := \Big\{
    q \in \mathfrak{D} \mid \left\langle D\cldmap(q), S\right\rangle = 1,
        \forall i \in I\setminus I_F: q_i \ge 0
    \Big\}.
  $$
  In other words, we may replace the constraint $D\cldmap (q) = \eta$, which
  was used to define the set $\mathbf{P}'$, with the (in general) much weaker
  constraint $\left\langle D\cldmap(q), S\right\rangle = 1$, as long as we only
  care about the image under the projection $\mathrm{pr}^F$.
  \begin{proof}
    Recall that $\mathbf{P}':= \{l+q \mid q\in \mathfrak{N}', \forall i \in I\setminus
    I_F: q_i \ge 0\}$, where $\mathfrak{N}'\subset \mathfrak{N}$ is a chosen
    and fixed complement of $\mathfrak{N}_F\subset \mathfrak{N}$. We will prove 
    \begin{equation}
      \label{eq:affine_spaces_same_image}
      \mathrm{pr}^F(l+ \mathfrak{N}') =  \mathrm{pr}^F\underbrace{\Big\{q'\in \mathfrak{D} \mid
      \left\langle D\cldmap(q'), S\right\rangle=1\Big\}}_{=:\mathcal{S}}.
    \end{equation}
    That is, we first ignore the inequalities $q_i \ge 0$ so that both sets
    become affine spaces in $\mathrm{pr}^F(\mathfrak{D})$. Once
    Eq.~\eqref{eq:affine_spaces_same_image} is proved, the original statement
    will follow immediately because $\mathrm{pr}^F(\mathbf{P}')$ and
    $\mathrm{pr}^F(\mathbf{P}'_{\mathrm{relax}})$ are further cut out 
    from $\mathrm{pr}^F(l+\mathfrak{N}')$ and $\mathrm{pr}^F(\mathcal{S})$
    respectively by the same set of inequalities, namely $q_i \ge 0$ for $i\in I\setminus I_F$.

    Take any $q  \in l+ \mathfrak{N}'$. Then $\braket{D\cldmap(q), S} =
    \braket{\eta, S} = 1$. This shows $\mathrm{pr}^F(l + \mathfrak{N}')\subset
    \mathrm{pr}^F(\mathcal{S})$.

    The strategy to show the inclusion in the other direction is to show that
    $\mathrm{pr}^F(l+\mathfrak{N}')$ and $\mathrm{pr}^F(\mathcal{S})$ have the
    same dimension as affine spaces. Clearly, $\dim
    (\mathrm{pr}^F(\mathcal{S})) = |I| - |I_F| - 1$.

    We claim that $\mathrm{pr}^F$ is injective on $\mathfrak{N}'$. Indeed,
    suppose $\mathrm{pr}^F(q) = 0$ for some $q\in \mathfrak{N}'$, then $q_i\neq
    0$ only when $i \in I_F$. But this means $q \in \mathfrak{N}_F$. Since $\mathfrak{N}_F
    \cap \mathfrak{N}'= 0$, we have $q =0$. This implies $\dim\mathrm{pr}^F(l +
    \mathfrak{N}') = \dim \mathrm{pr}^F(\mathfrak{N}') = \dim \mathfrak{N}'$.

    It remains to show that $\dim \mathfrak{N}'= |I| - |I_F| - 1$. But this follows
    directly from Proposition~\ref{thm:dim_of_facet_offfacet_null_spaces}, which asserts
    $$
      \begin{aligned}
        &\dim \mathfrak{N} = |I| - \dim \conv(\Omega) -  1\\
        &\dim \mathfrak{N}_F = |I_F| - \dim F -  1.
      \end{aligned}
    $$
    Taking the difference then gives $\dim \mathfrak{N}' = |I| - |I_F| - 1$.
    Therefore, $\dim(\mathrm{pr}^F(l+\mathfrak{N}')) = \dim
    (\mathrm{pr}^F(\mathcal{S}))$.
  \end{proof}
\end{lem}

\begin{cor}
\label{thm:replace_P_relax_2}
  $\mathrm{pr}^F(\mathbf{P}') = 
\left\{q\in \mathbb{R}^{I\setminus I_F} \mid q_i \ge 0, \sum_{i\in I\setminus I_F}
    q_i \braket{\omega_i-\gamma, S} = 1
    \right\}$ for any $\gamma\in \aff(F)$.
    \begin{proof}
      By Lemma~\ref{lem:replace_P_relax}, we can replace the left hand side by
      $\mathrm{pr}^F(\mathbf{P}'_{\mathrm{relax}})$.
      Take any $q\in \mathbf{P}_{\mathrm{relax}}'$, then $\sum_{i\in I} q_i
      \braket{\omega_i, S} = 1$. Since $\braket{\omega_i, S} = \braket{\gamma,
      S}$ for all $i\in I_F$, we have
      $$
        \sum_{i\in I_F}q_i \braket{\gamma, S} + \sum_{i\in I\setminus I_F}
        \braket{\omega_i, S} = \sum_{i\in I\setminus I_F}
        q_i\braket{\omega_i-\gamma, S} = 1,
      $$
      showing ``$\subset$''. Conversely, take any $q\in \mathbb{R}^{I\setminus
      I_F}\subset \mathbb{R}^I$ satisfying $\sum_{i\in I\setminus I_F}q_i
      \braket{\omega_i-\gamma, S}=1$. We can find $\tilde q \in
      \mathbb{R}^{I_F}$ such that $q+\tilde q \in \mathfrak{D}$. Then
      $\mathrm{pr}^F(q + \tilde q) = q$, and 
      $$
        \braket{D\cldmap(q+\tilde q), S} = \sum_{i\in I\setminus I_F}q_i \braket{\omega_i,S}
        + \sum_{i\in I_F} \tilde q_i \underbrace{\braket{\omega_i, S}}_{=\braket{\gamma, S}}
        = \sum_{i\in I\setminus I_F} q_i \braket{\omega_i - \gamma, S},
      $$
      so we also have ``$\supset$''.
    \end{proof}
\end{cor}

Finally, we prove Theorem~\ref{thm:boundary_force_formula}:

\begin{proof}[proof of Theorem~\ref{thm:boundary_force_formula}]

  Eq.~\eqref{eq:G_with_min_split_2} together with
  Corollary~\ref{thm:replace_P_relax_2} gives
\begin{equation}
  \label{eq:G_with_min_split_3}
  \begin{aligned}
    &\doverd{\sqrt{\epsilon}}\Big|_{\epsilon=0}\Fp(\rho_*+\epsilon\eta) 
      =-2 \max_{(p,\zeta)\in \tilde X_0} \max_{q\in \mathbf{P}'_{\mathrm{relax}}}
      \sum_{j\in I\setminus I_F}\sqrt{q_j}
      \left|\braket{\Phi(p, \zeta)|W|E_j}\right|\\
      &= -2 \max_{(p,\zeta)\in \tilde X_0} 
      \max_{\substack{s\in \mathbb{R}^{I\setminus I_F}, s_i \ge 0\\ 
      \sum_{j\in I\setminus I_F}s_j^2\braket{\omega_j-\gamma,S} = 1}
      }
      \sum_{j\in I\setminus I_F}s_j
      \left|\braket{\Phi(p, \zeta)|W|E_j}\right|.
  \end{aligned}
\end{equation}
This constrained optimization problem is elementary, with solution given by
$$
  s_j = \left[\sum_{j'\in I\setminus I_F}\frac{\left|\braket{\Phi(p,
  \zeta)|W|E_{j'}}\right|^2}{\braket{\omega_{j'}-\gamma, S}}\right]^{-\frac{1}{2}}
  \frac{\left|\braket{\Phi(p,\zeta)|W|E_j}\right|}{\braket{\omega_j - \gamma, S}}.
$$
Plugging this back into Eq.~\eqref{eq:G_with_min_split_3} finally yields
$$
\doverd{\sqrt{\epsilon}}\Big|_{\epsilon=0}\Fp(\rho_* + \epsilon \eta)
= -2\max_{(p,\zeta)\in \tilde X_0} 
    \left[\sum_{j\in I\setminus I_F}\frac{\left|\braket{\Phi(p,\zeta)|W|E_j}\right|^2}{\braket{\omega_j-\gamma, S}}
    \right]^{\frac{1}{2}}.
$$
To make the independence of the chosen basis $(\ket{E_i})_{\in I}$ manifest,
  note that the denominator of each summand only depends on $\omega_j$, so we
  can collect the terms corresponding to each weight $\omega \in \Omega$ so
  that the individual projectors $\ketbrap{E_j}$ in the numerators sum to the
  orthogonal projection $\Pi_\omega$ onto the weight space
  $\mathcal{H}_\omega$. The equation then becomes
$$
\doverd{\sqrt{\epsilon}}\Big|_{\epsilon=0}\Fp(\rho_* + \epsilon \eta)
= -2\max_{(p,\zeta)\in \tilde X_0} 
    \left[\sum_{\omega\in \Omega\setminus \Omega_F}
    \frac{\lVert \Pi_\omega W \ket{\Phi(p,\zeta)}\rVert^2}{\braket{\omega-\gamma, S}}
    \right]^{\frac{1}{2}}.
$$
\end{proof}

\chapter{Interlude: Momentum Maps}
\label{chap:interlude}

Although our development of abelian functional theories in
\cref{chap:abelian_functional_theory} is mathematically satisfactory,
interesting functional theories are often not abelian. RDMFT is such an
example, where $\iota(V)$ consists of all single-particle operators, which in
general do not commute. As another example, consider a system of $N$ two-level
spins (\cref{ex:nonabelian_spin_chain}), where we allow all local operators
to be external potentials. Certainly, the external potentials in this case also
do not commute. In \cref{chap:abelian_functional_theory}, we saw that the
properties of the functionals are heavily influenced by the geometry of their
domains. Before we can hope to derive any interesting properties of functional
theories with non-commuting external potentials, then, it is necessary to first
address the representability problem, that is, that of characterizing
$\iota^*(\mathcal{P})$ and $\iota^*(\mathcal{E})$. In this scenario, we can
no longer apply \cref{thm:abelian_nrep_set}: the notion of ``weights'' does not
even make sense in the first place if $\iota(V)$ is not commutative. As argued
already in \cref{chap:abelian_functional_theory} , it
would be overoptimistic to hope to solve the representability problem in full
generality. Therefore, it would be desirable to isolate a class of functional
theories which is not too large to involve for example the two-body
$N$-representability problem (see \cref{ex:2bodyqma}), but still large enough
to include
functional theories like RDMFT or the one for spin chains with local external
potentials.
As it turns out, the representability problem can be treated using 
techniques from symplectic geometry and representation theory
whenever the set of external potentials $V$ has the structure of a Lie algebra
and the potential map $\iota: V\rightarrow i\mathfrak{u}(\mathcal{H})$ is the
Lie algebra representation (up to $i$) corresponding to a unitary
representation of a compact Lie group on $\mathcal{H}$. In this case, the pure
state density map $\iota^*|_{\mathcal{P}}: \mathcal{P}\rightarrow V^*$ will be
a so-called \textit{momentum map} (up to $i$), the image of which has been
considerably studied in the symplectic geometry literature. 


Before diving into the details, we will briefly summarize the history of the problem.
Due to its application in quantum chemistry, quantum many-body theory and
quantum information theory, the problem of characterizing
$\iota^*(\mathcal{P})$ and $\iota^*(\mathcal{E})$ for a given potential map
$\iota: V\rightarrow i\mathfrak{u}(\mathcal{H})$, which is the same as that of
characterizing the set of all possible combinations of expectation values of a
given list of operators, is not new and has been previously studied in the
literature. It is a generalization of the \textit{quantum marginal problem},
which is concerned with determining whether a given list of reduced density
matrices (marginals) can originate from a pure or ensemble state in a
multipartite Hilbert space (see, for example,
Refs.~\cite{schillingQuantumMarginalProblem2014a,schillingQuantumMarginalProblem2014}).
An important special case of the quantum marginal problem is that of
characterizing the set of realizable 1RDMs, which is also known as the
\textit{$N$-representability problem}. It has long been known
\cite{colemanStructureFermionDensity1963} that for fermions, the necessary and
sufficient conditions for a 1RDM to arise from an $N$-particle
ensemble state are the usual Pauli constraints. In other words, the eigenvalues
of the 1RDM lie in $[0,1]$, and sum to $N$. The pure state $N$-representability
problem, on the other hand, is much more complicated. Borland and Dennis
\cite{borlandConditionsOnematrixThreebody1972} found a set of complete
constraints on the 1RDM spectrum for $N=3$ fermions with a $d=6$-dimensional
single-particle Hilbert space, which is the lowest combination of $(d,N)$ for
which the $N$-representability problem is not trivial (if $N=1$ or $N=d-1$, the
problem is completely trivial. If $N=2$ or $N=d-2$, the eigenvalues of the 1RDM
have to come in pairs, which is the only constraint apart from normalization).
As another variant of the quantum marginal problem, Higuchi et al.
\cite{higuchiOnequbitReducedStates2003} studied and solved the task of
identifying all collections of single-qubit reduced density matrix spectra
$(\lambda^{(1)}, \dotsb, \lambda^{(N)})$, $\lambda^{(i)}\in \mathbb{R}^2$, which
arise from some $N$-qubit pure state (see also
Ref.~\cite{bravyiRequirementsCompatibilityLocal2003}). Later, Klyachko
\cite{klyachkoQuantumMarginalProblem2004} provided a general solution to the
\textit{pure univariant} quantum marginal problem, which includes the preceding
two scenarios as special cases (see also
Refs.~\cite{knutsonSymplecticAlgebraicGeometry2000,daftuarQuantumStateTransformations2005,klyachkoQuantumMarginalProblem2006,christandlSpectraDensityOperators2006,altunbulakPauliPrincipleRepresentation2008,altunbulakPauliPrincipleRevisited2008}),
based on previous work by Berenstein and Sjamaar
\cite{berensteinCoadjointOrbitsMoment1999}, which is in turn built upon
previous results on momentum maps
\cite{heckmanProjectionsOrbitsAsymptotic1982,guilleminConvexityPropertiesMoment1982,guilleminConvexityPropertiesMoment1984,atiyahMomentMapEquivariant1984,kirwanConvexityPropertiesMoment1984,sjamaarConvexityPropertiesMoment1998}.


In this chapter, we lay out in a self-contained way the mathematical
fundamentals of momentum maps, which will be applied in
\cref{chap:momentum_map_functional_theories} to our discussion of a large class
of nonabelian functional theories. Beyond developing the machinery and language
needed for functional theories, the material that follows also serves as a
basic introduction to lines of investigation in quantum information theory to
understand the geometry of multipartite quantum states using momentum map
techniques (see, for example,
Refs.~\cite{sawickiSymplecticGeometryEntanglement2011,walterEntanglementPolytopesMultiparticle2013,sawickiConvexityMomentumMap2014}).
It is also of independent interest in other fields of physics, most notably
geometric quantization \cite{simmsLecturesGeometricQuantization1976}. The
contents of this chapter are mainly drawn from
Refs.~\cite{guilleminSymplecticTechniquesPhysics1990,guilleminConvexityPropertiesHamiltonian2005,walterMultipartiteEntanglement2017},
with
Refs.~\cite{leeIntroductionSmoothManifolds2012,tuDifferentialGeometry2017,leeIntroductionComplexManifolds2024,ortegaMomentumMapsHamiltonian2004,dasilvaLecturesSymplecticGeometry2008}
as general references for differential geometry and symplectic geometry and
Refs.~\cite{humphreysIntroductionLieAlgebras1972,fultonRepresentationTheory2004,hallLieGroupsLie2015}
for representation theory. We will keep following our convention that a
\textit{thing} always means a \textit{finite-dimensional thing} implicitly,
whenever applicable. In particular, ``Lie algebra'' without modifier will mean
``finite-dimensional Lie algebra''. Throughout this chapter, we will avoid
using the Dirac bra-ket notation, so a vector in $\mathcal{H}$ will simply be
written as $\psi$ instead of $\ket{\psi}$.


\section{Symplectic Geometry}
\label{sec:symplectic_basics}
Let $M$ be a smooth manifold. A \textit{symplectic form} on $M$ is a nondegenerate
2-form $\omega \in \Omega^2(M)$ that is closed. The pair $(M,\omega)$ is called
a \textit{symplectic manifold}. Given a smooth function $f\in C^\infty(M)$, the
\textit{symplectic gradient} is the vector field defined by $\sgrad f :=
(\dstraight f)^\sharp$, where $\sharp: \Omega^1(M)\rightarrow \mathfrak{X}(M)$
is the isomorphism induced by $\omega$. In other words, $\sgrad f$ is the
unique vector field satisfying
\begin{equation}
  \iota_{\sgrad f} \omega = \dstraight f.
\end{equation}
It is common to denote $\sgrad f$ by $X_f$, but we will avoid this notation. A
vector field $X$ is \textit{Hamiltonian} if $X = \sgrad{f}$ for some smooth
function $f$. If $X$ is Hamiltonian, then
$$
L_X\omega = \dstraight\iota_X\omega  + \iota_X \dstraight \omega =
\dstraight\dstraight f = 0,
$$
where $L_X(\cdot)$ and $\iota_X(\cdot)$ denote the Lie derivative along $X$ and
contraction with $X$ respectively.
A vector field $Y\in\mathfrak{X}(M)$ satisfying $L_Y\omega = 0$ is called
\textit{symplectic}. Hence, all Hamiltonian vector fields are symplectic.
Conversely, all symplectic vector fields are only locally Hamiltonian, but in
general there are symplectic vector fields that are not globally Hamiltonian.
Let $\mathfrak{X}_H(M)$ and $\mathfrak{X}_S(M)$ denote the sets of Hamiltonian
and symplectic vector fields respectively, then the situation is as follows:
$$
  \mathfrak{X}_H(M)\subset \mathfrak{X}_S(M)\subset \mathfrak{X}(M)
$$


\begin{propo}
\label{thm:comm_of_symplectic_vfs}
Let $X,Y\in \mathfrak{X}_S(M)$ be symplectic vector fields, then 
\begin{equation}
[X,Y] = \sgrad(\omega(Y,X)).
\end{equation}
In particular, $[X,Y]$ is Hamiltonian and therefore $[\mathfrak{X}_S(M), \mathfrak{X}_S(M)]\subset \mathfrak{X}_H(M)$.
\begin{proof}
$\iota_{[X,Y]} \omega = L_X\iota_Y \omega - \underbrace{\iota_YL_X\omega}_{=0}
= \dstraight(\iota_X\iota_Y\omega) + \underbrace{\iota_X\mathrm{d}(\iota_Y\omega)}_{=0}
= \mathrm{d}(\omega(Y,X))$.
\end{proof}
\end{propo}

The symplectic form $\omega$ induces a map $\{\cdot ,\cdot \}:
C^\infty(M)\times C^\infty(M)\rightarrow C^\infty(M)$ defined by $\{f, g\} =
\omega(\sgrad f, \sgrad g)$, called the \textit{Poisson bracket}, making
$(C^\infty(M), \{\cdot, \cdot \})$ into a Poisson algebra, in particular a Lie
algebra.

\begin{propo}
  The map $\sgrad: C^\infty(M)\rightarrow \mathfrak{X}(M)$ is a Lie algebra
  antihomomorphism. That is, 
  \begin{equation}
  [\sgrad f, \sgrad g] = -\sgrad\{f, g\}.
  \end{equation}
  \begin{proof}
  The vector fields $\sgrad f, \sgrad g$ are Hamiltonian, so in particular
    symplectic. By Proposition~\ref{thm:comm_of_symplectic_vfs}, we have 
  $[\sgrad f, \sgrad g] = \sgrad (\omega(\sgrad g, \sgrad f)) = -\sgrad \{f,g \}$.
  \end{proof}
\end{propo}
The kernel of the map $\sgrad$ is the locally constant functions, so we have an
exact sequence of Lie algebras
\begin{equation}
  \label{eq:symplectic_exact_sequence}
  0 \rightarrow \HdR^0(M) \rightarrow C^\infty(M) \xrightarrow{\sgrad} \mathfrak{X}_H(M) \rightarrow 0.
\end{equation}

\subsection{Momentum Maps}
Let $G$ be a Lie group acting on a symplectic manifold $(M,\omega)$ on the
left. For $g\in G$, let $l_g: M\rightarrow M$ denote the diffeomorphism $p
\mapsto g\cdot p$. Let $\mathfrak{g}$ denote the Lie algebra of $G$.
The $G$-action on $M$ defines a Lie algebra antihomomorphism $\phi:
\mathfrak{g}\rightarrow \mathfrak{X}(M)$, sending an element $A\in
\mathfrak{g}$ of the Lie algebra to its corresponding fundamental vector field
$A^*\subset\mathfrak{X}(M)$.
\begin{defn}
  The $G$-action on $M$ is \textbf{symplectic} if 
  \begin{equation}
    \label{eq:symplectic_action_condition}
    l_g^*\omega = \omega
  \end{equation}
    for all $g\in G$.
\end{defn}
If the $G$-action on $M$ is symplectic, then taking $g = \exp(tA)$ in
Eq.~\eqref{eq:symplectic_action_condition}, $A\in \mathfrak{g}$, and
differentiating with respect to $t$ gives $L_{A^*}\omega = 0$. That is, the
fundamental vector field $A^*$ is symplectic and therefore $\phi(\mathfrak{g})
\subset \mathfrak{X}_S(M)$.

\begin{defn}
  \label{def:momentum_map}
  Let $G$ be a connected Lie group acting on a symplectic manifold
  $(M,\omega)$, not necessarily preserving the symplectic form.
  A \textbf{comomentum map} (for the $G$-action on
  $M$) is a Lie algebra homomorphism $\mu: \mathfrak{g}\rightarrow C^\infty(M)$
  such that the following diagram commutes:
  \begin{equation}
    \label{diag:momentum_map_def_lift}
  \begin{tikzcd}
    C^\infty(M) \arrow{r}{\sgrad} & \mathfrak{X}(M) \\
    & \mathfrak{g}\arrow[swap]{u}{\phi} \arrow[dashed]{ul}{\mu}
  \end{tikzcd}
  \end{equation}
  That is, $\mu$ is a lift of $\phi: \mathfrak{g}\rightarrow \mathfrak{X}(M)$
  to $\mathfrak{g}\rightarrow C^\infty(M)$. The corresponding \textbf{momentum
  map} is the map $\Phi: M\rightarrow \mathfrak{g}^*$ defined by
  $\braket{\Phi(p), A} = \mu(A)(p)$.
\end{defn}


\begin{rem}
  In Diagram~\ref{diag:momentum_map_def_lift}, the maps $\sgrad$ and $\phi$
  are antihomomorphisms, so it makes sense to expect $\mu$ to be a (usual) homomorphism.
\end{rem}

\begin{rem}
  There is a more general definition of momentum maps for not necessarily
  connected Lie groups, but we will always work with connected ones.
\end{rem}

\begin{rem}
  \label{rem:momentum_map_def_symplectic}
  The image of $\sgrad$ in $\mathfrak{X}(M)$ is, by definition, the Hamiltonian
  vector fields $\mathfrak{X}_H(M)\subset \mathfrak{X}(M)$. It follows that if
  $\phi(\mathfrak{g})\not\subset\mathfrak{X}_H(M)$, then there will be no
  momentum maps. In other words, if there is a momentum map, then
  $\phi(\mathfrak{g})\subset \mathfrak{X}_H$, so the $G$-action must be at
  least symplectic.
\end{rem}

\subsection{Existence and Uniqueness}
In this section, we will discuss the existence and uniqueness of momentum maps
given a $G$-action on $M$. For applications in functional theories, we will
simply show that a given map $M\rightarrow \mathfrak{g}^*$ is a momentum map,
and we will not be concerned with its uniqueness. Thus, this section is not
relevant for \cref{chap:momentum_map_functional_theories}.

From now on we will always assume that $G$ is connected. From the definition
alone, it is not clear whether a (co)momentum map should exist or, if one
exists, whether it is unique. For example, as pointed out in
Remark~\ref{rem:momentum_map_def_symplectic}, if the image of the homomorphism
$\phi: \mathfrak{g} \rightarrow \mathfrak{X}(M)$ is not contained in
$\mathfrak{X}_H(M)$, then there would not exist even a map of \textit{sets}
$\mu$ that makes Diagram~\ref{diag:momentum_map_def_lift} commute. If
$\phi(\mathfrak{g})\subset \mathfrak{X}_H(M)$, then we can always find a
\textit{linear} map from $\mathfrak{g}$ to $C^\infty(M)$ making
Diagram~\ref{diag:momentum_map_def_lift} commute, but it is not clear whether
we can always find a Lie algebra homomorphism.

The conditions for the existence and uniqueness of momentum maps will be
phrased in terms of the Lie algebra cohomology $H^\bullet(\mathfrak{g})$ of
$\mathfrak{g}$, which is defined as the cohomology of the differential complex 
$$
  0\xrightarrow{\delta} \mathbb{R}\xrightarrow{\delta} \mathfrak{g}^*
  \xrightarrow{\delta} \exterior^2\mathfrak{g}^* \xrightarrow{\delta} 
\exterior^3\mathfrak{g}^*
\xrightarrow{\delta} 
  \dotsb,
$$
where $\delta: \exterior^k\mathfrak{g}^*\rightarrow \exterior^{k+1}\mathfrak{g}^*$ is defined as
$$
\delta\omega(A_1, \dotsb, A_{k+1})
 = \sum_{1\le i < j \le k+1} (-1)^{i+j}\omega([A_i, A_j], A_1, \dotsb,
 \widehat{A_i}, \dotsb, \widehat{A_j}, \dotsb, A_{k+1})
$$
and $\delta\omega =0 $ for $\omega \in\exterior^0\mathfrak{g}^* = \mathbb{R}$.
An important basic result is \textit{Whitehead's lemma}, which asserts that
$H^1(\mathfrak{g})=H^2(\mathfrak{g})=0$ if $\mathfrak{g}$ is semisimple.

\begin{propo}
  Assume that the $G$-action on $M$ is symplectic.
If $H^1(\mathfrak{g})=0$ or $\HdR^1(M)=0$, then $\phi(\mathfrak{g})\subset \mathfrak{X}_H(M)$.
\begin{proof}
    Let $j: \mathfrak{g}\rightarrow \Omega^1(M)$ denote the map $A\mapsto
    \iota_{A^*} \omega \in \Omega^1(M)$. Then $j$ is the composition 
    of the inverse of $\sharp$, which we denote by
    $\flat:
    \mathfrak{X}(M)\rightarrow \Omega^1(M)$, and $\phi$. Hence, showing that
    $\phi$ sends everything to Hamiltonian vector fields is equivalent to
    showing that $j$ sends everything to exact $1$-forms.
    $$
      \begin{tikzcd}
        & \mathfrak{X}(M)\arrow{dr}{\flat} & \\
        \mathfrak{g}\arrow{ur}{\phi}\arrow[swap]{rr}{j} & & \Omega^1(M)\\
      \end{tikzcd}
    $$
    Take any $A\in \mathfrak{g}$. By Cartan's magic formula,
    $$
    \mathrm{d}j(A) = \mathrm{d}(\iota_{A^*}\omega) = L_{A^*}\omega - \iota_{A^*}\mathrm{d}\omega = 0,
    $$
    so $j(A)$ is closed. If $\HdR^1(M)= 0$, then $j(A)$ is also exact and we are done.

    The above argument shows that we can define a map
    $$
    \begin{aligned}
        j': &\;\mathfrak{g} \rightarrow \HdR^1(M)\\
        & A \mapsto [j(A)],
      \end{aligned}
    $$
    where $[j(A)]$ denotes the cohomology class of $j(A)$. Hence, $j'$ can be
    viewed as an element in $\mathfrak{g}^*\otimes \HdR^1(M)$. For any
    $A,B\in \mathfrak{g}$, we then have
    $$
      \delta j' (A,B) = -j'([A,B]) = -[j([A,B])] = [\iota_{[A^*, B^*]}\omega] = [\dstraight(\omega(B^*, A^*))]=0,
    $$
    where we have used Proposition~\ref{thm:comm_of_symplectic_vfs} in the penultimate 
    equality. It follows that $\delta j' = 0$. Hence, if $H^1(\mathfrak{g})=0$,
    then $j' = 0$.
\end{proof}
\end{propo}

\begin{thm}
  \label{thm:existence_uniqueness_momentum_map}
  Assume $M$ is connected and $\phi(\mathfrak{g})\subset \mathfrak{X}_H(M)$ (in particular, the $G$-action is symplectic).
  \begin{enumerate}[label=(\arabic*)]
    \item If $H^2(\mathfrak{g})=0$, then a momentum map exists.
    \item If a momentum map exists, then the set of all momentum maps is an affine space for $H^1(\mathfrak{g})$.
  \end{enumerate}
  \begin{proof}
    $M$ being connected implies $\HdR^0(M) = \mathbb{R}$. Hence, the short exact
    sequence~\eqref{eq:symplectic_exact_sequence} becomes
    $$
      0 \rightarrow \mathbb{R} \rightarrow C^\infty(M) \xrightarrow{\sgrad} \mathfrak{X}_H(M) \rightarrow 0,
    $$
    where $\mathbb{R}\rightarrow C^\infty(M)$ is the inclusion of constant
    functions into $C^\infty(M)$. The statements are now consequences of the following lemma.
  \end{proof}
\end{thm}

\begin{lem}
  Let 
  \begin{equation}
    \label{eq:lie_algebra_central_extension}
    0\rightarrow \mathbb{R}\rightarrow  \mathfrak{h} \xrightarrow{\Dstraight} \mathfrak{k}\xrightarrow{} 0
  \end{equation}
  be a central extension of Lie algebras (that is, the sequence is exact and
  $\mathbb{R}\subset Z(\mathfrak{h})$), with $\mathfrak{h}$ and $\mathfrak{k}$
  possibly infinite-dimensional. Let $\mathfrak{g}$ be a Lie algebra and $\phi:
  \mathfrak{g}\rightarrow
  \mathfrak{k}$ a homomorphism. 
  \begin{enumerate}[label=(\arabic*)]
    \item If $H^2(\mathfrak{g}) = 0$, then a lift $\mu:
    \mathfrak{g}\rightarrow \mathfrak{h}$ exists.
  \item If at least one lift exists, the space of all lifts
    $\mathfrak{g}\rightarrow \mathfrak{h}$ is an affine space for
      $H^1(\mathfrak{g})$.
  \end{enumerate}

  $$
    \begin{tikzcd}
      \mathfrak{h} \arrow{r}{\Dstraight} & \mathfrak{k}\\
      & \mathfrak{g}\arrow[swap]{u}{\phi} \arrow[dashed]{ul}{\mu\;?}
    \end{tikzcd}
  $$

  \begin{proof}
    Since $\Dstraight$ is surjective, we can find a \textit{linear} map
    $\nu:\mathfrak{g}\rightarrow\mathfrak{h}$ such that $\Dstraight\circ \nu =
    \phi$. In general, however, the map $\nu$ will not be a Lie algebra
    homomorphism, so we might hope to find a ``correction'' linear map $\beta:
    \mathfrak{g}\rightarrow \mathfrak{h}$ so that $\nu + \beta$ is a Lie
    algebra homomorphism. If $\Dstraight\circ \beta = 0$, then $\nu+\beta$ will
    still be a lift of $\phi$. By the exactness of
    \eqref{eq:lie_algebra_central_extension}, this is equivalent to requiring
    $\beta(\mathfrak{g})\subset \mathbb{R}$, i.e., $\beta \in \mathfrak{g}^*$.

    The requirement that $\nu + \beta$ be a Lie algebra homomorphism then becomes
    \begin{equation}
      \label{eq:_beta_condition}
      [(\nu + \beta)(A), (\nu + \beta)(B)] = [\nu(A), \nu(B)] \overset{!}{=} (\nu + \beta)[A,B]
    \end{equation}
    for all $A, B\in \mathfrak{g}$. Define $\epsilon: \mathfrak{g}\times \mathfrak{g}\rightarrow \mathfrak{h}$
    by $\epsilon(A,B):= [\nu(A), \nu(B)] - \nu([A,B])$. Then
    $$
      \begin{aligned}
        &\Dstraight (\epsilon(A,B)) = \Dstraight([\nu(A), \nu(B)]) - \Dstraight\circ \nu([A,B])\\
        &=[\Dstraight\circ \nu(A), \Dstraight\circ \nu(B)] - \Dstraight\circ \nu ([A,B])\\
        &=[\phi(A), \phi(B)] - \phi([A,B]) = 0,
      \end{aligned}
    $$
    where we have used the fact that $\Dstraight$ and $\phi$ are homomorphisms.
    By the exactness of \eqref{eq:lie_algebra_central_extension}, the image of
    $\epsilon$ is contained in $\mathbb{R}\subset \mathfrak{h}$. Since
    $\epsilon$ is clearly skew-symmetric, we have $\epsilon \in \exterior^2\mathfrak{g}^*$.
    Condition \eqref{eq:_beta_condition} then reads
    \begin{equation}
      \label{eq:_beta_condition_2}
      -\delta\beta \overset{!}{=} \epsilon.
    \end{equation}
    We check that $\epsilon$ is closed:
    $$
    \begin{aligned}
      -\delta\epsilon(A,B,C) &= \epsilon([A,B], C) + \epsilon([B,C], A) + \epsilon([C,A], B)\\
      &=[\nu([A,B]), \nu(C)] - \nu([[A,B], C])  + (\text{cyclic permutations})\\
      &\overset{\text{\tiny Jacobi}}{=}[\nu([A,B]), \nu(C)] + (\text{cyclic permutations})\\
      &\overset{\text{\tiny Jacobi}}{=}\big[\nu([A,B]) - [\nu(A), \nu(B)], \nu(C)\big] + (\text{cyclic permutations})\\
      &=\big[-\epsilon(A,B), \nu(C)\big]  + (\text{cyclic permutations}) = 0,
    \end{aligned}
    $$
    where we have used $\mathrm{im}(\epsilon)\subset \mathbb{R}$ in the last line.
    If $H^2(\mathfrak{g}) = 0$, then $\epsilon$ is also exact, so we can
    find a $\beta \in \mathfrak{g}^*$ satisfying Eq.~\eqref{eq:_beta_condition_2}, proving (1).

    Now suppose $\mu: \mathfrak{g}\rightarrow \mathfrak{h}$ is any homomorphism
    such that $\Dstraight\circ \mu = \phi$. Take $\beta \in H^1(\mathfrak{g}) =
    \ker(\delta: \mathfrak{g}^*\rightarrow \exterior^2\mathfrak{g}^*)$. Then $\Dstraight \circ \beta = 0$, so $\Dstraight \circ
    (\mu+\beta)= \phi$. Furthermore, 
    $$
      [(\mu+\beta)(A), (\mu+\beta)(B)] = (\mu + \beta)([A,B])
    $$
    since $\beta([A,B]) = -\delta\beta(A,B) = 0$ and $\beta(A), \beta(B)\in
    \mathbb{R}$, showing that $\mu + \beta$ is a homomorphism.

    If $\mu': \mathfrak{g}\rightarrow \mathfrak{h}$ is another homomorphism
    such that $\Dstraight\circ \mu' = \phi$, then $\mu' -\mu \in
    \mathfrak{g}^*$, and 
    $$
    \begin{aligned}
      &(\mu' - \mu)([A,B]) = [\mu'(A), \mu'(B)] - [\mu(A), \mu(B)]\\
      &=[\underbrace{\mu'(A) - \mu(A)}_{\in \mathbb{R}}, \mu'(B)] + [\mu(A), \underbrace{\mu'(B)- \mu(B)}_{\in \mathbb{R}}]\\
      &= 0,
    \end{aligned}
    $$
    so $\delta(\mu'-\mu) = 0$
    and therefore $\mu'-\mu \in H^1(\mathfrak{g})$. This proves (2).
  \end{proof}
\end{lem}

If $\mathfrak{g}$ is semisimple, then $H^1(\mathfrak{g})=H^2(\mathfrak{g})=0$
by Whitehead's lemma. Consequently, we have:

\begin{cor}
  Let $G$ be a connected Lie group acting on a connected symplectic manifold
  $M$. If $\mathfrak{g}$ is semisimple, then there is a unique momentum map for
  the $G$-action on $M$.
\end{cor}

\subsection{Properties of Momentum Maps}
In this section, we collect various results on properties of momentum maps.
Among them, Kirwan's theorem (\cref{thm:kirwan}) is the most important one and
will be relevant for \cref{chap:momentum_map_functional_theories}. Although the
remaining results will not all be used directly in this thesis, we include them
here because we expect them to be useful for future developments of nonabelian
functional theories.

Let $(M,\omega)$ be a connected symplectic manifold with a symplectic action of
a connected Lie group $G$. Suppose $\Phi: M\rightarrow \mathfrak{g}^*$ is a
momentum map.

\begin{propo}
  \label{thm:momentum_map_composition}
  Let $H$ be a connected Lie group and $F: H\rightarrow G$ a Lie group
  homomorphism, making $H$ act on $M$ as well. Suppose $\Phi: M\rightarrow
  \mathfrak{g}^*$ is a momentum map for the $G$-action on $M$. Then $F^*\circ
  \Phi: M\rightarrow \mathfrak{h}^*$ is a momentum map for the $H$-action.
  \begin{proof}
    Let $\mu: \mathfrak{g}\rightarrow C^\infty(M)$ denote the comomentum map. 
    The following diagram commutes (where $\phi_G(A) = A^*$ etc.):
    \begin{equation}
    \begin{tikzcd}
      C^\infty(M)\arrow{rr}{\sgrad} & & \mathfrak{X}(M) \\
      & \mathfrak{g} \arrow[swap]{ul}{\mu} \arrow{ur}{\phi_G} \\
      & \mathfrak{h} \arrow{u}{F_*} \arrow[swap]{uur}{\phi_H} \arrow{uul}{\mu \circ F_*}
    \end{tikzcd}
    \end{equation}
    So $\mu\circ F_*$ is a comomentum map for the $H$-action. It is easy to
    check that the momentum map corresponding to $\mu\circ F_*$ is $F^*\circ
    \Phi$.
  \end{proof}
\end{propo}

\begin{lem}
  \label{lem:mu_bracket_omega}
  For all $A,B\in \mathfrak{g}$, $X\in T_pM$:
  \begin{enumerate}[label=(\arabic*)]
    \item $\braket{\Phi, [A,B]} = \omega(A^*, B^*)$
    \item $\braket{D_p\Phi(X), A} = \braket{(A^*)_p^\flat, X} = \omega(A^*_p, X)$.
  \end{enumerate}
  \begin{proof}
    \;\\
    (1)
    $
    \braket{\Phi, [A,B]} = \{\braket{\Phi, A}, \braket{\Phi, B}\}
    = \omega\big(\sgrad\braket{\Phi, A}, \sgrad\braket{\Phi, B}\big)
    = \omega(A^*, B^*)
    $
    \\\noindent
    (2) $\braket{D_p\Phi(X), A} = \dstraight_p(\braket{\Phi,A})(X) =
    (\sgrad\braket{\Phi, A})^\flat_p(X) = \braket{(A^*)^\flat_p, X} = \omega(A^*_p, X)$.
  \end{proof}
\end{lem}

\begin{cor}
  \label{thm:derivative_dual_map}
  The derivative $D_p\Phi: T_pM \rightarrow \mathfrak{g}^*$ is the dual map of
  the map $\mathfrak{g}\rightarrow T_p^*M$, $A\mapsto (A^*_p)^\flat$.
\end{cor}

Now, we are ready to establish an important fact about momentum maps: they are
$G$-equivariant. Recall that the Lie group $G$ acts naturally on
$\mathfrak{g}^*$ via the \textit{coadjoint action} $\Ad^*$, which is
defined by $\braket{\Ad^*_g(\alpha), A} = \braket{\alpha, \Ad_{g^{-1}}(A)}$
for all $\alpha\in\mathfrak{g}^*$ and $A\in \mathfrak{g}$.

\begin{propo}
  The map $\Phi: M\rightarrow \mathfrak{g}^*$ is $G$-equivariant with respect to the coadjoint action. That
    is,
    \begin{equation}
      \label{eq:mu_equivariant}
      \mu(g\cdot p) = \mathrm{Ad}_g^*\Phi(p).
    \end{equation}
    \begin{proof}
      Since $G$ is connected, we can write $g = \exp(tA_1)\exp(tA_2)\dotsb
      \exp(tA_N)$ for $A_i \in \mathfrak{g}$. Hence, it suffices to show
      Eq.~\eqref{eq:mu_equivariant} for $g= \exp(tA)$. 

      Take any $A\in \mathfrak{g}$. Define two curves in $\mathfrak{g}^*$ by
      \begin{equation}
        \begin{aligned}
           \gamma_1(t) := \Phi(\exp(tA)p) 
          \hspace{4em} \gamma_2(t) := \mathrm{Ad}_{\exp(tA)}^*\Phi(p).
        \end{aligned}
      \end{equation}
      The goal is then to show $\gamma_1(t) {=} \gamma_2(t)$ for all
      $t$. The trick is to show that $\gamma_1$ and $\gamma_2$ satisfy the same first-order
      differential equation, since we already have $\gamma_1(0) = \gamma_2(0)=
      \Phi(p)$. Take any $B\in \mathfrak{g}$, then
      $$
      \begin{aligned}
        & \braket{\dot\gamma_1(t_0), B} = \doverd{t}\Big|_{t=0}\braket{\Phi(\exp(tA)\exp(t_0A)p), B}
        = \braket{D_{\exp(t_0A)p}\Phi(A^*_{\exp(t_0A)p}), B}\\
        &\overset{\text{Lem. \ref{lem:mu_bracket_omega}}}{=}  \omega\left(B^*_{\exp(t_0A)p}, A^*_{\exp(t_0A)p}\right)
        \overset{\text{Lem. \ref{lem:mu_bracket_omega}}}{=} \braket{\Phi(\exp(t_0A)p), [B,A]} \\
        &= 
        \braket{-\gamma_1(t_0) \circ \ad_A, B}\\[1em]
        &\braket{\dot\gamma_2(t_0), B} = \doverd{t}\Big|_{t=0}\braket{\Ad_{\exp((t_0+t)A)}^*\Phi(p), B}\\
          &=\doverd{t}\Big|_{t=0}\braket{\Ad_{\exp(t_0A)}^*\Phi(p), \Ad_{\exp(-tA)}B}
          =\braket{\Ad_{\exp(t_0A)}^*\Phi(p), [-A,B]}\\
          &=\braket{-\gamma_2(t_0)\circ \ad_{A}, B}.
        \end{aligned}
      $$
      So both curves satisfy the differential equation $\dot \gamma = -\gamma \circ \ad_A$.
    \end{proof}
\end{propo}

Recall that if $E\subset \mathfrak{g}$ is any subset, the \textit{annihilator}
of $E$ in $\mathfrak{g}^*$ is the set $E^\circ := \{\eta \in \mathfrak{g}^*
\mid \forall A\in E: \braket{\eta, A} = 0\}$.

\begin{propo}
  \label{thm:image_of_derivative_of_moment_map}
  $D_p\Phi(T_pM) = \mathfrak{g}_p^\circ$, where $\mathfrak{g}_p$ is the stabilizer subalgebra of $p\in M$.
  \begin{proof}
    Take any vector $X \in T_pM$ and $A\in \mathfrak{g}_p$. Then
    $$
     \braket{D_p\Phi(X), A}  = \omega(A^*_p, X) = 0,
    $$
    since $A_p^*=0$. This shows $D_p\Phi(T_pM)\subset \mathfrak{g}_p^\circ$.
    By Corollary~\ref{thm:derivative_dual_map}, we have
    $$
    \dim D_p\Phi(T_pM) = \dim \mathrm{im}(A\mapsto A_p^*) = \dim G - \dim
    \mathfrak{g}_p = \dim \mathfrak{g}_p^\circ,
    $$
    so $D_p\Phi(T_pM) = \mathfrak{g}^\circ$.
  \end{proof}
\end{propo}

Recall that a smooth map $f:M\rightarrow N$ is said to meet a submanifold
$Z\subset N$ \textit{transversally} if $D_pf (T_pM) + T_{f(p)}Z = T_{f(p)}N$
for all $p\in f^{-1}(Z)$. A generalization of the regular value theorem asserts
that in this case $f^{-1}(Z) \subset M$ is a submanifold.

Let $\mathfrak{t}^*_{>0}\subset \mathfrak{t}^*_+$ denote the open Weyl chamber.
The following will be useful for the ``Selection Rule''
(\cref{thm:selection_rule}) discussed in the next section. 
\begin{propo}
  The momentum map $\Phi: M\rightarrow \mathfrak{g}^*$ meets the open Weyl
  chamber $\mathfrak{t}_{>0}^*$ transversally.
  \begin{proof}
    Take $p\in M$ such that $\Phi(p)\in \mathfrak{t}_{>0}^*$. By definition, we
    need to show
    \begin{equation}
      \label{eq:momentum_map_transversal}
      D_{p}\Phi(T_pM) + \mathfrak{t}^* {=} \mathfrak{g}^*.
    \end{equation}
    By Proposition~\ref{thm:image_of_derivative_of_moment_map}, the first
    summand is $\mathfrak{g}_p^\circ$, the annihilator of the stabilizer
    subalgebra of the $G$-action at $p\in M$. 

    $\mathfrak{g}_p \subset \mathfrak{g}_{\Phi(p)}$ because $\Phi$ is $G$-equivariant.
    But $\mathfrak{g}_{{\Phi(p)}} = \mathfrak{t}$ because $\Phi(p)\in
    \mathfrak{t}^*_{>0}$. Hence $\mathfrak{g}_{p}\subset \mathfrak{t}$, or
    $\mathfrak{g}_{p}^\circ \supset \mathfrak{t}^\circ$. It follows that
    $$
    D_{p}\Phi(T_{p}M) + \mathfrak{t}^* =
    \mathfrak{g}_{p}^\circ + \mathfrak{t}^* \supset 
    \mathfrak{t}^\circ + \mathfrak{t}^* = \mathfrak{g}^*.
    $$
  \end{proof}
\end{propo}

\begin{cor}
  \label{thm:preimage_of_open_weyl_submanifold}
  $\Phi^{-1}(\mathfrak{t}^*_{>0})$ is a submanifold of $M$.
\end{cor}

For our purpose in \cref{chap:momentum_map_functional_theories}, the most
relevant properties of momentum maps are those of the image $\Phi(M)$. For the
following two theorems, we assume that $M$ is compact and connected.

\begin{thm}[Atiyah-Guillemin-Sternberg
  \cite{atiyahConvexityCommutingHamiltonians1982,guilleminConvexityPropertiesMoment1982}] 
  Suppose a torus $T= (S^1)^n$ acts on $M$ with momentum map $\Phi:
  M\rightarrow \mathfrak{t}^* =\mathbb{R}^n$. Then $\Phi(M)$ is the convex hull
  of the images of all $T$-fixed points in $M$.
\end{thm}

\begin{rem}
  When $M= \mathbb{P}(\mathcal{H})$ is the projective Hilbert space with the
  appropriate symplectic structure (see
  \cref{sec:projective_space_symplectic}), we already know this from
  \cref{thm:abelian_nrep_set}.
\end{rem}

If $G$ is not necessarily abelian, there is a weaker convexity result due to
Kirwan. In this case, choose any Cartan subalgebra $\mathfrak{t}\subset
\mathfrak{g}$ and a Weyl chamber $\mathfrak{t}_+^* \subset \mathfrak{t}^*$. After
fixing a $G$-invariant inner product on $\mathfrak{g}$, we may think of
$\mathfrak{t}^*$ as a subspace of $\mathfrak{g}^*$.
Because $\Phi$ is $G$-equivariant, it is enough to know how $\Phi(M)$
intersects with $\mathfrak{t}^*_+$.

\begin{thm}[Kirwan \cite{kirwanConvexityPropertiesMoment1984}]
  \label{thm:kirwan} Let $G$ be a compact connected Lie group acting on $M$
  with momentum map $\Phi: M\rightarrow \mathfrak{g}^*$. The image $\Phi(M)$
  intersects $\mathfrak{t}^*_+$ in a convex polytope.
\end{thm}

\section{The Projective Hilbert Space}
\label{sec:projective_space_symplectic}

Let $\mathcal{H}$ be any complex vector space. Recall that the projective space
$\mathbb{P}(\mathcal{H})$ is the complex manifold whose points are
equivalence classes $[\psi]$ of nonzero vectors in
$\mathcal{H}$, where $\psi\sim \psi'$ if $\psi = \lambda \psi'$ for some
$\lambda \in \mathbb{C}\setminus \{0\}$.
Any choice of basis for $\mathcal{H}$
induces a biholomorphism $\mathbb{P}(\mathcal{H})\rightarrow \mathbb{CP}^{\dim
\mathcal{H}-1}$. If $\mathcal{H}$ is endowed with a Hermitian inner product
$\braket{\cdot, \cdot}$, it is convenient to embed $\mathbb{P}(\mathcal{H})$ into
$i\mathfrak{u}(\mathcal{H})$, the real vector space of Hermitian endomorphisms, via the
map
\begin{equation}
  \label{eq:projective_space_embedding}
  \begin{aligned}
    \tilde \Phi: & \;\mathbb{P}(\mathcal{H}) \rightarrow i\mathfrak{u}(\mathcal{H})\\
    & [\psi] \mapsto \frac{\psi\braket{\psi,\cdot}}{\braket{\psi,\psi}}.
  \end{aligned}
\end{equation}
It is easy to check that $\tilde \Phi$ is indeed an embedding. Under
$D_{[\psi]}\tilde \Phi: T_{[\psi]} \mathbb{P}(\mathcal{H}) \rightarrow
T_{\tilde \Phi[\psi]}i\mathfrak{u}(\mathcal{H}) = i\mathfrak{u}(\mathcal{H})$,
a tangent vector in $T_{[\psi]}\mathbb{P}(\mathcal{H})$ is mapped to a
Hermitian endomorphism on $\mathcal{H}$. For notational simplicity, we will
denote by $\proj{\psi}$ the orthogonal projection onto the subspace spanned by
$\psi$ for $\psi\neq 0$. That is, $\tilde \Phi[\psi] = \proj{\psi}$. 
The map $\tilde \Phi$ then gives, up to the identification of
$i\mathfrak{u}(\mathcal{H})$ with its dual, an identification between
$\mathbb{P}(\mathcal{H})$ and the set of pure states $\mathcal{P}\subset
i\mathfrak{u}(\mathcal{H})^*$.



\subsection{Unitary Group Action}
There is a natural action of $\mathrm{U}(\mathcal{H})$ on
$\mathbb{P}(\mathcal{H})$ given by $g\cdot [\psi] = [g\psi]$. 
The
following proposition gives information about the fundamental vector fields
$A^*$ for $A\in \mathfrak{u}(V)$. (Hereafter, we will adopt the convention that
we always pick representatives $\psi\in[\psi]$ which are normalized, i.e.,
$\lVert \psi\rVert=1$.)

\begin{propo}
  \label{thm:image_of_fundamental_vector}
  Let $A \in \mathfrak{u}(\mathcal{H})$. Then
  \begin{equation}
    D_{[\psi]} \tilde\Phi (A^*_{[\psi]}) = A\psi\braket{\psi,\cdot} - \psi\braket{\psi,A(\cdot)} = [A, \proj{\psi}],
  \end{equation}
  where $A^*\in\mathfrak{X}(\mathbb{P}(\mathcal{H}))$ is the fundamental vector
  field of $A$ for the $\mathrm{U}(\mathcal{H})$-action on $\mathbb{P}(\mathcal{H})$.
  \begin{proof}
    $$
    \begin{aligned}
      &D_{[\psi]}\tilde\Phi(A^*_{[\psi]}) = \doverd{t}\Big|_{t=0}\tilde\Phi\Big(\exp(tA)[\psi]\Big)
      =\doverd{t}\Big|_{t=0} (\tilde\Phi\circ \pi) (\exp(tA) \psi)\\
      &=\doverd{t}\Big|_{t=0}\frac{\exp(tA)\psi\braket{\exp(tA)\psi, \cdot}}{\braket{\exp(tA)\psi, \exp(tA)\psi}}
      =A\psi\braket{\psi,\cdot} + \psi\braket{A\psi, \cdot}\\
      &=A\psi\braket{\psi,\cdot} - \psi\braket{\psi, A(\cdot)},
    \end{aligned}
    $$
    where $\pi: \mathcal{H}\setminus \{0\}\rightarrow \mathbb{P}(\mathcal{H})$
    is the canonical projection and we have used the skew-Hermiticity of
    $A$ in the last line.
  \end{proof}
\end{propo}
\begin{rem}
  In the physics notation, on writes $\ketbrap{\psi}$ instead of
  $\psi\braket{\psi,\cdot}$ (which is equal to $\proj{\psi}$ if $\lVert
  \psi\rVert =1$).
\end{rem}

\begin{rem}
  \label{rem:unitary_group_tangent_space}
  At any point $[\psi]\in \mathbb{P}(\mathcal{H})$ in the projective space, the map
  $\mathfrak{u}(\mathcal{H})\rightarrow T_{[\psi]}\mathbb{P}(\mathcal{H})$, $A \mapsto A^*_{[\psi]}$ gives
  us a very convenient way of thinking about the tangent space at $[\psi]$. To be
  specific, since the $\mathrm{U}(\mathcal{H})$-action on $\mathbb{P}(\mathcal{H})$ is transitive,
  the map $A\mapsto A^*_{[\psi]}$ is surjective onto $T_{[\psi]}\mathbb{P}(\mathcal{H})$. In
  other words, every tangent vector at $[\psi]$ is the value of some fundamental
  vector field at $[\psi]$. However, this is not a ``perfect'' parametrization of the
  tangent space in the sense that the map $A\mapsto A^*_{[\psi]}$ is not injective.
\end{rem}




\subsection{The Momentum Map}
The projective space $\mathbb{P}(\mathcal{H})$ has the structure of a K\"ahler
manifold. In Appendix~\ref{app:projective_kaehler}, we construct explicitly
the Fubini-Study metric $h^{\mathrm{FS}}$ and the Fubini-Study $2$-form
$\omega^{\mathrm{FS}}$ on $\mathbb{P}(\mathcal{H})$ and show that they are
compatible with the almost complex structure $J$.
Of course, what is relevant for us is the symplectic structure, 
for which we have an explicit formula suitable for computation
(\cref{lem:omega_explicit_formula}):
\begin{equation}
  \label{eq:omega_explicit_formula_main}
  \hspace{3em}\omega^{\mathrm{FS}}(A^*_{[\psi]}, B^*_{[\psi]})
   = i \braket{\psi, [A,B]\psi}\hspace{3em} (A,B\in \mathfrak{u}(\mathcal{H})).
\end{equation}


\begin{thm}
  The map
\begin{equation}
  \begin{aligned}
    \Phi: &\;\mathbb{P}(\mathcal{H})\rightarrow \mathfrak{u}(\mathcal{H})^*\\
    & [\psi] \mapsto i\Tr(\tilde\Phi[\psi](\cdot)) = i\Tr(\proj{\psi}(\cdot)) = i\braket{\psi, (\cdot) \psi}
  \end{aligned}
\end{equation}
  is a momentum map for the $\mathrm{U}(V)$-action on
  $(\mathbb{P}(V), \omega^{\mathrm{FS}})$.
  \begin{proof}
    Define $\mu: \mathfrak{g}\rightarrow C^\infty(\mathbb{P}(\mathcal{H}))$ by $\mu(A)[\psi] =
    \braket{\Phi[\psi], A}$.
    The definition of momentum maps (Definition~\ref{def:momentum_map})
    requires us to check two things: 
    \begin{enumerate}[label=(\arabic*)]
      \item $\sgrad\circ \mu = \phi$
      \item $\mu$ is a Lie algebra homomorphism, i.e., $\mu([A,B])=\{\mu(A), \mu(B)\}$
    \end{enumerate}
    where $\phi: \mathfrak{g}\rightarrow \mathfrak{X}(\mathbb{P}(\mathcal{H}))$ is the
    map $A\mapsto A^*$.

    The first condition is equivalent to $\dstraight \mu(A) =
    \iota_{A^*}\omega^{\mathrm{FS}}$ for all $A\in \mathfrak{g}$. Take any
    $B\in \mathfrak{g}$, then
    $$
    \begin{aligned}
      &\braket{\dstraight \mu(A), B^*_{[\psi]}} 
      =\left\langle\dstraight ([\psi]\mapsto i\braket{\psi, A\psi}),B^*_{[\psi]}\right\rangle\\
      &= \doverd{t}\Big|_{t=0}i\left\langle{\exp(tB)\psi, A\exp(tB)\psi}\right\rangle\\
      &=i\braket{\psi, [A,B]\psi}
      =\omega^{\mathrm{FS}}(A^*_{[\psi]}, B^*_{[\psi]})\\
      &= \left(\iota_{A^*}\omega^{\mathrm{FS}}\right)(B^*_{[\psi]}),
    \end{aligned}
    $$
    where we have used Eq.~\eqref{eq:omega_explicit_formula_main} for the
    penultimate equality. It follows that $\dstraight\mu(A) =
    \iota_{A^*}\omega^{\mathrm{FS}}$.

    The second condition also holds:
    $$
    \begin{aligned}
      &\{\mu(A), \mu(B)\} = \omega^{\mathrm{FS}}(\sgrad \mu(A), \sgrad \mu(B))\\
      & = \omega^{\mathrm{FS}}(A^*, B^*)
      = ([\psi]\mapsto i\braket{\psi, [A,B]\psi})\\
      &= \mu([A,B]),
    \end{aligned}
    $$
    where we have used Eq.~\eqref{eq:omega_explicit_formula_main}
     again.
  \end{proof}
\end{thm}

\begin{cor}
  \label{thm:group_rep_momentum_map}
  Let $G$ be a connected Lie group and $\tau: G\rightarrow
  \mathrm{U}(V)$ a unitary representation, 
  so that $G$ acts on $\mathbb{P}(\mathcal{H})$ by $(g, [\psi]) \mapsto
  [\tau(g)\psi]$. 
  Let $\tau_*: \mathfrak{g}\rightarrow
  \mathfrak{u}(\mathcal{H})$ denote the induced Lie algebra homomorphism. Then the map
  \begin{equation}
    \begin{aligned}
      \Phi_G: &\;\mathbb{P}(\mathcal{H}) \rightarrow \mathfrak{g}^*\\
      & [\psi] \mapsto i\Tr(\proj{\psi}\tau_*(\cdot)) = (A \mapsto i\braket{\psi, \tau_*(A)\psi})
    \end{aligned}
  \end{equation}
  is a momentum map for the $G$-action on $\mathbb{P}(\mathcal{H})$.
  \begin{proof}
    Clearly $\Phi_G = \tau^*\circ \Phi$. The statement follows from applying
    Proposition~\ref{thm:momentum_map_composition} to $\tau$.
  \end{proof}
\end{cor}

%
%
%

Let $G$ be a compact connected Lie group and $\tau:G\rightarrow
\mathrm{U}(\mathcal{H})$ a unitary representation with momentum map $\Phi_G:
\mathbb{P}(\mathcal{H})\rightarrow\mathfrak{g}^*$, $[\psi]\mapsto (A\mapsto
i\braket{\psi, \tau_*(A)\psi})$. Pick a Cartan subalgebra
$\mathfrak{t}\subset\mathfrak{g}$ and a $G$-invariant inner product on
$\mathfrak{g}$, together with a Weyl chamber $\mathfrak{t}_+^*$.


Kirwan's theorem (Theorem~\ref{thm:kirwan}) states that
$\Phi_G(\mathbb{P}(\mathcal{H}))\cap \mathfrak{t}_+^*$ is a convex polytope, so
it is characterized by a finite set of inequalities $\braket{\cdot, S}\ge c$, where
$S\in \mathfrak{t}$ and $c\in \mathbb{R}$. 
In the abelian case, any $[\psi]\in
\mathbb{P}(\mathcal{H})$ mapping to a facet must be spanned by the weight
vectors whose weights are on the facet. This turns out to be true even when
$\mathfrak{g}$ is not abelian, a result called the ``Selection Rule''
\cite{klyachkoPauliExclusionPrinciple2009} which we now state and prove.


\begin{thm}[Selection Rule (Lemma 2.13 of Ref.~\cite{walterMultipartiteQuantumStates2014})]
  \label{thm:selection_rule}
  Let $\braket{\cdot, S}\ge c$ be an inequality corresponding to a facet of the
  convex polytope $\Phi_G(\mathbb{P}(\mathcal{H}))\cap \mathfrak{t}^*_+$. 
  Suppose $\psi \in \mathcal{H}$ ($\lVert \psi \rVert=1$) is such that
  $\Phi_G[\psi]\in \mathfrak{t}_{>0}^*$ and $\braket{\Phi_G[\psi], S}=c$ (i.e.,
  $\Phi_G[\psi]$ is lying on the facet). Then $i\tau_*(S)\psi = c\psi$ and
  $\dstraight_{[\psi]}\braket{\Phi_G,S}=0$.
  \begin{proof}
    Take any $Y\in D_{[\psi]}\Phi_G(T_{[\psi]}\mathbb{P}(\mathcal{H}))\cap \mathfrak{t}^*$.
    Since $\Phi_G^{-1}(\mathfrak{t}_{>0}^*)\subset \mathbb{P}(\mathcal{H})$ is a
    submanifold (\cref{thm:preimage_of_open_weyl_submanifold}), we can find a smooth curve
    $\gamma:(-\epsilon,\epsilon)\rightarrow \Phi_G^{-1}(\mathfrak{t}_{>0}^*)$
    such that $\gamma(0) = [\psi]$ and $D_{[\psi]}\Phi_G(\dot \gamma(0)) = Y$. Then
    $$
    \doverd{t}\Big|_{t=0}\braket{\Phi_G(\gamma(t)), S} =\braket{Y, S}.
    $$
    But $\braket{\Phi_G(\gamma(t)),S}\ge c$ is smooth and satisfies
    $\braket{\Phi_G(\gamma(0)), S}=\braket{\Phi_G[\psi], S}=c$, so the
    derivative at $t=0$ must be zero, implying $\braket{Y,S}=0$.

    It follows that $\mathfrak{g}_{[\psi]}^\circ \cap \mathfrak{t}^*\subset
    S^\circ$, so $\mathbb{R} S\subset \mathfrak{g}_{[\psi]} +
    \mathfrak{t}^\perp$. 
    We must have $S\in \mathfrak{g}_{[\psi]}$ because $S \notin
    \mathfrak{t}^\perp$, implying
    \begin{equation}
    \label{eq:H_is_stabilizer}
      \exp(t\tau_*(S))[\psi]  = [\psi]
    \end{equation}
    for all $t$. So $\exp(t\tau_*(S))\psi = k(t)\psi$ for some smooth function
    $k(t)$. Differentiating with respect to $t$ at $t=0$ yields $\tau_*(S)\psi
    = \dot k(0)\psi$. But, by assumption,
    $$
    c = \braket{\Phi_G[\psi], S} = i\braket{\psi, \tau_*(S)\psi} = i\dot k(0),
    $$
    so $\dot k(0) = -ic$. This proves the first part of the theorem.

    To show the second part, note that by definition $(\sgrad\braket{\Phi_G,
    S})_{[\psi]} = S^*_{[\psi]}$, but Eq.~\eqref{eq:H_is_stabilizer} implies
    $S^*_{[\psi]}=0$, so $(\sgrad\braket{\Phi_G, S})_{[\psi]} =
    (\dstraight_{[\psi]}\braket{\Phi_G,S})^\sharp=0$. Since $\sharp$ is an isomorphism, we conclude
    $\dstraight_{[\psi]}\braket{\Phi_G,S}=0$.
  \end{proof}
\end{thm}

Finally, we will quote a result from
Ref.~\cite{walterMultipartiteQuantumStates2014} for determining the facets of
the convex polytope $\Phi_G(\mathbb{P}(\mathcal{H}))\cap \mathfrak{t}_+^*$. 
We need some preparation for the statement of the theorem. Let $G$ be a maximal
compact subgroup of a complex reductive algebraic group $G'$, so that $\mathfrak{g}' =
\mathfrak{g}_{\mathbb{C}}$, with a representation $\tau':
G'\rightarrow \mathrm{GL}(\mathcal{H})$ such that $\tau'(G)\subset
\mathrm{U}(\mathcal{H})$. Let $$ \mathfrak{g}' = \mathfrak{t}_{\mathbb{C}}
\oplus \bigoplus_{\alpha\in R} \mathfrak{g}'_\alpha $$ be the root space
decomposition of $\mathfrak{g}'$, where $R\subset i\mathfrak{t}^*$ is the set
of roots. Pick a base for $R$, which fixes the positive Weyl chamber
$\mathfrak{t}_+^*$ and the decomposition of $R$ into the disjoint union of
positive and negative roots, which we denote by $R_+$ and $R_-$ respectively.
For $S\in \mathfrak{t}, c\in \mathbb{R}$, define
$$
\begin{aligned}
  &\mathcal{H}(S< c) := \bigoplus_{\omega\in \Omega: i\braket{\omega, S} < c}\mathcal{H}_\omega\\
  &\mathcal{H}(S= c) := \bigoplus_{\omega\in \Omega: i\braket{\omega, S} = c}\mathcal{H}_\omega\\
  &\mathfrak{n}_-(S< 0) := \bigoplus_{\alpha\in R_-: i\braket{\alpha, S} < 0}\mathfrak{g}'_{\alpha},
\end{aligned}
$$
where $\Omega\subset i\mathfrak{t}^*$ is the set of weights of the
representation $\tau'$ relative to $\mathfrak{t}_{\mathbb{C}}$.
Note that $\tau'(\mathfrak{n}_-(S < 0)) \mathcal{H}(S=c)\subset
\mathcal{H}(S < c)$.

\begin{thm}[Theorem 3.14 of Ref.~\cite{walterMultipartiteQuantumStates2014}]
  \label{thm:characterization_of_kirwan_polytope}
  If the convex polytope $\Phi_G(\mathbb{P}(\mathcal{H}))\cap \mathfrak{t}_+^*$
  has full dimension in $\mathfrak{t}^*$, then it is characterized by
  (in addition to the inequalities defining the positive Weyl chamber)
  $$
  \braket{\cdot, S} \ge c,
  $$
  for all $S\in\mathfrak{t}$, $c\in \mathbb{R}$ such that 
    \begin{enumerate}[label=(\arabic*)]
      \item the hyperplane $\braket{\cdot, S}= c$ is the affine hull of a subset of
            $i\Omega$, and 
      \item there exists a
            $\psi \in \mathcal{H}(S= c)$ for which the map 
            \begin{equation}
              \label{eq:ressayre_inequality_condition}
            \begin{aligned}
              \mathfrak{n}_-(S<0) &\rightarrow \mathcal{H}(S<c)\\
              A &\mapsto  \tau'_*(A) \psi
          \end{aligned}
            \end{equation}
          is an isomorphism.
    \end{enumerate}
\end{thm}

\begin{ex}
  \label{ex:2t3_moment_map_computation}
  Take $\mathcal{H} = \mathbb{C}^2\otimes \mathbb{C}^3$, $G=
  \mathrm{SU}(2)\times \mathrm{SU}(2)$, $G' = \mathrm{SL}(2,\mathbb{C})\times
  \mathrm{SL}(2,\mathbb{C})$, with $G'$ acting on $\mathcal{H}$ by the tensor
  product of the fundamental representation $\tau^{(2)}$
  and the adjoint representation $\tau^{(3)}$ of
  $\mathrm{SL}(2,\mathbb{C})$. The momentum map is
  $$
  \begin{aligned}
    \Phi: &\;\mathbb{P}(\mathbb{C}^2\otimes \mathbb{C}^3) \rightarrow \mathfrak{su}(2)^* \oplus \mathfrak{su}(2)^*\\
    & [\psi\otimes \chi] \mapsto 
    (A\mapsto i\braket{\psi, \tau^{(2)}_*(A)\psi}, B\mapsto i\braket{\chi, \tau^{(3)}_*(A) \chi}).
  \end{aligned}
  $$
  For the Cartan subalgebra, we choose
  $$
  \mathfrak{t}_{\mathbb{C}} = \mathbb{C} Z_1 \oplus \mathbb{C} Z_2,
  $$
  where $Z_1 = (Z,0)$ and $Z_2 = (0, Z)$ with $Z = \mathrm{diag}(1,-1)$.
  Relative to $\mathfrak{t}_{\mathbb{C}}$, the four roots are 
  $$
  \begin{aligned}
    \phantom{-}\alpha = \;& (Z \mapsto 2, 0)\\
    -\alpha = \;& (Z \mapsto -2, 0)\\
    \beta= \;& (0, Z \mapsto 2)\\
    -\beta= \;& (0, Z \mapsto 2).
  \end{aligned}
  $$
  Similarly, the weights of the representation $\tau^{(2)}\otimes \tau^{(3)}$
  are
  $$
  \begin{array}{ll}
    \omega_{-1,2} = (Z\mapsto -1, Z\mapsto 2) & \omega_{1,2} = (Z\mapsto 1, Z\mapsto 2) \\
     \omega_{-1,0} = (Z\mapsto -1, Z\mapsto 0) & \omega_{1,0} = (Z\mapsto 1, Z\mapsto 0)  \\
     \omega_{-1,-2} = (Z\mapsto -1, Z\mapsto -2) & \omega_{1,-2} = (Z\mapsto 1, Z\mapsto -2), 
  \end{array}
  $$
  each with multiplicity one. Note that all the roots and weights belong to $i\mathfrak{t}^*$.
  We will choose $\alpha$ and $\beta$ to be the
  positive simple roots, yielding the positive Weyl chamber
  $$
  \mathfrak{t}_+^* =\{\lambda i\alpha + \mu i\beta \mid \lambda,\mu \ge 0\}\subset \mathfrak{t}^*.
  $$
  We will now use \cref{thm:characterization_of_kirwan_polytope} to determine
  the polytope $\Lambda:=\Phi_G(\mathbb{P}(\mathcal{H})) \cap
  \mathfrak{t}_+^*$.

  First we show that $\Lambda$ satisfies the hypothesis that $\dim\Lambda =
  \dim \mathfrak{t}^* = 2$. To see this, consider the images of the vectors
  $\phi_{1,2}, \phi_{1,0}, \psi:=\frac{1}{\sqrt{2}}(\phi_{1,2}+\phi_{-1,-2})$,
  where we label the (unit-norm) weight vectors by the eigenvalues of $Z_1$
  and $Z_2$. Clearly, we have
  $$
  \begin{aligned}
    &\Phi[\phi_{1,2}] = i\omega_{1,2} \;\;\;\; \Phi[\phi_{1,0}] = i\omega_{1,0}\;\;\;\;
    &\Phi[\psi] = 0.
  \end{aligned}
  $$
  The points $i\omega_{1,2}, i\omega_{1,0}, 0$ all belong to $\mathfrak{t}^*_+$
  and are not collinear, so $\dim \Lambda$ must be two and the hypothesis of
  \cref{thm:characterization_of_kirwan_polytope} is satisfied.

  To obtain the inequalities, we only have to check lines (codimension one
  affine spaces) $\braket{\cdot, S} = c$ passing through (at least) two weights.
  Looking at \cref{fig:2t3_inequality_candidates}, we see that there are twelve
  possibilities (six lines in total): 

  \begin{center}
    \begin{tabular}{ccc}
        line \# & $S$ & $c$\\
        \hline
        1 & $\pm iZ_2$ & $\mp 2$\\
        2 & $\pm (iZ_1-iZ_2)$ & $\pm 1$ \\ 
        3 & $\pm (2iZ_1 -iZ_2)$ & $0$\\
        4 & $\pm iZ_1$ &  $\mp 1$\\
        5 & $\pm (iZ_1+iZ_2)$ & $\mp 1$\\
        6 & $\pm (iZ_1-iZ_2)$ & $\mp 1$\\
    \end{tabular}
    \end{center}


For each candidate $(S, c)$, we have to show whether there exists a
$\psi\in\mathcal{H}(S=c)$ such that the map
\eqref{eq:ressayre_inequality_condition} is an isomorphism. This already rules
out most candidates due to dimension reasons:  only those $(S,c)$ for which
$\dim\mathfrak{n}_-(S<0) = \dim\mathcal{H}(S<c)$ have to be considered.

\begin{figure}[h]
  \centering
  \begin{minipage}{.4\textwidth}
    \centering
    \includegraphics[width=.8\textwidth]{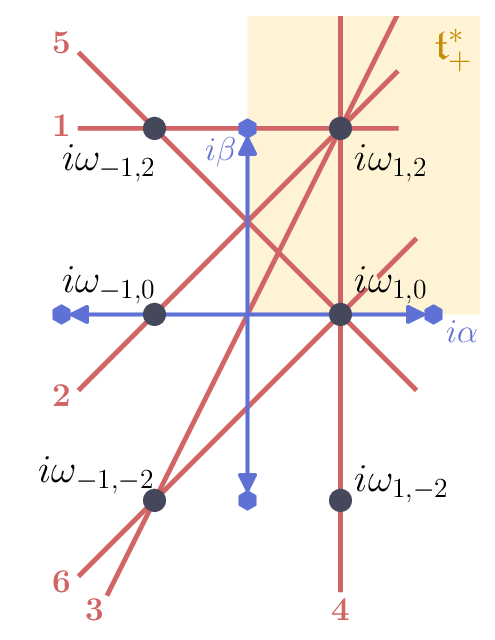}
    \captionof{figure}{Candidate inequalities.}
    \label{fig:2t3_inequality_candidates}
  \end{minipage}
  \hspace{.1\textwidth}
  \begin{minipage}{.4\textwidth}
    \centering
    \includegraphics[width=.8\textwidth]{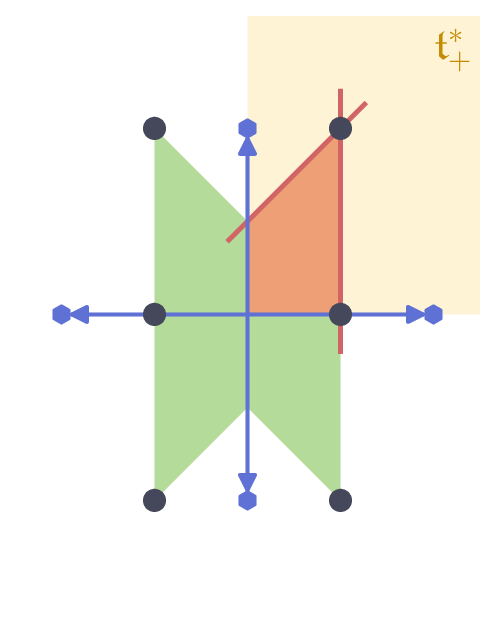}
    \captionof{figure}{$\Phi(\mathbb{P}(\mathcal{H}))\cap \mathfrak{t}^*$
    (green and orange) and $\Lambda=\Phi(\mathbb{P}(\mathcal{H}))\cap \mathfrak{t}^*_+$ (orange).}
    \label{fig:2t3_image}
  \end{minipage}
\end{figure}

In \cref{tab:2t3_ineq_candidates}, we list all twelve candidate inequalities,
together with the corresponding subspaces $\mathfrak{n}_-(S<0)$,
$\mathcal{H}(S=c)$, and $\mathcal{H}(S<c)$, where a filled hexagon or circle
indicates that the corresponding root space or weight space is included in the
respective subspaces. Each inequality is labeled by a pair $(m,s)$, where $m$
is the index of the line on which the inequality is saturated, and $s$ is the
choice of sign (choice of one of the two half spaces bounded by the line). For
example, inequality $(1,+)$ is $\braket{\cdot, iZ} \ge -2$. For inequalities
$(1,-), (2,+), (3,+), (3,-), (4,-), (5,+), (5,-), (6,-)$, the dimensions of
$\mathfrak{n}_-(S<0)$ and $\mathcal{H}(S<c)$ differ, so they cannot be valid
inequalities.

This leaves us with inequalities $(1,+), (2,-), (4, +), (6,+)$. For $(1,+)$ and
$(4,+)$, the isomorphism criterion~\eqref{eq:ressayre_inequality_condition} 
is trivially satisfied since $\dim\mathfrak{n}_-(S<0) =
\dim\mathcal{H}(S<c)=0$, so both are valid inequalities. Moreover, these
already imply $(6,+)$, so we only have to check criterion~\eqref{eq:ressayre_inequality_condition}
for $(2,-)$, in which case 
$$
\begin{aligned}
  &\mathfrak{n}_-(S<0) = \mathfrak{g}'_{-\alpha}= \mathbb{C}(X_1-iY_1)\\
  &\mathcal{H}(S=c) = \mathbb{C}\phi_{-1,0} + \mathbb{C}\phi_{1,2}\\
  &\mathcal{H}(S<c) = \mathbb{C}\phi_{-1,2}.
\end{aligned}
$$
Since $\tau_*'(X_1-iY_1)\phi_{1,2} = \lambda \phi_{-1, 2}$, $\lambda \neq 0$,
the map $\mathfrak{n}_-(S<0)\rightarrow\mathcal{H}(S<0), A\mapsto
\tau'_*(A)\phi_{1,2}$ is bijective. It follows that $\braket{\cdot, -iZ_1+iZ_2}\ge
-1$ is a bounding inequality of the polytope $\Lambda$.

We conclude that the polytope $\Lambda$ is bounded by the inequalities $(1,+),
(4,+),$ and $(2,-)$ (see \cref{fig:2t3_image}).


\begingroup
\newcommand{\mcheightscalefactor}{.052}

\begin{table}[h]
  \centering
  \begin{tabular}{cccM{2cm}M{2cm}M{2cm}}
    \makecell{candidate inequality\\(line \#, sign)}&$S$ & $c$ & $\mathfrak{n}_-(S<0)$ & $\mathcal{H}(S=c)$ & $\mathcal{H}(S<c)$\\
    \hline
    $1, +$ & $iZ_2$ & $-2$ 
      & \includegraphics[height=\mcheightscalefactor\pdfpageheight]{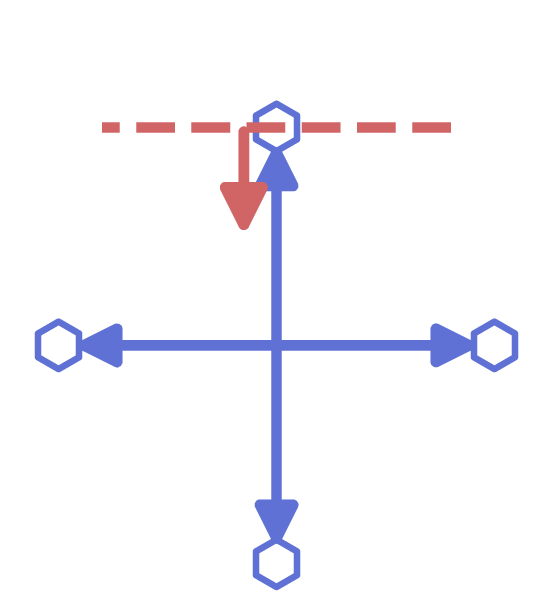}
      & \includegraphics[height=\mcheightscalefactor\pdfpageheight]{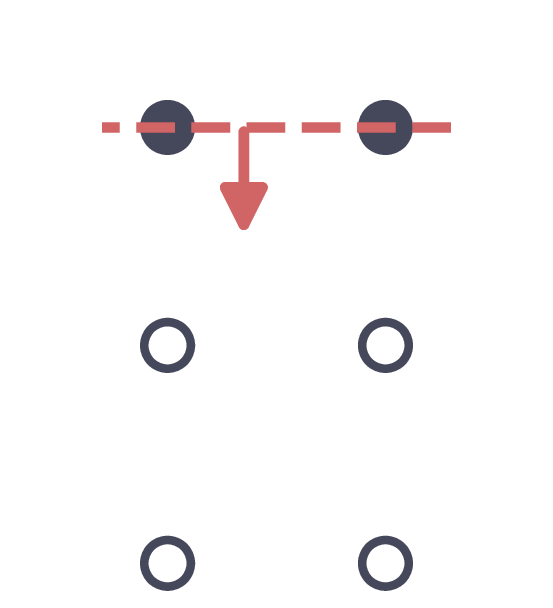}
      & \includegraphics[height=\mcheightscalefactor\pdfpageheight]{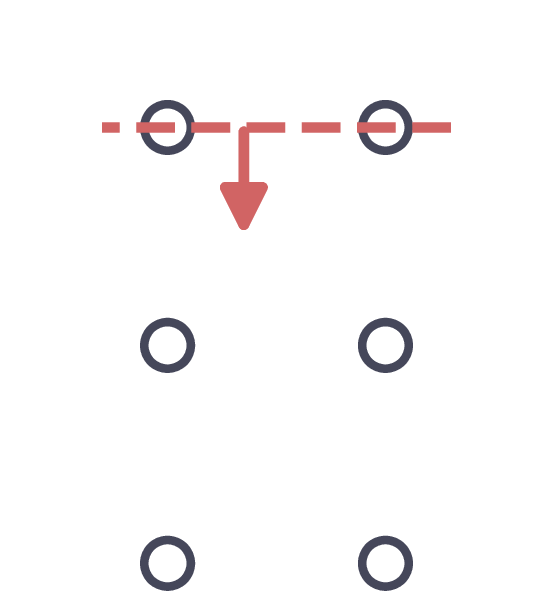}\\
    \hline
    $1, -$ & $-iZ_2$ & $2$
      & \includegraphics[height=\mcheightscalefactor\pdfpageheight]{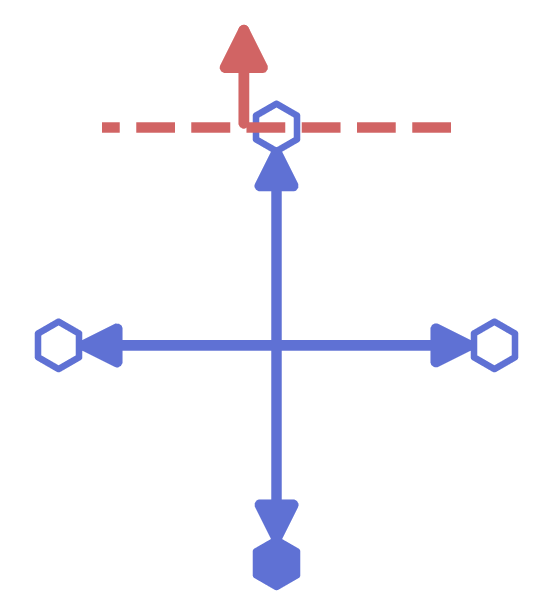}
      & \includegraphics[height=\mcheightscalefactor\pdfpageheight]{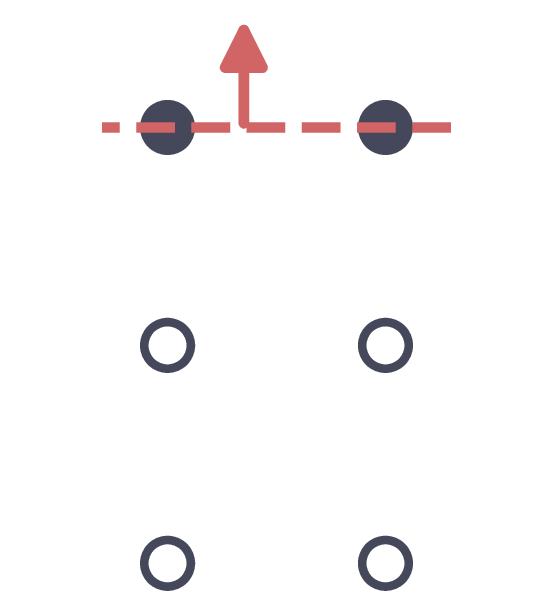}
      & \includegraphics[height=\mcheightscalefactor\pdfpageheight]{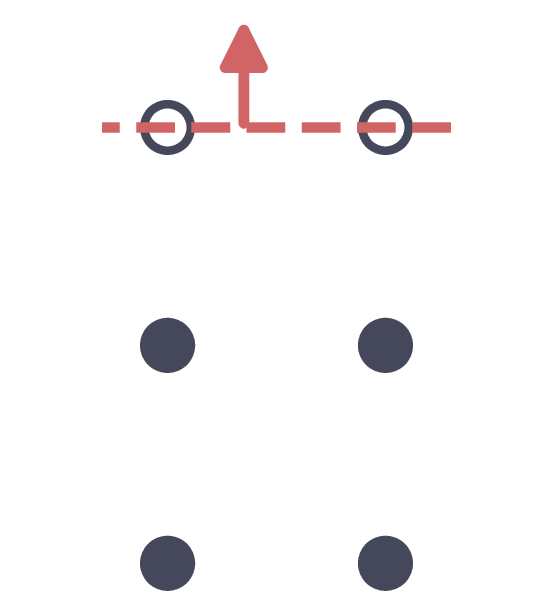}\\
    \hline

    $2, +$ & $iZ_1- iZ_2$ & $1$ 
      & \includegraphics[height=\mcheightscalefactor\pdfpageheight]{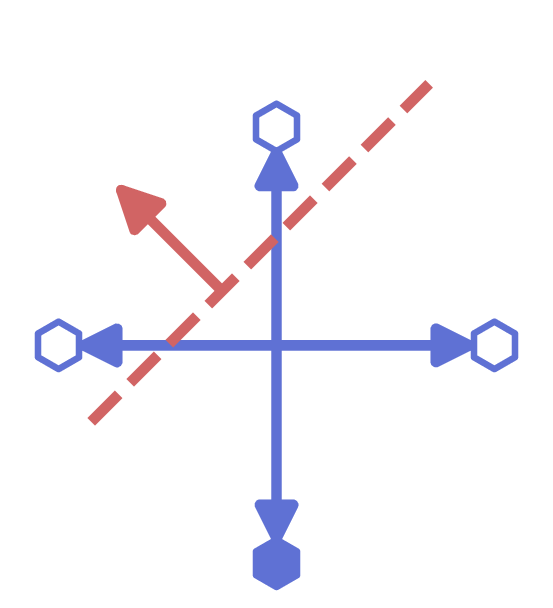}
      & \includegraphics[height=\mcheightscalefactor\pdfpageheight]{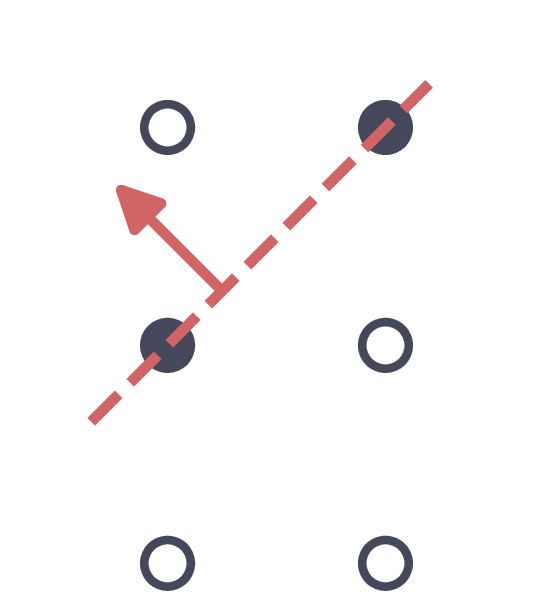}
      & \includegraphics[height=\mcheightscalefactor\pdfpageheight]{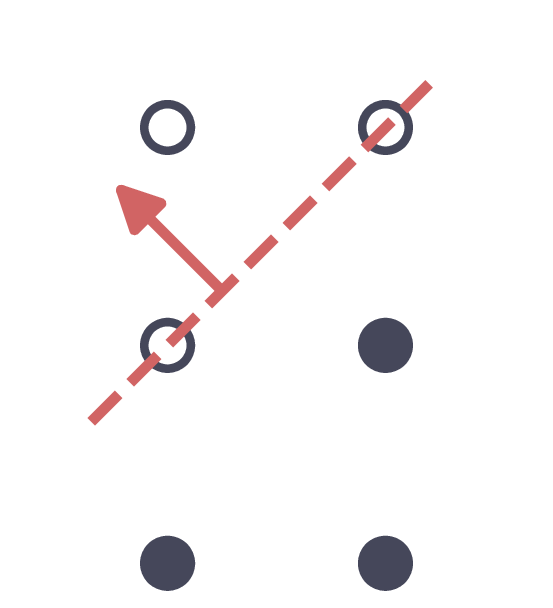}\\
    \hline

    $2, -$ & $-iZ_1+ iZ_2$ & $-1$ 
      & \includegraphics[height=\mcheightscalefactor\pdfpageheight]{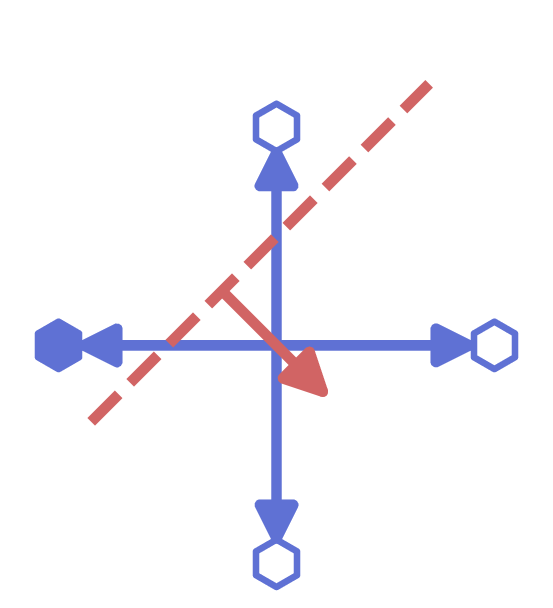}
      & \includegraphics[height=\mcheightscalefactor\pdfpageheight]{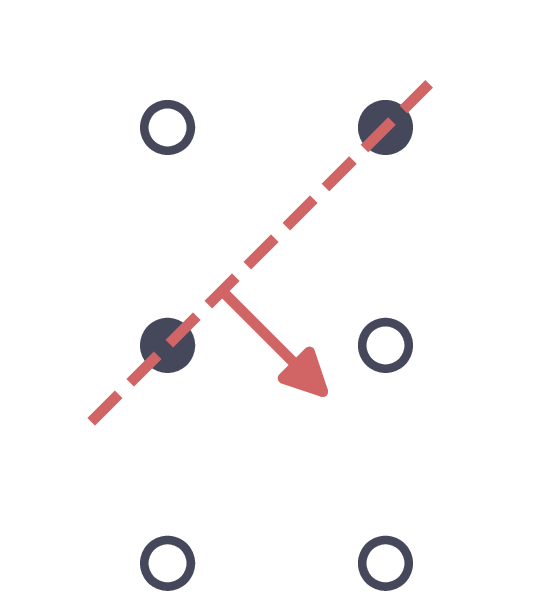}
      & \includegraphics[height=\mcheightscalefactor\pdfpageheight]{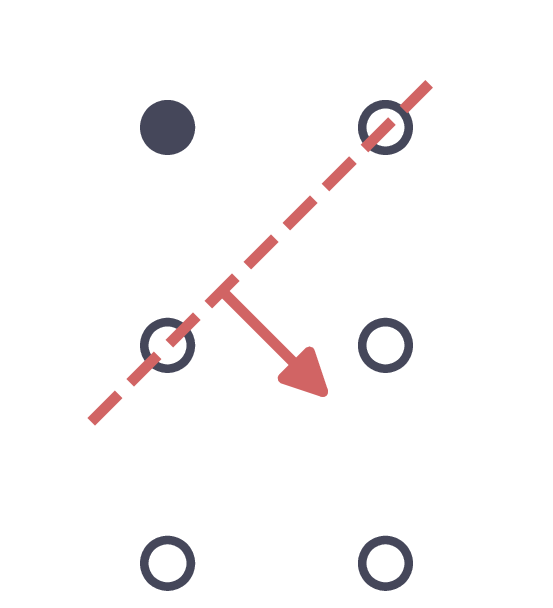}\\
    \hline

    $3, +$ & $2iZ_1 - iZ_2$ & $0$ 
      & \includegraphics[height=\mcheightscalefactor\pdfpageheight]{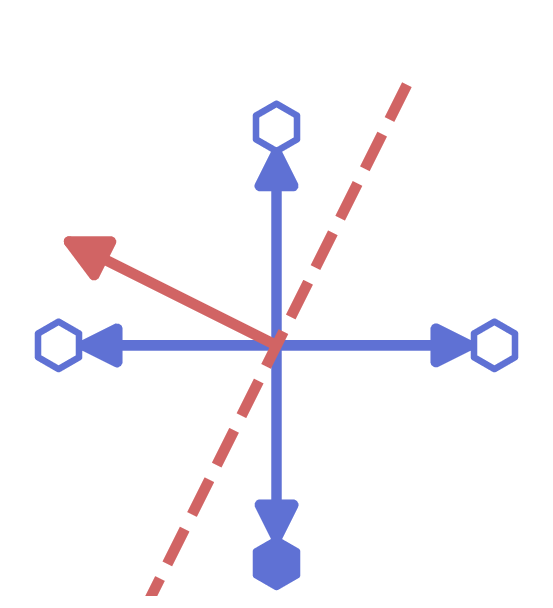}
      & \includegraphics[height=\mcheightscalefactor\pdfpageheight]{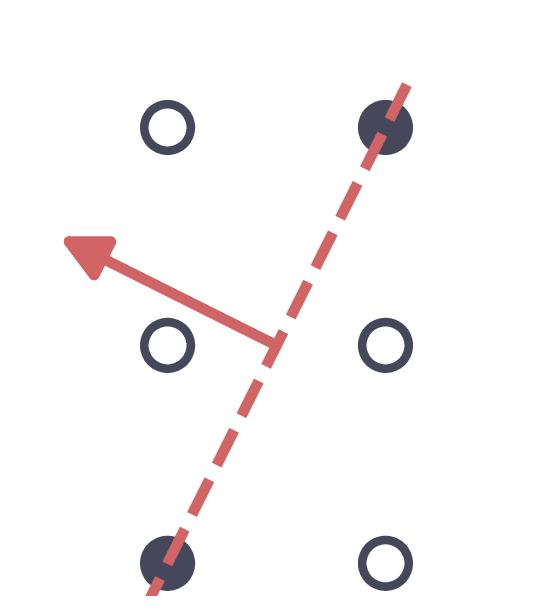}
      & \includegraphics[height=\mcheightscalefactor\pdfpageheight]{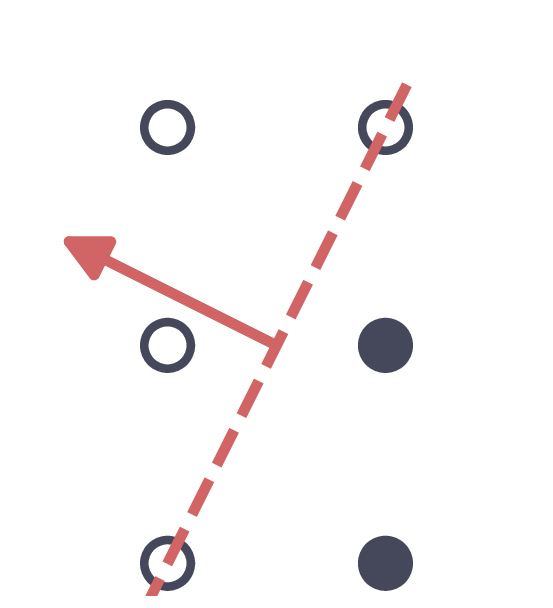}\\
    \hline

    $3, -$ & $-2iZ_1 + iZ_2$ & $0$ 
      & \includegraphics[height=\mcheightscalefactor\pdfpageheight]{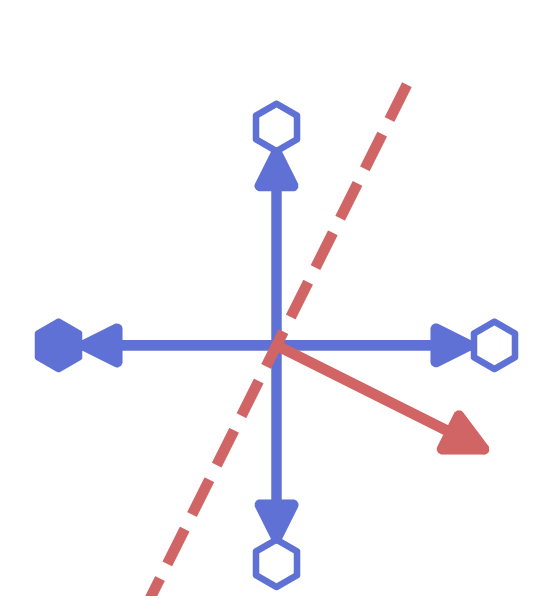}
      & \includegraphics[height=\mcheightscalefactor\pdfpageheight]{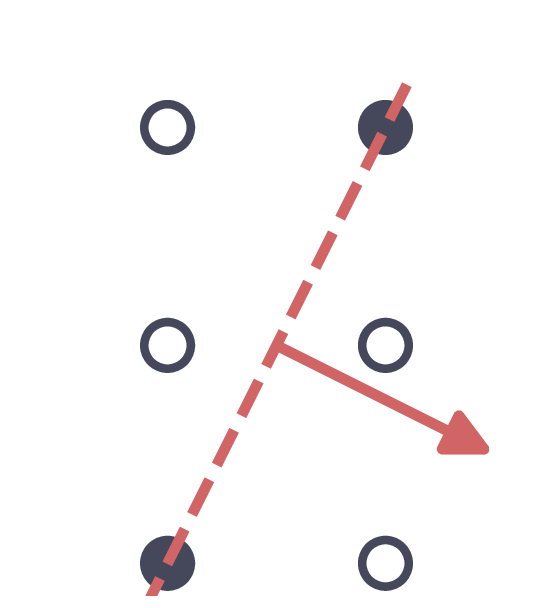}
      & \includegraphics[height=\mcheightscalefactor\pdfpageheight]{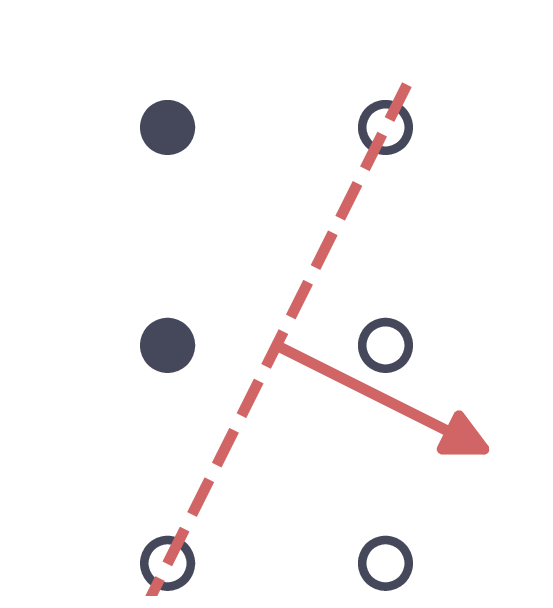}\\
    \hline

    $4, +$ & $iZ_1$ & $-1$
      & \includegraphics[height=\mcheightscalefactor\pdfpageheight]{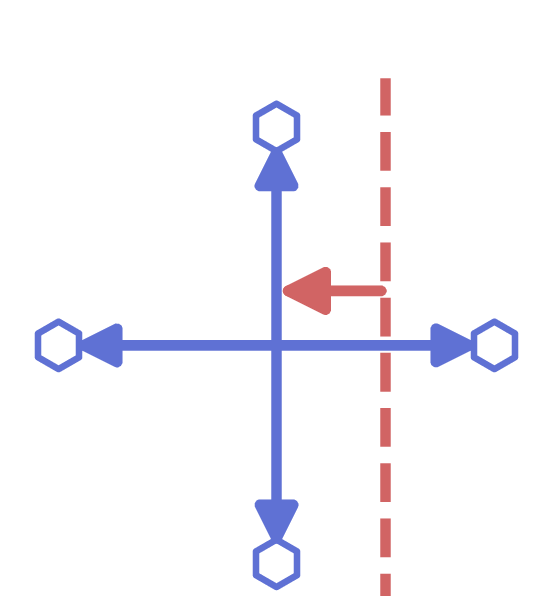}
      & \includegraphics[height=\mcheightscalefactor\pdfpageheight]{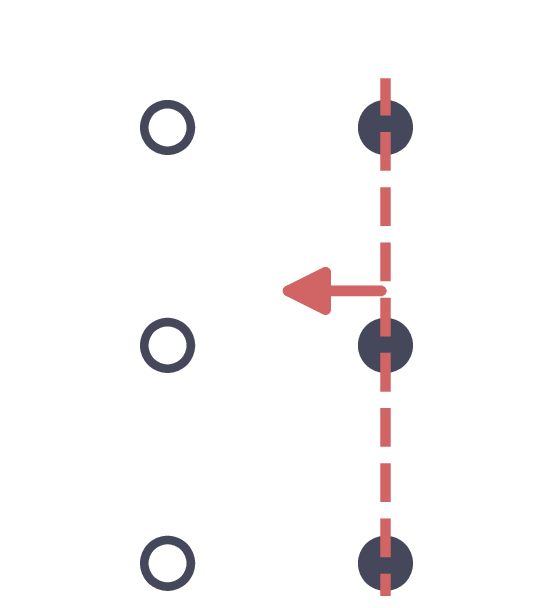}
      & \includegraphics[height=\mcheightscalefactor\pdfpageheight]{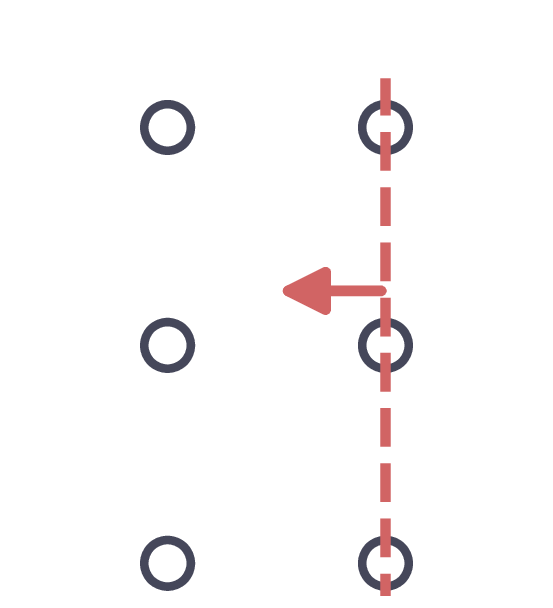}\\
      \hline

    $4, -$ & $-iZ_1$ & $1$ 
      & \includegraphics[height=\mcheightscalefactor\pdfpageheight]{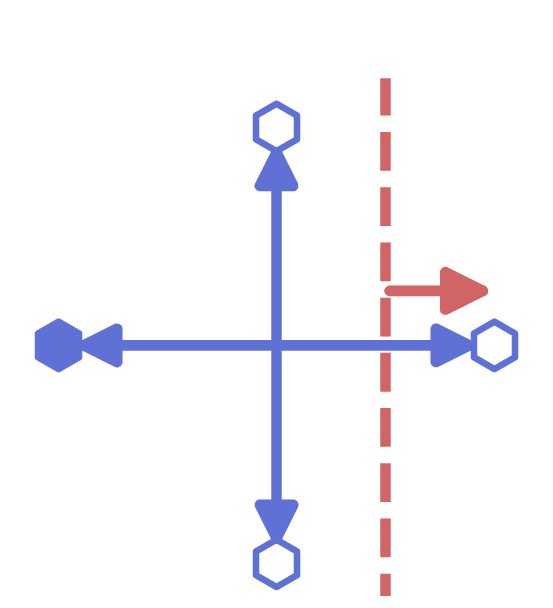}
      & \includegraphics[height=\mcheightscalefactor\pdfpageheight]{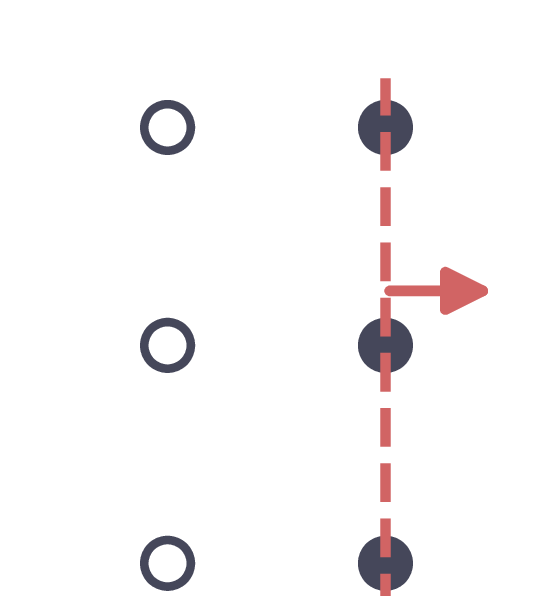}
      & \includegraphics[height=\mcheightscalefactor\pdfpageheight]{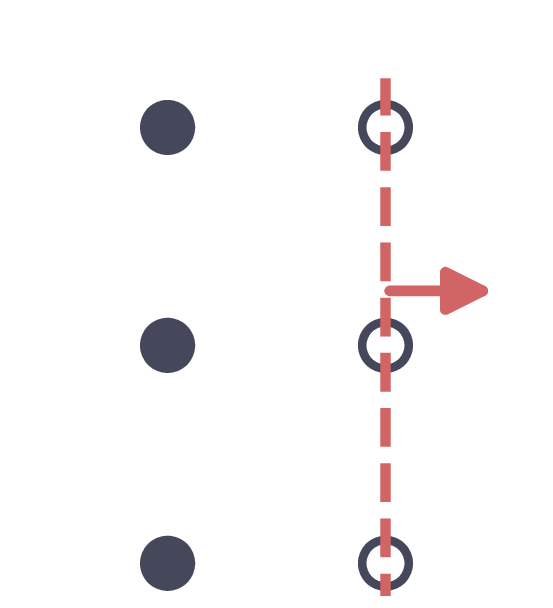}\\
      \hline

    $5, +$ & $iZ_1 + iZ_2$ & $-1$
      & \includegraphics[height=\mcheightscalefactor\pdfpageheight]{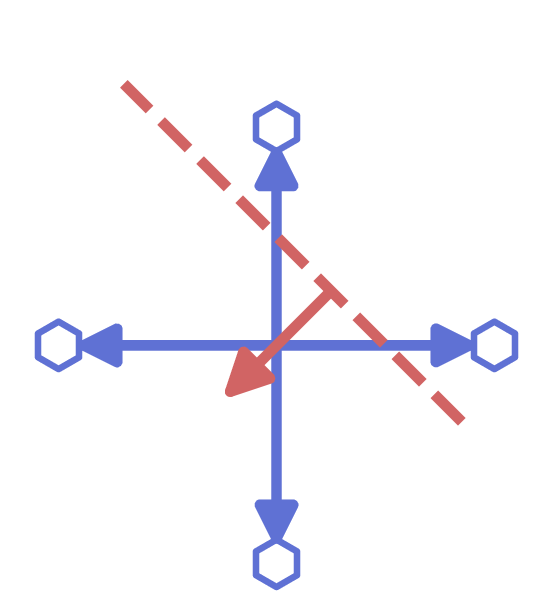}
      & \includegraphics[height=\mcheightscalefactor\pdfpageheight]{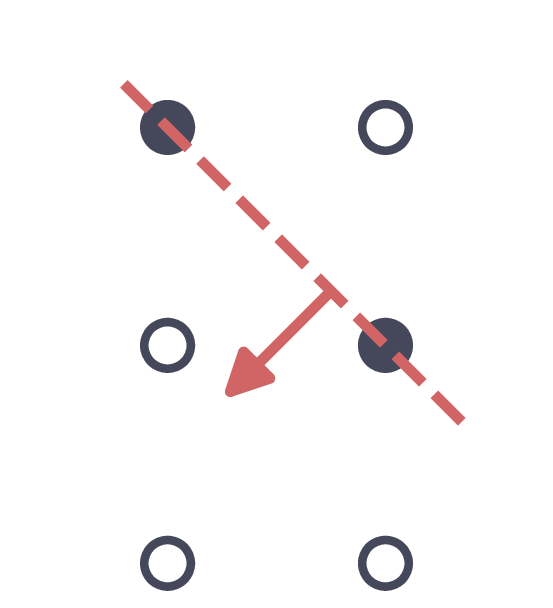}
      & \includegraphics[height=\mcheightscalefactor\pdfpageheight]{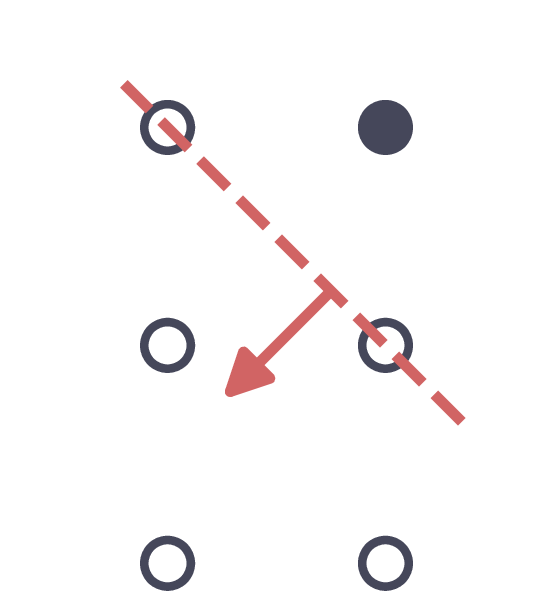}\\
      \hline

    $5, -$ & $-iZ_1 - iZ_2$ & $+1$
      & \includegraphics[height=\mcheightscalefactor\pdfpageheight]{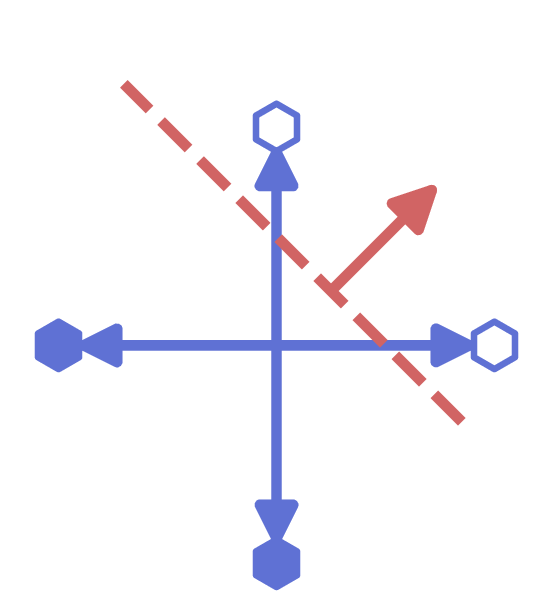}
      & \includegraphics[height=\mcheightscalefactor\pdfpageheight]{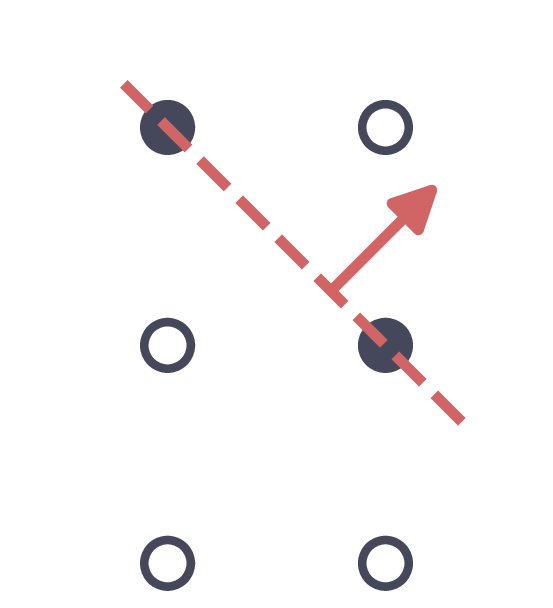}
      & \includegraphics[height=\mcheightscalefactor\pdfpageheight]{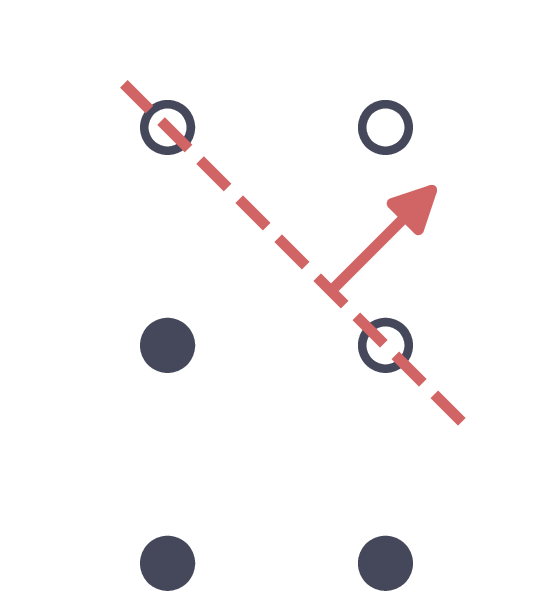}\\
      \hline

    $6, +$ & $iZ_1 - iZ_2$ & $-1$
      & \includegraphics[height=\mcheightscalefactor\pdfpageheight]{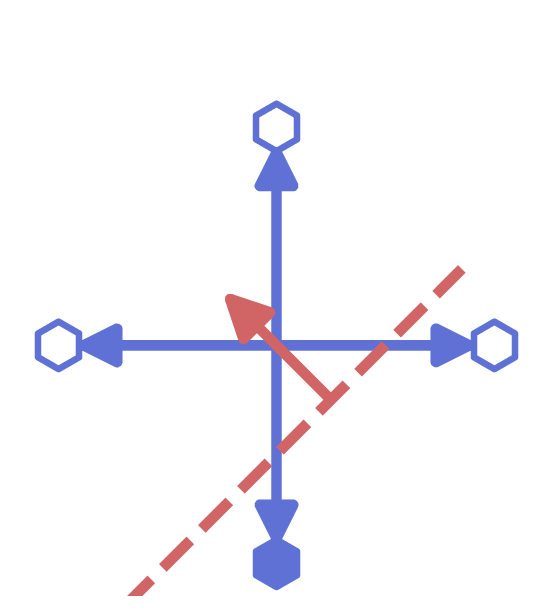}
      & \includegraphics[height=\mcheightscalefactor\pdfpageheight]{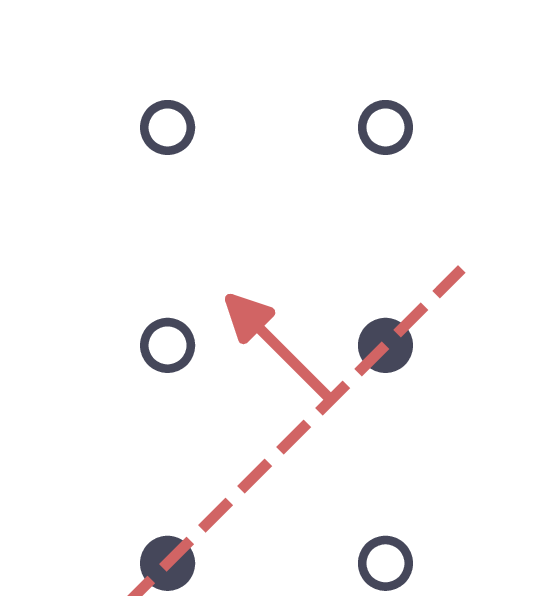}
      & \includegraphics[height=\mcheightscalefactor\pdfpageheight]{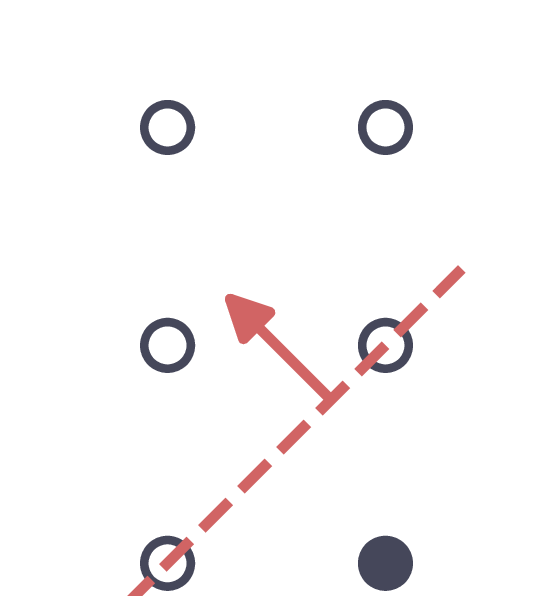}\\
      \hline

    $6, -$ & $-iZ_1 +iZ_2$ & $1$
      & \includegraphics[height=\mcheightscalefactor\pdfpageheight]{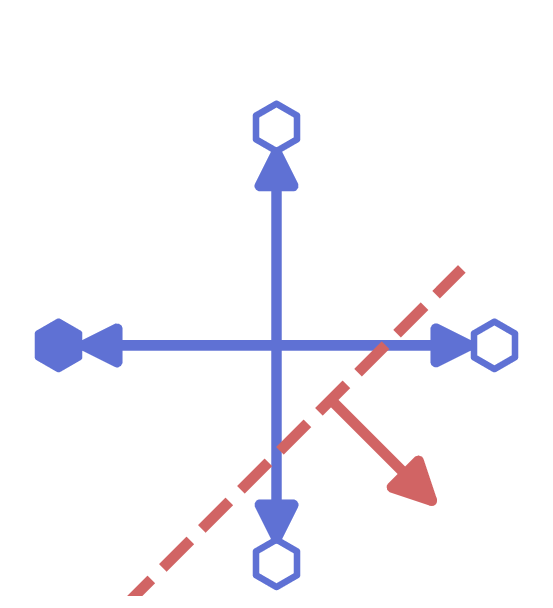}
      & \includegraphics[height=\mcheightscalefactor\pdfpageheight]{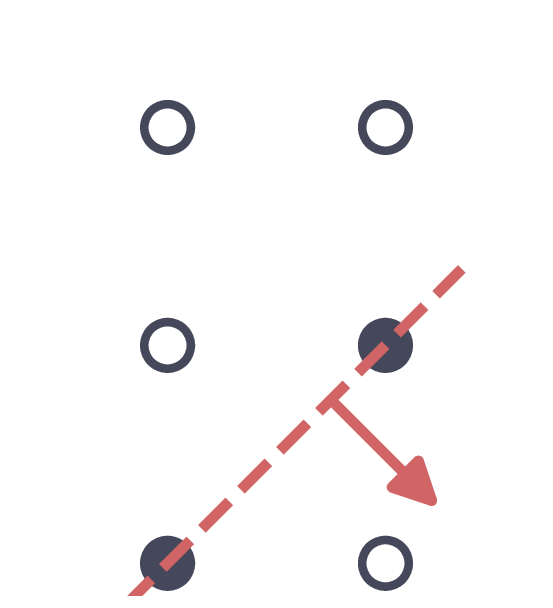}
      & \includegraphics[height=\mcheightscalefactor\pdfpageheight]{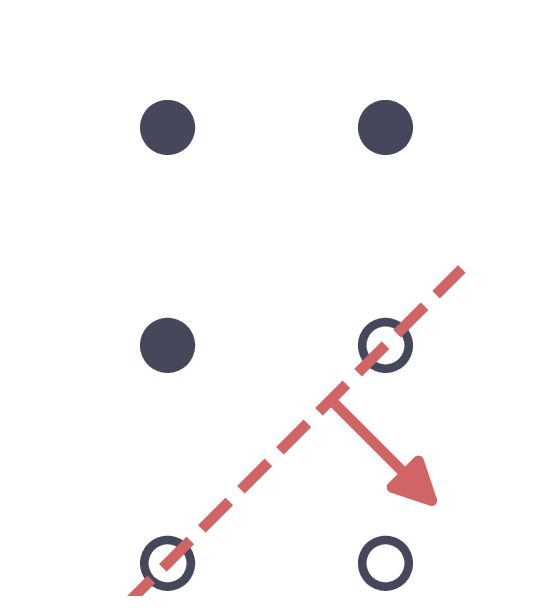}\\

  \end{tabular}

  \caption{Candidate inequalities for the action of $\mathrm{SU}(2)\times \mathrm{SU}(2)$ on   
  $\mathbb{C}^2\otimes \mathbb{C}^3$ (\cref{ex:2t3_moment_map_computation}).}
  \label{tab:2t3_ineq_candidates}

\end{table}

\endgroup

\end{ex}

\begin{ex}
  \label{ex:su3_adjoint_moment_map_computation}
  Take $\mathcal{H} = \mathfrak{sl}(3,\mathbb{C})$, $G= \mathrm{SU}(3)$, $G' =
  \mathrm{SL}(3,\mathbb{C})$, with $G'$ acting on $\mathcal{H}$ by the adjoint
  action $\tau' = \mathrm{Ad}$. We choose the subalgebra of diagonal matrices
  to be the Cartan subalgebra $\mathfrak{t}_{\mathbb{C}}$, which has a basis
  given by $Z := \mathrm{diag}(1,-1,0)$, $Z' := \mathrm{diag}(0,1,-1)$.
  The six roots are
  $$
  \begin{array}{lll}
    \alpha_1 = (Z\mapsto 2; Z'\mapsto -1)& \alpha_2 = (Z\mapsto -1; Z'\mapsto 2)& \alpha_3 = (Z\mapsto 1; Z'\mapsto 1)\\
    -\alpha_1= (Z\mapsto -2; Z'\mapsto 1)& -\alpha_2 = (Z\mapsto 1; Z'\mapsto -2)& -\alpha_3 = (Z\mapsto -1; Z'\mapsto -1),
  \end{array}
  $$
  which, together with $0\in i\mathfrak{t}^*$ with multiplicity two, are the
  weights of $\tau'$. After carrying out the same procedure as in
  \cref{ex:2t3_moment_map_computation}, one obtains the polytope
  $\Phi(\mathbb{P}(\mathcal{H}))\cap\mathfrak{t}_+^*$, which is shown 
  in \cref{fig:su3_adjoint_image}. In this case, there are two inequalities,
  which are given by
  $$
  \begin{aligned}
    &\braket{\cdot, iZ} \ge -1\\
    &\braket{\cdot, iZ'} \ge -1.
  \end{aligned}
  $$

  \begin{figure}[h]
    \centering
    \includegraphics[width=.45\textwidth]{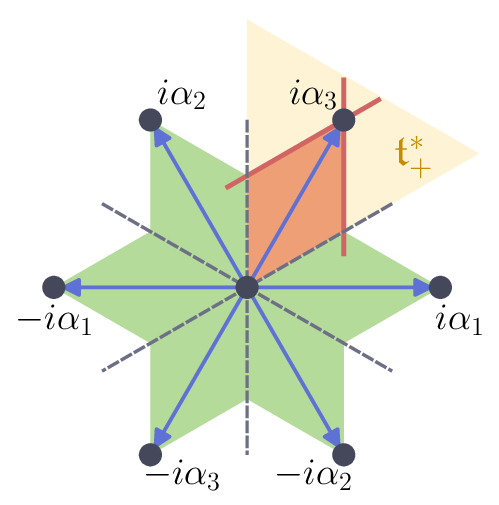}
    \caption{The image of the momentum map for the adjoint action of
    $\mathrm{SU}(3)$ on $\mathfrak{sl}(3,\mathbb{C})$
    intersected with $\mathfrak{t}^*$ (green and orange)
    and with $\mathfrak{t}^*_+$ (orange).
    The gray dashed lines are the reflection hyperplanes 
    of the Weyl group. The two red lines are the bounding 
    hyperplanes of $\Phi(\mathbb{P}(\mathcal{H}))\cap \mathfrak{t}_+^*$.
    }
    \label{fig:su3_adjoint_image}
  \end{figure}

\end{ex}

%
%
%
%
%
%
%
%

\chapter{Boundary Forces in Nonabelian Functional Theories}
\label{chap:momentum_map_functional_theories}

\section{Definitions}
\label{sec:nonab_ft_intro}

As remarked in the beginning of \cref{chap:interlude}, investigating all
nonabelian functional theories in the most general setting might be overly
ambitious. Therefore, we will limit ourselves to the case where the potential
map comes from a unitary representation of a compact Lie group on the Hilbert
space. This motivates the following definition:

\begin{defn}
  \label{defn:momentum_map_ft}
  A \textbf{momentum map functional theory} is a tuple $(G, \mathcal{H}, \tau,
  W)$, where $G$ is a compact connected Lie group, $\mathcal{H}$ is a Hilbert
  space, $\tau: G\rightarrow \mathrm{U}(\mathcal{H})$ is a representation, and
  $W\in i\mathfrak{u}(\mathcal{H})$ is any Hermitian operator.
\end{defn}

Let $\mathfrak{g}$ denote the Lie algebra of $G$ and $\tau_*:
\mathfrak{g}\rightarrow \mathfrak{u}(\mathcal{H})$ the Lie algebra
representation. Then the tuple $(i\mathfrak{g}, \mathcal{H}, \iota, W)$ is a
generalized functional theory in the sense of \cref{defn:generalized_ft}, where
the potential map $\iota: i\mathfrak{g}\rightarrow i\mathfrak{u}(\mathcal{H})$
is defined by $\iota(iA) = i \tau_*(A)$. Instead of writing $\iota$, we will
often simply use $\tau_*$ to denote the potential map whenever there is no
ambiguity. To put it another way, the potential map is the restriction of the
complexified representation $(\tau_*)_{\mathbb{C}}:
\mathfrak{g}_{\mathbb{C}}\rightarrow \mathfrak{gl}(\mathcal{H})$ to
$i\mathfrak{g}\subset \mathfrak{g}_{\mathbb{C}}$.

The density map is then the restriction of the dual map $\tau^*:
i\mathfrak{u}(\mathcal{H})^* \rightarrow i\mathfrak{g}^*$ to the set of
ensemble states $\mathcal{E}\subset i\mathfrak{u}(\mathcal{H})^*$ (recall our
convention of viewing an ensemble state as a linear functional on
$i\mathfrak{u}(\mathcal{H})$). As before, we will not indicate the restriction
to $\mathcal{E}$ explicitly, so the density map is simply denoted as $\tau^*$.

\begin{ex}
  RDMFT is a momentum map functional theory. Indeed, let $\mathcal{H}_1$ be any
  Hilbert space (interpreted as the single-particle Hilbert space), and let
  $\mathcal{H} = \exterior^N\mathcal{H}_1$ or $\mathrm{Sym}^N\mathcal{H}_1$.
  There is a natural homomorphism $\tau: \mathrm{U}(\mathcal{H}_1)\rightarrow
  \mathrm{U}(\mathcal{H})$, and the corresponding potential map is
  $$
  \begin{aligned}
    \tau_*: &\;i\mathfrak{u}(\mathcal{H}_1) \rightarrow i\mathfrak{u}(\mathcal{H})\\
    & h \mapsto  h\otimes \mathbbm{1}^{\otimes (N-1)} + \dotsb +
    \mathbbm{1}^{\otimes (N-1)}\otimes h.
  \end{aligned}
  $$
  The density map $\tau^*:
  i\mathfrak{u}(\mathcal{H})^*\rightarrow i\mathfrak{u}(\mathcal{H}_1)^*$ is
  just $N$ times the partial trace over $N-1$ particles.
\end{ex}

\begin{ex}
  Let $\mathcal{H}= (\mathbb{C}^d)^{\otimes N}$ be the Hilbert space of $N$
  qudits. Let $G = \prod_{i=1}^N\mathrm{U}(d)$, and
  $$
  \begin{aligned}
    \tau: & \;G\rightarrow \mathrm{U}(\mathcal{H})\\
    & (g_1, \dotsb, g_N) \mapsto g_1\otimes \dotsb \otimes g_N.
  \end{aligned}
  $$
  So $\tau(G)\subset \mathrm{U}(\mathcal{H})$ consists of the local unitaries.
  The Lie algebra is $\mathfrak{g} = \bigoplus_{i=1}^N \mathfrak{u}(d)$, and
  $\tau_*(i\mathfrak{g})\subset i\mathfrak{u}(\mathcal{H})$ is the Lie algebra
  of local observables. That is, $\tau_*(\mathfrak{g})\subset
  \mathfrak{u}(\mathcal{H})$ consists of operators of the form $h_1\otimes
  \mathbbm{1}^{\otimes (N-1)} + \dotsb + \mathbbm{1}^{\otimes (N-1)}\otimes
  h_N$, where $h_i \in \mathfrak{u}(d)$. The density map $\tau^*:
  i\mathfrak{u}(\mathcal{H})^*\rightarrow i\mathfrak{g}^*$ sends a state
  $\Gamma \in i\mathfrak{u}(\mathcal{H})^*$ to the list of partial traces
  $$
    (\Tr_{2,\dotsb, N}\Gamma, \Tr_{1,3,\dotsb, N}\Gamma, \dotsb, \Tr_{1,\dotsb, N-1}\Gamma).
  $$
\end{ex}

The advantage of studying momentum map functional theories is that the pure and
ensemble state representability problems are not completely out of reach as
compared to, for example, the two-body problem (see \cref{ex:2bodyqma}).
This is because we can apply the machinery developed in \cref{chap:interlude}
to investigate the density map $\tau^*$:

\begin{propo}
  \label{thm:density_map_is_momentum_map}
  The composition
  $$
  \mathbb{P}(\mathcal{H}) \rightarrow \mathcal{P}
  \xrightarrow{\tau^*} i\mathfrak{g}^* \xrightarrow{\times i} \mathfrak{g}^*
  $$
  is a momentum map.
  \begin{proof}
    See \cref{thm:group_rep_momentum_map}.
  \end{proof}
\end{propo}

Let $(G, \mathcal{H}, \tau, W)$ be a momentum map functional theory. Fix a
Cartan subalgebra $\mathfrak{t}\subset \mathfrak{g}$ and a $G$-invariant inner
product $\braket{\cdot,\cdot}$ on $\mathfrak{g}$, so that we can think of
$\mathfrak{t}^*$ as a subspace of $\mathfrak{g}^*$. Pick a Weyl chamber
$\mathfrak{t}_+^*\subset \mathfrak{g}^*$. 
Let $\Omega\subset i\mathfrak{t}^*$ denote the set of weights of the
representation $\tau: \mathfrak{g}\rightarrow \mathfrak{u}(\mathcal{H})$
relative to $\mathfrak{t}$. For each weight $\omega\in \Omega$, let
$\mathcal{H}_\omega$ stand for the weight space, 
which is defined as
$\{\ket{\Psi}\in \mathcal{H} \mid \forall A\in \mathfrak{t}: \tau(A)\ket{\Psi}
= \braket{\omega,A}\ket{\Psi}\}$. Then we have the weight space decomposition
$$
\mathcal{H} = \bigoplus_{\omega \in \Omega}\mathcal{H}_\omega.
$$

\begin{propo}
  The set of pure state representable densities $\tau^*(\mathcal{P})\subset
  i\mathfrak{g}^*$ intersects $i\mathfrak{t}^*_+$ in a convex polytope
  $\Lambda$, which we will refer to from now on as the \textbf{Kirwan polytope}.
  \begin{proof}
    This is a straightforward corollary to \cref{thm:density_map_is_momentum_map}
    and Kirwan's theorem (\cref{thm:kirwan}).
  \end{proof}
\end{propo}

\begin{ex}
  In the case of RDMFT, this means that the spectra of the 1RDMs which come
  from $N$-particle pure states form a convex polytope.
\end{ex}

The ensemble state representability problem in this case is still easy:

\begin{propo}
  The intersection of $i\mathfrak{t}^*$ and the set of ensemble state
  representable densities $\tau^*(\mathcal{E})$ is exactly $\conv(\Omega)$.
  \begin{proof}
    Clearly, $\tau^*(\mathcal{E})\cap i\mathfrak{t}^*$ is convex and contains
    $\Omega$, so $\tau^*(\mathcal{E}) \cap i\mathfrak{t}^*\supset
    \conv(\Omega)$. Conversely, suppose 
    $$
      \rho = \tau^*\left(\sum_i t_i \ketbrap{\Psi_i}\right) \in i\mathfrak{t}^*.
    $$
    For each $i$, decompose $\ket{\Psi_i} = \sum_{\omega\in\Omega}\ket{\Psi_i^\omega}$ , then
    $$
    \begin{aligned}
      \braket{\rho, v} = \sum_{i} \sum_{\omega,\omega'\in \Omega} 
      t_i\braket{\Psi_i^\omega| \tau(v) |\Psi_i^{\omega'}}
      = \sum_i \sum_{\omega\in \Omega}t_i\lVert\Psi_i^\omega\rVert^2 
      \braket{\omega, v} 
    \end{aligned}
    $$
    for any $v\in i\mathfrak{t}$, so $\rho \in \conv(\Omega)$.
    Hence $\tau^*(\mathcal{E})\cap i\mathfrak{t}^*\subset \conv(\Omega)$.
  \end{proof}
\end{propo}

In the following, we will always assume that the Kirwan polytope $\Lambda:=
\tau^*(\mathcal{P})\cap i\mathfrak{t}^*_+$ has full dimension in
$i\mathfrak{t}^*$, which implies that $\conv(\Omega)$ has full dimension in
$i\mathfrak{t}^*$. Note that the converse is not true in general. Indeed,
consider the ``trivial'' functional theory on $\mathcal{H}=\mathbb{C}^N$ (the
interaction $W$ is irrelevant here), where $G = \mathrm{SU}(N)$ and $\tau:
\mathrm{SU}(N)\hookrightarrow \mathrm{U}(N)$ is the inclusion. A Cartan
subalgebra for $\mathfrak{g}=\mathfrak{su}(N)$ is the set $\mathfrak{t}$ of
imaginary diagonal traceless matrices, and we can take $\mathfrak{t}_+^*$ to be
the subset of matrices whose diagonal entries times $i$ are in nonincreasing
order. Then 
$$
  \Lambda= \tau^*(\mathcal{P})\cap i\mathfrak{t}^*_+ = \left\{\mathrm{diag}
  \left(\frac{N-1}{N}, -\frac{1}{N}, \dotsb, -\frac{1}{N}\right)\right\}
$$
is zero-dimensional, yet $\conv(\Omega)$ clearly has full dimension in
$i\mathfrak{t}^*$.

Let $R\subset i\mathfrak{t}^*$ be the set of roots, for which we have the root
space decomposition
$$
\mathfrak{g}_{\mathbb{C}} = \mathfrak{t}_{\mathbb{C}} \oplus
\bigoplus_{\alpha\in R} \mathfrak{g}_\alpha'.
$$
For each $\alpha \in R$, let $(L^+_\alpha, L^-_\alpha, H_\alpha)$ be an
$\mathfrak{sl}(2,\mathbb{C})$-tuple, meaning in particular that $L^\pm_\alpha
\in \mathfrak{g}_{\pm \alpha}'$, 
$H_\alpha \in \mathfrak{t}_{\mathbb{C}}$, and
$[L^+_\alpha, L^-_\alpha] = H_\alpha$. Let $R_+$ be the set of positive roots
relative to $\mathfrak{t}_+^*$, and write
$$
\mathfrak{g} = \mathfrak{t} \oplus 
\bigoplus_{\alpha\in R_+} (i\mathbb{R} X_\alpha)
\oplus \bigoplus_{\alpha\in R_+} (i\mathbb{R} Y_\alpha),
$$
where $X_\alpha:= L^+_\alpha + L^-_\alpha$ and 
$Y_\alpha := -iL^+_\alpha + iL^-_\alpha$.

As in the abelian case, we would like to understand the behavior of the pure
functional near the boundary of its domain. Ideally, we would prove a
generalization of \cref{thm:boundary_force_formula}, but it is not at all clear
how one would adapt the proof we presented in
\cref{chap:abelian_functional_theory} for nonabelian $\mathfrak{g}$. For
example, it is not obvious anymore whether the fiber of the density map can be
understood by first studying the classical density map
(Definition~\ref{defn:classical_density}), as we did in
\cref{chap:abelian_functional_theory} for the proof of the boundary force using
constrained search. To see what kind of complications might arise, consider a
state $\ket{\Psi}= \sum_{\omega\in\Omega} c_\omega\ket{E_\omega}$, where
$(\ket{E_\omega})_{\omega \in \Omega}$ is a basis for $\mathcal{H}$ (assuming
that all weights have multiplicity one for simplicity), satisfying $\sum_{\omega
\in \Omega} |c_\omega|^2\omega = \rho \in i\mathfrak{t}^*$. We still
have $\braket{\tau^*(\ketbrap{\Psi}), v} = \braket{\rho, v}$ for all $v \in
i\mathfrak{t}$. However, we cannot conclude immediately anymore that
$\tau^*(\ketbrap{\Psi}) = \rho$ as in the abelian setting, because
$\tau^*(\ketbrap{\Psi})$ might not vanish on $X_\alpha$ or $Y_\alpha$. In
other words, $\tau^*(\ketbrap{\Psi})$ might not even be in $i\mathfrak{t}^*$!

Nevertheless, it is still possible to obtain a formula analogous to
Eq.~\eqref{eq:boundary_force_formula} using perturbation theory, in a manner similar to
the argument presented in \cref{sec:abelian_boundary_force_perturbation}
before the rigorous proof in \cref{sec:boundary_force_rigorous_proof}. As we
will see, however, the perturbative method works only for the so-called
\textit{nice} facets, which we will now define.

First, note that we should not expect the boundary force to be present near all
facets of the convex polytope $\Lambda\subset i\mathfrak{t}^*_+$. Intuitively, a
facet that is entirely contained in $i\mathfrak{t}^*_{+} \setminus
i\mathfrak{t}^*_{>0}$ (the latter being the open Weyl chamber) has nothing to do with the ``true boundary'' of the domain
of the pure functional. Following
Ref.~\cite{walterMultipartiteQuantumStates2014}, we will call such a facet $F\subset
\Lambda$ \textit{trivial}. 

\begin{defn}
  \label{defn:nice_facet}
  A nontrivial facet $F\subset \Lambda$ is said to be \textbf{nice} if $F$ is
  contained in a facet of $\conv(\Omega)$.
\end{defn}

\begin{figure}[h]
  \centering
  \subfloat[]{
    \includegraphics[width=.3\textwidth]{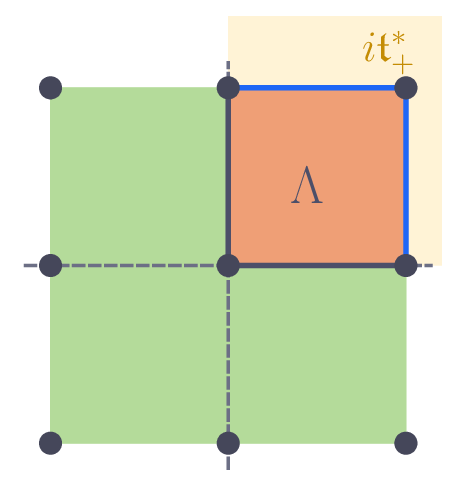}
  }
  \subfloat[\label{fig:nice_facets_a}]{
    \includegraphics[width=.3\textwidth]{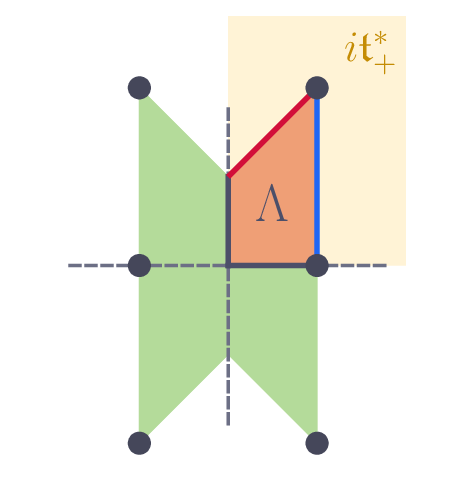}
  }
  \subfloat[]{
    \includegraphics[width=.3\textwidth]{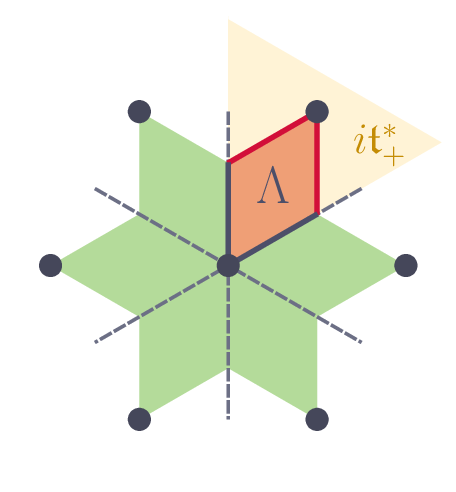}
  }

  \caption{Trivial (black), nice (blue), and nontrivial not nice (red) facets.
  (a): $G= \mathrm{SU}(2)\times \mathrm{SU}(2)$, $\mathcal{H} = \mathbb{C}^3\otimes \mathbb{C}^3$.
  (b): $G= \mathrm{SU}(2)\times \mathrm{SU}(2)$, $\mathcal{H} = \mathbb{C}^2\otimes \mathbb{C}^3$.
  (c): $G= \mathrm{SU}(3)$, $\mathcal{H} = \mathfrak{sl}(3,\mathbb{C})\cong\mathbb{C}^8$.
  All representations are the obvious ones. The gray dashed lines are the Weyl
  group reflection hyperplanes.
  }
  \label{fig:nice_facets}
\end{figure}

\cref{fig:nice_facets} shows several examples of trivial, nontrivial and nice,
and nontrivial and not nice facets.
Informally speaking, a nice facet of $\Lambda$ is one which remains after convexifying
$\tau^*(\mathcal{P})\cap i\mathfrak{t}^*$, which is the union of $\Lambda$ and
its images under the Weyl group action. If a facet $F'$ of $\conv(\Omega)$ is
the union of a nice facet of $\Lambda$ and its images under the Weyl group, we
will also call $F'$ nice.

If $F\subset \Lambda$ is a nice facet, we will let $\Omega_F$ denote the subset
of weights that lie on the affine hull of $F$. Thus, in general, $F$ will be a
proper subset of $\conv(\Omega_F)$. Similarly, $\mathcal{H}_F\subset \mathcal{H}$
denotes the subspace spanned by the weight spaces of weights in $\Omega_F$, and
$\Pi_F: \mathcal{H}\rightarrow \mathcal{H}_F$ is the orthogonal projection.

As in the case of abelian functional theories, a notion of the restricted functional
theory to a facet $F$ will be useful for the boundary force derivation (see
\cref{def:facet_functional_theory} for the definition in the abelian case),
although it is not immediately clear what the correct definition should be when
$\mathfrak{g}$ is not abelian. In the abelian setting, all we had to do was
restrict the Hilbert space to the subspace spanned by the weight spaces on the
facet. However, doing so is not good enough for nonabelian functional theories:
for arbitrary $g\in G$, the operator $\Pi_F^{\phantom{\dagger}}\tau(g)\Pi_F^\dagger\in
\mathrm{End}(\mathcal{H}_F)$ might fail to be unitary. In fact, $g\mapsto
\Pi_F^{\phantom{\dagger}}\tau(g)\Pi_F^\dagger$ might fail to be a representation at all! Hence,
naively applying \cref{def:facet_functional_theory} will in general result in a
functional theory which is not a momentum map functional theory anymore. What
follows is a definition which turns out to play well with boundary force
calculations presented in \cref{sec:general_momentum_perturbation}.

\begin{defn}
  \label{def:facet_functional_theory_nonabelian}
  Let $F\subset\conv(\Omega)$ be a nice facet, and let $\braket{\cdot, S}\ge c$ be
  a corresponding inequality, where $S\in i\mathfrak{t}$ and $c\in\mathbb{R}$.
  The \textbf{restricted functional theory on the facet} $F$ (or simply
  \textbf{facet functional theory}) is the momentum map functional theory given
  by $(G_F, \mathcal{H}_F, \tau_F, \Pi_F^{\phantom{\dagger}} W \Pi_F^\dagger)$, where 
  $G_F$ is the identity component of $C_{G}(S)$ (the centralizer) and 
  $$
    \mathcal{H}_F := \bigoplus_{\omega \in \Omega_F} \mathcal{H}_\omega
    \hspace{3.5em}
    \tau_F(g) := \Pi_F^{\phantom{\dagger}} \tau(g) \Pi_F^\dagger,
  $$
\end{defn}

Note: (1) $G_F$ is compact because $C_G(S)$ is; (2) If $\ket{\Psi}\in
\mathcal{H}_F$, then $\tau_*(S)\tau(g)\ket{\Psi} = \tau(g)\tau_*(S)\ket{\Psi}
= c\tau(g)\ket{\Psi}$ for all $g\in G_F$, so $\tau(g)\ket{\Psi}\in
\mathcal{H}_F$. This shows that $\mathcal{H}_F$ is stable under the $G$-action.

\begin{ex}
  If $\mathfrak{g}$ is abelian, then $G_F = G$ because $G$ is connected and
  compact, so we recover \cref{def:facet_functional_theory}.
\end{ex}

We will also define
$$
\begin{aligned}
  & \mathfrak{g}^\nparallel := \bigoplus_{\alpha\in R_+: \braket{\alpha, S}\neq 0}
  (i\mathbb{R}X_\alpha \oplus i\mathbb{R} Y_\alpha)\\
  & \mathfrak{g}^\parallel := \bigoplus_{\alpha\in R_+: \braket{\alpha, S}=0}
  (i\mathbb{R}X_\alpha \oplus i\mathbb{R} Y_\alpha).
\end{aligned}
$$
Note that $\mathfrak{g}^\nparallel = 0$ if $\mathfrak{g}$ is abelian.

\begin{propo}
  \label{propo:facet_algebra_decomposition}
  $\mathfrak{t} \oplus\mathfrak{g}^\parallel = \mathfrak{g}_F$,
  where $\mathfrak{g}_F$ is the Lie algebra of $G_F$.
  \begin{proof}
    Suppose $\alpha\in R$ is parallel to the facet, i.e., $\braket{\alpha, S}=0$.
    Then $[L^\pm_\alpha, S] = \mp\braket{\alpha, S}L^\pm_\alpha = 0$, so
    $L^\pm_\alpha\in C_{\mathfrak{g}}(S) = \mathfrak{g}_F$ and consequently 
    $iX_\alpha, iY_\alpha \in \mathfrak{g}_F$.
    On the other hand, if $B\in \mathfrak{t}$, then $[B, S]=0$ trivially. This
    shows ``$\subset$''.

    Now take any $B\in \mathfrak{g}$ such that $[B,S]=0$. Decompose $B$ in
    $\mathfrak{g}' = \mathfrak{g}_{\mathbb{C}}$ into 
    $B = B_0 + \sum_{\alpha\in R} c_{\alpha} L^+_\alpha$ 
    with $B_0\in \mathfrak{t}$. Then $[B,S]=0$ implies
    $\sum_{\alpha\in R} c_\alpha\braket{\alpha,S}L^+_\alpha =0$. 
    That is, $c_\alpha=0$ whenever $\braket{\alpha, S}\neq 0$. This shows $B\in
    \mathfrak{t} \oplus\mathfrak{g}^\parallel$.
  \end{proof}
\end{propo}

\begin{ex}
  \label{ex:2t3_facet_dft}
  Consider the action of $\mathrm{SU}(2)\times \mathrm{SU}(2)$ on
  $\mathbb{C}^2\otimes \mathbb{C}^3$ as in \cref{ex:2t3_moment_map_computation}.
  We will take $F$ to be the only nice facet of $\Lambda$ (see
  \cref{fig:nice_facets_a}), for which $S$ can be taken to be $-Z_1 = (-Z,
  0)\in i\mathfrak{su}(2)\oplus i\mathfrak{su}(2)$.
  Then 
  $$
  \begin{aligned}
    \mathfrak{g}_F = &C_{\mathfrak{g}}(-Z_1) = 
    \mathrm{span}_{\mathbb{R}}\{iZ_1, iX_2, iY_2, iZ_2\}\\
    &\mathfrak{g}^\nparallel = \mathrm{span}_{\mathbb{R}}\{iX_1, iY_1\}\\
    &\mathfrak{g}^\parallel = \mathrm{span}_{\mathbb{R}}\{iX_2, iY_2\}.
  \end{aligned}
  $$
  The connected Lie group of the facet functional theory is then $G_F =
  \mathrm{U}(1)\times \mathrm{SU}(2)$, where $\mathrm{U}(1)$ is embedded into
  $\mathrm{SU}(2)$ by $g \mapsto \mathrm{diag}(g, g^{-1})$.
  The weights lying on the affine hull of $F$ are 
  $$
  \begin{aligned}
    &\omega_{1,2} = (Z\mapsto 1, Z\mapsto 2)\\
    &\omega_{1,0} = (Z\mapsto 1, 0)\\
    &\omega_{1,-2} = (Z\mapsto 1, Z\mapsto-2),
  \end{aligned}
  $$
  so the restricted Hilbert space $\mathcal{H}_F$ is the subspace of
  $\mathcal{H}$ spanned by $\ket{1,2}, \ket{1,0},$ and $\ket{1,-2}$ 
  (where $\ket{m_1, m_2}$ is what we called $\phi_{m_1,m_2}$ in \cref{ex:2t3_moment_map_computation}). 
  Hence, we
  end up with an action of $\mathrm{U}(1)\times \mathrm{SU}(2)$ on
  $\mathcal{H}_F\cong \mathbb{C}^3$, with $\mathrm{U}(1)$ acting 
  by $e^{i\varphi}\mapsto e^{i\varphi}$ and $\mathrm{SU}(2)$
  by the three-dimensional irreducible representation.
\end{ex}

For the nice facets, we have a boundary force formula:
\begin{conj}
  \label{conj:boundary_force_formula_nonabelian}
  Let $F\subset \conv(\Omega)$ be a nice facet with a corresponding inequality
  $\braket{\cdot, S}\ge c$. 
  Let $\rho_* \in F$ be a density which is ``sufficiently
  nice''. Let $\eta \in i\mathfrak{t}^*$ be inward-pointing with $\braket{\eta,
  S}=1$. Then
  \begin{equation}
    \label{eq:boundary_force_formula_nonabelian}
    \doverd{\sqrt{\epsilon}}\Big|_{\epsilon=0}
    \Fp(\rho_* + \epsilon \eta)
     = -2 \min_{v\in \mathfrak{g}^{\nparallel}}\left(\sum_{\omega \in \Omega\setminus \Omega_F}
     \frac{\lVert \Pi_\omega (\tau_*(v) + W) \ket{\Phi}\rVert^2}{\braket{\omega,S} - c}
     \right)^{\frac{1}{2}}.
  \end{equation}
\end{conj}
This result is labeled as a conjecture because we only have a perturbative
derivation, which relies heavily on various strong assumptions, just like
what was necessary in \cref{sec:abelian_boundary_force_perturbation} for the
perturbative derivation of the abelian boundary force
(Eq.~\eqref{eq:boundary_force_formula}). Currently, we do not have a proof of
\cref{conj:boundary_force_formula_nonabelian} with constrained search.
In the statement of \cref{conj:boundary_force_formula_nonabelian}, a density
$\rho_*\in F$ being ``nice'' is an unspecified condition that we hope to make
more precise in the process of finding a rigorous proof.
Intuitively, this condition should at least resemble the regularity condition
in the abelian case (see the statement of \cref{thm:boundary_force_formula}).

The reason for isolating the nice facets for boundary forces is that the bad
facets of $\Lambda$ are exactly those which cannot be reached, in loose terms,
by ``sending the external potential to infinity in the correct direction''. It
is conceivable that the pure functional might behave quite differently near bad
facets, although there are analytical arguments (see the second half of \cref{chap:conclusion})
which suggest that a boundary force may still exist.

In the rest of this chapter, we will first show a concrete derivation of the
boundary force for the bosonic Hubbard dimer in the next section, followed by a
perturbative calculation in the general setting for obtaining
Eq.~\eqref{eq:boundary_force_formula_nonabelian} in the last section.


\section{The Bosonic Hubbard Dimer}

Before showing Eq.~\eqref{eq:boundary_force_formula_nonabelian} for general
momentum map functional theories, we will work out in this section the boundary
force of the bosonic Hubbard dimer, for which a derivation can be found in
Ref.~\cite{benavides-riverosReducedDensityMatrix2020}. However, the approach
here will be easier to adapt to general functional theories. The reason
for discussing the dimer first is that it captures one of the essential
differences between abelian and nonabelian functional theories, namely that
$\mathfrak{g}^\nparallel$ is not trivial, resulting in, as we will see, an additional
optimization over the subspace spanned by the nonparallel root vectors.
Nevertheless, the dimer is still slightly too simple to be representative of all
momentum map functional theories. For one, the polytope $\Lambda$ is
one-dimensional, and has only one trivial facet and one nice facet. More
importantly, the facet functional theory has a one-dimensional Hilbert space,
which eliminates some conceptual difficulties encountered in the general
derivation presented in \cref{sec:general_momentum_perturbation}.

The bosonic Hubbard dimer is a system consisting of $N$ identical (spinless)
bosons on two lattice sites interacting via the Hubbard interaction. To be
more precise, the single-particle Hilbert space is $\mathbb{C}^2$, and the
$N$-particle Hilbert space is $\mathcal{H} = \mathrm{Sym}^N(\mathbb{C}^2)\cong
\mathbb{C}^{N+1}$, which has an orthonormal basis given by
$$
	\ket{N,0}, \ket{N-1,1}, \dotsb , \ket{0, N},
$$
with the state $\ket{m, N-m}$ having $m$ bosons on the first site and $N-m$ on the second
one. The boson-boson interaction is taken to be
$$
W = n_1^2 + n_2^2 = (b_1^\dagger b_1^{\phantom{\dagger}})^2 + (b_2^\dagger b_2^{\phantom{\dagger}})^2,
$$
where $n_i, b_i^{\phantom{\dagger}}, b_i^\dagger$ are the number operator, annihilation operator,
and creation operator of the $i$-th site respectively.

To define a momentum map functional theory, we need to specify a compact
connected Lie group $G$ and a unitary representation $\tau: G\rightarrow
\mathrm{U}(\mathcal{H})$. We take $G= \mathrm{SU}(2)$, with $\tau: \mathrm{SU}(2)
\rightarrow \mathrm{U}(\mathcal{H})$, $g\mapsto g\otimes \dotsb\otimes g$,
identifying $\mathrm{Sym}^N(\mathbb{C}^2)$ with a subspace of
$(\mathbb{C}^2)^{\otimes N}$. The potential map is then
$$
\begin{aligned}
	\tau_*: &\;i\mathfrak{su}(2) \rightarrow i\mathfrak{u}(\mathcal{H})\\
	& v \mapsto v\otimes \mathbbm{1}^{\otimes (N-1)} + \dotsb + \mathbbm{1}^{\otimes (N-1)}\otimes v.
\end{aligned}
$$
In the following, we will identify $i\mathfrak{su}(2)$ with $\mathbb{R}^3$ via
the basis consisting of the Pauli matrices $X, Y, Z$, which also gives an
identification $i\mathfrak{su}(2)^*\subset \mathbb{R}^3$, so a density $\rho
\in i\mathfrak{su}(2)^*$ will be thought of as a vector in $\mathbb{R}^3$. One
should keep in mind that the Hilbert-Schmidt inner product on
$i\mathfrak{su}(2)$ is twice the standard inner product on $\mathbb{R}^3$, and
we still use the latter to identify $(\mathbb{R}^3)^*\cong
\mathbb{R}^3$. Under this identification, the potential and density maps are
explicitly given by
\begin{equation}
	\label{eq:boson_dimer_potential_map}
	\begin{aligned}
		& \hspace{0em}\tau_*: \mathbb{R}^3 \rightarrow i\mathfrak{u}(\mathcal{H}) \\
		 &v \mapsto 
		 v_1(b_1^\dagger b_2^{\phantom{\dagger}} + b_2^\dagger b_1^{\phantom{\dagger}})
		 + v_2(-ib_1^\dagger b_2^{\phantom{\dagger}} + ib_2^\dagger b_1^{\phantom{\dagger}})
		 + v_3(b_1^\dagger b_1^{\phantom{\dagger}}- b_2^\dagger b_2^{\phantom{\dagger}})
	\end{aligned}
\end{equation}
\vspace{.3em}
\begin{equation}
	\label{eq:boson_dimer_density_map}
	\begin{aligned}
		& \hspace{0em}\tau^*: \mathcal{E} \rightarrow \mathbb{R}^3 \\
		 &\Gamma \mapsto
				\left(
				\braket{\Gamma,b_1^\dagger b_2^{\phantom{\dagger}} + b_2^\dagger b_1^{\phantom{\dagger}}},
				\braket{\Gamma,-ib_1^\dagger b_2^{\phantom{\dagger}} + ib_2^\dagger b_1^{\phantom{\dagger}}},
				\braket{\Gamma,b_1^\dagger b_1^{\phantom{\dagger}} - b_2^\dagger b_2^{\phantom{\dagger}}}
				\right).
	\end{aligned}
\end{equation}
If $N=1$, then $\mathcal{H} = \mathbb{C}^2$ and 
$b_1^\dagger b_2^{\phantom{\dagger}} + b_2^\dagger b_1^{\phantom{\dagger}} = X$, 
$-ib_1^\dagger b_2^{\phantom{\dagger}} + ib_2^\dagger b_1^{\phantom{\dagger}} = Y$, 
$b_1^\dagger b_1^{\phantom{\dagger}} - b_2^\dagger b_2^{\phantom{\dagger}} =Z$. 
Consequently, $\tau^*(\mathcal{P})\subset \mathbb{R}^3$ is
just the unit sphere (Bloch sphere). If $N>1$, consider the state
$$
\ket{\Psi} := \frac{1}{\sqrt{2}} (\ket{N,0}+ \ket{0, N}),
$$
for which $\braket{\Psi|b_1^\dagger b_2^{\phantom{\dagger}}|\Psi} = \braket{\Psi|b_2^\dagger
b_1^{\phantom{\dagger}}|\Psi} = \braket{\Psi|b_1^\dagger b_1^{\phantom{\dagger}} - b_2^\dagger b_2^{\phantom{\dagger}}|\Psi}= 0$, so
$\tau^*(\ketbrap{\Psi})= 0$. Since $\tau^*(\ketbrap{N,0}) = (0,0,N)$, we have
shown that both $0$ and $N\alpha_3$ belong to $\tau^*(\mathcal{P})$, 
where $(\alpha_i)_{i=1}^3$ is the standard basis on $\mathbb{R}^3$.
By Kirwan's theorem,
$\tau^*(\mathcal{P})\cap \mathbb{R}_{\ge 0}\alpha_3$ is a convex polytope, so 
$$
	\tau^*(\mathcal{P})\cap \mathbb{R}_{\ge 0}\alpha_3 \supset \{t\alpha_3 \mid t \in [0,N]\}.
$$
It is easy to show the inclusion in the other direction. By the
$\mathrm{SU}(2)$-equivariance of $\tau^*$, then, we have $\tau^*(\mathcal{P}) =
\{\sum_{i=1}^3 t_i\alpha_i \mid \sum_{i=1} t_i^2 \le N^2\}$.

To summarize the discussion above, the solution to the pure state
representability problem for the bosonic dimer is given by
\begin{equation}
	\dom \Fp = \tau^*(\mathcal{P})
	= \begin{cases}
		\text{sphere of radius } 1 & N=1\\
		\text{ball of radius } N & N>1.
	\end{cases}
\end{equation}
From now on, we will assume $N>1$. The boundary of $\dom \Fp$ is then the
sphere of radius $N$. 

\begin{figure}[h]
	\centering
	\includegraphics[width=0.48\textwidth]{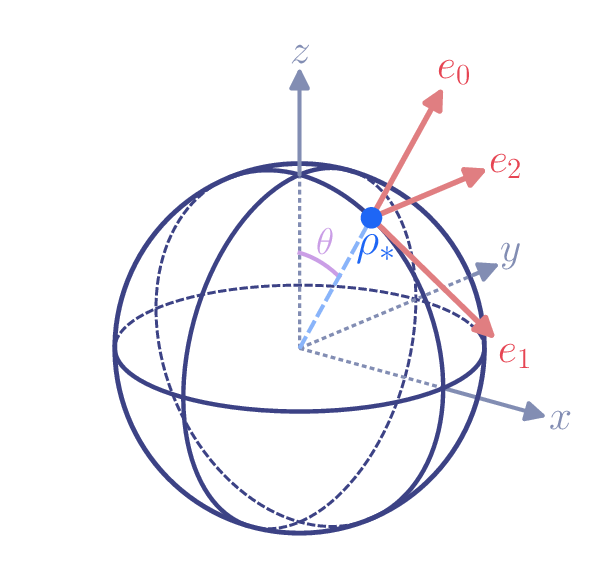}
	\caption{The chosen density $\rho_*$ (blue) on the boundary and 
	the corresponding basis vectors $e_i$ (red). The dark blue sphere represents the
	boundary of $\dom\Fp$, which is the ball of radius $N$.
	}
	\label{fig:hubbard_dimer_boundary}
\end{figure}

Fix an angle $\theta$. Consider the density
\begin{equation}
	\rho_* = (N\sin\theta, 0, N\cos\theta)
\end{equation}
and the basis vectors for $\mathbb{R}^3$ (see Fig.~\ref{fig:hubbard_dimer_boundary})
\begin{equation}
	\begin{aligned}
		 & e_0 := (\sin\theta, 0, \cos\theta)  \\
		 & e_1 := (\cos\theta, 0, -\sin\theta) \\
		 & e_2 := (0,1,0).
	\end{aligned}
\end{equation}
Let $\varepsilon_0, \varepsilon_1, \varepsilon_2 \in (\mathbb{R}^3)^*$ be the
corresponding dual basis (which are the same as the $e_i$).

\begin{figure}[H]
	\centering
	\includegraphics[width=0.48\textwidth]{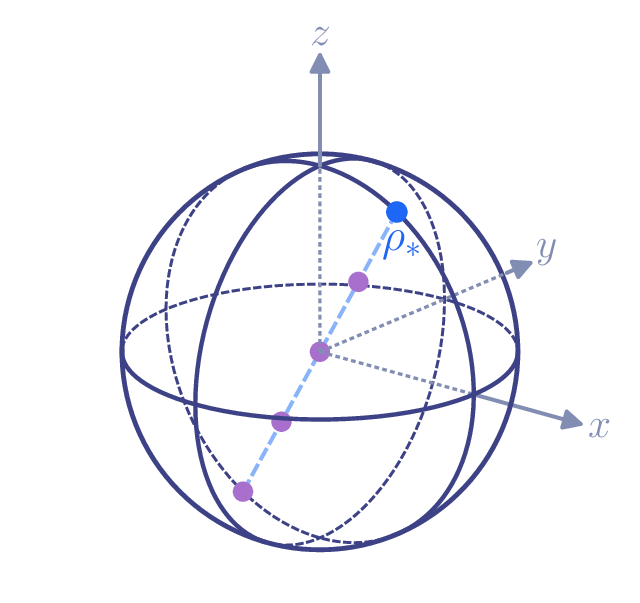}
	\caption{Images of the pure states $\ketbrap{\Phi_m}$ under $\tau^*$
	(purple: $m >0$, blue: $m = 0$).
	}
	\label{fig:hubbard_dimer_weight_diagram}
\end{figure}

For fixed $x_1, x_2\in \mathbb{R}$, consider a one-parameter family of Hamiltonians
\begin{equation}
	\label{eq: hamil_family}
	H_g := 
	 \tau_*(-e_0) + g(\tau_*(x_1e_1+x_2e_2) + W).
\end{equation}
It is easy to check that $\tau_*(-e_0) = -\sin\theta (b_1^\dagger b_2^{\phantom{\dagger}} +
	b_2^\dagger b_1^{\phantom{\dagger}}) - \cos\theta(b_1^\dagger b_1^{\phantom{\dagger}} - b_2^\dagger b_2^{\phantom{\dagger}})$ 
	has a
nondegenerate ground state given by
\begin{equation}
	\ket{\Phi_0}= \frac{1}{\sqrt{N!}}(a^\dagger)^N \ket{},
\end{equation}
where 
\begin{equation}
		  a^\dagger := b_1^\dagger\cos\frac{\theta}{2} + b_2^\dagger\sin\frac{\theta}{2}      
			\hspace{3em}
		  a_\perp^\dagger := b_1^\dagger \sin\frac{\theta}{2} - b_2^\dagger\cos\frac{\theta}{2}
\end{equation}
and $\ket{}$ is the zero-boson state in the Fock space. In general, the $m$-th excited
state of $\tau_*(-e_0)$ is
\begin{equation}
	\ket{\Phi_m} = \frac{1}{\sqrt{(N-m)!}}\frac{1}{\sqrt{m!}}(a^\dagger)^{N-m} (a^\dagger_\perp)^m\ket{},
\end{equation}
with eigenvalue $\lambda_m = \braket{\Phi_m|\tau_*(-e_0)|\Phi_m} = 2m-N$. In
the language of representation theory, the vectors $\ket{\Phi_m}$ are the
weight vectors of the representation $\tau_*$ relative to the one-dimensional
Cartan subalgebra $\mathfrak{t}$ spanned by $e_0 \in i\mathfrak{su}(2)$. The
appropriate Weyl chamber here is $i\mathfrak{t}^*_+ = \mathbb{R}_{\ge
0}\varepsilon_0 \subset i\mathfrak{t}^*$, and the Kirwan polytope $\Lambda :=
\tau^*(\mathcal{P})\cap i\mathfrak{t}^*_+$ is the line segment $[0, N]\cdot
\varepsilon_0$, which has $0$ as the only trivial facet and $\rho_*$ as the
only nice facet.
The images of $\ketbrap{\Phi_m}$ under the density map $\tau^*$ are shown in 
Fig.~\ref{fig:hubbard_dimer_weight_diagram}.

For small $|g|$, the lowest eigenvalue of $\eqref{eq: hamil_family}$ up to
second order in $g$ is
\begin{equation}\label{eq: energy_perturb}
	\lambda_0(g) \approx \lambda_0 + gE_1(x_1, x_2) + g^2E_2(x_1, x_2),
\end{equation}
with
\begin{equation}
	\label{eq: e1e2}
	\begin{aligned}
		 & E_1(x_1, x_2) = \braket{\Phi_0|\tau_*(x_1e_1+x_2e_2) + W|\Phi_0}                                                    \\
		 & E_2(x_1, x_2) = \sum_{m = 1}^N \frac{\left|\braket{\Phi_m |\tau_*(x_1e_1+x_2e_2) + W|\Phi_0}\right|^2}{\lambda_0 - \lambda_m}
		 = -\sum_{m=1}^N \frac{\left|\braket{\Phi_m|\tau_*(x_1e_1 + x_2e_2)+W|\Phi_0}\right|^2}{2m}.
	\end{aligned}
\end{equation}
In other words, the value of the energy function $E$ at $v\in \mathbb{R}^3$,
which is by definition the ground state energy of the Hamiltonian
\begin{equation}
	\begin{aligned}
		&H(v) = \tau_*(v) + W 
		= \sum_{i=1}^3\braket{\varepsilon_i, v} \tau_*(e_i) + W\\
		& = -\braket{\varepsilon_0, v} \Big[-\tau_*(e_0) + 
			\frac{1}{-\braket{\varepsilon_0, v}}\Big(\tau_*(\braket{\varepsilon_1,v}e_1 + \braket{\varepsilon_2,v}e_2)+ W\Big)\Big],
	\end{aligned}
\end{equation}
is given by, due to Eq.~\eqref{eq: energy_perturb},
\begin{equation}
	\begin{aligned}
		E(h) \approx -\braket{\varepsilon_0,v} \Bigg[ 
			\lambda_0 &+ \frac{1}{-\braket{\varepsilon_0,v}}
			\braket{\Phi_0|\tau_*\left(\braket{\varepsilon_1,v}e_1 + \braket{\varepsilon_2,v}e_2\right)+W|\Phi_0} \\
		  & + \left(\frac{1}{-\braket{\varepsilon_0,v}}\right)^2E_2(\braket{\varepsilon_1,v}, \braket{\varepsilon_2,v})
			\Bigg]. 
	\end{aligned}
\end{equation}
Rearranging, we get
\begin{equation}
	\label{eq: e_of_h}
	\begin{aligned}
		E(v) \approx & -\braket{\varepsilon_0,v} \lambda_0 +
		\braket{\Phi_0|\tau_*(\braket{\varepsilon_1,v}e_1+\braket{\varepsilon_2,v}e_2)+W|\Phi_0}
		-\frac{1}{\braket{\varepsilon_0,v}}E_2(\braket{\varepsilon_1,v}, \braket{\varepsilon_2,v})\\
		= & \braket{\Phi_0|\tau_*(v)+W|\Phi_0}
		-\frac{1}{\braket{\varepsilon_0, v}}E_2(\braket{\varepsilon_1,v}, \braket{\varepsilon_2,v}).
	\end{aligned}
\end{equation}
It is important to remember that Eq.~\eqref{eq: e_of_h} only holds when
$\braket{\varepsilon_0,v} < 0$ and $\left|\braket{\varepsilon_0,v}\right|$ is large enough. 
The derivative is
\begin{equation}
	\begin{aligned}
		\dstraight_v E  \approx  \tau^*(\ketbrap{\Phi_0})
		&+\frac{\varepsilon_0}{\braket{\varepsilon_0,v}^2}E_2(\braket{\varepsilon_1,v}, \braket{\varepsilon_2,v})\\
		& -\frac{1}{\braket{\varepsilon_0, v}}\Big(
			\varepsilon_1\partial_1 E_2(\braket{\varepsilon_1, v}, \braket{\varepsilon_2, v})
		 +\varepsilon_2\partial_2 E_2(\braket{\varepsilon_1, v}, \braket{\varepsilon_2, v})
		 \Big).
	\end{aligned}
\end{equation}
Note that the first term $\tau^*(\ketbrap{\Phi_0})$ is nothing but $\rho_*$.
Now, we are interested in the derivative of the functional at $\rho_* -
\epsilon \varepsilon_0$ with $\epsilon$ small. The plan is the following: if we
can find $v$ so that $\dstraight_v E = \rho_* - \epsilon \varepsilon_0$, then
we will have $\dstraight_{\rho_*-\epsilon \varepsilon_0}\mathcal{F} = -v$. The condition
$\dstraight_v E =\rho_* - \epsilon\varepsilon_0$ gives
\begin{equation}
	\label{eq:conditions}
	\begin{cases}
		-\epsilon = \frac{1}{\braket{\varepsilon_0,v}^2}E_2(\braket{\varepsilon_1,v}, \braket{\varepsilon_2,v}) \\
		0  = -\frac{1}{\braket{\varepsilon_0,v}}\partial_1 E_2(\braket{\varepsilon_1,v}, \braket{\varepsilon_2,v}) \\
		0  = -\frac{1}{\braket{\varepsilon_0,v}}\partial_2 E_2(\braket{\varepsilon_1, v}, \braket{\varepsilon_2,v})
	\end{cases}
	\Rightarrow
	\begin{cases}
		& \braket{\varepsilon_0,v} = -\frac{1}{\sqrt{\epsilon}}\sqrt{-E_2(\braket{\varepsilon_1,v}, \braket{\varepsilon_2,v})} \\
		 & \partial_1 E_2(\braket{\varepsilon_1, v}, \braket{\varepsilon_2,v}) = 0                                  \\
		 & \partial_2 E_2(\braket{\varepsilon_1,v}, \braket{\varepsilon_2,v}) = 0,
	\end{cases}
\end{equation}
where we ignored the higher order terms altogether. To proceed, note that
\begin{equation}
	\begin{aligned}
		  \tau_*(e_1) = a_\perp^\dagger a + a^\dagger a_\perp  
		  \hspace{4em}
		 \tau_*(e_2) = -ia_\perp^\dagger a +ia^\dagger a_\perp.
	\end{aligned}
\end{equation}
Therefore, $\braket{\Phi_m|\tau_*(e_1)|\Phi_0}$ is real and
$\braket{\Phi_m|\tau_*(e_2)|\Phi_0}$ is imaginary. Moreover, it is easy to see
that $\braket{\Phi_m|W|\Phi_0}$ is real. For Eq.~\eqref{eq:conditions} to hold,
then, it must be that $\braket{\varepsilon_2,v}=0$. We work out explicitly the
matrix elements of $\tau_*(e_1)$ and $W$:
\begin{equation}
	\braket{\Phi_m|\tau_*(e_1)|\Phi_0} = \braket{\Phi_m|a_\perp^\dagger a + a^\dagger a_\perp|\Phi_0}
	=\braket{\Phi_m|a_\perp^\dagger a|\Phi_0} = \delta_{m,1}\sqrt{N}
\end{equation}
\begin{equation}
	\begin{aligned}
		 & \braket{\Phi_1|W|\Phi_0} = (N-1)\sqrt{N}\sin\theta \cos\theta         \\
		 & \braket{\Phi_2|W|\Phi_0} = \frac{\sqrt{N(N-1)}}{\sqrt{2}}\sin^2\theta \\
		 & \braket{\Phi_{m > 2}|W|\Phi_0} = 0.
	\end{aligned}
\end{equation}
Therefore
\begin{equation}
	\label{eq:bose_dimer_second_order_energy}
	E_2(\braket{\varepsilon_1, v}, 0) = 
	- \frac{1}{2}\left\vert\braket{\varepsilon_1,v}\sqrt{N} + (N-1)\sqrt{N}\sin\theta\cos\theta\right\vert^2
	- \frac{1}{4}\left\vert\frac{\sqrt{N(N-1)}}{\sqrt{2}}\sin^2\theta\right\vert^2.
\end{equation}
According to Eq.~\eqref{eq:conditions}, the value of $\braket{\varepsilon_1,
v}$ is fixed by requiring that $E_2$ be stationary. Clearly, then, we should
choose $\braket{\varepsilon_1, v}$ so that the first term of
Eq.~\eqref{eq:bose_dimer_second_order_energy} vanishes, so the
stationary value of $E_2$ is $-\frac{1}{8}N(N-1)\sin^4\theta$. The
derivative of the functional is then
\begin{equation}
	\begin{aligned}
		& \dstraight_{\rho_*-\epsilon \varepsilon_0} \mathcal{F} = -v = -\braket{\varepsilon_0,v} e_0 -\braket{\varepsilon_1, v} e_1  \\
		& \approx\frac{e_0}{\sqrt{\epsilon}}\sqrt{\frac{1}{8}N(N-1)\sin^4\theta}
		+ e_1(N-1)\sin\theta\cos\theta                                             \\
		 & =\frac{e_0}{\sqrt{\epsilon}}\frac{\sqrt{N(N-1)}}{2\sqrt{2}}\sin^2\theta
		+ e_1\frac{N-1}{2}\sin 2\theta.
	\end{aligned}
\end{equation}
Finally, we obtain the directional derivative along $-\varepsilon_0$:
\begin{equation}
	\doverd{\epsilon}\mathcal{F}(\rho_* - \epsilon\varepsilon_0)
	=\braket{-\varepsilon_0,\dstraight_{\rho_*-\epsilon\varepsilon_0}\mathcal{F}}
	\approx-\frac{1}{\sqrt{\epsilon}}\frac{\sqrt{N(N-1)}}{2\sqrt{2}}\sin^2\theta.
\end{equation}
This agrees with Eq.~(12) of Ref.~\cite{benavides-riverosReducedDensityMatrix2020}.


\section{Perturbation Theory in General Momentum Map Functional Theories}
\label{sec:general_momentum_perturbation}

We will now derive Eq.~\eqref{eq:boundary_force_formula_nonabelian} by
perturbation theory. After making all the necessary niceness assumptions about
the functional theory as we did in
\cref{sec:abelian_boundary_force_perturbation} (see the actual assumptions
therein), we will not distinguish between $\Fp$ and $\FHK$, and will use $\mathcal{F}$ to denote both
in the following.

Let $(G, \mathcal{H}, \tau, W)$ be a momentum map functional theory with a
chosen Cartan subalgebra $\mathfrak{t}$ and Weyl chamber $i\mathfrak{t}_+^*$
such that the convex polytope $\Lambda = \tau^*(\mathcal{P})\cap
i\mathfrak{t}_+^*$ has full dimension in $i\mathfrak{t}^*$. Let $\Omega\subset
i\mathfrak{t}^*$ denote the set of weights of the representation $\tau_*:
\mathfrak{g}\rightarrow \mathfrak{u}(\mathcal{H})$. Fix a nice facet $F\subset
\conv(\Omega)$, with associated inequality $\braket{\cdot, S}\ge c$, where
$S\in i\mathfrak{t}$ and $c\in \mathbb{R}$, along with a density $\rho_*\in
\relint(F)$ and an inward vector $\eta \in i\mathfrak{t}^*$. We can scale $S$
and $c$ by a positive number so that $\braket{\eta, S} = 1$.

Recall from \cref{sec:nonab_ft_intro} that the facet $F$ induces a decomposition of vector spaces
$$
\mathfrak{g} = \mathfrak{t} \oplus \mathfrak{g}^\parallel \oplus \mathfrak{g}^\nparallel,
$$
where $\mathfrak{t}\oplus\mathfrak{g}^\parallel = C_{\mathfrak{g}}(S) =
\mathfrak{g}_F$ is the space of external potentials (up to $i$) of the facet
functional theory. Define $U := \ker(\eta)\oplus i\mathfrak{g}^\parallel$, where
$\ker(\eta):= \{w\in i\mathfrak{t} \mid \braket{\eta, w} = 0\}\subset
i\mathfrak{t}$. We then have the decomposition
$$
i\mathfrak{g} = \underbrace{\mathbb{R} S \oplus U}_{i\mathfrak{g}_F} \oplus  i\mathfrak{g}^\nparallel.
$$
With respect to this decomposition, define projections $\pi_{S}, \pi_U,
\pi_\nparallel: i\mathfrak{g}\rightarrow i\mathfrak{g}$. That is, 
$\pi_{S}$ is identity on $\mathbb{R} S$ and zero on
$U\oplus\mathfrak{g}^\nparallel$ etc. 
These satisfy $\pi_S + \pi_U +
\pi_\nparallel = \mathrm{Id}_{i\mathfrak{g}}$. Furthermore, it is easy to check
that $\pi_S(\cdot) = \braket{\eta, \cdot}S$. We will also denote the projection
onto the space $i\mathfrak{g}_F$ of external potentials of the facet functional
theory by $\pi_F := \pi_S + \pi_U$.

Now consider the family of Hamiltonians
\begin{equation}
  \label{eq:hamiltonian_decomposed}
  H(v) = \tau_*(v) + W = \braket{\eta, v}\tau_*(v) + \tau_*(\pi_U(v)) 
  + \tau_*(\pi_\nparallel(v)) + W.
\end{equation}
We will compute the ground state energy $E(v)$ in the limit
$\braket{\eta,v}\rightarrow \infty$ using perturbation theory. Let us write
Eq.~\eqref{eq:hamiltonian_decomposed} in a more suggestive form:
\begin{equation}
  H(v) = \braket{\eta, v} \left[
    \tau_*(S) + \frac{1}{\braket{\eta, v}} \left(\tau_*(\pi_U(v)) + \tau_*(\pi_\nparallel(v))  + W\right)
    \right]
\end{equation}
We treat everything inside of $\frac{1}{\braket{\eta, v}}(\dotsb )$ as a small
perturbation, and $\frac{1}{\braket{\eta, v}}$ will be the expansion parameter.

Since $F$ is a nice facet, all eigenvalues of $\tau_*(S) - c\mathbbm{1}$ are
nonnegative, and the zero eigenspace is just $\mathcal{H}_F$, the Hilbert space
of the facet functional theory, which is spanned by the weight vectors whose
weights are on $F$. Therefore, at zeroth order, any state in $\mathcal{H}_F$ is
a ground state of $H(v)$, with energy $c\braket{\eta, v}$.

The energy up to second order in $\braket{\eta, v}^{-1}$ is then
\begin{equation}
  E(v) \approx \braket{\eta, v} \left[
    c + \frac{1}{\braket{\eta, v}} E_1(v) + \frac{1}{\braket{\eta, v}^2} E_2(v) 
    \right],
\end{equation}
where $E_1$ and $E_2$ are the first and second order energy
corrections respectively. Although we write $E_1(v),
E_2(v)$, each of these corrections actually does not
depend on all of $v$, but only on certain components of $v$ selected by some of
the projections. For example, the first order energy correction is
\begin{equation}
  \label{eq:first_order_energy}
  E_1(v) = \braket{\Phi(v)|\tau_*(\pi_U(v) + \pi_\nparallel(v)) + W|\Phi(v)},
\end{equation}
where the term coming from $\pi_\nparallel(v)$ vanishes because
$\braket{\Phi(v)|\tau_*(\pi_\nparallel(v))|\Phi(v)} = 0$. In
Eq.~\eqref{eq:first_order_energy}, $\ket{\Phi(v)}$ is the ground state in
$\mathcal{H}_F$ of the Hamiltonian 
$$
  \Pi_{F}^{\phantom{\dagger}}\big(\braket{\eta, v}\tau_*(S) + \tau_*(\pi_U(v)) +  W\big)\Pi_F^\dagger , 
$$
which is nothing but the Hamiltonian $H_F(v)$ of the facet functional theory
(see \cref{def:facet_functional_theory_nonabelian}). This implies that the
present derivation of the boundary force only works near those points on $F$
that are pure state $v$-representable in the facet functional theory.

For each $\omega \in \Omega$, let $\Pi_\omega: \mathcal{H}\rightarrow
\mathcal{H}_\omega$ be the orthogonal projection onto the weight space
$\mathcal{H}_\omega$. The second order correction is then
\begin{equation}
  E_2(v) = -\sum_{\omega \in \Omega\setminus \Omega_F}
  \frac{\lVert\Pi_\omega(\tau_*(\pi_\nparallel(v)) + W))\ket{\Phi(v)}\rVert^2}
  {\braket{\omega, S}-c},
\end{equation}
where we used $\Pi_\omega \tau_*(\pi_U(v))\ket{\Phi(v)}=0$.

Note that $E_2(v)$ depends on $v$ in two ways: through
$\pi_\nparallel(v)$ explicitly, and through $\ket{\Phi(v)}$, 
which depends only on $\pi_U(v)$.

We proceed by employing the same method as in \cref{sec:abelian_boundary_force_perturbation} to compute
$\dstraight_{\rho_*+\epsilon \eta}\mathcal{F}$: we first determine
$\dstraight_v E$, and then solve $\dstraight_v E = \rho_* + \epsilon \eta$ for
$v$. Collecting everything, we get
\begin{equation}
  \begin{aligned}
    &E(v) \approx \braket{\eta, v}
    \Bigg[c \;+ 
      \frac{1}{\braket{\eta, v}}\braket{\Phi(v)|\tau_*(\pi_U(v)) + W|\Phi(v)}\\
      &\hspace{5em}- \frac{1}{\braket{\eta, v}^2}\sum_{\omega\in \Omega\setminus \Omega_F}
        \frac{\lVert \Pi_\omega(\tau_*(\pi_\nparallel(v))+W)\ket{\Phi(v)}\rVert^2}{\braket{\omega, S}-c}
    \Bigg]\\
    &=E_F(\underbrace{\pi_S(v) + \pi_U(v)}_{= \pi_F(v)\in i\mathfrak{g}_F}) - \frac{1}{\braket{\eta, v}}
      \sum_{\omega\in \Omega\setminus \Omega_F}
      \frac{\lVert \Pi_\omega(\tau_*(\pi_\nparallel(v))+W)\ket{\Phi(v)}\rVert^2}{\braket{\omega, S}-c},
  \end{aligned}
\end{equation}
where $E_F: i\mathfrak{g}_F \rightarrow \mathbb{R}$ is the ground state energy
function of the facet functional theory. Hence
\begin{equation}
  \label{eq:energy_derivative}
  \begin{aligned}
    &\dstraight_vE \approx  \dstraight_{\pi_F(v)} E_F  + 
    \frac{\eta}{\braket{\eta, v}^2}
      \sum_{\omega\in\Omega\setminus\Omega_F}
      \frac{\lVert \Pi_\omega(\tau_*(\pi_\nparallel(v))+W)\ket{\Phi(v)}\rVert^2}{\braket{\omega, S}-c}\\
      &-\frac{1}{\braket{\eta, v}}\dstraight_v
      \Bigg\{\sum_{\omega\in \Omega\setminus \Omega_F}
        \frac{\lVert \Pi_\omega(\tau_*(\pi_\nparallel(v))+W)\ket{\Phi(v)}\rVert^2}{\braket{\omega, S}-c}
      \Bigg\}.
  \end{aligned}
\end{equation}
Note that $\dstraight_{\pi_F(v)}E_F$ is the ground state density corresponding
to $\pi_F(v)\in i\mathfrak{g}_F = iC_{\mathfrak{g}}(S)$ in the facet functional
theory. We will assume $\dstraight_{\pi_F(v)}E_F\in F$. Fix a density
$\rho_*\in F$, which we assume to be pure state $v$-representable. We now want
to find a path $v(\epsilon) =:v_\epsilon$ such that $\dstraight_{v_\epsilon}E =
\rho_* + \epsilon \eta$. Using Eq.~\eqref{eq:energy_derivative}, this condition
reads
\begin{equation}
  \label{eq:energy_derivative_2}
  \begin{aligned}
    &\rho_*
      + \epsilon\eta \\
    & \approx
    \dstraight_{\pi_F(v)}E_F
       +
    \frac{\eta}{\braket{\eta, v_{\epsilon}}^2}
      \sum_{\omega\in \Omega\setminus \Omega_F}
      \frac{\lVert \Pi_\omega(\tau_*(\pi_\nparallel(v_\epsilon))+W)\ket{\Phi(v_\epsilon)}\rVert^2}{\braket{\omega, S}-c}\\
      &\hspace{8em}-\frac{1}{\braket{\eta, v_\epsilon}}\dstraight_{v_\epsilon}\Bigg\{\sum_{\omega\in\Omega\setminus \Omega_F}
        \frac{\lVert \Pi_\omega(\tau_*(\pi_\nparallel(v_\epsilon))+W)\ket{\Phi(v_\epsilon)}\rVert^2}{\braket{\omega, S}-c}
      \Bigg\}.
  \end{aligned}
\end{equation}
Evaluate both sides on $S$ to get
$$
\epsilon \approx  \frac{1}{\braket{\eta, v_\epsilon}^2}
    \sum_{\omega\in \Omega\setminus\Omega_F}
      \frac{\lVert \Pi_\omega(\tau_*(\pi_\nparallel(v_\epsilon))+W)\ket{\Phi(v_\epsilon)}\rVert^2}{\braket{\omega, S}-c},
$$
where we have used $\braket{\rho_*, S} = \braket{\dstraight_{\pi_F(v)}E_F, S}$,
which holds because both $\rho_*$ and $\dstraight_{\pi_F(v)}E_F$ are densities
on the facet $F$. The derivative of the functional is then 
\begin{equation}
  \label{eq:dF_nonabelian_intermediate}
\doverd{\epsilon}\mathcal{F}(\rho_* + \epsilon\eta)
= \braket{\eta, -v_\epsilon}\approx 
-\frac{1}{\sqrt{\epsilon}} \left[
    \sum_{\omega\in\Omega\setminus\Omega_F}
      \frac{\lVert \Pi_\omega(\tau_*(\pi_\nparallel(v_\epsilon))+W)\ket{\Phi(v_\epsilon)}\rVert^2}{\braket{\omega, S}-c}
  \right]^{\frac{1}{2}}.
\end{equation}
We may replace $\ket{\Phi(v_\epsilon)}$ with $\ket{\Phi} = \ket{\Phi(v_0)}$, which results
in a higher-order error. To determine $\pi_\nparallel(v_\epsilon)$, evaluate
Eq.~\eqref{eq:energy_derivative_2} on any $w \in i\mathfrak{g}^\nparallel$
to get
\begin{equation}
0 = \left\langle 
  \dstraight_{v_\epsilon}
  \Bigg\{\sum_{\omega\in\Omega\setminus\Omega_F}
        \frac{\lVert \Pi_\omega(\tau_*(\pi_\nparallel(v_\epsilon))+W)\ket{\Phi}\rVert^2}{\braket{\omega, S}-c}
      \Bigg\}
  , w\right\rangle,
\end{equation}
which means that $\pi_\nparallel(v_\epsilon)$ is a stationary point. Upon
closer examination, one finds that $\pi_\nparallel(v_\epsilon)$ minimizes the
square root in Eq.~\eqref{eq:dF_nonabelian_intermediate}. It follows that
\begin{equation}
  \label{eq:boundary_force_formula_nonabelian_2}
\doverd{\epsilon}\mathcal{F}(\rho_* + \epsilon\eta)
\approx  
-\frac{1}{\sqrt{\epsilon}} \min_{v\in i\mathfrak{g}^\nparallel}\left[
    \sum_{\omega\in \Omega\setminus\Omega_F}
      \frac{\lVert \Pi_\omega(\tau_*(v)+W)\ket{\Phi}\rVert^2}{\braket{\omega, S}-c}
  \right]^{\frac{1}{2}},
\end{equation}
which is the claimed boundary force formula Eq.~\eqref{eq:boundary_force_formula_nonabelian}.

\begin{ex}
  We will apply Eq.~\eqref{eq:boundary_force_formula_nonabelian_2} to the
  functional theory defined by 
  $$
    (G=\mathrm{SU}(2)\times \mathrm{SU}(2),
  \tau=\tau^{(2)}\otimes\tau^{(3)}, \mathbb{C}^2\otimes \mathbb{C}^3, W),
  $$
  where $\tau^{(2)}$ and $\tau^{(3)}$ are the two- and three-dimensional irreducible
  representations of $\mathrm{SU}(2)$ respectively.

  In \cref{ex:2t3_moment_map_computation}, we worked out the set of pure state
  representable densities $\iota^*(\mathcal{P})$, whose intersection with
  $i\mathfrak{t}^*_+$ is shown in \cref{fig:nice_facets_a}. For the
  interaction $W$, we will take
  $$
  \begin{aligned}
    & \braket{1,2|W| {-}1,2} = \braket{{-}1,2|W|1,2} =  u_1\\
    & \braket{1,0|W| {-}1,0} = \braket{{-}1,0|W|1,0} =  u_2\\
    & \braket{1,{-}2|W| {-}1,{-}2} = \braket{{-}1,{-}2|W|1,{-2}} =  u_3\\
    & \braket{1,2|W| 1,2} =  k_1\\
    & \braket{1,{-}2|W| 1,{-}2} =  k_3\\
    & \braket{1,2|W|1,0} = \braket{1,0|W|1,2} = 1\\
    & \braket{1,{-}2|W|1,0} = \braket{1,0|W|1,{-}2} = 1,
  \end{aligned}
  $$
  where $u_1,u_2, u_3, k_1, k_3$ are real numbers, and $\braket{i,j|W|k,l}=0$ for all
  matrix elements not listed above (see \cref{fig:2t3_interaction}).
  
  Take $\rho_*=(Z\mapsto 1, Z\mapsto1) \in F$, where $F\subset \Lambda$ is the
  unique nice facet. The inward-pointing vector $\eta\in i\mathfrak{t}^*$ and
  the normal vector $S\in i\mathfrak{t}$ are chosen as
  $$
    \eta = (Z\mapsto -1, 0) \;\;\;\; S = -Z_1,
  $$
  which satisfy $\braket{\eta, S}=1$.

  First, we try to understand the preimage of the density $\rho_*$ under the
  density map $\tau^*$. If $\tau^*(\ketbrap{\Phi}) = \rho_*$, then we must have
  $$
  \ket{\Phi} = a\ket{1,2} + b\ket{1,0} + c\ket{1,-2},
  $$
  where $b$ is real and nonnegative without loss of generality. The condition
  that $\tau^*(\ketbrap{\Phi})\in i\mathfrak{t}^*$ is equivalent to
  $$
  \begin{aligned}
    &\braket{\Phi|\tau_*(X_1)|\Phi} = \braket{\Phi|\tau_*(Y_1)|\Phi} = 0\\
    &\braket{\Phi|\tau_*(X_2)|\Phi} = \braket{\Phi|\tau_*(Y_2)|\Phi} = 0.
  \end{aligned}
  $$
  The first condition is trivially satisfied, while the second one is
  equivalent to $\bar a b + b c=0$. Hence, $b=0$ or $a = -\bar c$. If $a =
  -\bar c$, then $\tau^*(\ketbrap{\Phi})$ would be $(Z\mapsto 1, 0)$, which is not
  equal to $\rho_*$. Thus, we have $b=0$. It follows that
  $$
  \ket{\Phi} = \frac{\sqrt{3}}{2}\ket{1,2} + \frac{1}{2}e^{i\phi} \ket{1,-2},
  $$
  where $e^{i\phi}$ is any phase. Clearly, $\ket{\Phi}$ is a minimizer of 
  the constrained search for any $\phi$.

  Eq.~\eqref{eq:boundary_force_formula_nonabelian_2} now reads
  $$
  \begin{aligned}
    &\doverd{\epsilon}\mathcal{F}(\rho_*+\epsilon\eta)\approx\\
    &\frac{-1}{\sqrt{\epsilon}}
    \min_{v\in i\mathfrak{g}^\nparallel}
      \left[
        \frac{\left|\braket{-1,2|(\tau_*(v)+W)|\Phi}\right|^2}{2}
        +\frac{\left|\braket{-1,0|(\tau_*(v)+W)|\Phi}\right|^2}{2}
        +\frac{\left|\braket{-1,-2|(\tau_*(v)+W)|\Phi}\right|^2}{2}
  \right]^{\frac{1}{2}}.
  \end{aligned}
  $$
  Now, $i\mathfrak{g}^\nparallel = \mathrm{span}\{X_1, Y_1\}$ (see
  \cref{ex:2t3_facet_dft}), so
  \begin{equation}
    \label{eq:2t3_boundary_force}
  \begin{aligned}
    \doverd{\epsilon}\mathcal{F}(\rho_*+\epsilon\eta)&\approx
    -\frac{1}{\sqrt{\epsilon}}
    \min_{v\in i\mathfrak{g}^\nparallel}
      \left[
        \frac{3\left|\braket{-1,2|(\tau_*(v)+W)|1,2}\right|^2}{8}
        +\frac{\left|\braket{-1,-2|(\tau_*(v)+W)|1,-2}\right|^2}{8}
  \right]^{\frac{1}{2}}\\
      &=-\frac{1}{2\sqrt{2}\sqrt{\epsilon}}\min_{x,y\in\mathbb{R}}
    \sqrt{3\left|u_1 + x +iy\right|^2
      + \left|u_3 + x+iy\right|^2
      } \\
      &= -\frac{1}{2\sqrt{2}\sqrt{\epsilon}} \min_{x\in \mathbb{R}}
      \sqrt{3(u_1+x)^2 + (u_3+x)^2}\\
      &= -\frac{1}{2\sqrt{2}\sqrt{\epsilon}}\sqrt{\frac{3}{4}(u_1-u_3)^2}
      = -\frac{1}{\sqrt{\epsilon}}\cdot \frac{\sqrt{6} }{8} \left|u_1-u_3\right|.
  \end{aligned}
  \end{equation}
  Finally, we integrate Eq.~\eqref{eq:2t3_boundary_force} to get an
  approximation to the functional itself:
  \begin{equation}
    \label{eq:2t3_boundary_force_integrated}
    \begin{aligned}
      &\mathcal{F}(\rho_* +\epsilon\eta) \approx
          \mathcal{F}(\rho_*) - \int_0^\epsilon 
        \frac{1}{\sqrt{\epsilon'}}\cdot  \frac{\sqrt{6} }{8} \left|u_1-u_3\right| \dstraight \epsilon'
        = \mathcal{F}(\rho_*) - \sqrt{\epsilon} \frac{\sqrt{6} }{4} \left|u_1-u_3\right|,
      \end{aligned}
  \end{equation}
  where
  $$
    \mathcal{F}(\rho_*) = \braket{\Phi|W|\Phi} = \frac{3}{4}k_1 + \frac{1}{4} k_3.
  $$
  In \cref{fig:2t3_boundary_force_comparison}, we compare
  Eq.~\eqref{eq:2t3_boundary_force_integrated} with numerically obtained values
  of the pure functional $\mathcal{F}_p$. The latter is achieved by executing the pure state
  constrained search in the following way: $2\times 10^6$ states in the Hilbert
  space are sampled in a specific way so that their images under the density
  map all lie in $i\mathfrak{t}^*$ and are close to $\rho_*$. For each
  $\epsilon$, the expectation value $\braket{\Psi|W|\Psi}$ is minimized over
  all states $\ket{\Psi}$ mapping sufficiently close to $\rho_* + \epsilon
  \eta$.

  \begin{figure}[htb]
    \centering
    \subfloat[Schematic illustration of the matrix elements of the interaction $W$. 
    Two weights (black filled circles) are connected by a red line if the matrix element of $W$ between 
    the respective weight vectors is not zero.
    \label{fig:2t3_interaction}]{
      \includegraphics[width=.24\textwidth,valign=c]{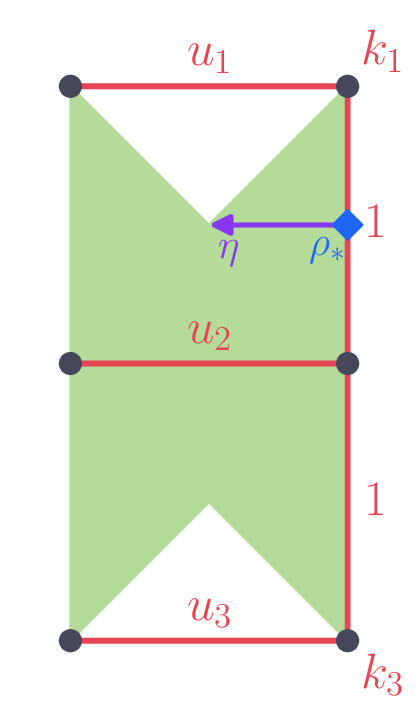}
      \vphantom{
        \includegraphics[width=.65\textwidth,valign=c]{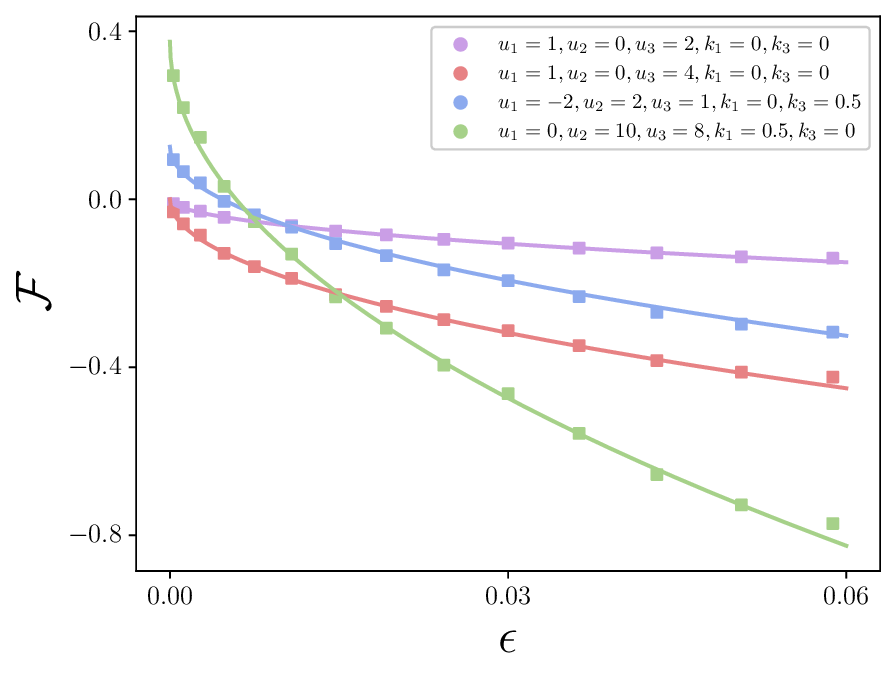}
      }
    }
    \hspace{1.5em}
    \subfloat[Comparison of numerical values of $\mathcal{F}_p$ (square dots)
    with
    Eq.~\eqref{eq:2t3_boundary_force_integrated} (solid lines). 
    \label{fig:2t3_boundary_force_comparison}]{
      \includegraphics[width=.65\textwidth,valign=c]{figs/chap6/twotensorthree_boundaryforce.eps}
    }
    \caption{}
    \label{fig:2t3_boundary_force}
  \end{figure}
\end{ex}

\chapter{Conclusion and Outlook}
\label{chap:conclusion}

In this thesis, inspired by recent observations in the RDMFT literature (namely
Refs.~\cite{schillingDivergingExchangeForce2019,benavides-riverosReducedDensityMatrix2020,maciazekRepulsivelyDivergingGradient2021,liebertFunctionalTheoryBoseEinstein2021})
regarding the behavior of the universal functional near the boundary of its
domain, we carried out a systematic investigation of the interplay between the
functional and its domain's geometry in a way agnostic to physical details of
individual systems, facilitated by the abstract notion of a generalized
functional theory defined in \cref{chap:generalized_ft}. We model a functional
theory by the following data: a Hilbert space $\mathcal{H}$, a fixed
interaction operator $W$, and a linear parametrization map $\iota:V\rightarrow
i\mathfrak{u}(\mathcal{H})$ called the \textit{potential map}, the dual of
which is the \textit{density map}. In \cref{chap:generalized_ft}, we showed
that these data alone already result in rich mathematical structure, in
particular allowing us to establish the weak Hohenberg-Kohn theorem
(\cref{thm:weak_HK}) and to define the universal functionals (see
\cref{sec:universal_functionals}).

Following the introduction of bosonic momentum space functional theories in
\cref{chap:momentum_rdmft}, where a concrete functional form and a boundary
force formula were derived informally, we identified in
\cref{chap:abelian_functional_theory} a special class of generalized functional
theories, called \textit{abelian functional theories}, for which many results
from \cref{chap:momentum_rdmft} could be generalized. An abelian functional
theory is one whose external potentials all commute, permitting a
straightforward characterization of the set of representable densities in terms
of \textit{weights}, which are given by the eigenvalues of the external
potentials (see \cref{defn:abelian_weights}). In particular, we showed
(\cref{thm:abelian_nrep_set}) that the domains of the pure and ensemble
functionals both coincide with the convex hull of all weights, which is a convex polytope.
Our most important finding in \cref{chap:abelian_functional_theory}
is \cref{thm:boundary_force_formula}, which gives an explicit formula
(Eq.~\eqref{eq:boundary_force_formula}) for calculating the prefactor of the
diverging boundary force, strongly constraining the behavior of the functional
near a facet of the polytope. Unlike previous works on this topic, we are the
first to write down such a general formula and, moreover, to provide a rigorous
proof based on the Levy-Lieb constrained search, which we elaborated in
\cref{sec:boundary_force_rigorous_proof}.

Finally, we generalized some of our results on abelian functional theories from
\cref{chap:abelian_functional_theory} to the nonabelian setting. For the
representability problem to be tractable, we restricted ourselves to a class of
generalized functional theories called \textit{momentum map functional
theories} (\cref{defn:momentum_map_ft}), which still include many functional
theories of physical interest such as DFT, RDMFT, and spin chains with local
observables as external potentials. Roughly speaking, a momentum map functional
theory is one for which the space of external potentials is a Lie algebra and
the density map is a momentum map, a mathematical object whose properties were
discussed in \cref{chap:interlude}. Because of Kirwan's theorem
(\cref{thm:kirwan}), we know that the domain of the pure functional intersects
any Weyl chamber of any Cartan subalgebra in a convex polytope, for which
\cref{thm:characterization_of_kirwan_polytope} gives a complete
characterization. We applied these mathematical statements in
\cref{chap:momentum_map_functional_theories}, in particular to formulate
\cref{conj:boundary_force_formula_nonabelian}, which is the principal result of
the last part of the thesis. \cref{conj:boundary_force_formula_nonabelian}
claims that for a \textit{nice facet} of the Kirwan polytope
(\cref{defn:nice_facet}) the counterpart of
Eq.~\eqref{eq:boundary_force_formula}, the abelian boundary force formula, is
given by Eq.~\eqref{eq:boundary_force_formula_nonabelian}, which involves
an additional optimization over the span of the root spaces 
whose roots are not parallel to the facet. Although we do not have a rigorous
proof at the moment, a derivation of Eq.~\eqref{eq:boundary_force_formula_nonabelian} based on
perturbation theory was given in \cref{sec:general_momentum_perturbation}.

\subsubsection*{Towards a Proof of \cref{conj:boundary_force_formula_nonabelian} and Its Generalization Beyond Nice Facets}

Looking ahead, we outline a few ideas that could potentially lead to a rigorous
proof of \cref{conj:boundary_force_formula_nonabelian} by constrained search,
which may work equally well also for nontrivial facets that are not nice.

A natural strategy would be to mimic what we did for abelian functional
theories in \cref{sec:boundary_force_rigorous_proof}. Recall the setting: we
choose a density $\rho_*$ on a facet $F$ of the convex polytope
$\conv(\Omega)$, where $\Omega$ is the set of weights, and an inward-pointing
vector $\eta \in \overrightarrow{\aff(\Omega)}$. The goal is to understand how
$\Fp(\rho_* + \epsilon \eta)$ depends on $\epsilon$ for small $\epsilon$. A key
ingredient was a simple description of the fiber of the density map over
$\rho_*+\epsilon\eta$ in terms of the fiber over $\rho_*$, which was achieved
by \cref{thm:classical_density_fiber_splitting}. Roughly speaking, we were able
to show that, up to phases, the fiber over $\rho_*+\epsilon\eta$ is almost the
product of the fiber over $\rho_*$ and a certain convex polytope, with the
latter parametrizing the different ways a state can deviate in order to change
the density by $\epsilon\eta$.

For nonabelian momentum map functional theories, we then have to solve the
following problem about momentum maps: 

\begin{problem}
Let $G$ be a compact connected Lie
group acting unitarily on a complex inner product space $\mathcal{H}$, and let
$\Phi_G: \mathbb{P}(\mathcal{H})\rightarrow\mathfrak{g}^*$ be the natural
momentum map (see \cref{thm:group_rep_momentum_map}). 
Pick a Cartan subalgebra $\mathfrak{t}$ and Weyl chamber $\mathfrak{t}^*_{+}$,
and let $\Lambda := \Phi_G(\mathbb{P}(\mathcal{H}))\cap \mathfrak{t}^*_+$ be
the Kirwan polytope, which we assume to have full dimension in
  $\mathfrak{t}^*$. Pick a nontrivial facet $F\subset \Lambda$, a point $\rho_*
  \in \relint(F)$, and an inward-pointing vector $\eta \in
  i\mathfrak{t}^*$. For small $\epsilon$, describe the
  fiber $\Phi_G^{-1}(\rho_* + \epsilon \eta)$, preferably in terms of
  $\Phi_G^{-1}(\rho_*)$.
\end{problem}

We already know something useful about $\Phi_G^{-1}(\rho_*)$: by the Selection
Rule (\cref{thm:selection_rule}), $\Phi_G^{-1}(\rho_*)$ must be contained in
$\mathbb{P}(\mathcal{H}_F)$. Ideally, we would like to say that the fiber
$\Phi_G^{-1}(\rho_*+\epsilon \eta)$ is not ``too different'' from
$\Phi_G^{-1}(\rho_*)$, in the sense that every element in the former lies
within distance $\delta(\epsilon)$ of some element in the latter with respect
to, say, the trace norm. A corollary of Lemma~2.14 of
Ref.~\cite{walterMultipartiteQuantumStates2014}, which is essentially a
consequence of the compactness of $\mathbb{P}(\mathcal{H})$ and the Selection
Rule, provides a partial confirmation:

\begin{lem}
There exists a function $\delta(\epsilon)$ such that 
  $$
  \begin{aligned}
    &\forall \epsilon: \forall [\psi] \in  \Phi_G^{-1}(\rho_* + \epsilon \eta):\\
    &\min_{[\phi]\in \mathbb{P}(\mathcal{H}_F)}
    \Tr(|\proj{\phi} - \proj{\psi}|)\le \delta(\epsilon)
  \end{aligned}
  $$
  and $\lim_{\epsilon\rightarrow 0} \delta(\epsilon) = 0$.
\end{lem}
(Recall that $\mathbf{p}_\psi:= \psi\braket{\psi, \cdot}$ denotes the projector
onto $\mathbb{C}\psi$.)

However, this is too weak for the boundary force for two reasons. First, we do
not have control over the minimizer $[\phi]\in \mathbb{P}(\mathcal{H}_F)$.
Second, if we want to conclude $\Fp(\rho_*+\epsilon\eta)\approx \Fp(\rho_*)-
\mathcal{G}\sqrt{\epsilon}$, which we believe to hold
for nice facets for good reasons (with $-\mathcal{G}$ given by
Eq.~\eqref{eq:boundary_force_formula_nonabelian}), we expect
$\delta(\epsilon)\sim \sqrt{\epsilon}$, which is clearly true if $\mathfrak{g}$
is abelian. Again, partial evidence for the latter claim can be found in
Ref.~\cite{schillingReconstructingQuantumStates2017}, which deals with the case
of fermionic RDMFT, for which $\mathcal{H} =\exterior^N\mathcal{H}_1$, $G=
\mathrm{U}(\mathcal{H}_1)$, and $\Phi_G$ is $N\mathrm{Tr}_{N-1}$ up to a factor of $i$.
Fix an ordered basis for $\mathcal{H}_1$, which induces a choice of Cartan
subalgebra $\mathfrak{t}\subset \mathfrak{u}(\mathcal{H})$ and a Weyl chamber
$\mathfrak{t}^*_{+}\subset \mathfrak{t}^*$.

\begin{thm}
  Let $\braket{\cdot, S}\ge c$ be an inequality corresponding to a nontrivial
  facet $F$ of the convex polytope $N\mathrm{Tr}_{N-1}(\mathcal{P}) \cap
  i\mathfrak{t}^*_+$.
  There exists some constant $C>0$ such that
  for $\epsilon>0$ small enough and for any $N$-particle state $\ket{\Psi}$ 
  for which $\rho:= N\mathrm{Tr}_{N-1}(\ketbrap{\Psi})\in i\mathfrak{t}^*_{>0}$,
  $\braket{\rho, S} = \epsilon$, and $\rho$ is not too close to $i\mathfrak{t}^*_+ \setminus i\mathfrak{t}^*_{>0}$, 
  there exists an $N$-particle state
  $\ket{\Phi}$ such that the 1RDM $N\mathrm{Tr}_{N-1}(\ketbrap{\Phi})$, after
  conjugating by a suitable single-particle unitary, lies on $F$, and
  $$
  \lVert \Phi-\Psi \rVert \le C\sqrt{\epsilon}.
  $$
\end{thm}

However, with this theorem we do not have control over whether
$N\mathrm{Tr}_{N-1}(\ketbrap{\Phi})$ actually lies on the facet $F$, which is what we
need. In the quantum chemistry language, one says that the states $\ket{\Phi}$
and $\ket{\Psi}$ may not share the same natural orbitals.

\begin{appendices}
    \chapter{K\"ahler Structure of $\mathbb{P}(\mathcal{H})$}
\label{app:projective_kaehler}

This appendix will be dedicated to giving $\mathbb{P}(\mathcal{H})$ the
structure of a K\"ahler manifold, in particular making
$\mathbb{P}(\mathcal{H})$ into a symplectic manifold. 

%



Recall that a K\"ahler manifold is a Riemannian, symplectic and complex
manifold such that the three structures are compatible with each other.
More precisely, a complex manifold $M$ is K\"ahler if it is endowed
with a Riemannian metric $h$ and a symplectic form $\omega$
such that $h(\cdot, \cdot) = \omega(\cdot, J(\cdot))$, where $J: T_pM\rightarrow T_pM$
is the almost complex structure.

In our case, it turns out to be easier to first construct a Riemannian metric
$h$ and then (try to) define from it a symplectic form. In this case, the
following lemma is useful:
\begin{lem}
  \label{lem:kaehler_conditions}
  Let $M$ be a complex manifold and let $J: TM\rightarrow TM$ denote the almost
  complex structure. Let $h$ be a Riemannian metric on $M$ and define
  $\omega\in \Gamma(T^*M\otimes T^*M)$ by $\omega(X,Y) = h(JX, Y)$ for tangent
  vectors $X,Y$. Then the following are equivalent:
  \begin{enumerate}[label=(\arabic*)]
    \item $h(\cdot,\cdot) = \omega(\cdot, J(\cdot))$
    \item $\omega$ is skew-symmetric, i.e., $\omega(X,Y) = -\omega(Y,X)$
    \item $h$ is invariant under $J$, i.e., $h(J(\cdot), J(\cdot)) = h(\cdot,\cdot)$
    \item $\omega$ is invariant under $J$, i.e., $\omega(J(\cdot), J(\cdot)) = \omega(\cdot,\cdot)$
  \end{enumerate}
  \begin{proof}
    $(1)\Rightarrow(2)$: $\omega(X,Y) = h(JX, Y) = h(Y, JX) = \omega(Y, JJX) = -\omega(Y,X)$.\\
    $(2)\Rightarrow (3)$: $h(JX, JY) = \omega(X, JY) = -\omega(JY, X) = -h(JJY, X) = h(X,Y)$.\\
    $(3)\Rightarrow (4)$: $\omega(JX, JY) = h(JJX, JY) = h(JX,Y) = \omega(X,Y)$.\\
    $(4)\Rightarrow (1)$: $\omega(X, JY) = \omega(JX, JJY) = h(JJX, JJY)  = h(X,Y)$.
  \end{proof}
\end{lem}

Our plan of giving $\mathbb{P}(\mathcal{H})$ a K\"ahler structure is as follows:
  \begin{enumerate}[label=(\arabic*)]
    \item Understand the almost complex structure $J$ on $\mathbb{P}(\mathcal{H})$. 
  \item Give $\mathbb{P}(\mathcal{H})$ a Riemannian metric, called the \textit{Fubini-Study metric} $h^{\mathrm{FS}}$.
  \item With $J$ and $h^{\mathrm{FS}}$, define $\omega^{\mathrm{FS}}(\cdot,
    \cdot) := h^{\mathrm{FS}}(J(\cdot), \cdot)$.  
  \item Show that $\omega^{\mathrm{FS}}$ is skew-symmetric, and hence defines a 2-form.
  \item Show that $\omega^{\mathrm{FS}}$ is closed, thus concluding that
    $(\mathbb{P}(\mathcal{H}), h^{\mathrm{FS}}, \omega^{\mathrm{FS}})$ is a K\"ahler manifold.
\end{enumerate}

\section{Almost Complex Structure}

Recall the general strategy we employ to understand the tangent space
$T_{[\psi]}\mathbb{P}(\mathcal{H})$ (see Remark~\ref{rem:unitary_group_tangent_space}) using
the map $A\mapsto A_{[\psi]}^*$. For the almost complex structure $J:
T_{[\psi]}\mathbb{P}(\mathcal{H})\rightarrow T_{[\psi]}\mathbb{P}(\mathcal{H})$, this
means that it suffices to describe $JA_{[\psi]}^*$ for all $A\in
\mathfrak{u}(\mathcal{H})$. Since $D_{[\psi]}\tilde\Phi:
T_{[\psi]}\mathbb{P}(\mathcal{H})\rightarrow i\mathfrak{u}(\mathcal{H})$ is
injective, a formula for $D_{[\psi]}\tilde\Phi(JA_{[\psi]}^*)$ would be a
satisfactory description of $JA_{[\psi]}^*$.

\begin{propo}
  \label{thm:complex_structure_PV}
  Let $[\psi]\in \mathbb{P}(\mathcal{H})$ and take any normalized representative $\psi\in[\psi]$. Then
  \begin{equation}
    \label{eq:complex_structure_PV}
    D_{[\psi]}\tilde\Phi(JA^*_{[\psi]}) =
     i(A \proj{\psi} + \proj{\psi}A) - 2\braket{\psi, iA\psi} \proj{\psi}
    =i(A \proj{\psi} + \proj{\psi}A) - 2i\proj{\psi}A\proj{\psi}
  \end{equation}
  for all $A\in\mathfrak{u}(\mathcal{H})$.
  \begin{proof}
    Let $\tilde A\in \mathfrak{X}(\mathcal{H})$ denote the fundamental vector field of
    the action of $\mathrm{U}(\mathcal{H})$ on $\mathcal{H}$. That is,
    \begin{equation}
      \tilde A_{\psi} = \doverd{t}\Big|_{t=0}\exp(tA)\psi = A\psi \in T_\psi\mathcal{H} = \mathcal{H}.
    \end{equation}
    Clearly, we have $D_\psi\pi (\tilde A_\psi) = A^*_{[\psi]}$, where $\pi: \mathcal{H}\setminus
    \{0\}\rightarrow \mathbb{P}(\mathcal{H})$ is the canonical projection. The map $\pi$
    is holomorphic, so   
    $$
    JA^*_{[\psi]} = JD_\psi\pi \tilde A_\psi = D_\psi\pi J\tilde A_\psi = D_\psi\pi (i\tilde A_\psi) = D_\psi\pi (iA\psi).
    $$
    It follows that
    $$
    \begin{aligned}
      &D_{[\psi]}\tilde\Phi (JA^*_{[\psi]})
      = D_{[\psi]}\tilde \Phi \circ D_\psi\pi (iA\psi) = D_\psi(\tilde \Phi\circ \pi) (iA\psi).
    \end{aligned}
    $$
    But the composition $\tilde \Phi \circ \pi$ is nothing but the map $\phi
    \mapsto \phi\braket{\phi, \cdot}/\braket{\phi,\phi}$. Therefore
    $$
    \begin{aligned}
      &D_{[\psi]}\tilde\Phi (JA^*_{[\psi]})
      = \doverd{t}\Big|_{t=0} \frac{(\psi+iA\psi t)\braket{\psi+iA\psi t, \cdot}}{\braket{\psi+iA\psi t, \psi+iA\psi t}}\\
      &=iA\psi\braket{\psi,\cdot} + \psi\braket{iA\psi,\cdot} - (\braket{iA\psi, \psi} + \braket{\psi, iA\psi})\psi\braket{\psi,\cdot} \\
      &=i(A\psi\braket{\psi,\cdot}+ \psi\braket{\psi,\cdot}A) - 2\braket{\psi, iA\psi} \psi\braket{\psi,\cdot}.
    \end{aligned}
    $$
  \end{proof}
\end{propo}
\begin{ex}
Sanity check: take $A = -i\mathbbm{1}\in \mathfrak{u}(\mathcal{H})$, then $A^*_{[\psi]}=0$
  and so we expect $D_{[\psi]}\tilde \Phi(JA^*_{[\psi]})=0$. Indeed, the right hand
  side of Eq.~\eqref{eq:complex_structure_PV} gives $\proj{\psi}+
  \proj{\psi} -2\proj{\psi}=0$.
\end{ex}

\begin{cor}
\begin{equation}
  D_{[\psi]}\tilde \Phi(JA^*_{[\psi]}) =  i \left[D_{[\psi]}\tilde \Phi(A^*_{[\psi]}), \proj{\psi}\right]
\end{equation}
  \begin{proof}
    Apply Propositions~\ref{thm:image_of_fundamental_vector} and \ref{thm:complex_structure_PV}:
    $$
    \begin{aligned}
      &i\left[D_{[\psi]}\tilde \Phi(A^*_{[\psi]}), \proj{\psi}\right]
      =i[[A, \proj{\psi}], \proj{\psi}]
      = i[(A\proj{\psi}-\proj{\psi}A), \proj{\psi}]\\
      &= i (A\proj{\psi}- \proj{\psi}A\proj{\psi} - \proj{\psi}A\proj{\psi} + \proj{\psi}A)
      =i(A\proj{\psi} + \proj{\psi}) - 2i\proj{\psi}A\proj{\psi}\\
      &= D_{[\psi]}\tilde \Phi(JA^*_{[\psi]})
    \end{aligned}
    $$
  \end{proof}
\end{cor}

\section{The Fubini-Study Metric}
Now we will give $\mathbb{P}(\mathcal{H})$ a Riemannian metric. On the real
vector space $i\mathfrak{u}(\mathcal{H})$, there is a natural inner product
given by
\begin{equation}
  h^{\mathrm{HS}}(X, Y) := \Tr(XY)
\end{equation}
for $X,Y\in i\mathfrak{u}(\mathcal{H})$. Thus $h^{\mathrm{HS}}$ defines a Riemannian
metric on $i\mathfrak{u}(\mathcal{H})$ via the identification $T_Ai\mathfrak{u}(\mathcal{H}) =
i\mathfrak{u}(\mathcal{H})$ for all $A\in i\mathfrak{u}(\mathcal{H})$. 

\begin{defn}
  The \textbf{Fubini-Study metric} $h^{\mathrm{FS}}$ on $\mathbb{P}(\mathcal{H})$ is the
  pullback of $h^{\mathrm{HS}}$ by $\tilde\Phi$. That is,
  \begin{equation}
    h^{\mathrm{FS}}_{[\psi]}(X,Y) = \mathrm{Tr}\Big[(D_{[\psi]}\tilde\Phi(X))\circ(D_{[\psi]}\tilde\Phi(Y))\Big]
  \end{equation}
  for all $X,Y\in T_{[\psi]}\mathbb{P}(\mathcal{H})$.
\end{defn}
That $h^{\mathrm{FS}}$ is a Riemannian metric follows from the fact that $\tilde\Phi$
is an embedding. The following proposition relates $h^{\mathrm{FS}}$ to the
inner product on $\mathcal{H}$:

\begin{propo}
  Take any unit vector $\psi\in \mathcal{H}$ and $X,Y\in T_\psi\mathcal{H} =
  \mathcal{H}$ such that $\braket{\psi,X} = \braket{\psi, Y} = 0$. Then
  \begin{equation}
    h^{\mathrm{FS}}(D_\psi\pi(X), D_\psi\pi(Y)) = 2\Re\braket{X,Y}.
  \end{equation}
  \begin{proof}
      $$
      \begin{aligned}
        &h^{\mathrm{FS}}(D_\psi\pi(X), D_\psi\pi (Y))
        = h^{\mathrm{HS}}(D_{[\psi]}\tilde\Phi\circ D_\psi\pi(X), D_{[\psi]}\tilde\Phi\circ D_\psi\pi(Y))\\
        &= h^{\mathrm{HS}}(D_{\psi}(\tilde\Phi\circ \pi)(X), D_{\psi}(\tilde\Phi\circ \pi)(Y))
      \end{aligned}
      $$
      We have
      $$
      D_\psi(\tilde\Phi\circ \pi)(X) = \doverd{t}\Big|_{t=0}\frac{(\psi+tX)\braket{\psi+tX, \cdot}}{\braket{\psi+tX,\psi+tX}}
      = X\braket{\psi,\cdot} + \psi\braket{X,\cdot},
      $$
      where we used $\braket{\psi,\psi}=1$ and $\braket{\psi,X}=0$. Therefore,
      $$
      \begin{aligned}
        &h^{\mathrm{FS}}(D_\psi\pi(X), D_\psi\pi(Y)) \\
        &= \Tr\Big[(X\braket{\psi,\cdot} +
      \psi\braket{X,\cdot})\circ (Y\braket{\psi,\cdot} + \psi\braket{Y,\cdot})\Big]\\
        &= \mathrm{Tr}\Big[X\braket{Y,\cdot} + \braket{X,Y}\psi\braket{\psi,\cdot}\Big]\\
        &= \braket{Y,X} + \braket{X,Y}.
      \end{aligned}
      $$
  \end{proof}
\end{propo}

\begin{ex}
  The complex projective line $\mathbb{CP}^1$ can be identified with the
  two-sphere $S^2$ by (think of the Bloch sphere)
  \begin{equation}
    \begin{aligned}
      \alpha: &\;\mathbb{CP}^1\rightarrow S^2\\
      & [z_0:z_1] \mapsto \frac{1}{|z_0|^2+|z_1|^2}(2\Re(\bar z_0z_1), 2\Im(\bar z_0 z_1), |z_0|^2 - |z_1|^2).
    \end{aligned}
  \end{equation}
  Under this identification, the Fubini-Study metric on $\mathbb{CP}^1$
  coincides, up to a factor of $\frac{1}{2}$, with the standard round metric
  $h^\circ$ on $S^2$. That is,
  $h^{\mathrm{FS}} = \frac{1}{2}\alpha^*h^\circ$.
  \begin{proof}
    Consider the map 
    $$
    \begin{aligned}
      &j: S^2 \rightarrow i\mathfrak{u}(2)\\
      &(x,y,z)\mapsto
      \frac{1}{2}
    \begin{pmatrix}
      1 +z & x-iy\\
      x+iy & 1 - z
    \end{pmatrix}.
    \end{aligned}
    $$
    Then the following diagram commutes:
    \begin{equation}
      \begin{tikzcd}
        \mathbb{CP}^1 \arrow{rr}{\tilde\Phi} \arrow[swap]{dr}{\alpha}& & i\mathfrak{u}(2)\\
          & S^2 \arrow[swap]{ur}{j}
      \end{tikzcd}
    \end{equation}
    It is easy to check that $h^\circ = 2j^*h^{\mathrm{HS}}$. Pulling back by
    $\alpha$ yields $\alpha^*h^\circ = 2\alpha^*j^*h^{\mathrm{HS}} = 2(j\circ \alpha)^*
    h^{\mathrm{HS}} = 2\tilde \Phi^*h^{\mathrm{HS}} = 2h^{\mathrm{FS}}$.
  \end{proof}
\end{ex}

\section{The Fubini-Study Symplectic Form}
We are now ready to give $\mathbb{P}(\mathcal{H})$ a symplectic structure:
\begin{defn}
  The \textbf{Fubini-Study 2-form} $\omega^{\mathrm{FS}}$on $\mathbb{P}(\mathcal{H})$ is defined by
  \begin{equation}
    \omega^{\mathrm{FS}}(X, Y) = h^{\mathrm{FS}}(JX, Y)
  \end{equation}
  for all $X,Y\in T_{[\psi]}\mathbb{P}(\mathcal{H})$.
\end{defn}
Of course, for $\omega^{\mathrm{FS}}$ to be a symplectic form, we need to check
a few things. First of all, for $\omega^{\mathrm{FS}}$ to be a 2-form at all,
it has to be skew-symmetric. This can be shown by, for example, proving that
$h^{\mathrm{FS}}$ is invariant under $J$, and applying
Lemma~\ref{lem:kaehler_conditions}. Instead, we will show this by direct
computation, which has the advantage of yielding an explicit formula for
$\omega^{\mathrm{FS}}(A^*_{[\psi]}, B^*_{[\psi]})$. Secondly, we need to show that
$\omega^{\mathrm{FS}}$ is closed.

\begin{lem}
  \label{lem:omega_explicit_formula}
  Take $A,B\in \mathfrak{u}(\mathcal{H})$ and $[\psi]\in \mathbb{P}(\mathcal{H})$. Choose a
  normalized representative $\psi$ of $[\psi]$. Then
  \begin{equation}
    \label{eq:omega_explicit_formula}
    \omega^{\mathrm{FS}}(A^*_{[\psi]}, B^*_{[\psi]}) = i\braket{\psi, [A,B]\psi}.
  \end{equation}
\begin{proof}
  $$
  \begin{aligned}
    &\omega^{\mathrm{FS}}(A^*_{[\psi]}, B^*_{[\psi]}) = h^{\mathrm{FS}}(JA^*_{[\psi]}, B^*_{[\psi]})\\
    &= h^{\mathrm{HS}}\Big(D_{[\psi]}\tilde\Phi JA^*_{[\psi]}, D_{[\psi]}\tilde\Phi B^*_{[\psi]}\Big)\\
    &= h^{\mathrm{HS}}\Big(i\big(A\proj{\psi}+ \proj{\psi}A -2\proj{\psi}A\proj{\psi}\big), [B, \proj{\psi}]\Big)\\
    &= i\mathrm{Tr}\Big[\big(A\proj{\psi}+ \proj{\psi}A -2\proj{\psi}A\proj{\psi}\big)(B\proj{\psi} - \proj{\psi}B)\Big]\\
    &= i\mathrm{Tr}\Big[(A\proj{\psi}+ \proj{\psi}A)(B\proj{\psi} - \proj{\psi}B)\Big]\\
    &=i\mathrm{Tr}(A\proj{\psi}B\proj{\psi} - A\proj{\psi}B + \proj{\psi}AB\proj{\psi}-\proj{\psi}A\proj{\psi}B)\\
    &= i \mathrm{Tr}(\proj{\psi}[A,B])\\
    &= i\braket{\psi, [A,B]\psi}
  \end{aligned}
  $$
\end{proof}
\end{lem}

\begin{rem}
  $[A,B]\in\mathfrak{u}(\mathcal{H})$ is skew-Hermitian, so $i\braket{\psi, [A,B]\psi}$ is real.
\end{rem}
Equation~\eqref{eq:omega_explicit_formula} shows that $\omega^{\mathrm{FS}}$ is
indeed skew-symmetric, and hence a 2-form. Now we need to check that
$\omega^{\mathrm{FS}}$ is closed.

\begin{lem}
  \label{lem:Lie_deriv_of_omega} Let $A,B,C\in \mathfrak{u}(\mathcal{H})$ and
  take $[\psi]\in \mathbb{P}(\mathcal{H})$ with normalized representative
  $\psi\in \mathcal{H}$. Then
  \begin{equation}
    \label{eq:Lie_deriv_of_omega}
    A_{[\psi]}^*\omega^{\mathrm{FS}}(B^*, C^*) = i\braket{\psi, [[B,C], A]\psi}.
  \end{equation}
\end{lem}
Note: In Eq.~\eqref{eq:Lie_deriv_of_omega}, $B^*, C^*$ are vector fields, so
$\omega^{\mathrm{FS}}(B^*, C^*)$ is a smooth function on $\mathbb{P}(\mathcal{H})$. We
have used the standard notation where $Xf$ denotes the Lie derivative of a
smooth function $f$ along a vector field $X$. That is, $Xf = \dstraight f(X)$.
If $p$ is a point, then $X_pf = (Xf)_p = \dstraight f(X_p)$.
\begin{proof}
  $$
  \begin{aligned}
    &A^*_{[\psi]} \omega^{\mathrm{FS}}(B^*, C^*) = A^*_{[\psi]} \Big([\psi]\mapsto i\braket{\psi, [B,C]\psi}\Big)\\
    &=i\doverd{t}\Big|_{t=0}\braket{\exp(tA)\psi, [B,C]\exp(tA)\psi}\\
    &=i\braket{A\psi, [B,C]\psi} + i\braket{\psi, [B, C]A\psi}\\
    &= i\braket{\psi, [[B,C], A]\psi}
  \end{aligned}
  $$
\end{proof}

\begin{propo}
  The Fubini-Study 2-form $\omega^{\mathrm{FS}}$ is closed.
  \begin{proof}
    We will use the following fact: if $\omega$ is any 2-form on a manifold $M$
    and $X,Y,Z\in \mathfrak{X}(M)$ are vector fields, then
    $$
    \begin{aligned}
      &\dstraight \omega(X,Y,Z)\\
      &= X\omega(Y,Z) + Y\omega(Z,X) + Z\omega(X,Y) 
      - \omega([X,Y], Z) - \omega([Y,Z],X) - \omega([Z,X],Y).
    \end{aligned}
    $$
    We apply this to $\omega = \omega^{\mathrm{FS}}$. Let $A,B,C\in \mathfrak{u}(\mathcal{H})$, then
    $$
    \dstraight\omega^{\mathrm{FS}}(A^*_{[\psi]}, B^*_{[\psi]}, C^*_{[\psi]})
    = A_{[\psi]}^*\omega^{\mathrm{FS}}(B^*,C^*) + B^*_{[\psi]}\omega^{\mathrm{FS}}(C^*,A^*) + C^*_{[\psi]}\omega^{\mathrm{FS}}(A^*,B^*),
    $$
    where the last three terms canceled due to
    Lemma~\ref{lem:omega_explicit_formula} and the Jacobi identity. Using
    Lemma~\ref{lem:Lie_deriv_of_omega}, we see that the remaining three terms
    also cancel due to the Jacobi identity. Since the fundamental vector fields
  span $T_{[\psi]}\mathbb{P}(\mathcal{H})$ at any $[\psi]$, we conclude
      $\dstraight\omega^{\mathrm{FS}}=0$. 
    \end{proof}
\end{propo}

\end{appendices}

\printbibliography

\cleardoublepage
\pagestyle{plain}


\chapter*{Acknowledgements}
This exciting project would not have been possible without the countless
stimulating ideas from Christian, my thesis advisor, and fruitful
discussions with him throughout the year. His everlasting enthusiasm for
physics and sharp intuition have helped me overcome numerous scientific
obstacles.

I would also like to express my gratitude to my amazing colleagues: Julia,
Damiano, \begin{CJK}{UTF8}{bkai}晟霖\end{CJK}, Markus, Ludvík, Unik, and Paul.
I feel lucky to be around such brilliant people.


The two-year academic journey in TMP would not have been complete without my
friends -- Omkar, Thomas, Miquel, \begin{CJK}{UTF8}{bkai}正羽\end{CJK},
Afonso, Andr\'as, Dmytro, Vittorio, Matteo, Martin, and many more. I thank them for
their company and friendship.

Finally, I am deeply indebted to Laura for her constant and unfailing support
and encouragement, and to my parents for their unconditional love and
unwavering belief in my education.


\cleardoublepage

\pagestyle{plain}
\noindent Erklärung:
\vspace{1cm}

\noindent Hiermit erkläre ich, die vorliegende Arbeit selbständig verfasst zu
haben und keine anderen als die in der Arbeit angegebenen Quellen und
Hilfsmittel benutzt zu haben.
\vspace{3cm}

\noindent München, 5.10.2025
\vspace{2cm}

\noindent Chih-Chun Wang

\end{document}